\documentclass[12pt,a4paper,twoside]{report}
\usepackage{amsmath}
\usepackage{amsthm}
\usepackage{latexsym}
\usepackage{multibox}
\usepackage{amssymb}
\usepackage{fancyhdr}
\usepackage{array}

\newcommand{\tB}{\tilde B }
\newcommand{\bb}{\begin{equation}}
\newcommand{\ee}{\end{equation}}
\newcommand{\bqn}{\begin{eqnarray}}
\newcommand{\eqn}{\end{eqnarray}}
\newcommand{\pp}{\partial }
\newlength{\blength}
\renewcommand{\proof}[1]{\vspace{-.05cm}
\begin{list}{\bf
{Proof:}}{\listparindent=\parindent\parsep=0pt
\labelwidth=0cm
\labelsep=\parindent
\addtolength{\labelsep}{-\blength}
\addtolength{\labelsep}{1.2cm}
\itemindent=-\blength
\addtolength{\itemindent}{\parindent}
\leftmargin=1.2cm}
\item
#1~$\qedsymbol$\end{list}
\vspace{.0cm}}

\newtheorem{theorem}{Theorem}

\begin{document}
\bibliographystyle{BKstyle}
\begin{titlepage}
\begin{center}
{\large UNIVERSIT\'E LIBRE DE BRUXELLES \\
Facult\'e des Sciences \\
Service de Physique Th\'eorique et Math\'ematique}

\vspace{3cm}

{\LARGE\bf Cohomologie BRST locale des th\'eories de
$p$-formes}

\vspace{3cm}

Dissertation pr\'esent\'ee en vue de l'obtention \\
du grade de Docteur en Sciences \\

\vspace{3cm}
Bernard Knaepen\\
Aspirant FNRS\\
\vspace{3cm} 
Ann\'ee acad\'emique 1998--1999 \\
\end{center}

\end{titlepage}
\thispagestyle{empty}
\ 
\clearpage      
\setcounter{page}{3}
\ 
\vspace{12 cm}

\noindent
{\large \bf Titre de la th\`{e}se annexe}
\vspace{.5cm}

\noindent
L'utilisation des
m\'{e}thodes de simulation des grandes \'{e}chelles et de la
mo\-d\'{e}\-li\-sa\-tion des structures les plus petites permet de
caract\'{e}riser la turbulence magn\'{e}to-hydrodynamique.

\newpage
\thispagestyle{empty}
\ 
\clearpage    
\ 
\vspace{6 cm}

\noindent
{\LARGE \bf Remerciements}
\vspace{.5cm}

Je remercie vivement mon promoteur, Marc Henneaux, pour ses nombreuses
contributions \`{a} ce travail et pour m'avoir assist\'{e} tout au long
de sa r\'{e}alisation.

Mes remerciements s'adressent \'{e}galement \`{a} Christiane Schomblond
pour sa participation \`{a} cette th\`{e}se et pour les nombreux conseils
qu'elle m'a prodigu\'{e}s pendant sa r\'{e}daction.

J'exprime aussi ma gratitude \`{a} Daniele Carati et Olivier Agullo pour
m'avoir patiemment initi\'{e} aux m\'{e}thodes num\'{e}riques de
simulation en ma\-gn\'{e}to-hydrodynamique.

Merci \`{a} J. Antonio Garcia, Glenn Barnich, Andr\'{e} Wilch et
Abdelilah Barkallil pour les nombreuses discussions que nous avons eues.

J'ai eu la chance de travailler pendant ces quatre ann\'{e}es dans le 
Service de Physique Th\'{e}orique et Math\'{e}matique (anciennement
Service de Physique Statistique, Plasmas et Optique non-lin\'{e}aire) au
sein duquel r\`{e}gne  une ambiance \`{a} la fois stimulante et
d\'{e}contract\'{e}e. En particulier, je n'oublierai jamais les nombreux
r\'{e}cits de blagues et les discussions de la pause caf\'{e}.  Il serait
trop long de nommer tous les participants \`{a} ce rendez-vous
incontournable mais je profite de ces quelques lignes pour leur
t\'{e}moigner mon amiti\'{e}.

Enfin, je remercie affectueusement mes amis, toute ma famille et en
particulier mes parents et B\'{e}n\'{e}dicte.

\newpage

\tableofcontents

\chapter{Introduction}
At present, the most promising candidates of models unifying
the four known fundamental interactions consist of string
theories \cite{GSW,Polchinskibook}. As their name indicates, the
fundamental objects of these theories are strings and it is
believed that all the particles found in nature can be viewed as
their different vibration modes.
At the lowest end of their energy spectrum,
string theories contain states which can be
described by antisymmetric tensor fields. These fields are
therefore essential ingredients of various supergravity models
which represent low energy approximations of string theories.
The study of these fields is thus an important
question of theoretical and mathematical physics.

Antisymmetric tensor fields, denoted $B_{\mu_1\ldots
\mu_p}$, generalize the vector potential $A_\mu$ used to
describe Maxwell's theory of electromagnetism. They are
antisymmetric with respect to the permutation of any two of
their indices and are thus naturally viewed as the
components of a $p$-form,
\begin{equation}
B=\frac{1}{p!}B_{\mu_1\ldots\mu_n} dx^{\mu_1}\wedge\ldots
\wedge dx^{\mu_p}.
\end{equation}
As in the case of electromagnetism, the
action of massless
$p$-form gauge theories is invariant under gauge
transformations. For Maxwell's theory, the gauge
transformations read, $A_\mu \rightarrow A_\mu +
\partial_\mu \epsilon$ where $\epsilon$ is
an arbitrary spacetime function. In form notation, this
symmetry is written as
$A=A_\mu dx^\mu
\rightarrow A+ d\epsilon$ where $d=dx^\mu \partial_\mu$ is
the spacetime exterior derivative. The gauge transformations of $p$-form gauge fields
are straightforward generalization of the above transformations
and are given by, $B\rightarrow B+d\eta$ where $\eta$ is now
an arbitrary spacetime dependent 
$p-1$-form. Their is however a significant difference. The
gauge transformations of $p$-forms ($p>1$) are reducible, i.e,
they are not all independent. Indeed the gauge parameters
$\eta$ and $\eta'=\eta + d\rho$ where $\rho$ is a
$p-2$-form produce the same transformation of $B$ by virtue
of $d^2=0$. As we shall see later, this is one of the
particularities of
$p$-form gauge theories which makes their study
a rich problem.

The aim of this thesis is to study some of the properties of
$p$-form gauge theories by using the BRST field-antifield
formalism. The essential ingredient of this formalism is the
BRST differential $s$ which was first discovered in the context
of the quantization of Yang-Mills theory. As for
any gauge theory, it is necessary when quantizing Yang-Mills
theory to introduce a gauge fixing term in the action in
order to obtain well-defined propagators and
Green functions. Furthermore one must ensure that the measurable
quantities calculated from the gauge fixed action do not depend
on the choice of the gauge fixation and that the theory is unitary
in the physical sector. An elegant way of achieving this program for
Yang-Mills theory was devised by Faddeev and Popov
\cite{FaddeevPopov}. Their method requires the introduction of new
fields, called ghosts which only appear as intermediate states in
transition amplitudes. 

Becchi, Rouet, Stora \cite{BRS1, BRS2} and Tyutin
\cite{Tyutin} then discovered that the gauge-fixed Yang-Mills
action incorporating the ghosts was invariant under a
nilpotent symmetry, the BRST symmetry, which was then
extended to theories with more general gauge structures
\cite{Kallosh1,WitHol1,BV1,BV2,BV3} and in particular to reducible
gauge theories. The most important property of the BRST symmetry is
that for any gauge theory it allows one to construct an
``extended action" which guarantees at the quantum level all the
conditions recalled in the preceding paragraph.

Another benefit of the BRST symmetry is
that one can study various problems of classical and quantum field
theory by calculating the local cohomology
$H(s\vert d)$ of its nilpotent generator $s$; this amounts to
the classification of the ``non-trivial'' solutions of the
{\em Wess-Zumino consistency condition} \cite{WZ}.

In quantum field theory, the local BRST cohomology is
especially useful in renormalization theory. Indeed, 
one can
show that in ``ghost number" 1 (see Section
{\bf\ref{BRSTformalismgen}}),
$H(s\vert d)$ contains all the candidate gauge anomalies which
can arise upon quantization of a gauge theory. 
The existence of these BRST cocycles does
not imply that the theory is anomalous. However, a calculation of
their coefficients indicates if the theory is consistent or not.
Other important
cohomological groups in renormalization theory are found in ghost
number $0$ and contain the counterterms needed for renormalization.
The calculation of these BRST cocycles is therefore useful to
examine the stability of the action and to determine if the
divergences can be absorbed in a redefinition of the fields,
masses, coupling constants or other parameters present in the
classical action. 

The calculation of the BRST cohomology is also relevant to
address some questions of classical field theory. Indeed, the
cohomological groups in ghost number $0$ which restrict the form of
the counterterms also determine the consistent interaction terms
that can be added to a lagrangian. By consistent it is meant
vertices which preserve gauge invariance and therefore the number
of degrees of freedom. The BRST cohomology is also very powerful to
study the gauge invariant nature of the conserved currents of a
theory. Indeed, as explained in Section {\bf\ref{cocudef}}, this
problem is related to the calculation of the cohomological
groups in ghost number $-1$.

To obtain the BRST cohomology we will closely follow the
approach developed for Yang-Mills theory\footnote{A
complete account of the analysis of the local BRST cohomology
in Yang-Mills theory and a list of the original works
on the subject can be found in
\cite{theseGlenn,BBHreport}.}. This is possible mainly because
the BRST differential of
$p$-form gauge theories splits as in the Yang-Mills case as
the sum of the Koszul-Tate differential $\delta$ and the
longitudinal exterior derivative $\gamma$: $s=\delta +
\gamma$ (for Chapline-Manton models (Chapter {\bf\ref{chapCM}}),
this form of
$s$ is obtained after a change of variables). The r\^{o}les of the
Koszul-Tate differential and the longitudinal exterior derivative
are respectively to implement the equations of motion and to take
into account the gauge invariance of the theory. 

Our analysis will be performed in
the presence of antifields; in the terminology of
path integral quantization, they are the sources of the BRST
variations of the fields and ghosts. They are
useful in renormalization theory to control how gauge
invariance survives renormalization and they also provide a
convenient way to take into account the dynamics of the model
when one addresses the problems of classical field theory
raised above.

As in the Yang-Mills case, our study of the BRST cohomology
is composed of two parts. In the first one, we
calculate the solutions of Wess-Zumino consistency condition
which do not depend on the antifields. Our analysis relies
on the resolution of the so-called ``descent equations" from
which one obtains the cohomology
$H(\gamma\vert d)$ by relating it to the cohomology $H(\gamma)$.

In the second part of our study, we calculate the
antifield dependent BRST cocycles. Their existence is
closely tied to the presence of non-vanishing classes in the
``characteristic cohomology" $H(\delta\vert d)$. This cohomology is
defined as the set of $q$-forms $w$ (see Section
{\bf\ref{mathframe}}) which satisfy the conditions:
\begin{equation}
dw \approx 0; \quad \quad w\not \approx d\psi.\label{ritiyu}
\end{equation}
In \eqref{ritiyu} $d$ is the spacetime exterior derivative and the
symbol $\approx$ means ``equal when the equations of motion hold".
The calculation of the characteristic cohomology is an interesting
problem on its own but is also related to important questions of
field theory (see Chapter {\bf\ref{CharaCohosec}}). 

Our analysis will show that the BRST cocycles naturally fall into
two categories. In the first category, one has the cocycles $a$
which are strictly annihilated by
$s$. They consist of the gauge invariant
functions of the theory and are easily constructed. In second
category, one has the cocycles which are $s$-closed only modulo the
spacetime exterior derivative $d$, i.e., which satisfy $sa+db=0$. A
priori, any cocycle can depend on the components of the
antisymmetric tensors, the corresponding ghosts and antifields as
well as their derivatives. However, a remarkable property of the
solutions of the Wess-Zumino consistency condition belonging to the
second category is that they only depend on the exterior forms
build up from the antisymmetric tensors, the ghosts and the
antifields and not on the individual components of these forms.
This property explains a posteriori why the calculation of the BRST
cohomology in the algebra of forms and their exterior derivatives
does not miss any of the cocycles of the second
category\footnote{This statement is true except for the 
antifield dependent BRST cocycles related to the conserved currents
of the theory (see Theorem {\bf\ref{bigtheo1}}).}.

From the analysis of the BRST
cohomology, we will obtain for the models considered all the
candidate anomalies, counterterms, consistent interactions and we
will analyze the gauge structure of the conserved currents.  Our
main result will be the classification of all the
first-order vertices which can be added to the action of a system of
free $p$-forms
$(p>1)$. Their construction has attracted a lot of interest in the
past
\cite{DeserJT,FT1,Teitelboim1,BergRooWitNieu,ChaplineManton,
Baulieu1, BrandtD1} but systematic results were
only obtained recently \cite{HK1}. 
We will show that besides the strictly gauge invariant ones,
the first-order vertices are necessarily of the Noether form
``conserved antisymmetric tensor" times ``$p$-form potential" and
exist only in particular spacetime dimensions.
The higher-order vertices will also be studied for a system
containing forms of two different degrees. This analysis will unveil
only one type of consistent interacting theory which is not of the
forms described in \cite{DeserJT,FT1,BergRooWitNieu,ChaplineManton}.

Our thesis is organized as follows. In Chapter 2, we
expose briefly the BRST field-antifield construction and
describe how it addresses the problems of classical and
quantum theory we have outlined. We will also define the
mathematical framework in which the calculations will be done.

In Chapter 3, we perform the BRST construction for an
arbitrary system of free $p$-forms, i.e, we introduce all the
necessary antifields, ghosts and define their
transformation under the BRST differential in terms of the
Koszul-Tate differential and the longitudinal exterior
derivative. We then give results concerning the cohomologies
of the Koszul-Tate differential and the spacetime exterior
derivative  (Poincar\'e Lemma). By making use of Young
diagrams we also calculate the cohomology
$H(\gamma)$ of the longitudinal exterior derivative which
represents an essential ingredient in the analysis of the
antifield independent solutions of the Wess-Zumino consistency
condition.

In Chapter 4 we begin our study of the BRST cohomology for the
system of $p$-forms described in Chapter 3. Chapter 4 is
divided into three sections and begins with an outline of
the methods used to obtain the antifield (in)dependent BRST
cocycles. In Section {\bf\ref{Sec:Antidep1}} we analyze the
Wess-Zumino consistency condition in the absence of
antifields. We begin by reviewing the technic of the ``descent
equations": we illustrate it with simple examples and then
describe the general theory. We then show that for free
$p$-forms, one can investigate the Wess-Zumino consistency
condition in the so-called ``small algebra". After that we
introduce the ``Universal algebra" and we formulate the {\em
Generalized transgression lemma} from which one is able to
obtain all the non-trivial cocycles. The results of the
analysis are presented in Section {\bf\ref{results}} and we
summarize them in Section {\bf\ref{cafree}} by listing the
various (antifield independent) counterterms and anomalies. In
Section {\bf{\ref{adssec}}} we turn our attention to the
antifield dependent solutions of the Wess-Zumino consistency
condition and we prove some preliminary results. In particular we
show that the existence of antifield dependent BRST cocycles is
related to the presence of non-vanishing classes in the
``characteristic cohomology". 

This cohomology is studied in Chapter
{\bf{\ref{CharaCohosec}}} which begins by a general description of
the subject and a summary of our results. 

In Chapter 6, we continue the analysis of the Wess-Zumino
consistency condition in the presence of antifields. Using the
results of Chapter {\bf{\ref{CharaCohosec}}} we obtain in
Section {\bf\ref{matho}} all the antifield dependent BRST
cocycles from which we infer in Section {\bf\ref{cfova}} and
{\bf{\ref{conscurent}}} the corresponding counterterms, first-order
vertices, anomalies and the gauge structure of the conserved
currents of the theory. The last Section of Chapter 6 is dedicated
to the construction of interacting
$p$-form theories which are consistent to all orders in the
coupling constant.

In Chapter 7 we study the BRST cohomology for Chapline-Manton
models. The analysis is similar to the one of the free theory since
after a redefinition of the antifields the BRST differential is
brought to the standard form $s=\delta+ \gamma$. Thus we begin by
studying the antifield independent cocycles, then the characteristic
cohomology and finally the antifield dependent cocycles. The
important lesson of the analysis of Chapline-Manton models is that
the results gathered for the free theory are useful in the analysis
of interacting theories since they allow to obtain the
cohomological groups by  ``perturbative arguments".
 
Finally, in Chapter 8 we conclude this dissertation with a few
comments.

\chapter{BRST formalism}

One of the benefits of the BRST field-antifield formalism is that
it allows one to formulate in a purely algebraic manner certain
questions concerning the (classical or quantum) theory of fields.
In this section we briefly describe how this formalism is
introduced and how it deals with the following questions:
\begin{enumerate}
\item
determination of the consistent deformations of a classical theory,
\item
analysis of the gauge invariant nature of conserved currents of a
classical theory.
\item
classification of possible counterterms occurring in the
renormalization of quantum theory,
\item determination of candidate anomalies which can occur as a
result of the quantization of a classical theory.

\end{enumerate}

We will also describe the mathematical framework in which the
calculations will be done.

\section{BRST action and master equation}
\label{BRSTformalismgen}

Let us start by introducing rapidly the various ingredients of
the BRST formalism which are relevant for our discussion of the
questions raised above. 

The starting point of the analysis
is a local classical action
$I$,
\begin{equation}
I=\int d^nx {\cal L}(\phi^i,\partial_\mu
\phi^i,\ldots).\label{Istart}
\end{equation}
By {\em local} we mean that ${\cal L}$ depends on the
fields and their derivatives up to a finite order $k$. 

The first step in the BRST construction is to associate with each
field $\phi^i$ present in \eqref{Istart} another field called an
{\em antifield}. These antifields, which we label $\phi^*_i$, are
assigned a degree called the antighost number which is equal to
$1$.

Besides being invariant under
some global symmetries acting on the fields, the action $I$ can be
invariant under some {\em gauge transformations}.
These have the property that their parameters,
$\epsilon^\alpha$, can be set independently at each
point in spacetime, i.e., $\epsilon^\alpha=\epsilon^\alpha (x)$.
For each of these {\em gauge parameters}, one introduces another
antifield $\phi^*_\alpha$ of antighost number $2$, as well as a
field called a ghost, which we label $C^\alpha$. This ghost is
assigned a degree called the pureghost number which is equal to
$1$. The pureghost number of the antifields and the antighost
number of the ghosts are set equal to zero. One also introduces a
third degree called the {\em ghost number} which is equal to the
difference between the pureghost number and the antighost number.

A gauge theory is said to be {\em reducible} when the gauge
parameters are not all independent. In that case, the gauge
transformations are themselves invariant under
certain variations of the gauge parameters $\epsilon^\alpha$. For
each parameter involved in the transformations of 
$\epsilon^\alpha$, one again introduces a new antifield
$\phi^*_\Gamma$ of antighost number $3$ and a ghost $C^\Gamma$ of
pureghost number $2$.

This construction continues if the transformations of the gauge
parameters are themselves reducible and so on. We will suppose
here that the construction stops after a finite number of steps so
that the number of antifields and ghosts introduced is finite. The
last ghosts introduced are called the {\em ghosts of ghosts}.

Once all the necessary antifields and ghosts have been defined,
one can introduce the BRST differential $s$ and define its action
on all the variables. The general expression of $s$ is of the
form,
\begin{equation}
s=\delta + \gamma  + s_1 + s_2 + \ldots .\label{difs}
\end{equation}
In \eqref{difs} the first part of $s$ is the {\em Koszul-Tate
differential}
$\delta$. Its r\^{o}le is to implement the equations
of motion in cohomology, i.e., its cohomology is given by the
on-shell functions and in particular, any 
function which vanishes on-shell is $\delta$-exact.  The second
operator present in
\eqref{difs} is the {\em longitudinal exterior derivative} $\gamma$.
Its r\^{o}le is to take into account the gauge invariance of the
theory; $\gamma$ is thus constructed in such a way that
any {\em observable} is annihilated by $\gamma$ on-shell.
The general expressions for $\delta$ and 
$\gamma$ may be found in
\cite{HenneauxTeitelboim} and will be given later for the various
models considered in this thesis. Let us only note here that: 
$\delta$ lowers by one unit the antighost number while $\gamma$
leaves it unchanged; $\delta$ is
nilpotent and $\gamma$ is nilpotent modulo $\delta$, i.e,
$\gamma$ is a differential in the cohomology $H(\delta)\equiv
\frac{\text{Ker }
\delta}{\text{Im } 
\delta}$; $\delta$ and $\gamma$ anticommute ($\delta \gamma
+\gamma\delta=0$). 

The extra terms $s_j$ present in $s$ are of antighost
number $j$ (they raise the antighost number by $j$ units).
Their action on the variables is obtained by requiring that
$s^2=0$. 

For all the models we consider in this thesis, we will see
that $s$ reduces to the sum of $\delta$ and $\gamma$:
$s=\delta+\gamma$.\footnote{For
Chapline-Manton models this form of $s$ is obtained by a
redefinition of the antifields.} In the rest of the text we will assume that $s$
has this particular form.

Once the action of $s$ on all the fields, ghosts and antifields
has been obtained, one can construct a new action $S$ of ghost
number $0$ called the BRST action which generates the BRST symmetry
$s$. It is defined by\footnote{This form of the BRST differential
is not the most general one; indeed there are some theories for
which
$S$ contains terms which are not linear in the antifields.
However, the models examined in this thesis all admit a BRST
action of the form
\eqref{BRSTact}.},
\begin{equation}
S=\int d^n x ({\cal L} + (-)^{\epsilon_A} \phi^*_A
s\phi^A),\label{BRSTact}
\end{equation}
where $\epsilon_A$ is the Grassmann parity of $\phi^A$. We
have denoted all the antifields by
$\phi^*_A
\equiv (\phi^*_i, \phi^*_\alpha, \phi^*_\Gamma, \ldots )$ and the
fields and ghosts by
$\phi^A \equiv (\phi^i, C^\alpha, C^\Gamma, \ldots)$. By
construction the BRST action is such that,
\begin{align}
&s\phi^A = (S,\phi^A)=-\frac{\delta^R 
S}{\delta \phi^*_A(x)},\\
&s\phi^*_A = (S,\phi^*_A) = \frac{\delta^R 
{S}}{\delta \phi^A(x)}.
\end{align}
Furthermore, $S$ satisfies the {\em master equation},
\begin{equation}\label{nilpot}
(S,S)=0,
\end{equation}
where the {\em antibracket} $(.,.)$ of two local functionals $A$ and
$B$ is defined by,
\begin{equation}
(A,B)=\int d^n x \left( \frac{\delta^R {A}}{\delta \phi^A(x)}
\frac{\delta^L {B}}{\delta \phi^*_A(x)}- \frac{\delta^R 
{A}}{\delta \phi^*_A(x)}
\frac{\delta^L {B}}{\delta \phi^A(x)}\right).
\end{equation}
The master equation \eqref{nilpot} is a direct consequence of
$s^2=0$.

An important feature is that by construction, the BRST action
contains all the information about the action and its gauge
structure. For instance one can directly read
off from its expression the gauge transformations, their algebra,
the reducibility identities etc.

In the following sections we quickly review how the BRST formalism
addresses the questions raised at the beginning of this
section.

\section{Consistent deformations}
\label{ckgkdfo}
Let us illustrate the problem of consistent deformations for an
irreducible gauge theory (for a recent and more complete
review see
\cite{henneauxSC}).

Let $I_0$ be a ``free"\footnote{Although the action $I_0$ can
already contain interaction vertices, we call it {\em free}
as it is the action we want to deform.} action invariant under
some gauge transformations $\delta_{\epsilon} \phi^i=R^i_{0
\alpha}\epsilon^\alpha$, i.e.,
\begin{equation}
\int d^n x \frac{\delta I_0}{\delta
\phi^i}R^i_{0 \alpha}\epsilon^\alpha=0.
\end{equation}
A deformation of the free action consists in adding to
$I_0$ interaction vertices,
\begin{equation}
I_0\rightarrow I=I_0 + gI_1 + g^2 I_2+\ldots,
\end{equation}
where $g$ is a coupling constant. In order to be {\em consistent},
we require the deformation to preserve gauge invariance. Therefore,
the action $I$ must remain gauge invariant, possibly under modified
gauge transformations:
\begin{equation}
\int d^n x \frac{\delta I}{\delta
\phi^i}R^i_{\ \alpha}\epsilon^\alpha=0.\label{gitux}
\end{equation}
with,
\begin{equation}\label{ria}
R^i_{\ \alpha}=R^i_{0 \alpha} +g R^i_{1 \alpha} +g^2 R^i_{2 \alpha}
+ \ldots.
\end{equation}
Eq. \eqref{gitux} has to be satisfied at each order in the coupling
constant.

The deformations of an action can be of two types. In the first
one, gauge invariant terms are added to the original lagrangian
and therefore no modification of the gauge
transformations is required. In Maxwell theory, Chern-Simons terms
produce such deformations. The second type of deformations are
obtained by adding vertices which are not invariant
under the original gauge transformations and which therefore
require a modification in them. In such a case, the algebra of the
new gauge transformations can also be altered. A famous example is
the Yang-Mills theory in which the abelian $U(1)^N$ symmetry is
replaced by the non-abelian
$SU(N)$ symmetry.  

For reducible theories, one also imposes the gauge
transformations \eqref{ria} to remain reducible and all higher
order reducibility identities to hold (possibly in a
deformed way). These consistency requirements guarantee that both
the deformed theory and the free theory contain the same number of
degrees of freedom.

The problem of
consistent interactions among fields can be elegantly
formulated within the BRST field-antifield formalism
\cite{BH1}. Indeed, let
$S_0$ be the solution of the master equation for a free theory, 
\begin{equation}
(S_0,S_0)=0.
\end{equation}
If one deforms this free theory,
the lagrangian, its
gauge transformations, reducibility identities etc.\;are modified.
Therefore, the BRST action $S_0$ which encodes all this information
it is also modified and becomes,
\begin{equation}
S=S_0 +g S_1 + g^2 S_2 + \ldots,
\end{equation}
where all the terms are of ghost number $0$. 
All our consistency requirements will be met if the new BRST action
continues to satisfy the master equation \eqref{nilpot}. Indeed,
$(S,S)=0$ guarantees in particular that condition \eqref{gitux} is
fulfilled but this is also true for all the higher order
reducibility identities.

As we already stressed, the BRST action encodes all the
information about the gauge structure of the theory. In its
expansion 
\eqref{BRSTact} according to the antighost number,
one encounters three terms of special interest in the present
discussion:
\begin{enumerate}
\item
the antifield independent term which corresponds to the
full interacting lagrangian; 
\item
a term of the form $\phi^*_i R^i_{\ \alpha} C^\alpha$ (where
$\phi^*_i$ are the antifields of antighost number $1$ and the
$C^\alpha$ are the ghosts of pureghost number $1$). From this 
term one deduces the modified gauge transformations of the
fields $\delta_\epsilon \phi^i = R^i_{\ \alpha}
\epsilon^\alpha$. Note that if $S_1, S_2, \ldots$ do not
depend on the antifields then the gauge transformations
leaving the deformed action invariant are identical to those of the
free theory. 
\item
a term of the form $C^\alpha_{\beta\gamma}\phi^*_{\alpha}
C^\beta C^\gamma$ (where $\phi^*_\alpha$ are the antifields of
antighost number 2). This term is present when the gauge
transformations are not abelian even on-shell; the
$C^\alpha_{\beta\gamma}$ are the structure ``constants" of the
algebra of the gauge transformations.
\end{enumerate}
The advantage of using the BRST
formalism to address the problem of consistent deformations is that
one can make use of the cohomological techniques. Indeed, as Eq.
$(S,S)=0$ must be fulfilled at each order in the coupling constant,
one gets a tower of equations which reads,
\begin{align}
(S_0,S_0)&=0, \\
2(S_0,S_1)&=0,\label{roqnx}\\
2(S_0,S_2)+(S_1,S_1)&=0,\label{dddfff}\\
&\ \ \vdots \nonumber
\label{ordreg}
\end{align}
The first equation is satisfied by hypothesis. The second one
implies that $S_1$ is a cocycle of the BRST differential $s$ of the
{\em free} theory. This condition, known as the {\em Wess-Zumino
consistency condition}, imposes very strong restrictions on $S_1$
and thus provides a convenient way to determine all the first
order vertices. 

The interesting $S_1$ are in fact elements of the cohomology
$H^0(s)$.  To see this, suppose that $S_1$ is of the form
$S_1=s T_1$ where $T_1$ is of ghost number
$-1$. Since $s^2=0$, $S_1$ is automatically a BRST
cocycle. However, such solutions
correspond to ``trivial" deformations because they amount
to field redefinitions in the original action
\cite{BH1,henneauxSC}. Therefore, the non-trivial deformations of
the BRST action are represented by the cohomological classes of
$H^0(s)\equiv \frac{\text{Ker}\ s}{\text{Im}\ s}$.

At order $g^2$, condition
\eqref{dddfff} indicates that the antibracket of $S_1$ with
itself must be
$s$-exact. This is a new constraint for $S_1$ but it also defines
$S_2$ up to an $s$-exact term. If locality (as defined above) is not
imposed to the functionals, one can show that condition
\eqref{dddfff} is always satisfied because the antibracket of
two closed functionals is always exact. This is also true for all
the other equations beneath \eqref{dddfff}. It is thus only when
one requires the functionals to be local that one
can meet an obstruction in the construction of the higher order
vertices corresponding to $S_1$.

To properly take into account locality one reformulates all the
equations in terms of their integrands. For instance, Eq.
\eqref{roqnx} can be written,
\begin{equation}
sS_1=0 \Leftrightarrow s\big(\int {\cal S}_1
\big)=0\Leftrightarrow
\int (s {\cal S}_1)=0.
\end{equation}
In terms of the integrand ${\cal S}_1$ the last equation reads,
\begin{equation}
s{\cal S}_1 +d {\cal M}_1=0,\label{SQSQSQ}
\end{equation}
where ${\cal M}_1$ is a local form of degree $n-1$ and ghost
number $1$ and $d$ is the spacetime exterior
derivative. Again one can show that BRST-exact terms modulo $d$ are
trivial solutions of \eqref{SQSQSQ} and correspond to trivial
deformations. In order to implement locality, the proper
cohomology to evaluate is thus $H^n_0(s\vert d)$ where the
superscript and subscript respectively denote the form degree and
ghost number.

The local equivalent of
\eqref{dddfff} is,
\begin{equation}
s{\cal S}_2+({\cal S}_1,{\cal S}_1)+d{\cal M}_2=0,\label{lococn}
\end{equation}
where we have set $S_2=\int {\cal S}_2$ while $({\cal S}_1,{\cal
S}_1)$ is defined  by
$(S_1,S_1)=\int ({\cal S}_1,{\cal S}_1)$ up to an irrelevant
$d$-exact term. Contrary to the previous situation in which one did
not impose locality, Eq.
\eqref{lococn} is not automatically fulfilled if one requires
${\cal S}_2$ and ${\cal M}_2$ to be local functions. 
It indicates that $({\cal S}_1,{\cal S}_1)$, which is of form
degree $n$ and ghost number $1$,  must be BRST-exact modulo $d$.
Therefore, the construction of $S$ can be obstructed at this stage
if
$H^n_1(s\vert d)$ does not vanish.

The consistency conditions at higher order in the coupling constant
are also non-trivial for local functionals and it is immediate to
check that further obstructions to the construction of $S$ can
arise only if $H^n_1(s\vert d)$ is not zero.

To summarize, the problem of consistent deformations is elegantly
captured by the BRST-field antifield formalism. The non-trivial
first order vertices are representatives of $H^n_0(s\vert d)$ and
the construction of the higher order vertices can be obstructed if
$H^n_1(s\vert d)$ is non-vanishing. These observations motivate the
study of these two cohomological groups.

\section{Gauge invariant conserved currents and
global symmetries}\label{cocudef} 
In this section we discuss the r\^ole of the BRST cohomology in the
study of the gauge structure of global symmetries and
conserved currents \cite{BBH4}.

A {\em global
symmetry} of a field theory is a transformation of the fields
depending on constant parameters,
\begin{equation}
\Delta \phi^i
= a^i (\phi^j,\partial_\mu \phi^j,\ldots),
\end{equation}
which leaves the action invariant or in other words, which leaves
the lagrangian invariant up to a total derivative, i.e.,
\begin{equation}
\frac{\delta {\cal L}}{\delta \phi^i}\Delta \phi^i + \partial_\mu
j^\mu =0.\label{symaaa}
\end{equation}
In \eqref{symaaa}, $j^\mu$ is the conserved current associated with
the symmetry $\Delta \phi^i$.

If we set $a_1= \phi^*_i a^i d^n x$, \eqref{symaaa} can be rewritten
as,
\begin{equation}
\delta a_1 + dj=0,
\end{equation}
where $j$ is the $n-1$ form
dual to $j^\mu$. We have used the fact that by {\em definition},
$\delta \phi^*_i
= -\frac{\delta {\cal L}}{\delta \phi^i}$. The $n$-form $a_1$ of
antighost number $1$ therefore defines an element of the
cohomological group
$H^n_1(\delta
\vert d)$.

A global symmetry is said to be trivial if the corresponding $a_1$
defines a trivial element in $H^n_1(\delta\vert d)$, i.e.,
\begin{equation}
a_1 = \delta r_2 + dc_1.
\end{equation}
This is the case if and only if the global symmetry reduces on-shell
to a gauge transformation \cite{HenneauxTeitelboim,BBH1}. Two global
symmetries are said to be equivalent if they differ by a trivial
symmetry. Therefore, the equivalence classes of non-trivial
symmetries correspond to the classes of $H^n_1(\delta\vert d)$.

From Eq. \eqref{symaaa}, we see that there is a well defined map
between the cohomological classes of $H^n_1(\delta\vert
d)$ and the classes of conserved currents, where two currents are
said to be equivalent if they differ by a term of the form: $\delta
e_1 + dm$ where $m$ is a $n-2$ form. Such a term reduces on-shell
to an identically conserved current and will be called a {\em
trivial current}. This does not mean that trivial currents are
devoid of physical interest but rather that their expression is
easily obtained.

The question we want to answer is: can $a_1$ and $j$ be redefined
(``improved") by the addition of trivial terms in order to make them
gauge invariant. In other words, does one have
$\gamma a'_1=\gamma j'=0$ with $a'_1=a_1+\delta r_2+dc_1$ and
$j'=j+\delta e_1+dm$? 

Using the following isomorphism theorem, one
can relate this question to the calculation of the local BRST
cohomology
$H(s\vert d)$:
\begin{theorem}(valid in particular for the $p$-form gauge theories
considered in this thesis)\label{isosym}
\begin{align}
H_k(s|d)\simeq
\begin{cases}
H_k(\gamma|d,H_0(\delta))&\quad
 k\geq 0 \\
H_{-k}(\delta|d)& \quad k<0 
\end{cases}
\end{align}
\end{theorem}
\proof{The proof of this result can be found in \cite{BBH2}.}
Using the case $k<0$ of Theorem {\bf\ref{isosym}} we see that to
any class of  $H^n_1(\delta\vert d)$ representing a global symmetry
we can associate a class of $H^n_{-1}(s\vert d)$ of ghost
number
$-1$. 
The map $q: H^n_{-1}(s\vert d) \rightarrow H^n_1(\delta\vert d)$ is
realized in the following manner. Let $a$ be a representative of a
class of $H^n_{-1}(s\vert d)$. The expansion of
$a$ according to the antighost number is of the form:
\begin{equation}\label{devconsym}
a={\overline a_1}+{\overline a_2}+\ldots +{\overline a_k}.
\end{equation}
Since $a$ satisfies the Wess-Zumino consistency condition $sa+db=0$,
the term
${\overline a_1}$ satisfies $\delta {\overline a_1}+db_0=0$ and
therefore defines an element of $H^n_1(\delta\vert d)$. It is
easy to see that the map
$q: H^n_{-1}(s\vert d) \rightarrow H^n_1(\delta\vert d): [a]
\rightarrow [{\overline a_1}]$ is well defined in cohomology
since a change of representative in $H^n_{-1}(s\vert d)$
($a\rightarrow a+ sr+dc$) does not change the class of ${\overline
a_1}$ in
$H^n_1(\delta \vert d)$ (${\overline a_1}\rightarrow {\overline
a_1}+
\delta r_2 + d c_1$).

The BRST cocycles \eqref{devconsym} can be classified according to
the value of $k$ at which their expansion (according to the
antighost number) stops. This expansion genuinely stops at order $k$
if it is not possible to remove the term ${\overline a_k}$ by the
addition of trivial terms.

Because $a$ is a BRST cocycle, the last term ${\overline a_k}$
satisfies the equation $\gamma {\overline a_k}+db_k=0$. In
Section {\bf\ref{adssec}} however we will show that in each class of
$H(s\vert d)$ there is a representative which satisfies the
simpler equation $\gamma {\overline a_k}=0$. This means that the
only global symmetries which cannot be made gauge
invariant are those for which the term ${\overline a_k}$ in the
corresponding BRST cocycle is genuinely of order $k$. Indeed, if
the terms in
$a$ of antighost numbers $j>1$ can be removed by the addition of
trivial terms then the corresponding class of $H^n_1(\delta\vert
d)$ has a representative which obeys $\gamma a'_1=0$ and which
is therefore gauge invariant.

Our analysis of Section {\bf\ref{conscurent}} will show that when
one excludes an explicit dependence on the spacetime coordinates,
the only BRST cocycles in ghost number
$-1$ which cannot be assumed to stop at order $1$ are those which
correspond to the following global symmetries:
\begin{equation}\label{jkhlgm}
\delta B^a_{\mu_1\ldots\mu_p}=k^a_{\
b}B^b_{\mu_1\ldots\mu_p}, \quad \text{with $k^a_{\ b}$ an
antisymmetric matrix}.
\end{equation}
Since the $k^a_{\ b}$ are antisymmetric matrices these
global symmetries correspond to rotations of the $p$-forms among
themselves. Only these are not equivalent to a
gauge invariant global symmetry.

Once this result has been obtained, one may investigate which
conserved currents can or cannot be made gauge
invariant. Our analysis establishes that the only currents that
cannot be made gauge invariant are those associated
to the global symmetries \eqref{jkhlgm}.

The crucial part of our study of the gauge invariant nature of
the global symmetries and the conserved currents is the
classification of the solution of the Wess-Zumino consistency
condition in ghost number $-1$.

\section{Counterterms and Anomalies}\label{coan}
In this section we analyze the r\^ole of the BRST field-antifield
formalism in the theory of renormalization. Our presentation is
based on \cite{theseGlenn}. 

In order to quantize a gauge theory by
the method of path integral it is necessary to fix the
gauge in order to obtain
properly defined Green functions. A convenient way of achieving
this is to introduce {\em auxiliary fields} and to define a
gauge-fixing fermion. 

The auxiliary fields consist of ghosts and
antifields and will respectively be denoted by $C^i, b^i$ and
$C^*_i, b^*_i$. They are associated in pairs by the BRST
differential,
\begin{align}
sC^i&=b^i,\quad sb^i=0,\label{s1}\\
sb^*_i&=C^*_i,\quad sC^*_i=0,\label{s2}
\end{align}
and their ghost numbers and Grassmann parities are related by,
\begin{align}
ghost\ C^i = ghost\ b^*_i = ghost\ b^i -1=ghost\ C^*_i -1,\\
\epsilon (C^i) = \epsilon (b^*_i) = \epsilon (b^i)-1= \epsilon
(C^*_i)-1.
\end{align}
The number of auxiliary fields required depends on the theory
considered and in particular on its order of reducibility.

By virtue of \eqref{s1} and \eqref{s2}, the BRST action generating
the differential $s$ in presence of auxiliary fields is still given
by,
\begin{equation}
S=\int d^n x ({\cal L} + (-)^{\epsilon_A} \phi^*_A
s\phi^A),
\end{equation}
where the $\phi^A$ and $\phi^*_A$ now include  the auxiliary
fields.

The extended BRST action $S_{ext}$ is defined as,
\begin{equation}
S_{ext}[\phi^A,{\tilde \phi}^*_A]=S[\phi^A,{\tilde
\phi}^*_A+\frac{\delta^R \psi}{\delta \phi^A}].\label{Sext}
\end{equation}
In \eqref{Sext}, $\psi$ is the gauge-fixing fermion. It is of ghost
number $-1$ and depends only on the $\phi^A$. It is chosen in such
a way that the Feynman rules obtained from $S_{ext}[\phi^A,{\tilde
\phi}^*_A=0]$ are well defined. It is easy to see that the extended
BRST action satisfies the master equation
$(S_{ext},S_{ext})=0$ where the antibracket is now expressed in
terms of the variables $\phi^A$ and ${\tilde \phi}^*_A$.
From \eqref{Sext} one can {\em formally} calculate all the Green
functions of the theory. 

The effective action $\Gamma$ is defined as
the generating functional for the connected one particle
irreducible, amputated vertex functions,
\begin{equation}\label{1PI}
\Gamma [\tilde \varphi]=\sum_{m=2}^{\infty}{1\over m!}\int
d^nx_1...d^nx_m
\tilde \varphi (x_1)...\tilde \varphi (x_m){<0\vert T\varphi (x_1)...\varphi (x_m) 
\vert 0>}^{1PI,amp}.
\end{equation}
In \eqref{1PI} all the fields, ghosts and antifields have been
collectively denoted $\varphi$. The ${\tilde \varphi}$ are the 
corresponding smooth, fast decreasing test functions.

As is well known, when one tries to calculate the
effective action by using the Feynman rules, one encounters
difficulties arising from ill-defined (infinite) quantities.
Therefore, in order for the theory to make sense one needs a
regularization scheme which eliminates those infinities. 

To be more precise, $\Gamma$ admits an expansion in powers of
$\hbar$,
\begin{equation} \label{expeff}
\Gamma =\Gamma^{(0)}+\hbar \Gamma^{(1)}+ \hbar^2
\Gamma^{(2)}+\ldots .
\end{equation}
In \eqref{expeff} one can show that the first term is a tree diagram
which actually coincides with the extended BRST action $S_{ext}$.
The other terms are loop diagrams, the number of loops being given
by the power in $\hbar$. The problems usually begin when one tries
to calculate $\Gamma^{(1)}$. In order to eliminate the infinite
quantity present in $\Gamma^{(1)}$ one needs to regularize the
theory by giving a meaning to divergent integrals. This can be done
for example but introducing a cut-off $\Lambda$. Eq. \eqref{expeff}
then becomes,
\begin{equation}
\Gamma^{reg}({\tilde \phi}^A,{\tilde \phi}^*_A) =S_{ext}+\hbar
\Gamma^{(1)}_{div}(\Lambda)+\hbar \Gamma^{(1)}_{f}(\Lambda) + O(\hbar^2),
\end{equation}
where $\Gamma^{(1)}_{div}(\Lambda)$
is the sum of the one loop contributions to $\Gamma$ which diverge
when $\Lambda\rightarrow \infty$.

Problems can occur when the regularization introduced fails to preserve
at the quantum level the local symmetries of the classical action.
Indeed, the {\em quantum
action principle}
\cite{PiguetSor1} states that the following identity holds,
\begin{equation}\label{qap}
\frac12 (\Gamma^{reg},\Gamma^{reg})(\tilde \phi)=\hbar \Delta
(\tilde \phi) + O(\hbar^2),
\end{equation}
where the antibracket is taken with respect to the sources ${\tilde
\phi}^A$ and ${\tilde \phi}^*_A$. The term $\Delta (\tilde \phi)$
is an integrated polynomial of ghost number $1$. It represents at
order $\hbar$ the obstruction for $\Gamma^{reg}$ to satisfy the
equation,
\begin{equation}\label{Zinn}
(\Gamma^{reg},\Gamma^{reg})=0.
\end{equation}
If a term $\Delta(\tilde \phi)$ is present and cannot
be removed by adding local counterterms to the classical action, the
theory is said to be anomalous.
Indeed, Eq.
\eqref{Zinn} reflects gauge invariance and has
to be satisfied in order to guarantee the consistency of the
quantum theory and in particular its unitarity.

By taking the antibracket of Eq. \eqref{qap} with $\Gamma^{reg}$
and using the Jacobi identity
$(\Gamma^{reg},(\Gamma^{reg},\Gamma^{reg}))\equiv 0$ one obtains a
consistency requirement for the anomaly
$\Delta(\tilde\phi)$ \cite{WZ}:
\begin{equation} \label{WZumnio}
(S_{ext},\Delta)=0.
\end{equation}
This implies that the anomalies have to satisfy the Wess-Zumino
consistency condition  for the differential
$s_{ext}=(S_{ext},.\;)$. The problem of finding the possible
anomalies of the theory is in fact of cohomological nature since
trivial solutions of \eqref{WZumnio} can be easily eliminated
by modifying the original action $S_{ext}$. Indeed
if $\Delta=(S_{ext}, \Sigma)$, the redefinition
$S_{ext}\rightarrow S_{ext}-\hbar \Sigma$ yields at order $\hbar$,
$(\Gamma^{reg},\Gamma^{reg})=0$. Therefore, the non-trivial
anomalies are elements of $H^n_1(s_{ext})$.

If the theory can be shown to be free from anomalies, then the
BRST cohomology is also useful to determine the 
counterterms required to renormalize the theory. Indeed, in that case Eq.
\eqref{qap} implies at order
$\hbar$ for the divergent contributions to $\Gamma$,
\begin{equation}
(S_{ext}, \Gamma^{(1)}_{div})=0.\label{coun}
\end{equation}
The counterterms are thus solutions of the Wess-Zumino consistency
condition in ghost number $0$. As in the case of anomalies, there
are some trivial solutions to \eqref{coun} given by
$\Gamma^{(1)}=(S_{ext}, \Omega)$ where $\Omega$ is of ghost number
$-1$. They are called trivial because they amount to field
redefinitions in the action $S_{ext}$.

The above discussion shows that two important questions concerning
renormalization can be examined by the calculation of the BRST
cohomology of the differential $s_{ext}$. One can in fact show that
this cohomology is isomorphic to $H(s)$ where $s$ is the BRST
differential of the {\em free theory}. The isomorphism is
implemented by making the change of variables
${\tilde \phi}^*_A\rightarrow {\tilde
\phi}^*_A+\frac{\delta^R \psi}{\delta \phi^A}$ in the
representatives of $H(s)$. 

This concludes our survey of the use of the BRST cohomology in
classical and quantum field theory. In the next section we will
formulate precisely the mathematical framework in which the 
Wess-Zumino consistency condition will be studied.

\section{General definitions}\label{mathframe}

In the previous sections we have emphasized the r\^ole of the BRST
cohomology in field theory. However, in order to focus the
attention on the physical ideas we deliberately postponed the precise
definition of the space in which the calculations
have to be done.

The standard way to impose locality in the BRST formalism is to
work in the so-called ``jet spaces". To that end, we define a local
function $f$ as a function which depends on the spacetime
coordinates
$x^\mu$, the fields (including the ghosts and  the antifields) and
their derivatives up to a finite order $k$,
\begin{equation}
f=f(x^\mu,\Phi,\partial_\mu
\Phi,\ldots,\partial_{{\mu_1}\ldots{\mu_k}}\Phi).
\end{equation}
A local function is thus a function over a {\em finite dimensional
space} $V^k$ called a jet space which is 
parameterized by the spacetime coordinates
$x^\mu$, the fields $\Phi$ and their subsequent derivatives
$\partial_{\mu_1\ldots\mu_k}\Phi$ up to a finite order
$k$. In the sequel we will always assume that spacetime has the 
topology of $\mathbb{R}^n$.

A local functional $F$ is then defined as the integral of local
function:
\begin{equation}
F=\int d^n x f(x^\mu,\Phi,\partial_\mu
\Phi,\ldots,\partial_{{\mu_1}\ldots{\mu_k}}\Phi).
\end{equation}
By definition, the local BRST cohomology $H(s)$ 
is the set of local functionals which satisfy the
equation,
\begin{equation}
sF=s\int d^n x\, f =0,\label{wdsazqs}
\end{equation}
(for all allowed field configurations) and which are not of the
form,
\begin{equation}
F=sG=s\int d^n x\, g,\label{bngk}
\end{equation}
with $g$ a local function.

The next step in the analysis is to remove the integral
sign. This is done by recalling that Eq. \eqref{wdsazqs} is
satisfied if and only if
there exist a local 
$n-1$ form $m$ such that,
\begin{equation}
sf+dm=0,\label{tuyifov}
\end{equation}
with $\oint m=0$; a local $q$-form $a$ is defined as a spacetime
form with coefficients belonging to the algebra of local
functions:
\begin{equation}
a=a_{\mu_1\ldots\mu_q}(x^\mu,\Phi,\partial_\mu
\Phi,\ldots,\partial_{{\mu_1}\ldots{\mu_k}}\Phi)
dx^{\mu_1}\wedge\ldots\wedge dx^{\mu_q}.
\end{equation}
In
\eqref{tuyifov}
$d$ is the algebraic spacetime exterior derivative defined on local
forms as,
\begin{align}
dm&=dx^\mu \partial_\mu m, \text{\ and} \\
\partial_\mu m &=\frac{\partial m}{\partial
x^\mu}+\partial_\mu \Phi\frac{\partial^L m}{\partial \Phi}+
\partial_{\mu\nu} \Phi\frac{\partial^L m}{\partial(\partial_\nu
\Phi)}+\ldots +
\partial_{\mu\nu_1\ldots\nu_s} \Phi\frac{\partial^L
m}{\partial(\partial_{\nu_1\ldots\nu_s}
\Phi)}.
\end{align}

Similarly, the
coboundary condition
\eqref{bngk} is equivalent to,
\begin{equation}
f=sg+dl,\label{dhfkvzaa}
\end{equation}
with $l$ a local $n-1$ form such that $\oint l=0$.

Accordingly, the cohomology $H(s)$ in the space of local
functionals is isomorphic to the cohomology $H(s\vert d)$ in the
space of local functions. 

The conditions $\oint m=\oint l=0$ are rather 
awkward to take into account
(in particular they depend on the precise conditions imposed on the 
fields at the boundaries).
Therefore, one usually investigates 
$H(s|d)$ without restrictions 
on the $(n-1)$-forms at the boundary. 
One thus allows elements of
$H(s|d)$ that {\it do not define} $s$-closed local functionals 
because of non-vanishing 
surface terms.

Because we are interested in the physical problems described in
Section {\bf\ref{ckgkdfo}}-{\bf\ref{coan}} we will further restrict
the algebra of local functions. Indeed, the
counterterms and anomalies are in fact local integrated polynomials.
We will also impose this condition on the consistent deformations
(at each order in the coupling constant) and on the conserved
currents. The
local BRST cohomology $H(s\vert d)$ will therefore be investigated
in the space of local $q$-forms with coefficients that
are polynomials in fields and their derivatives. 
This algebra will be denoted by ${\cal P}$. Thus, $a$ belongs
to ${\cal P}$ if and only if
\begin{equation}
a = \alpha_{\nu_1 \ldots \nu_q}\label{formfff}
(x^\mu,[\Phi]) dx^{\nu_1} \ldots dx^{\nu_q},
\end{equation} 
with $\alpha$ a polynomial. The notation $f([y])$ means that
$f$ depends on
$y$ and its successive derivatives. Note that in \eqref{formfff} we
have dropped the $\wedge$ symbol between the $dx^\mu$.

We will also perform many of the calculations in the algebra of
local $q$-forms which do not depend explicitly on the spacetime
coordinates $x^\mu$. This algebra will be denoted ${\cal P}_{-}$.

\chapter{Free theory: BRST construction}
\section{Lagrangian, equa\-tions of mo\-tion and gau\-ge
in\-va\-ri\-ance} 
The free action of a system of $p$-forms
$B^a_{\mu_1 \ldots
\mu_{p_a}}$ is given by,
\begin{align}
I&=\int d^n x {\cal L}\\
&=\int d^n x\sum_a \big({-1 \over 2(p_{a} +1)!}H^a_{\mu_1
\ldots 
\mu_{p_{a}+1}} H^{a \mu_1
\ldots \mu_{p_{a}+1}}\big), \label{Lagrangian} 
\end{align}
where the field strengths $H^{a \mu_1\ldots \mu_{p_{a}+1}}$ are
defined by,
\begin{align}
H^a&={1\over (p_{a}+1)!} 
H^a_{\mu_1 \ldots \mu_{p_{a}+1}}dx^{\mu_1} \ldots 
dx^{\mu_{p_{a}+1}} =dB^a, \label{FieldStrength}\\
B^a&={1\over p_{a}!} B^a_{\mu_1 \ldots \mu_{p_{a}}}
dx^{\mu_1} \ldots dx^{\mu_{p_a}}.
\end{align}
Otherwise stated, we will always suppose in the sequel that
$p_a \geq 1$ and $n>p_a+1$ for all values of $a$. The second
condition is necessary in order for the $p$-forms to have local
degrees of freedom.

From \eqref{Lagrangian} one easily obtains the equations of motion
by varying the fields $B^a_{\mu_1 \ldots
\mu_{p_a}}$. They can be written equivalently as
\begin{equation}
\partial_\rho H^{a \rho \mu_1 \ldots \mu_{p_a}}=0
\Leftrightarrow d \overline{H}^a=0, \label{FE1}
\end{equation}
where $\overline{H}^a$ is the $n-p_a-1$ form dual to $H^{a \rho
\mu_1 \ldots \mu_{p_a}}$. 

The main feature of theories involving $p$-form gauge fields
is that their gauge symmetries
are {\em reducible}.  More precisely, in the present case, the 
Lagrangian (\ref{Lagrangian}) is invariant under the gauge
transformations,
\begin{equation}
B^a \rightarrow B^{'a} = B^a + d\Lambda^a  \label{GaugeSym}
\end{equation}
where $\Lambda^a$ are arbitrary ($p_a-1$)-forms.  
Now, if $\Lambda^a = d\epsilon^a$,
then, the variation of $B^a$ vanishes identically.  Thus, the gauge 
parameters
$\Lambda^a$ do not all provide independent gauge symmetries:  
the gauge transformations (\ref{GaugeSym}) are reducible.
In the same way, if $\epsilon^a$ is equal to $d\mu^a$,
then, it yields a vanishing $\Lambda^a$.
There is ``reducibility of reducibility"
unless $\epsilon^a$ is a zero form.  If $\epsilon^a$ is not
a zero form, 
the process keeps going until one reaches $0$-forms. For 
the theory with Lagrangian (\ref{Lagrangian}), there
are thus 
$p_M-1$ stages of reducibility of the
gauge transformations ($\Lambda^a$ is a ($p_a-1$)-form),
where $p_M$ is the
degree of the form of highest degree occurring in (\ref{Lagrangian})
\cite{Siegel,Thierry1,Thierry2,Baulieu1}. 
One says 
that the  theory
is a  reducible gauge theory of reducibility order
$p_M-1$.

\section{BRST differential}
\label{BRSTdiff}

We define the action of the BRST differential $s$ along the lines
recalled in Section {\bf \ref{BRSTformalismgen}}. Therefore we
need to introduce besides the fields, some antifields and some
ghosts. Because the theory is reducible (of order $p_a -1$ for
each $a$), the following set of antifields
\cite{HenneauxTeitelboim,BV2,BV3} is required:
\begin{equation} B^{*a \mu_1 \ldots \mu_{p_a}}, B^{*a\mu_1
\ldots \mu_{p_a-1}},\ldots, B^{*a\mu_1},B^{*a}.
\label{antifieldlist}
\end{equation} The Grassmann parity and the {\it antighost}
number of  the antifields $B^{*a \mu_1
\ldots \mu_{p_a}}$ associated with the fields $B^a_{\mu_1
\ldots \mu_{p_a}}$ are equal to $1$.  The Grassmann parity
and the {\it antighost} number of the other antifields is
determined according to the following rule.  As one moves
from one term to the next one to its right in
(\ref{antifieldlist}),  the Grassmann parity changes and
the antighost number increases by one unit. Therefore the
parity and the antighost number of a given antifield
$B^{*a \mu_1 \ldots \mu_{p_a-j}}$ are respectively $j+1$
modulo $2$ and $j+1$.

Reducibility also imposes the following set of
ghosts \cite{HenneauxTeitelboim,BV2,BV3}:
\begin{equation} C^a_{\mu_1 \ldots
\mu_{p_a-1}},\ldots,C^a_{\mu_1
\ldots
\mu_{p_a-j}},\ldots, C^a.
\label{ghosts}
\end{equation} These ghosts carry a degree called the pure
ghost number. The pure  ghost number of $C^a_{\mu_1
\ldots\mu_{p_a-1}}$ and its Grassmann parity are equal to
1.   As one moves from one term to the next one to its right
in  (\ref{ghosts}), the Grassmann parity changes and the
ghost  number increases by one unit up to $p_{a}$.

For each field, the {\em ghost number} is defined as the
difference between the pureghost number and the antighost number.

The following table summarizes for each field the various values of
the gradings we have introduced:
\vspace{.1cm}

\begin{center}
\setlength{\doublerulesep}{0pt}
\setlength{\extrarowheight}{5pt}
\begin{tabular}{!{\vrule width 1pt depth 10pt}c!{\vrule width
1pt}c|c|c|c!{\vrule width 1pt}}
\hline\hline\hline
Fields & parity mod 2& pureghost & antighost
& ghost
\\
\hline\hline\hline
$B_{a \mu_1 \ldots \mu_{p_a}}$ & 0 & 0 & 0 & 0 
\\
\hline
$B^{*a \mu_1 \ldots \mu_{p_a}}$ & 1 & 0 & 1 & $-1$ \\ \hline
$B^{*a \mu_1 \ldots \mu_{p_a-1}}$ & 0 & 0 & 2 & $-2$ \\ \hline
$\ldots$ &$\ldots$ &$\ldots$&$\ldots$&$\ldots$ \\ \hline
$B^{*a \mu_1 \ldots \mu_{p_a-j}}$ & $j+1$& 0 & $j+1$ &
$-j-1$
\\
\hline
$\ldots$ &$\ldots$ &$\ldots$&$\ldots$&$\ldots$ \\
 \hline
$B^{*a}$ & $p_a+1$ & 0 & $p_a+1$ & $-p_a-1$ \\ \hline
$C^a_{\mu_1\ldots \mu_{p_a-1}}$ & 1 & 1 & 0 & 1
\\
\hline
$C^a_{\mu_1\ldots \mu_{p_a-2}}$ & 0 & 2 & 0 & 2 \\ \hline
$\ldots$ &$\ldots$ &$\ldots$&$\ldots$&$\ldots$ \\ \hline
$C^a_{\mu_1\ldots \mu_{p_a-j}}$ & $j$  & $j$ & 0 & $j$ \\
\hline
$\ldots$ &$\ldots$ &$\ldots$&$\ldots$&$\ldots$ \\ \hline
$C^a$ & $p_a$  & $p_a$ & 0 & $p_a$ \\
\hline $x^\mu$ & $0$ & $0$ & 0 & $0$ \\
\hline $dx^\mu$ & $1$ & $0$ & 0 & $0$ \\
\hline\hline\hline
\end{tabular}
\end{center}
\vspace{.5cm}

As explained in Section {\bf\ref{mathframe}}, we denote by ${\cal
P}$ the algebra of spacetime forms depending explicitly on the
spacetime coordinates
$x^\mu$ with coefficients that are polynomials in the fields,
antifields, ghosts and their derivatives. The corresponding
algebra without explicit dependence on the spacetime coordinates
is denoted ${\cal P}_{-}$.

The action of $s$ in ${\cal P}$ is the sum of two parts,
namely, the 'Koszul-Tate differential $\delta$' and the
'longitudinal exterior derivative $\gamma$':
\begin{equation}
s=\delta +\gamma,
\end{equation}
and by definition we have,
\begin{eqnarray}
\delta B^a_{\mu_1 \ldots \mu_{p_a}}&=&0, \\
\delta C^a_{\mu_1 \ldots \mu_{p_a-j}}&=&0, \\
\delta {\overline B}^{*a}_1 +d{\overline H}^a &=&0,
\nonumber \\
\delta {\overline B}^{*a}_2 +d{\overline B}^{*a}_1
&=&0,\nonumber \\ &\vdots& \label{defduaux} \\
\delta {\overline B}^{*a}_{p_a+1}+d{\overline
B}^{*a}_{p_a} &=& 0,
\nonumber
\end{eqnarray}
and,
\begin{eqnarray}
\gamma {{B}^{*a\mu_1 \dots \mu_{p_a+1-j}}}&=&0,\\
\gamma B^a + dC^a_1 &=&0 ,\\
\gamma C^a_1 + dC^a_2 &=&0,\\
&\vdots& \nonumber \\
\gamma C^a_{p_a-1} + dC^a_{p_a} &=&0,\\
\gamma C^a_{p_a} & = &0.
\end{eqnarray}
In the above equations,
$C^a_{j}$ is the
$p_a-j$ form whose components are $C^a_{\mu_1 \ldots
\mu_{p_a-j}}$. Furthermore, we have denoted the duals by an
overline (to avoid confusion with the *-notation of
the antifields); for instance, ${\overline B}^{*a}_1$ is the dual
of the antifield of antighost number $1$, $B^{*a \mu_1 \ldots
\mu_{p_a}}$. 

The actions of $\delta$ and $\gamma$ are extended to the rest of
the generators of
${\cal P}$ and ${\cal P}_{-}$ by using the rules,
\begin{equation}
\delta\partial_\mu=\partial_\mu \delta,\quad
\gamma\partial_\mu=\partial_\mu
\gamma,
\end{equation}
and
\begin{equation}
\delta(x^\mu)=\gamma(x^\mu)=0.
\end{equation}

By definition,
we choose both
$\delta$ and
$\gamma$ to act like left-antiderivations, e.g.,
\begin{equation}
\delta (ab)= (\delta a) b + (-)^{\epsilon_a} a (\delta b),
\end{equation}
where $\epsilon_a$ is the Grassmann parity of $a$.

With these conventions we have:
\vspace{.1cm}

\begin{center}
\setlength{\doublerulesep}{0pt}
\setlength{\extrarowheight}{5pt}
\begin{tabular}{!{\vrule width 1pt depth 10pt}c!{\vrule width
1pt}c|c|c|c|c!{\vrule width 1pt}}
\hline\hline\hline
Fields & parity mod 2& form-degree & pureghost & antighost
& ghost
\\
\hline\hline\hline
$ {\overline H}^{a} $ & $n-p_a-1$  & $n-p_a-1$ & 0 & 0 & 0 
\\
\hline
${\overline B}^{*a}_{1}$ & $n-p_a-1$  & $n-p_a$ & 0 & 1 & $-1$
\\
\hline
${\overline B}^{*a}_{2} $ & $n-p_a-1$  & $n-p_a+1$ & 0 & 2 &
$-2$
\\
\hline
$\ldots$ &$\ldots$ &$\ldots$&$\ldots$&$\ldots$&$\ldots$ \\ \hline
${\overline B}^{*a}_j$ & $n-p_a-1$  & $n-p_a-1+j$ & 0 & $j$ &
$-j$
\\
\hline
$\ldots$ &$\ldots$ &$\ldots$&$\ldots$&$\ldots$ &$\ldots$\\
 \hline
${\overline B}^{*a}_{p_a+1}$ & $n-p_a-1$  & $n$ & 0 & $p_a+1$
&
$-p_a-1$
\\
\hline
$C^a_{1}$ & $p_a$  & $p_a-1$ & 1 & 0 & 1
\\
\hline
$C^a_{2}$ & $p_a$  & $p_a-2$ & 2 & 0 & 2 \\ \hline
$\ldots$ &$\ldots$ &$\ldots$&$\ldots$&$\ldots$ &$\ldots$ \\ \hline
$C^a_{j}$ & $p_a$  & $p_a-j$ & $j$ & 0 & $j$ \\
\hline
$\ldots$ &$\ldots$ &$\ldots$&$\ldots$&$\ldots$&$\ldots$ \\ \hline
$C^a$ & $p_a$  & 0 & $p_a$ & 0 & $p_a$ \\
\hline $x^\mu$ & $0$ & $0$ & $0$ & 0 & $0$ \\
\hline $dx^\mu$ & $1$ & 1 & $0$ & 0 & $0$ \\
\hline\hline\hline
\end{tabular}
\end{center}
\vspace{.5cm}

The r\^ole of the Koszul-Tate differential is to implement the
equations of motion in cohomology. This statement is
contained in the following theorem:
\begin{theorem} (Valid in ${\cal P}$ and ${\cal P}_{-}$)
Let $i$ be the antighost number. Then,
$H_i(\delta)=0$ for $i>0$, i.e., the cohomology of $\delta$
vanishes in antighost number strictly greater than zero.
Furthermore, the cohomology of $\delta$ in antighost number
zero is given by the algebra of ``on-shell spacetime
forms''.
\label{propkoszul}
\end{theorem}
\proof{See \cite{FHST,FH,HenneauxTeitelboim,HenneauxCMP1} and also
\cite{BBH1} for the explicit proof in the case of $1$-forms and
$2$-forms.} 
While
$\delta$ controls the dynamics of the theory, the  r\^ole of the
exterior longitudinal derivative is to take care of the gauge
invariance. Therefore, the only combinations of the
fields $B^a_{{\mu_{1}\ldots
\mu_{p_{a}}}}$ and their derivatives appearing in the
cohomology of
$\gamma$ should be the field strengths and their
derivatives. Indeed, those are the only gauge invariant
objects that can be formed out of the fields and their
derivatives. The following theorem shows that the definition
of $\gamma$ is consistent with this requirement.
\begin{theorem} (Valid in ${\cal P}$ and ${\cal P}_{-}$)
The cohomology of $\gamma$ is given by,
\begin{equation}
H(\gamma) = {\cal I} \otimes {\cal C},
\end{equation}
where ${\cal C}$ is the algebra generated by the
{\em undifferentiated} ghosts $C^a_{p_a}$, and ${\cal I}$ is
the algebra of spacetime forms with coefficients that are
polynomials in the fields strengths, their
derivatives, the antifields and their derivatives. In particular,
in antighost and pure ghost numbers equal to zero, one can take as
representatives  of the cohomological classes the gauge invariant
forms.

\noindent {\rm \underline{rem}: From now on, the variables
generating
${\cal I}$ will be denoted by $\chi$.}
\label{hgamma}
\end{theorem}
\proof{One follows the standard method
which consists in separating the variables into
three sets obeying respectively $\gamma x^i=0,\ \gamma y^\alpha=z^\alpha$,
$\gamma z^\alpha=0$. The
variables $y^\alpha$ and $z^\alpha$ form ``contractible pairs" and
the cohomology is then generated by the (independent) variables $x^i$.
In our case, the $x^i$ are given by $x^\mu$, $dx^\mu$, the fields
strength components, the antifields and their
derivatives as well as the
last (undifferentiated) ghosts of ghosts (in ${\cal P}_{-}$ the
$x^\mu$ are absent). Of course, for the proof to work, any identity
verified by the
$y^\alpha$ should also be verified by the $z^\alpha$, so that the
independent
$z^\alpha$ are all killed in cohomology.

Let us first note that $x^\mu$, $dx^\mu$, the antifields and their
derivatives are automatically part of the $x^i$ since they are
$\gamma$-closed and do not appear in the $\gamma$-variations.

To show that indeed the rest of the generators of ${\cal P}$, which
are
\begin{equation}
C^a_{\mu_1\ldots
\mu_l},\partial_{\alpha_1\ldots\alpha_k}C^a_{\mu_1\ldots
\mu_l},\ldots,C^a,\partial_{\alpha_1\ldots\alpha_k}C^a,
\ldots,
\end{equation}
(with the convention 
$B^a_{\mu_1\ldots\mu_p}=C^a_{\mu_1\ldots\mu_p}$)
split as indicated,
we decompose the
$\partial_{\alpha_1\ldots\alpha_k}C^a_{\mu_1\ldots \mu_l}$
into irreducible tensors under the full linear group
$GL(n)$. Since the
$\partial_{\alpha_1\ldots\alpha_k}C^a_{\mu_1\ldots \mu_l}$
are completely symmetric in $\alpha_1\ldots \alpha_k$ and
completely antisymmetric in $\mu_1\ldots \mu_l$ they
transform under $GL(n)$ as the variables of the tensor
product representation symbolically denoted by

\begin{picture}(0,105)(0,0)
\multiframe(0,50)(15.5,0){5}(15,15){$\alpha_1$}...{$\alpha_k$}
\put(85,55){$\bigotimes$}
\multiframe(105,20)(0,15.5){5}(15,15){$\mu_l$}...{$\mu_1$}
\put(135,55){$\simeq$}
\multiframe(170.5,82)(15.5,0){5}(15,15){$\alpha_1$}...{$\alpha_k$}
\multiframe(155,20)(0,15.5){5}(15,15){$\mu_l$}...{$\mu_1$}
\put(255,55){$\bigoplus$}
\multiframe(290.5,82.5)(15.5,0){4}(15,15){$\alpha_2$}..{$\alpha_k$}
\multiframe(275,5)(0,15.5){6}(15,15){$\alpha_1$}{$\mu_l$}...{$\mu_1$}
\put(215,45){1)}
\put(330,45){2)}
\end{picture}
in \cite{hamermesh}\nocite{Dragon1}.

Convenient generators for the irreducible spaces
corresponding to diagram 1) and 2) are respectively,
\begin{equation}
\partial_{(
{\alpha_1\ldots\alpha_k}}C^a_{[\mu_1{)}_1\ldots
\mu_l{]}_2}\hbox{\ \ \ and \ \ \ }
\partial_
{\alpha_2\ldots\alpha_{k}}H^a_{\alpha_1\mu_1\ldots
\mu_l},
\end{equation}
with $H^a_{\mu_1\ldots
\mu_l}=\partial_{[\mu_1}C^a_{\mu_2\ldots\mu_l]}$. $[\ ]$
and $(\ )$ mean respectively antisymmetrization and
symmetrization; the subscript indicates the order in which
the operations are done.

A direct calculation shows that,
\begin{eqnarray}
\gamma
C^a_{\mu_1\ldots\mu_l}&=&H^a_{\mu_1\ldots\mu_l} \hbox{\ for\
} 2 \leq l\leq p, \\ \gamma
C^a_{\mu_1}&=&\partial_{\mu_1}C^a,
\\ \gamma
\partial_{(
{\alpha_1\ldots\alpha_k}}C^a_{[\mu_1{)}_1\ldots
\mu_l{]}_2}&=&c\partial_{\alpha_1\ldots\alpha_{k}}
H^a_{\mu_1\ldots
\mu_l}\hbox{\ for\
} 2 \leq l\leq p
,\label{defc} \\
\gamma \partial_{(
{\alpha_1\ldots\alpha_k}}C^a_{\mu_1{)}}&=&\partial_{\alpha_1
\ldots\alpha_k\mu_1}C^a, \\
\gamma
H^a_{\mu_1\ldots\mu_{p+1}}&=&0,\\ \gamma
\partial_{\alpha_1\ldots\alpha_{k}}
H^a_{\mu_1\ldots
\mu_{p+1}}&=&0, \\
\gamma C^a&=&0,
\end{eqnarray}
with $c$ in \eqref{defc} given by $c=\frac{k+l}{l(k+1)}$.
All the generators are are now split according to the
rule recalled at the beginning of the subsection. The
cohomology is therefore generated by
\begin{equation}
C^a, \ H^a_{\mu_1\ldots\mu_{p+1}}
\hbox{\ \ and\ \ }
\partial_{\alpha_1\ldots\alpha_{k}}
H^a_{\mu_1\ldots
\mu_{p+1}}.
\end{equation}
This ends the proof of the theorem.
Note that the generators are not independent but restricted by
the Bianchi identity $dH^a=0$.
}

The last two theorems remain valid if we restrict ourselves to
${\cal P}_{-}$ except that the spacetime forms are now independent
of the coordinates $x^\mu$.

\section{Poincar\'e Lemma}
In the sequel, we will also need the following result on the
cohomology of $d$:
\begin{theorem}
The cohomology of $d$ in the algebra of local forms ${\cal P}$ is
given by,
\begin{eqnarray}
H^0(d) \simeq R, \\
H^k(d)=0 \hbox{ for } k\not = 0, k\not=n,  \label{dCohomo1} \\
H^n(d) \simeq \hbox{ space of equivalence classes of local forms,}
\label{dCohomo2}
\end{eqnarray}
where $k$ is the form degree and $n$ the spacetime dimension.
In (\ref{dCohomo2}), two local forms are said to be 
equivalent if and only if
they have identical Euler-Lagrange derivatives with respect to all
the fields and the antifields.
\label{Poincare}
\end{theorem}
\proof{This theorem is known as the algebraic Poincar\'e Lemma. It
differs from the usual Poincar\'e lemma because here we only work
with {\em local} forms in the various fields and their
derivatives. For various proofs, see
\cite{Vinogr1,Vinogr2,DeWilde1,Tulczyjew1,Tulczyjew2,Tsujishita1,
BonoraCoRa1,Olver1,Wald1,Dickey1,BDK3,DVHTV1}.}
It should be mentioned that the theorem holds as such only in
${\cal P}$.  In ${\cal P}_{-}$, 
(\ref{dCohomo1}) would have to be amended as
\begin{equation}
H^k(d) \simeq \{\hbox{constant forms} \}
 \hbox{ for } k\not = 0, k\not=n,
\end{equation}
where the constant forms are by definition
the local exterior forms
with constant coefficients. Indeed, the explicit
$x$-dependence enables one to remove the constant $k$-forms ($k>0$)
from the cohomology, since these are exact, $c_{i_1 i_2 \dots
i_k} dx^{i_1} dx^{i_2} \dots dx^{i_k} = d(c_{i_1 i_2 \dots
i_k} x^{i_1}$ $dx^{i_2} \dots dx^{i_k})$. Note that the constant exterior forms 
can be alternatively eliminated without introducing
an explicit $x$-dependence, but by imposing
Lorentz invariance (there is no Lorentz-invariant
constant $k$-form for $0<k<n$). 

We shall denote in the sequel
the algebra of constant forms by $\Lambda^*$ and the subspace of
constants forms of degree $k$ by $\Lambda^k$.  

The following formulation of the Poincar\'e lemma is also
useful.
\begin{theorem}
Let $a$ be a local, closed $k$-form ($k<n$) 
that vanishes when the fields, the ghosts and the antifields are
set equal to zero.  Then, $a$ is $d$-exact.
\label{poincarebis}
\end{theorem}
\proof{The condition that $a$ vanishes when the 
fields and the antifields are
set equal to zero eliminates the constants.}   
This form 
of the Poincar\'e lemma holds in  both the algebras
of $x$-de\-pen\-dent and
$x$-independent local exterior forms.

\section{Mixed forms}\label{mixedforms}
In our analysis of the BRST cohomology, in turns out that two
combinations of the fields and antifields play a central
r\^ole. The first one groups the field strengths
and the duals of the antifields and is denoted $\tilde H^a $,
\begin{equation}
\tilde H^a = {\overline H}^a + \sum_{j=1}^{p_a+1} {\overline
B}^{*a}_j.\label{totalB}
\end{equation}
The second one combines the $p_a$-forms and their associated
ghosts and is denoted $\tilde B^a$,
\begin{equation}
\tB^a = B^a + C^a_1 + \ldots + C^a_{p_a}.
\end{equation}
The regrouping of physical fields with ghost-like
variables is quite standard in BRST theory
\cite{Stora2}.  Expressions similar (but not
identical) to
(\ref{totalB}) have appeared in the analysis of the Freedman-Townsend
model and of string field theory \cite{Thorn1,BauBergSezg1}, as
well as in the context of
topological models \cite{CarVilSasSor1,Baulieu4}.  Note
that for a one-form, expression (\ref{totalB}) reduces
to relation (9.8) of \cite{BBH2}.

It is easy to see that both $\tilde H^a$ and $\tB^a$ have
a definite Grassmann parity respectively given by $n-p_a+1$
and $p_a$ modulo $2$. On the other hand, products of $\tilde
H^a$ or
$\tilde B^a$ are not homogeneous in form degree and ghost
number. 
To isolate a
component of a given form degree $k$ and ghost number $g$ we
enclose the product in brackets $[\ldots]^k_g$. 

Since
products of $\tB^a$ very frequently appear  in the rest of
the text, we introduce the following notations,
\begin{equation}
{\cal Q}^{a_1 \ldots a_m}=\tB^{a_1}\ldots \tB^{a_m} \quad
\hbox{and} \quad {\cal Q}^{a_1\ldots
a_m}_{k,g}=[\tB^{a_1}\ldots
\tB^{a_m}]^k_g.
\end{equation}
For example, using these conventions, we write
the most general representative of $H(\gamma)$ as,
$
a=\sum_m \alpha_{a_1\ldots a_m}(\chi) [\tB^{a_1} \ldots
\tB^{a_m}]^0_l
= \sum_m \alpha_{a_1\ldots a_m}(\chi) 
{\cal Q}^{a_1 \ldots a_m}_{0,l},
$
with $l=\sum_m p_{a_m}$.

We also define the three ``mixed operators": $\Delta =
\delta + d$, $\tilde \gamma =\gamma + d$ and $\tilde s=s+d$.

Using those definitions we have the following relations:
\begin{align}
\Delta {\tilde H}^a &=0, \quad \Delta {\tilde B}^a
 =0,\\
\tilde \gamma {\tilde H}^a &=0,  \quad\tilde \gamma
{\tilde B}^a =H^a, \label{formulerusse} \\
\tilde s {\tilde H}^a &=0,  \quad \tilde s {\tilde
B}^a =H^a.
\end{align}
Eq. $\tilde \gamma {\tilde
B}^a=H^a$ is known in the literature as the ``horizontality
condition" \cite{Baulieu1}.

\chapter{Free theory: BRST cohomology (Part I)}
\section{Antifield dependence of solutions}
In the previous chapters, we have introduced the various
ingredients of the BRST field-antifield for the system
of free $p$-form gauge fields
\eqref{Lagrangian}. We have defined all the necessary
fields, antifields, ghosts and their transformation law
under the BRST differential which in this case reduces to the
sum of the Koszul-Tate differential and the longitudinal exterior
derivative. We now start the analysis of
the BRST cohomology. 

The equation we want to solve is,
\begin{equation}
s a^k_g + d a^{k-1}_{g+1}=0,
\label{WZ2}
\end{equation}
where $a^k_g$ is a
polynomial in
${\cal P}_{(-)}$ of ghost number $g$ and form degree $k$. 

Historically, solutions of \eqref{WZ2} were first sought in the
absence of antifields. This approach is incomplete since it does
only take into account the gauge invariance of the theory but not
its dynamics. For example, by not investigating the BRST
cohomology in the presence of antifields, one would miss the
Yang-Mills vertex when computing the various consistent
interactions of Maxwell's theory. We will nevertheless start by
determining the antifield independent solution of \eqref{WZ2} for
the following reasons. First, they are of course
interesting on their own. But also,  their knowledge is required in
order to apply our method of analysis of the antifield dependent
solutions.

We will analyze the antifield independent solutions of the
Wess-Zumino consistency condition by studying the so-called
descent equations \cite{Stora1,Stora2,Zumino1}. They were first
studied in the Yang-Mills theory context. They are however useful
for a large class of theories for which the longitudinal exterior
derivative is nilpotent. This is the case for the system
\eqref{Lagrangian} but also for the ``Chapline-Manton" theories as
we shall see later. All the relevant details of
these descent equations will be recalled in the next section.

The descent equations for the antifield dependent solutions also
exist. However, they cannot be used straightforwardly to obtain the
general solution of \eqref{WZ2}. The main difficulty lies in the
fact that the antifields are of negative ghost number. Therefore,
one cannot easily formulate a {\em Generalized ``transgression"
lemma} when they are present (see Theorem {\bf{\ref{gentrlm}}}).

In order to analyze the antifield dependent solutions we therefore
follow another approach. First,
we decompose the solutions of
(\ref{WZ2})  according to the antighost number,
$a_g^k=a_{g,0}^k+ a_{g,1}^k+\ldots+ a_{g,q}^k$. When performing
such a decomposition, we suppose that the term of highest antighost
number $(a_{g,q}^n)$ cannot be eliminated by adding trivial
terms to $a^n_g$, i.e., $a^k_{g,q} \not = 
se^k_{g-1} +d c^{k-1}_g$; otherwise, the expansion would
stop at
$q-1$. If $q\not =0$ then $a^k_g$ is said to depend non-trivially
on the antifields.
To proceed, we will first determine the most general form for the
term $a^n_{g,q}$. Then, substituing this form in \eqref{WZ2} we
will try to compute recursively the components $a^n_{q,q-1},
a^n_{g,q-2}$ etc. of lower antighost numbers in $a^n_g$. As we
shall see, this construction is not obstructed only for a small
number of $a^n_{q,q}$. This way of analyzing the antifield
dependent solutions of the Wess-Zumino consistency condition has
some resemblance with the method of the descent equations and we
shall later comment on this. 

\section{Antifield independent solutions}\label{Sec:Antidep1}

As stated, we start by calculating the antifield independent
solution of the Wess-Zumino consistency condition by solving the
so-called ``descent equation". We first recall the general
procedure of the analysis and then apply it to free $p$-forms.

\subsection{Descent equations}

Because we work here in the absence of antifields, Eq. \eqref{WZ2}
becomes,
\begin{equation}
\gamma a^k_g + d a^{k-1}_{q+1} = 0,\label{lfrtsx}
\end{equation} 
(where $g$ now denotes the  pureghost number). This is permissible
since both $\gamma$ and $d$ are of antighost number $0$. To a given
solution
$a^k_g$ of \eqref{lfrtsx}, one can associate another solution
of the Wess-Zumino consistency condition, namely,
$a^{k-1}_{g+1}$.  Indeed, the triviality of
$d$ (Theorem {\bf\ref{poincarebis}})
implies,
\begin{equation}
\gamma a^{k-1}_{g+1} + d a^{k-2}_{g+2} =0,
\end{equation} for some $a^{k-2}_{g+2}$.  There are ambiguities in
the choice of
$a^{k-1}_{g+1}$ given the class $[a^k_g]$ of
$a^k_g$ in
$H^{(k,g)}(\gamma \vert d)$, but it is easy to verify that
the map
$\partial : H^{(k,g)}(\gamma \vert d) \rightarrow
H^{(k-1,g+1)}(\gamma \vert d)$ is well defined.

The map $\partial$ is in general not injective.  There are
non trivial classes of $H^{*,*}(\gamma \vert d)$ that are
mapped on zero through the descent.  For instance, if one
iterates $\partial$, one gets from
$a^k_g$ a chain of cocycles in $H^{*,*}(\gamma \vert
d)$,
$[a^k_g] \mapsto \partial [a^k_g] \in
H^{(k-1,g+1)}(\gamma
\vert d)
\mapsto \partial^2 [a^k_g] \in H^{(k-2,g+2)} \mapsto
\cdots \mapsto \partial^l [a^k_g] \in H^{(k-l,g+l)}
\mapsto 0$ which must eventually end on zero since there
are no forms of negative form degree.  

The equations
defining the successive images of $[a^k_g]$ are
\begin{eqnarray}
\gamma a^k_g + d a^{k-1}_{g+1} &=& 0, \\
\gamma a^{k-1}_{g+1} + d a^{k-2}_{g+2} &=& 0, \\ &\vdots&
\nonumber \\
\gamma a^{k-l}_{g+l} + da^{k-l-1}_{g+l+1} &=& 0,
\end{eqnarray} 
and are known as the ``descent equations"
\cite{Stora1,Stora2,Zumino1}.  

If 
$a^{k-l-1}_{g+l+1}$ is trivial in
$H^{(k-l-1,g+l+1)}(\gamma \vert d)$, i.e.,
$a^{k-l-1}_{g+l+1}=
\gamma  b^{k-l-1}_{g+l} +  db^{k-l-2}_{g+l+1}$,   one may
redefine
$a^{k-l}_{g+l}
\rightarrow a^{k-l}_{g+l} - db^{k-l-1,g+l} =
{a'}^{k-l,g+l)}$ in such a way that we have $\gamma
{a'}^{k-l}_{g+l} = 0$, i.e.,
${a'}^{k-l-1}_{g+l+1}=0$.  Conversely, if $a^{k-l}_{g+l}$ is
annihilated by $\gamma$, then $\partial [a^{k-l}_{g+l}]
=0$.  Thus, the last non-trivial element $a^{k-l}_{g+l}$,
or ``bottom", of the descent is a $\gamma$-cocycle that is
not exact in
$H^{*,*}(\gamma \vert d)$.  The non-injectivity of
$\partial$ follows precisely from the existence of such
cocycles.

The length of the descent associated with $[a^k_g]$ is
the integer $l$ for which $\partial^l [a^k_g]$ is the
last non-trivial cocycle occurring in the chain.  One says
that a descent is non trivial if it has length $\geq 1$.
The idea of
\cite{DVTV0,DVTV1} is to classify the elements of
$H^{*,*}(\gamma \vert d)$ according to the length of the
associated descent.

In order to achieve this, one must determine the possible
bottoms, i.e., the elements of $H(\gamma)$ which are not
trivial in
$H(\gamma \vert d)$ and which can be lifted $l$ times.

\subsection{Lifts of elements of $H(\gamma)$ - An example}
\label{exlift}
The difficulty in the analysis of the lift is that contrary
to the descent which carries no ambiguity in cohomology,
the lift is ambiguous because
$H(\gamma)$ is not trivial.  Furthermore, for the same
reason, the lift can be obstructed, i.e., given $a \in
H(\gamma)$, there may be no descent (i) which has
$a$ as bottom; and (ii) which starts with a solution
$b$ of the Wess-Zumino consistency condition such that $db
=0$ (while any descent ends always with an $a$ such that
$\gamma a = 0$). The ``first" $b$ may be such that $db
\not= 0$ or even $db \not=
\gamma$(something). In that case, there is no element $c$
above $b$ such that $\gamma c + db =0$ (while there is
always an element $e$ below $a$ such that $\gamma a + de =
0$, namely $e=0$: the descent effectively stops at $a$ but
is not obstructed at $a$).

In this subsection, we shall illustrate these features on a
specific example: that of a free $1$-form $A$ and a free
$2$-form $B$, with BRST algebra
\begin{eqnarray}
\gamma A + d A_1 &=& 0, \; \gamma A_1 =
0, \\
\gamma B + d B_1 &=& 0, \; \gamma B_1 +
d B_2 = 0, \; \gamma B_2 = 0.
\end{eqnarray} The curvatures are $F= dA$ and $H=dB$, with
$\gamma F=
\gamma H =0$.

Consider the $\gamma$-cocycle $A_1 B_2$.  It has
form-degree zero and ghost number three.  The descent that
ends on this bottom has length one, and not the maximum
length three. Indeed, the $\gamma$-cocycle
$A_1 B_2$ can be lifted once, since there exists
$a \in {\cal P}$ such that $\gamma a +
d(A_1 B_2) = 0$.  One may take $a = A
B_2 + A_1 B_1$. Of course, $a$ has
form-degree one and ghost number two.  If one tries to lift
the given $\gamma$-cocycle
$A_1 B_2$ once more, one meets an obstruction. 
Namely, there is no $b$ such that $\gamma b + da = 0$. This
is because $da$ is in the same $\gamma$-class as $F
B_2$, which is non-trivial, i.e., which cannot be
written as a $\gamma$-variation. It is easy to verify that
one cannot remove the obstruction by adding to $a$ a
$\gamma$-cocycle (which would not change
$\gamma a$).  This provides an example of a
$\gamma$-cocycle for which the lift is obstructed after one
step.

Consider now the $\gamma$-cocycle $\frac{1}{2} F
(B_2)^2$ with ghost number four and form-degree two. 
This cocycle can be lifted a first time, for instance $F
B_1 B_2$ is above it,
\begin{equation}
\gamma [F B_1 B_2] + d [\frac{1}{2} F
(B_2)^2] = 0.
\label{firstliftof}
\end{equation} It can be lifted a second time to  $ \frac{1}{2}
(B_1)^2 + F B B_2$. However, if one
tries to lift it once more, one meets apparently the
obstruction $F H B_2$, since the exterior derivative
of
$ \frac{1}{2} (B_1)^2 + F B B_2$
differs from the
$\gamma$-cocycle $F H B_2$ by a $\gamma$-exact term.
It is true that $F H B_2$ is a non-trivial
$\gamma$-cocycle. However, the obstruction to lifting three
times $\frac{1}{2} F (B_2)^2$ is really absent.  What
happens is that we made a ``wrong" choice for the term above the
$\gamma$-cocycle $\frac{1}{2} F (B_2)^2$ and should
have rather taken a term
that differs from $F B_1 B_2$ by an appropriate
$\gamma$-cocycle.  This is because $F H B_2$ is in fact the true
obstruction to lifting twice the $\gamma$-cocycle
$A_1 H B_2$.  Thus if one replaces
(\ref{firstliftof}) by
\begin{equation}
\gamma [F B_1 B_2 - A_1 H B_2] +
d[\frac{1}{2} F (B_2)^2]= 0,
\end{equation} which is permissible since $\gamma(A_1 H
B_2) = 0$, one removes the obstruction to further lifting
$\frac{1}{2} F (B_2)^2$.  This shows that the
obstructions to lifting $k$ times a $\gamma$-cocycle are
not given by elements of $H(\gamma)$, but rather, by
elements of $H(\gamma)$ that are not themselves
obstructions of shorter lifts.  The ambiguity in the
choice of the lifts plays accordingly a crucial r\^ole in
the analysis of the obstructions.

In fact, the given $\gamma$-cocycle
$\frac{1}{2} F (B_2)^2$ is actually trivial in
$H(\gamma \vert d)$
\begin{align}
\frac{1}{2} F (B_2)^2 =&  d[\frac{1}{2} A
(B_2)^2 + A_1 B_1 B_2] \nonumber \\
& +
\gamma[A B_1 B_2 + \frac{1}{2}
A_1(B_1)^2 + 
A_1B B_2]
\end{align} and therefore, its lift can certainly never
be obstructed.

\subsection{Lifts of elements of $H(\gamma)$ - The first
two steps}

In order to control these features, it is
convenient to introduce new differential algebras
\cite{DVTV0,DVTV1}.  Let $E_0 \equiv H(\gamma)$. We define a
map $d_0: E_0 \rightarrow E_0$ as follows:
\begin{equation} d_0[a] = [da],
\end{equation} where $[\ ]$ is here the class in $H(\gamma)$.  This
map is well defined because $\gamma da = -d \gamma a = 0$ (so
$da$ is a
$\gamma$-cocycle) and $d(\gamma m) = -\gamma (dm)$ (so $d$
maps a
$\gamma$-coboundary on a $\gamma$-coboundary). Now, $d_0$
is a derivation and $d_0^2=0$, so it is a differential.
Cocycles of $d_0$ are elements of $H(\gamma)$ that can be
lifted at least once since
$d_0[a] = 0 \Leftrightarrow da+ \gamma b =0$ for some $b$,
so $b$ descends on $a$.  By contrast, if
$d_0[a] \not= 0$, then $a$ cannot be lifted and, in
particular, $a$ is not exact in $H(\gamma \vert d)$ (if it
were, $a = \gamma m + dn$, one would have $da = - \gamma
dm$, i.e., $da = 0$ in $H(\gamma)$).  Let $F_0$ be a
subspace of $E_0$ supplementary to $\text{Ker} \, d_0$.  One has
the isomorphism (as vector spaces),
\begin{equation} E_0 \simeq \text{Ker} \, d_0 \oplus F_0.
\label{iso1}
\end{equation}

The next step is to investigate cocycles that can be lifted
at least twice. In order to be liftable at least once,
these must be in $\text{Ker}
\, d_0$ . Among the elements of $\text{Ker} \, d_0$, those that
are in $\text{Im}
\, d_0$ are not interesting, because they are elements of
$H(\gamma)$ that are trivial in $H(\gamma \vert d)$ ($[a] =
d_0[b] \Leftrightarrow a = db + \gamma m$).  Thus the
relevant space is $E_1 \equiv H(d_0, E_0)$.  One has,
\begin{equation} \text{Ker} \, d_0 \simeq \text{Im} \, d_0 \oplus E_1.
\label{iso1'}
\end{equation} One then defines the differential $d_1: E_1 \rightarrow
E_1$,
\begin{equation} d_1 [[a]] = [[db]],
\label{defd1}
\end{equation} where $b$ is defined through $da + \gamma b = 0$ --
recall that
$d_0[a] = 0$ -- and where $[[a]]$ is the class of $[a]$ in
$E_1$.  It is easy  to see that (\ref{defd1}) provides a
well-defined differential in $E_1$\footnote{Proof: $d_0[a]
= 0 \Rightarrow da+\gamma b= 0 \Rightarrow \gamma db=0$. 
Hence, $db$ is a
$\gamma$-cocycle, which is clearly annihilated by $d_0$,
$d_0 [db]= [d^2 b] = 0$. Furthermore, the class of $db$ in
$E_1$ does not depend on the ambiguity in the definition of
$b$, since if $b$ is replaced by $b+dm+ u$ with $\gamma
u=0$, then $db$ is replaced by $db + du$ which is
equivalent to $db$ in
$E_1$ (the class of $du$ in $E_0$ is equal to $d_0[u]$
since $\gamma u = 0$, and this is zero in $E_1$). The
derivation property is also easily verified, $d_1 (ab) =
(d_1a) b + (-1)^{\epsilon_a}a d_1 b$.}.

If $[[a]] \in E_1$ is a $d_1$-cocycle, then it can be
lifted at least twice since $[[db]]=0$ in $E_1$ means $db
=d u + \gamma$(something) with
$\gamma u = 0$.  Thus one has $da + \gamma b' = 0$ with
$b' = b - u$ and $d b' = \gamma$(something). If on the
contrary, $d_1 [[a]] \not= 0$, then the corresponding
elements in $H(\gamma)$ cannot be lifted twice, $d_1 [[a]]$
being the obstruction to the lift.  More precisely, the
inequality $d_1 [[a]] \not= 0$ in $E_1$ means $[db]
\not= d_0 [c]$ in $E_0$. Thus, $db$ cannot be written as a
$\gamma$-variation, even up to the exterior derivative of a
$\gamma$- closed term (ambiguity in the definition of $b$). It is
straightforward to see that in that case both $a$ and $b$ are
non-trivial in $H(\gamma \vert d)$.

Analogous to the decomposition (\ref{iso1}) one has,
\begin{equation} E_1 \simeq \text{Ker} \, d_1 \oplus F_1,
\end{equation} where $F_1$ is a subspace of $E_1$ supplementary to
$\text{Ker} \, d_1$. The elements in $\text{Im} \, d_1$ are trivial in
$H(\gamma \vert d)$ and thus of no interest from the point
of view of the Wess-Zumino consistency condition.

To investigate the (non-trivial) $\gamma$-cocycle that can
be lifted at least three times one defines,
\begin{equation} E_2 = H(d_1, E_1),
\end{equation} and the differential
$d_2$ through,
\begin{equation} d_2 : \; E_2 \rightarrow E_2 , \; d_2 [[[a]]]  =
[[[dc]]],
\end{equation} where the triple brackets denote the classes in $E_2$
and where
$c$ is defined through the successive lifts
$da + \gamma b = 0$, $db + \gamma c = 0$ (which exist since
$d_1 [[a]] =0$).  It is  straightforward to
verify that $d_2$ is well-defined in $E_2$, i.e., that the
ambiguities in $b$ and $c$ play no r\^ole in $E_2$. 
Furthermore, a $\gamma$-cocycle $a$ such that
$d_0 [a] = 0$ (so that $[[a]] \in E_1$ is well-defined) and
$d_1 [[a]] = 0$ (so that $[[[a]]] \in E_2$ is well-defined)
can be lifted a third time if and only if $d_2 [[[a]]] =0$.
Indeed, the relation $d_2 [[[a]]] =0$ is equivalent to
$[[[dc]]] = 0$, i.e. $dc = \gamma u + d v + d w$, with
$\gamma v = 0$ (this is the $d_0$-term) and $\gamma w + dt
= 0$,
$\gamma t = 0$ (this is the $d_1$-term).  Thus, by
redefining $b$ as
$b - t$ and $c$ as $c -v - w$, one gets, $dc_{Redefined} =
\gamma u$. If $a$ cannot be lifted a third time then it is
non-trivial in $H(\gamma \vert d)$; this is also true for
$b$ and $c$. 

Note that we have again the decomposition,
\begin{equation}
E_2 \simeq \text{Ker} d_2 \oplus F_2,
\end{equation}
where $F_2$ is a subspace of $E_2$ supplementary to $\text{Ker} d_2$.

To summarize, the above discussion shows that: 1) the elements of
$F_0$ are the non-trivial $\gamma$-cocycles in $H(\gamma \vert d)$
that cannot be lifted at all; 2) the elements of $F_1$ are the 
non-trivial $\gamma$-cocycles in $H(\gamma \vert d)$ that can be
lifted once; and 3) the elements of $F_2$ are the 
non-trivial $\gamma$-cocycles in $H(\gamma \vert d)$ that can be
lifted twice. Furthermore, the successive lifts of elements of
$F_1$ and $F_2$ are non-trivial solution of the Wess-Zumino
consistency condition.

\subsection{Lifts of elements of $H(\gamma)$ - General
theory}

One can  proceed in the same way for the next lifts. One
finds in that manner a sequence of spaces $E_r$ and
differentials $d_r$ with the properties:

\begin{enumerate}
\item $E_{r} = H(E_{r-1}, d_{r-1})$.
\item There exists an antiderivation $d_r : E_r \rightarrow
E_r$ defined by $d_r [[\dots [ X ] \dots ]]$ $= [[\dots
[db]\dots ]]$ for
$[[\dots [X] \dots ]] \in E_r$ where $[[\dots [db]
\dots ]]$ is the class of the $\gamma$-cocycle $db$ in
$E_r$ and where $b$ is defined through
$dX+ \gamma c_1 = 0$, $dc_1 + \gamma c_2 = 0$, ...,
$d c_{r-1} + \gamma b = 0$. Similarly, $[[\dots [X] \dots
]]$ denotes the class of the
$\gamma$-cocycle $X$ in $E_r$ (assumed  to fulfill the
successive conditions $d_0 [X] = 0$, $d_1 [[X]] =0$ etc ...
so as to define an element of $E_r$).
\item $d_r^2 = 0$. 
\item A $\gamma$-cocycle $X$ can be lifted $r$ times if and
only if
$d_0 [X] = 0$,
$d_1 [[X]]=0$,
$d_2 [[[X]]]=0$,
 ..., $d_{r-1} [[\dots[X]\dots]]=0$.  If $d_r [[\dots[ X
]\dots]]
\not=0$, the $\gamma$-cocycle $X$ cannot be lifted
$(r+1)$ times and it is, along with its successive lifts, not exact
in
$H(\gamma
\vert d)$.
\item A necessary and sufficient condition for an element
$Y$ in
$H(\gamma)$ to be exact in
$H(\gamma \vert d)$ is that there exists a $k$ such that
$d_i [\dots[Y]\dots]=0$, ($i=1,2, \dots, k-1$) {\em and}
$[\dots[Y]\dots] = d_k [\dots[Z]\dots]$. This implies in
particular
$d_j [\dots[ Y ]\dots] = 0$ for all $j$'s.
\item Conversely, if a $\gamma$-cocycle $Y$ fulfills
$d_i [\dots[Y]\dots]=0$ for $i=0, 1, ..., k-1$ and
$d_k [\dots[Y\dots] \not= 0$, then, it is not exact in
$H(\gamma \vert d)$.  The condition is not necessary,
however, because there are elements of $H(\gamma)$ that are
non trivial in
$H(\gamma \vert d)$ but which are annihilated by all
$d_i$'s.  This is due to the fact that there are no
exterior form of degree higher than the spacetime
dimension.  We shall come back to this point below.
\end{enumerate}

The meaning of the integer $k$ for which $ Y = d_k Z$ in
item 5 (with
$Y \not= d_i$(something) for $i<k$) is as follows (we shall
drop the multiple brackets when no confusion can arise). 
If the
$\gamma$-cocycle $a$ is  in $\text{Im} \, d_0$, then $a = db +
\gamma c$, where $b$ is also a $\gamma$-cocycle. If $a$ is
a non-zero element of
$E_1$ in the image of $d_1$, then again $a = db + \gamma
c$, but $b$ is now {\em not} a cocycle of $\gamma$ since
$a$ would then be in $\text{Im}
\, d_0$ and thus zero in $E_1$. Instead, one has $\gamma b
+ d \beta = 0$ where
$\beta$ {\em is} a cocycle of $\gamma$ ($\gamma \beta = 0$)
which is not trivial in $H(\gamma
\vert d)$.  More generally, $k$ characterizes the length of
the descent below
$b$ in $a = db + \gamma c$, $\gamma b + d \beta = 0$ etc.

The proof of items 1 to 4 proceeds recursively.  Assume
that the differential algebras $(E_i,d_i)$ have been
constructed up to order
$r-1$, with the properties 2 through 4.  Then, one defines
the next space $E_r$ as in 1.  Let $x$ be an element of
$E_r$, and let
$X$ be one of the $\gamma$-cocycles such that the class
$[[\dots [ X]
\dots ]]$ in $E_r$ is precisely $x$. Since $X$ can be
lifted $r$ times, one has a sequence  $dX + \gamma c_1 =0$,
..., $dc_{r-1} +
\gamma b =0$. The ambiguity in $X$ is $X \rightarrow X +
\gamma a +du_0 + du_1 + \cdots + du_{r-1}$, where $u_0$ is
a $\gamma$-cocycle (this is the $d_0$-exact term), $u_1$ is
the first lift of a
$\gamma$-cocycle (this is the $d_1$-exact term) etc. 
Setting $u = u_0 + u_1 + \cdots u_{r-1}$, one sees that the
ambiguity in $X$ is of the form $X \rightarrow X +
\gamma a + du$.  On the other hand, the ambiguity in the
successive lifts takes the form $c_1 \rightarrow c_1 +
m_1$, where $m_1$ is a
$\gamma$-cocycle that can be lifted $r-1$ times, $c_2
\rightarrow c_2 + n_1 + m_2$, where $n_1$ descends on $m_1$
and $m_2$ is a
$\gamma$-cocycle that can be lifted
$r-2$ times, ..., and finally $b \rightarrow b + a_1 + a_2
+ \cdots + a_{r-1} + a_r$, where $a_1$ descends $(r-1)$
times, on $m_1$, $a_2$ descends $(r-2)$ times, on $m_2$,
etc, and $a_r$ is a
$\gamma$-cocycle.

The element $X_r \equiv db$ is clearly a cocycle of
$\gamma$, which is annihilated by $d_0$ and the successive
derivations $d_k$ because
$dX_r=0$ exactly and not just up to $\gamma$-exact terms. 
The ambiguity in the successive lifts of $X$ plays no
r\^ole in the class of $X_r$ in $E_r$, since it can
(suggestively) be written $db
\rightarrow db + d_{r-1} m_1 + d_{r-2} m_2 + \cdots + d_1
m_{r-1} + d_0 a_r$.  Thus, the map
$d_r$ is well-defined as a map from $E_r$ to $E_r$.  It is
clearly nilpotent since $dX_r=0$.  It is also a derivation,
because one may rewrite the lift equations for $X$ as
$\tilde{\gamma}(X+ c_1 + c_2 +
\dots + b)=d_r X$ where
\begin{equation}
\tilde{\gamma} = \gamma + d.
\end{equation} Let $Y$ be another element of
$E_r$ and $e_1$,  $e_2$, ...$\beta$ its successive lifts.
Then,
$\tilde{\gamma} (Y+ e_1 + e_2 + \dots + \beta) = d_r Y$.
Because
$\tilde{\gamma}$ is a derivation, one has
$\tilde{\gamma}[(X+c_1 + \dots + b)(Y+e_1+\dots
+\beta)]=(d_rX)Y +(-1)^{\epsilon_X} X d_rY + $ forms of
higher form-degree, which implies
$d_r(XY)=(d_rX)Y + (-1)^{\epsilon_X} X d_r Y$: $d_r$ is
also an odd derivation and thus a differential.  This
establishes properties 2 and 3.

To prove property 4, one observes that $X$ can be lifted
once more if and only if one may choose its successive
lifts such that $db$ is
$\gamma$-exact. This is equivalent to stating that $d_r X$
is zero in
$E_r$. Properties 5 and 6 are rather obvious: if $a$ is a
$\gamma$-cocycle which is exact in $H(\gamma \vert d)$, $a=
db+
\gamma c$, then $a=d_k m$ where $k$ is the length of the
descent associated with $\gamma b + dn =0$, which has
bottom $m$.

As shown in \cite{DVTV0,DVTV1}, the above construction may be
elegantly captured in an exact couple \cite{Massey}.  The
detailed analysis of this exact couple and the proof of the
above results using directly the powerful tools offered by
this couple may be found in
\cite{DVTV1,DVTV2,Talon1}.

One has, for each $r$, the vector space isomorphisms,
\begin{equation} E_r \simeq \text{Ker} \, d_r \oplus F_r \simeq \text{Im}
\, d_r
\oplus E_{r+1}
\oplus F_r,
\label{iso2}
\end{equation} where $F_r$ is a subspace supplementary to $\text{Ker} \,
d_r$ in $E_r$. Thus,
\begin{equation} E_0 \simeq \oplus_{k=0}^{k=r-1} F_k
\oplus_{k=0}^{k=r-1} \text{Im} \, d_k
\oplus E_r.
\end{equation} Because there is no form of degree higher than the
spacetime dimension,
$d_n =0$ ($d_n a$ has form-degree equal to
$FormDeg(a)+n+1$). Therefore, $E_{n} = E_{n+1} = E_{n+2} =
\dots$. This implies
\begin{equation} E_0 \simeq \oplus_{k=0}^{k=n-1} F_k
\oplus_{k=0}^{k=n-1} \text{Im} \, d_k
\oplus E_n.
\label{isofinal}
\end{equation} The elements in any of the $F_k$'s are non trivial
bottoms of the descent which can be lifted exactly $k$
times.  All the elements above them in the descent are also
non trivial solutions of the Wess-Zumino consistency
condition. The elements in $\text{Im} \, d_k$ are bottoms which
are trivial in $H(\gamma \vert d)$ and which therefore define
trivial solutions of the Wess-Zumino consistency
condition. Finally, the elements in $E_n$ are bottoms that
can be lifted all the way up to form degree $n$.  These are
non trivial in $H(\gamma \vert d)$, since they are not
equal to $d_i m$ for some $i$ and $m$. The difference
between the elements in $\oplus F_k$ and those in
$E_n$ is that the former cannot be lifted all the way
up to form-degree $n$: one meets an obstruction before,
which is $d_{k} a$ (if $a \in F_k$). By contrast, the
elements in $E_n$ can be lifted all the way up to form
degree
$n$.  This somewhat unpleasant distinction between
$\gamma$-cocycles that are non-trivial in $H(\gamma \vert
d)$ will be removed below, where we shall assign an
obstruction to the elements of $E_n$ in some appropriate
higher dimensional space.

In order to solve the Wess-Zumino consistency condition,
our task is now to determine explicitly the spaces $E_r$
and $F_r$.

\subsection{Invariant Poincar\'e lemma -- Small algebra --
Universal Algebra}
\label{smallA}

To that end, we first work out the cohomology of $d_0$ in
$E_0
\equiv H(\gamma)$.  Let $u$ be a $\gamma$-cocycle.  Without
loss of generality, we may assume that $u$  takes the form
\begin{equation} 
u = \sum P_I \omega^I
\end{equation} 
where the $\omega^I$ are polynomials in the last
undifferentiated ghosts of ghosts $C^a_{p_a}$ and
where the $P_I$ are polynomials in the  field
strength components and their derivatives, with
coefficients that involve $dx^\mu$ and $x_\mu$.  The $P_I$ are
called  ``gauge-invariant polynomials". A direct
calculation using the fact that the $d$-variation of the
last ghosts is $\gamma$-exact yields $du = \sum (dP_I)
\omega^I +
\gamma v'$.  The r.h.s of the previous equation is 
$\gamma$-exact if and only if $dP_I =0$. 

The first consequence is
that $F_0$ is isomorphic to the space of polynomials of the form,
\begin{equation}
 F_0 \simeq \{u = \sum P_I
\omega^I \text{with\ } dP_I\not = 0\ \text{and\ } P_I
\sim P_I + dQ_I\}.\label{eqpourf0}
\end{equation}
$F_0$ exhausts all the solutions of the Wess-Zumino consistency
condition in form degree $<n$ which descend trivially.

Now, if $P_I
= dR_I$ where
$R_I$ is also a gauge invariant polynomial, then $u$ is $d$-exact
modulo
$\gamma$,
$u=da +
\gamma b$, with
$\gamma a = 0$.  Conversely, if $u =da + \gamma b$ with
$\gamma a = 0$, then
$P_I$ is $d$-exact in the space of invariant polynomials. 
Thus, the class of $u$ (in $E_0$) is a non trivial cocycle
of $d_0$ if and only if $P_I$ is a non trivial cocycle of
the {\em invariant} cohomology of $d$ which we now calculate.

Since we are interested in lifts of $\gamma$-cocycles from
form-degree $k$ to form-degree $k+1$, we shall investigate
the
$d$-invariant cohomology only in form-degree strictly
smaller than the spacetime dimension $n$.  This will be
assumed throughout the remainder of this section. [In
form-degree $n$, there is clearly further invariant
cohomology since any invariant $n$-form is annihilated by
$d$, even when it cannot be written as the $d$ of an
invariant form].

\subsubsection{Invariant Poincar\'e Lemma}
In the literature, the result covering the invariant cohomology of
$d$ is known as the invariant Poincar\'e Lemma. It is contained in
the following theorem which we formulate in the presence of
antifields because we will also apply it further on the
antifield dependent solutions of the Wess-Zumino consistency
condition.
\begin{theorem}
Let ${\cal H}^k$ be the subspace of form degree $k$
of the finite dimensional algebra
${\cal H}$ of polynomials in the curvature ($p_a+1$)-forms
$H^a$, ${\cal H} = \oplus_k {\cal H}^k$.
One has 
\begin{equation}
H^{k,inv}_j(d) = 0, \; k<n, \; j>0
\end{equation}
and
\begin{equation}
H^{k,inv}_0(d) = {\cal H}^k, \; k<n. \label{invpoincare2}
\end{equation}
Thus, in particular, 
if $a=a(\chi)$ with $da=0$, antighost $a>0$ and deg $a<n$
then $a=db$ with $b=b(\chi)$.  And if $a$ has antighost number zero,
then $a = P(H^a) +db$, where $P(H^a)$ is a polynomial in the
curvature forms and $b = b([H])$.
\label{invpoincare}
\end{theorem}
This generalizes the result
established for $1$-forms in \cite{BDK1,BDK2,DVHTV1}.

\proof{Let us first show that in antighost number $>0$ the
invariant cohomology of $d$ is trivial. Let $\alpha$ be a
solution of $d\alpha=0$ with $\alpha=\alpha(\chi)$. We decompose
$\alpha$ according to the number of derivatives of the antifields:
\begin{equation}
\alpha= \alpha_0 +\ldots +\alpha_k.
\end{equation}
With $d$ written as $d={\overline
d} +{\overline{\overline d}}  $ where $\overline{d}$ acts only on
the antifields and ${\overline{\overline d}}$ on all the others
variables, equation
$d\alpha=0$ then implies $\overline{d}\alpha_k=0$. According to
Theorem {\bf{\ref{poincarebis}}} we have, $\alpha_k=
\overline{d}\beta_{k-1}(\chi)$ (the fact that all the antifields
and their derivatives appear in the $\chi$ is crucial here) and
therefore by redefining in
$\alpha$ the terms of order less than $k$, one can get rid of
$\alpha_k$, so that 
$\alpha=
\alpha_0+\ldots+\alpha_{k-1}+ d\beta_{k-1}$. In the same way, one
shows that all the $\alpha_i$ up to $\alpha_1$ can be removed from
$\alpha$ by adding the $d$-exact term $d\beta_{i-1}$. Finally,
$\alpha_0$ has to vanish because the condition
$\overline{d}\alpha_0=0$ implies
$\alpha_0=\alpha_0(x^\mu,[H])$ and we are in antighost $>0$.

We now prove that in antighost number $0$ the invariant cohomology
of $d$ is exhausted by the polynomials in the curvatures. We first
establish the result for one
$p$-form and then extend the analysis to an arbitrary system of
$p$-forms.

So let us start with one $p$-form. It can be either of odd or
even degree. Let us begin with the odd-case. Because we have
$dP_J=0$ and $\gamma P_J=0$ we can build a descent equation as
follows:
\begin{align} 
dP^k_J=0 \Rightarrow P^k_J&=da^{k-1}_0 + K\\
0&=\gamma a^{k-1}_0 + da^{k-2}_{1}\label{form321asub}\\ 
& \vdots \\
0&=\gamma a^{k-j}_{j-1} + da^{k-j-1}_{j}\label{form323asub} \\
0&=\gamma a^{k-j-1}_{j},
\end{align}
where $K$ is a constant. In the case of one $p$-form of odd degree,
the last equation of the descent and Theorem {\bf
\ref{hgamma}} tell us that $a^{k-j-1}_{j}=a^{k-p-1}_p = M_J C_p$
where
$M_J$ is a polynomial in the field strength components and their
derivatives. If we substitute this in \eqref{form323asub} we
obtain $dM_J C_P + \gamma (a^{k-j}_{j-1}-M_J C_{p-1})=0$. This
implies $dM_J=0$ (Theorem {\bf
\ref{hgamma}} again). Because the form degree of $M_J$ is strictly
less than
$k$ (the form degree of $P_J$), we make the recurrence
hypothesis that the theorem holds for $M_J$, i.e, $M_J=M_J(H)$.
$a^{k-p-1}_p$ then lifts with no ambiguity (except for $\gamma$
and $d$ exact terms which are irrelevant) up to
$a^{k-2}_1= M_J(H)C_1$. Equation \eqref{form321asub} then implies
$a^{k-1}_0 = M_J(H)B + R_J$ where $R_J$ is a polynomial in the field strength components and their
derivatives. Therefore we have  $P^k_J=M_J(H)H + dR_J$ which
proves the theorem for one $p$-form of odd degree.

In the case where the $p$-form is of even degree the proof proceed
as follows. We first construct the same descent as previously with 
$a^{k-j-1}_{j}$ this time of the form
$a^{k-j-1}_{j}=a^{k-pl-1}_{pl}=M_J C^l_p$. Just as in the the odd
case, \eqref{form323asub} implies $dM_J=0$. We thus make the
recurrence hypothesis $M_J=M_J(H) = \alpha H + \beta$ where
$\alpha$ and $\beta$ are constants. Therefore, $a^{k-pl-1}_{pl}=
(\alpha H +\beta)C^l_p$. We then note that $\alpha H C^l_p$ is
$\gamma$-exact modulo $d$. One can see this by using the
horizontality condition \eqref{formulerusse}. Indeed, $\tilde
\gamma (\alpha \tB^{l+1})=(l+1)\alpha H \tB^{l}$ which implies
$(l+1)\alpha H C^l_p=\gamma [\alpha
\tB^{l+1}]^{p+1}_{pl-1}+d[\alpha
\tB^{l+1}]^{p}_{pl}.$
Thus we may suppose that $a^{k-pl-1}_{pl}=a^0_{pl}=\beta
C^l_p$. Let us now show that if $l>1$ then $\beta=0$. In that
case, the bottom $a^0_{pl}$ can be lifted without any ambiguity up
to
$a^{p-1}_{p(l-1)+1}= [\beta \tB^l]^{p-1}_{p(l-1)+1}$ (using Eq. 
\eqref{formulerusse} again). The next equation of the descent then
yields:
$a^{p}_{p(l-1)}= [\beta \tB^l]^{p}_{p(l-1)} + R_J C^{l-1}_p$
where $R_J$ is a polynomial in the field strength components and
their derivatives. Substituting this into $\gamma
a^{p+1}_{p(l-1)-1}+d a^{p}_{p(l-1)}=0$ we get $\beta l H C^{l-1}_p
+ dR_J C^{l-1}_p + \gamma ({a'}^{p+1}_{p(l-1)-1})=0 \Rightarrow 
\beta l H+ dR_J=0$. Thus, using our recurrence hypothesis, we see
that for
$l>1$ we necessarily have $\beta=0$. If $l=1$, the above
obstruction is not present. The bottom $\beta C_p$ then yields
$P^k_J=k +\beta H+ dR_J$ which proves the theorem for the
even-case.

Let us finally prove the theorem for an arbitrary system of
$p$-forms. We label one of the $p$-form with $A$ and the rest
with $B$ and decompose $P_J$ according to the number of
derivatives of the field strengths of the $p$-form labeled by
$A$:
$P_J=P_0+\ldots +P_k$. Because
$dP_J=0$  we have
$d_A P_k=0$, where $d_A$ only acts on the fields of the sector
labeled by $A$. Using the theorem in the case of a single
$p$-form we get $P_k= d_A R_{k-1}+ V$ where $V$ is only present
for $k=0$ because it only depends on the sector $A$ through $H_A$
and on the sector
$B$ through the field strengths and their derivatives. Thus,
except when $k=0$, $P_k$ can be removed from $P_J$ by subtracting
the coboundary $dR_{k-1} \Rightarrow P_J= V$ up to trivial terms.
We now expand
$V$ in powers of
$H_A$:
$V=\sum (H_A)^k v_k$. The condition $dV=0$ then implies $dv_k=0$.
By induction on the number of $p$-forms occurring in $v_k$ we obtain
the desired result.}
If
the local forms are not taken  to be explicitly $x$-dependent,
Equation (\ref{invpoincare2}) must be replaced by
\begin{equation}
H^{k,inv}_0(d) = (\Lambda \otimes {\cal H})^k.\label{inp2}
\end{equation} 
\subsubsection{Small Algebra}
From now on, we restrict our attention to the algebra of
$x$-dependent spacetime forms ${\cal P}$. In Section
{\bf{\ref{respmoins}}} we comment on how the results are
affected when the analysis is pursued in ${\cal P}_{-}$.

Theorem {\bf{\ref{invpoincare}}} implies, according to the general
analysis of the descent equation given above, that the only
bottoms $u$ ($\gamma u = 0$) that can be lifted at least once can
be expressed in terms of exterior products of the curvature
forms $H^a$ and the last ghosts of ghosts (up to trivial
redefinitions). Out of the infinitely many generators of
$H(\gamma)$, only $H^a$ and
$C^a_{p_a}$ survive in $E_1$.

Because the objects that survive the first step in the lift
can be expressed in terms of forms, it is convenient to
introduce the so-called ``small algebra"
${\cal A}$ generated in the exterior product by the
exterior forms, $B^a$, $H^a$,
$C^a_{p_a-k}$ and $dC^a_{p_a-k}$ ($k=0, ..., p_a-1$).
This algebra is stable under
$\gamma$ and $d$.  Denoting by $E^{small}_0$ the
cohomology of
$\gamma$ in the small algebra, one finds
\begin{equation} E_0^{small} \equiv H(\gamma, {\cal A}) \simeq {\cal B}
\end{equation} where ${\cal B}$ is the subalgebra of ${\cal A}$
generated by the curvatures $H^a$ and the last ghosts of
ghosts $C^a_{p_a}$.

One also defines $E_1^{small}$ as $H(d_0^{small}, E_0^{small})$,
where
$d_0^{small}$ is the restriction of $d_0$ to $E_0^{small}$.
Because
$d H^a = 0$ and $dC^a_{p_a} = \gamma$(something), the
restriction
$d_0^{small}$ identically vanishes. Thus
\begin{equation} E_1^{small}
\simeq E_0^{small} \simeq {\cal B}.
\end{equation}

What is the relationship between $E_1^{small}$ and $E_1$? 
These two spaces are in fact isomorphic,
\begin{equation}
 E_1 \simeq E_1^{small}.
\label{E1E1small}
\end{equation}
Indeed, let $q$ be the map from $E_1^{small}$ to $E_1$
that assigns to a cohomological class in $E_1^{small}$ its
cohomological class in $E_1$ ($a \in  E_1^{small} \simeq
{\cal B}$ fulfills $\gamma a = 0$ and $d_0 a =0$ and so
defines of course an element of $E_1$). It follows from the
above theorem that the map $q$ is surjective since any
class in $E_1$ possesses a representative in the small
algebra.  The map $q$ is also injective because there is no
non trivial class in $E_1^{small}$ that becomes trivial in
$E_1$. If the small-algebra $\gamma$-cocycle
$r = \sum P_I \omega^I$ with $P_I, \omega^I \in {\cal B}$
can be written as $r = du + \gamma t$ with $u$ and $v$ 
in the big algebra and $u$ a $\gamma$-cocycle, then
$r$ is actually zero.  Indeed, if
$\gamma u=0$ we have, $u=Q_I \omega^I + \gamma m$. This implies,
$P_I\omega^I=dQ_I \omega^I + \gamma t'$ and thus $P_I =d Q_I$ which
is impossible according to Theorem {\bf{\ref{invpoincare}}}.

In fact, it is easy to see that we also have $E_k \simeq
E^{small}_k$ for each $k>1$. Indeed, suppose the result holds for
$E^{small}_{k-1}$ and $E_{k-1}$ and let $q$ be the bijective map
between these two spaces. If $a \in E^{small}_{k-1}$ than there
exists $c_1, \ldots, c_{k-1} \in {\cal A}$ such that $da+\gamma
c_1=0, \ldots, dc_{k-2}+dc_{k-1}=0$ and we have by definition,
\begin{equation}
d^{small}_{k-1} [a]_{E^{small}_{k-1}} =
[dc_{k-1}]_{E^{small}_{k-1}}.
\label{mapq1}
\end{equation}
Furthermore, because $E^{small}_{k-1} \simeq E_{k-1}$, $a$ also
represents a non trivial class of $E_{k-1}$ and we may choose as
its successive ${k-1}$ lifts the previous $c_1,\ldots, c_{k-1}$. So
by definition,
\begin{equation}
d_{k-1}[a]_{E_k}=[dc_{k-1}]_{E_k}.
\label{mapq2}
\end{equation}
Equation \eqref{mapq1} and \eqref{mapq2} imply that the
differentials $d_{k-1}^{small}$ and $d_{k-1}$ are mapped on each
other through the isomorphism, $qd^{small}_{k-1}=d_{k-1}q$.
Therefore $E^{small}_k = H(d_{k-1}^{small}, E^{small}_{k-1})
\simeq H(d_{k-1},E_{k-1}) = E_{k-1}$ which proves the result.

By virtue of this result, one can equivalently compute the spaces
$E^{small}_k$ instead of the spaces $E_k$ in order to obtain the
elements of $H(\gamma)$ which can be lifted $k$ times and which are
not
$d_k$-exact. 

What about the relationship between $F^{small}_k$ and
$F_k$ for $k>0$? Suppose $a \in F^{small}_k$. This means that even
when taking into account the ambiguities in the definitions of
$c_1,\ldots, c_k$ in the {\em small algebra} one does not have
$dc'_k + \gamma c_{k+1}=0$. One may ask whether or not the
obstruction to the lift of $c_k$ can be removed when the
ambiguities are not restricted to the small algebra? The answer is
negative for the following reason. The ambiguities in any of the
$c_i\ (i<k)$ have to be lifted at least once; so up to trivial
terms they can be supposed to be in the small algebra as well as
their successive lifts. This implies that the ambiguity in $c_k$ is
$c_k \rightarrow c_k + m + u_0$ with $m$  in 
${\cal A}$ and
$u_0$ a $\gamma$-closed term. $c_k$ can be lifted if it is
possible to adjust $m$ and $u_0$ so that $d(c_k + m) + d u_0
=\gamma r$. However, the same argument used in the proof of the
injectivity of the map $q$ from $E^{small}_1$ to $E_1$ shows
that this is impossible.

We can summarize the above discussion in the following theorem:
\begin{theorem} There is  no loss of generality in
investigating in the small algebra the solutions of the Wess-Zumino
consistency condition that descend non trivially.\label{smallbig}
\end{theorem}

\subsubsection{Universal Algebra}
The small algebra
${\cal A}$ involves only exterior forms, exterior products
and exterior derivatives.   It does ``remember" the
spacetime dimension since its generators are not free: any
product of generators with form-degree exceeding the
spacetime dimension vanishes.

It is useful to drop this relation and to work in the
algebra freely generated by the potentials, the 
ghosts and their exterior derivatives with the sole
condition that these commute or anti-commute (graded
commutative algebra) but without imposing any restriction
on the maximally allowed form degree
\cite{DVTV1,BonoraCoRaRiSta1}. This algebra is called the universal
algebra and denoted by ${\cal U}$.  In this algebra, the
cohomology of $d$ is trivial in all form-degrees and the
previous theorems on the invariant cohomology of $d$ are
also valid in form-degree
$\geq n$. Furthermore, one can sharpen the condition for a
cocycle in
$H(\gamma)$ to be non trivial in $H(\gamma \vert d)$.

\begin{theorem} A necessary and sufficient condition for $X
\in H(\gamma)$ to be non-trivial in $H(\gamma \vert d)$ is
that there exists
$r$ such that $d_rX \not=0$.  That is, the lift of $X$ must
be obstructed at some stage. (For the equation $d_rX
\not=0$ to make sense, $d_i X$ must vanish for $i<r$. 
Here also we denote by the same letter $X \in E_0$ and its
representative in
$E_r$).\label{suffnec}
\end{theorem}

\proof{The decomposition of $E_n$ is now non-trivial since
$da$ does not necessarily vanish even when $a$ is a
$n$-form. Thus,
$d_n$ is not necessarily zero and the procedure of lifting
can be pursued above form-degree $n$. Suppose that one 
encounters no obstruction in the lifting of $X$. That
is, one can go all the way up to ghost number zero, the
last two equations being $dc_k + \gamma b = 0$ (with $b$ of
ghost number zero) and
$db = 0$ (so $b$ lifts to zero).  Then, one can write $b =
d m$ since the cohomology of $d$ is trivial in any
form-degree in the universal algebra ${\cal U}$ (except for
the constants, which cannot arise here since $b$ involves
the fields). The triviality of the top-form $b$ implies the
triviality in $H(\gamma \vert d)$ of all the elements below
it.  Thus, a necessary condition for the bottom to be non
trivial in $H(\gamma \vert d)$ is that one meets an
obstruction in the lift at some stage. The condition is
also clearly sufficient.}
One can summarize our results as follows:

\begin{theorem} (Generalized ``transgression" lemma) Let $X
\in E_0$ be a non-trivial element of
$H(\gamma \vert d)$.  Then there exists an integer $r$ such
that $d_i X = 0$, $i<r$ and $d_rX = Y \not=0$.  The element
$Y$ is defined through the chain $dX+ \gamma c_1 = 0$, ...,
$dc_{r-1} + \gamma c_r=0, dc_r + \gamma c_{r+1}= Y$, where
the elements $c_i \in {\cal U}$ ($i=1, r+1$) are chosen so
as to go all the way up to $c_{r+1}$. One has $\gamma Y=0$
and $Y$ should properly viewed as an element of
$E_r$ (reflecting the ambiguities in the lift). One calls
the obstruction $Y$ to a further lift of $X$ the
(generalized) ``transgression" of $X$.  The element $X$ and
its transgression have opposite statistics. \label{gentrlm}
\end{theorem} This is the direct generalization of the
analysis of
\cite{DVTV1} to the case of $p$-forms.  ``Primitive
elements" of $E_0$ are those that have form-degree zero and
for which the transgression has ghost number zero, i.e.,
they are the elements that can be lifted all the way up to
ghost number zero (``that can be transgressed").  We refer
to \cite{DVTV1,GreubHalVan1,Cartan} for more background
information applicable to the Yang-Mills case.

Because the space $E_n$ and the next ones can be further
decomposed in the universal algebra,
\begin{equation} E_n \simeq \text{Im}\, d_{n} \oplus E_{n+1} \oplus F_n,
\; E_{n+1} \simeq \text{etc}
\end{equation} where the decomposition for a given
$\gamma$-cocycle ultimately ends at form-degree equal to
the ghost number, one has
\begin{equation} E_0 \simeq \oplus_{k=0}^\infty F_k \oplus_{k=0}^\infty
\text{Im}\, d_{k}.
\end{equation}

\subsection{Results}
\label{results}
We now compute the spaces $E_k$ for the system of free
$p$-forms.

Let $0<p_1<p_2< \dots <p_M$ be the form degrees of the gauge
potentials $B^a$.  We denote by $B_1^{a_1}$ the forms of
degree
$p_1$, $B_2^{a_2}$ the forms of degree
$p_2$ etc.

The first non-vanishing differential (in $E_0^{small}$) is
$d_{p_1} $ so that $E_0^{small} = E_1 = E_2 = ... =
E_{p_1}$. Any bottom in
$E_0^{small}$ can be lifted at least $p_1$ times.  In
$E_1$, the differential $d_{p_1}$ acts as follows,
\begin{equation} d_{p_1} C^{a_1}_{p_1} = H_1^{a_1}, \; d_{p_1}
H_1^{a_1} = 0,
\end{equation} in the sector of the forms of degree $p_1$ and,
\begin{equation} d_{p_1} C^{a_k}_{p_k} = 0, \; d_{p_1} H_k^{a_k}
= 0, \; k >1
\end{equation} in the other sectors.  The form of the differential
$d_{p_1}$ makes explicit the contractible part of
$(E_{p_1}, d_{p_1})$. The variables $C^{a_1}_{p_1}$ and
$H_1^{a_1}$  are removed from the cohomology, so that
$E_{p_1+1}$ is isomorphic to the algebra generated by the
curvatures $H_k^{a_k}$ of form-degree
$>p_1 +1$ and the last ghosts of ghosts of ghost number
$>p_1$.

A subspace $F_{p_1}$ complementary to $\text{Ker} \, d_{p_1}$ is
easily constructed.  In fact, a monomial in
$C^{a_1}_{p_1}$ and $H_1^{a_1}$ is defined by a tensor
$f_{a_1 \dots a_k b_1 \dots b_m}$ which is symmetric
(respectively antisymmetric) in $a_1, \dots, a_k$ and
antisymmetric (respectively symmetric) in $b_1, \dots, b_m$
if the last ghosts of ghosts are commuting (respectively
anticommuting).  For definiteness, suppose that the $H_1^{a_1}$
are anticommuting and the $C^{a_1}_{p_1}$ commuting. The
irreducible components of $f_{a_1 \dots a_k b_1 \dots b_m}$ are
then of the two following Young-symmetry types,

\begin{picture}(0,105)(0,0)
\multiframe(0,50)(15.5,0){5}(15,15){$a_1$}...{$a_k$}
\put(85,55){$\bigotimes$}
\multiframe(105,20)(0,15.5){5}(15,15){$b_m$}...{$b_1$}
\put(135,55){$\simeq$}
\multiframe(170.5,82)(15.5,0){5}(15,15){$a_1$}...{$a_k$}
\multiframe(155,20)(0,15.5){5}(15,15){$b_m$}...{$b_1$}
\put(255,55){$\bigoplus$}
\multiframe(290.5,82.5)(15.5,0){4}(15,15){$a_2$}..{$a_k$}
\multiframe(275,5)(0,15.5){6}(15,15){$a_1$}{$b_m$}...{$b_1$}
\put(215,45){1)}
\put(330,45){2)}
\end{picture}

\noindent so that the polynomial in $C^{a_1}_{p_1}$ and
$H_1^{a_1}$ can be written as,
\begin{align}
f_{a_1 \dots a_k b_1 \dots b_m}C^{a_1}_{p_1} \ldots C^{a_k}_{p_1}
H_1^{b_1}\ldots H_1^{b_m} =& f^{(1)}_{a_1 \dots a_k b_1 \dots
b_m}C^{a_1}_{p_1} \ldots C^{a_k}_{p_1} H_1^{b_1}\ldots H_1^{b_m}
\label{decomp2} \\ \nonumber  &+ f^{(2)}_{a_1 \dots
a_k b_1
\dots b_m}C^{a_1}_{p_1}
\ldots C^{a_k}_{p_1} H_1^{b_1}\ldots H_1^{b_m}.
\end{align}
The first term on the r.h.s \eqref{decomp2} is annihilated by
$d_{p_1}$ while the second term is not and therefore defines an
element of $F_{p_1}$.
The space
$F_{p_1}$ can be taken to be the space generated by the
monomials of this symmetry type (not annihilated by
$d_{p_1}$), tensored by the algebra generated by the
curvatures and last ghosts of ghosts of higher degree.
Together with their successive lifts, the elements in
$F_{p_1}$ provide all the non-trivial solutions of the
Wess-Zumino consistency condition which are involved in
descents whose bottoms can be lifted exactly $p_1$ times. 

In the
case where the $H_1^{a_1}$
are commuting and the $C^{a_1}_{p_1}$ anticommuting one simply
exchanges the r\^oles of $a_i$ and $b_j$ in the previous
discussion.

Similarly, one finds that the next non-vanishing
differential is
$d_{p_2}$.  The generators $C^{a_2}_{p_2}$ and
$H_2^{a_2}$ drop from the cohomology of $d_{p_2}$ while
those of higher degree remain.  A space $F_{p_2}$ can be
constructed along exactly the same lines as the space
$F_{p_1}$ above and characterizes the solutions of the
Wess-Zumino consistency condition involved in descents
whose bottoms can be lifted exactly $p_2$ times.

More generally, the non-vanishing differentials are
$d_{p_k}$. They are defined (in $E_{p_k}$, which is
isomorphic to the algebra generated by the curvatures of
form-degree $>p_{k-1}+1$ and the last ghosts of ghosts of
ghost number $>p_{k-1}$) through
\begin{equation} d_{p_k}C^{a_k}_{p_k}=H_k^{a_k}, \; d_{p_k}
H_k^{a_k} = 0
\end{equation} and
\begin{equation} d_{p_k} C^{a_j}_{p_j} = 0, \; d_{p_k}H_j^{a_j}
= 0,
\; j>k.
\end{equation} The generators $C^{a_k}_{p_k}$ and $H_k^{a_k}$
disappear in cohomology.  The subspace $F_{p_k}$ is again
easily constructed along the previous lines. Together with
their successive lifts, the elements in $F_{p_k}$ provide
all the non-trivial solutions of the Wess-Zumino
consistency condition which are involved in descents whose
bottoms can be lifted exactly $p_k$ times. 

Along with
$F_0$ (see
\eqref{eqpourf0}), the
$F_{p_i}$'s constructed in this section provide in form degree $<n$
all the non-trivial $\gamma$-cocycle remain non-trivial as
elements of $H(\gamma \vert d)$. Together with their successive
lifts they form a complete set of representatives of $H(\gamma
\vert d)$ involved in non-trivial descents.
\vspace{.3cm}

\noindent{\underline{Example:}} Let us illustrate this discussion
in the case of the simple model with one free $1$-form $A$ and one
free
$2$-form $B$ considered in Section {\bf\ref{exlift}}.  The
space $E_0^{small}$ is isomorphic to the space of
polynomials in the curvature-forms $F$, $H$ and the last
ghosts of ghosts $A_1$, $B_2$.  The differential
$d_0^{small}$ vanishes so $E_1 \simeq E_0^{small}$.  One next finds
 $d_1 A_1 = F$,
$d_1 F = 0$, $d_1 B_2 = 0$ and $d_1 H=0$.  The space
$E_2$ is isomorphic to the space of polynomials in
$B_2$ and $H$.  One may take for
$F_1$ the space of polynomials linear in $A_1$. 
These can be lifted exactly once, their lifts being linear
in
$A$ and $A_1$,
\begin{equation} a \in F_1 \Leftrightarrow a = A_1 \sum
(B_2)^l F^k H^m \; \; (m=0 \hbox{ or } 1).
\end{equation} Then, one gets
\begin{equation} da + \gamma b = 0,
\end{equation} with
\begin{equation} b = \sum \big( A (B_2)^l F^k H^m + l
A_1 B_1 (B_2)^{l-1} F^k H^m  \big).
\end{equation} They cannot be further lifted since
the obstruction $d_1 a =\sum (B_2)^l$ $F^{k+1} H^m$ does
not vanish.  The above $a$'s and
$b$'s are the most general solutions of the Wess-Zumino
consistency condition involved in descents of length $1$.

The differential $d_2$ in $E_2$ is given by
$d_2 B_2 = H$, $d_2 H = 0$.  Because $H^2=0$, one may
take for
$F_2$ the space of polynomials in $B_2$ only.  For
those, the descent reads,
\begin{eqnarray}
\alpha = (B_2)^l &,& \gamma \alpha = 0,
\nonumber \\
\beta = l B_1 (B_2)^{l-1} &,& d \alpha +\gamma
\beta = 0,
\nonumber \\
\lambda = l B (B_2)^{l-1} + \frac{l(l-1)}{2}
(B_2)^{l-2} (B_1)^2 &,& d \beta+\gamma \lambda 
= 0.
\end{eqnarray} The elements of the form $\alpha$, $\beta$
or $\lambda$ are the most general solutions of the
Wess-Zumino consistency condition involved in descents of
length $2$.  With the solutions involved in descents of
length $1$ and those that do not descend (i.e., which are
strictly annihilated by $\gamma$), they exhaust all the
(antifield-independent) solutions of the Wess-Zumino
consistency condition.
\vspace{.5cm}

A straightforward consequence of our discussion is the
following theorem, which will prove useful in the analysis of
antifield dependent solutions.
\begin{theorem}
\label{triviality} Let $\omega$ be a $\gamma$-cocycle of the form
\begin{equation}
\omega = \alpha(H_s^{a_s} \; C^{a_s}_{p_s})
\beta(C^{a_k}_{p_k}, H_k^{a_k}), \; k>s
\end{equation} where $\alpha$ vanishes if $H_s^{a_s}$ and 
$C^{a_s}_{p_s}$ are set equal to zero (no constant term) and which
fulfills,
\begin{equation} d_{p_s} \alpha = 0.
\end{equation} (i.e. the first possible obstruction in the lift of
$\omega$ is absent). Then, $\omega$ is trivial in $H(\gamma
\vert d)$.
\end{theorem} 
\proof{
The proof is direct: one has $\alpha =
d_{p_s} \mu$ for some
$\mu(H_s^{a_s}, C^{a_s}_{p_s})$ since $d_{p_s}$ is
acyclic in the space of the $\alpha(H_s^{a_s} C^{a_s}_{p_s})$ 
with no constant term. Thus $\omega$ is
$d_{p_s}$-exact, $\omega= d_{p_s} (\mu \beta)$; so $\omega$ is
the first obstruction to the further lift of $\mu \beta$
and as such, is trivial.}
The theorem applies in particular when $\alpha$ is an
arbitrary polynomial of strictly positive degree in the
curvatures
$H_s^{a_s}$.

\subsection{Results in ${\cal P}_{-}$}\label{respmoins}
In the algebra of $x$-independent spacetime forms ${\cal P}_{-}$,
the analysis proceeds similarly as in ${\cal P}$. 

Because of \eqref{inp2}, we now define the small algebra ${\cal
A}_{-}$ as the algebra generated by the exterior products of the
forms $H^a$, $B^a$, $C^a_{p_a-k}$, $dC^a_{p_a-k}$ ($k=0,\ldots
p_a-1$) but also
$dx^\mu$ since we now need the constant forms. By exactly the
same arguments used in ${\cal P}$, one shows that the
isomorphisms
$E_k^{small}\simeq E_k,\; k\geq 1$ still hold in ${\cal P}_{-}$
and that one cannot remove obstructions to lifts by going to
the ``big algebra''. Therefore,  Theorem {\bf{\ref{smallbig}}}
remains valid in ${\cal P}_{-}$.

The universal algebra is then defined as the algebra freely 
generated by the exterior products of the forms $H^a$, $B^a$,
$C^a_{p_a-k}$, $dC^a_{p_a-k}$ ($k=0,\ldots p_a-1$) and
$dx^\mu$ without any restriction on the maximally  allowed form
degree. In that algebra the cohomology of $d$ and the invariant
cohomology of $d$ are respectively given by the constant forms
and
\eqref{inp2} (in all form-degrees including form-degrees $\geq n$).
Theorems {\bf{\ref{suffnec}}} and {\bf{\ref{gentrlm}}} are then
proved as they are in ${\cal P}$.

Therefore, the calculation of the spaces $E_k^{small}$ in ${\cal
P}_{-}$ can proceed as in Section {\bf{\ref{results}}}. The only
difference is that the $f_{a_1\ldots a_k b_1\ldots b_m}$ in
\eqref{decomp2} are now constant forms instead of just constants.

Note that even in ${\cal P}_{-}$, the constant
forms can be eliminated by requiring Lorentz invariance.

\subsection{Counterterms and anomalies}
\label{cafree}
In this section we summarize the above results by giving
explicitly the anti\-field-inde\-pen\-dent counterterms and
anomalies:
$H^{(n,0)}(\gamma \vert d)$ and $H^{(n,1)}(\gamma \vert
d)$. These can be of two types: (i) those that descend
trivially (``type A") and can be assumed to be strictly
annihilated by $\gamma$; they are described by $H(\gamma)$ up
to trivial terms; and (ii) those that lead to a
non-trivial descent (``type B") and can be assumed to be
in the small algebra modulo solutions of the previous type.
For small ghost number, it turns out to be more convenient
to determine the solutions of ``type B" directly from the
obstructions sitting above them rather than from the
bottom. That this procedure, which works in the universal
algebra, yields all the solutions, is guaranteed by the analysis
of Section {\bf \ref{smallA}}.

The following results apply equally to ${\cal P}$ and ${\cal
P}_{-}$ since for the counterterms and anomalies we impose
Lorentz-invariance which
eliminates the constant forms.

\subsubsection{Counterterms and anomalies of type A}

The counterterms that lead to a trivial descent involve in
general the individual components of the gauge-invariant
field strengths and their derivatives and cannot generically be
expressed as exterior products of the forms $F$ or $H$. They are
the gauge-invariant polynomials introduced above and read
explicitly,
\begin{equation} a = a([H^a]) d^n x.\label{71}
\end{equation} 
In order to be non-trivial in $H(\gamma \vert d)$ the above
cocycles must satisfy the condition 
$a \not= db$ which is equivalent to the
condition that the variational derivatives of $a$ with
respect to the fields do not identically vanish.
We have assumed that the spacetime forms
$dx^\mu$ occur only through the product $dx^0 dx^1 \cdots
dx^{n-1}
\equiv d^n x$ as this is required by Lorentz-invariance.

The anomalies that lead to a trivial descent are sums of
terms of the form
$a = P \, C \, d^n x$ where $P$ is a gauge-invariant
polynomial and
$C$ is a last ghost of ghost with ghost number one.  These
anomalies exist only for a theory with $1$-forms. 
One has explicitly
\begin{equation} a = P_A([H^a]) C^A_1 \label{454}
\end{equation} where $A$ runs over the $1$-forms. $a$ will be
trivial if
and only if $P=dR$ where $R$ is an invariant polynomial or if
$P_A=P_A(H^a)$ with
$ P_A H^A=0$. Indeed, suppose that $a$ is trivial, i.e., $P_A C^A
= \gamma c + de$. By acting
with $\gamma$ on this equation we see that $e$ satisfies $\gamma e
+ dm=0$. We can thus decompose $e$ as the sum of an element of
$H(\gamma \vert d)$ which descends trivially and a term $v$ in the
small algebra which is the lift of a
$\gamma$-cocycle: $e=R_A([H^a]) C^A_1 + v$. This implies $de=d R_A
C^A_1 + Q_A(H^a)C^A_1 + \gamma u$ and thus $P_A([H^a])=dR_A([H^a])
+ Q_A(H^a)$ where $Q_A H^A=0$ (see Eq. \eqref{obstruction2} in the
section on Anomalies of type B).

The existence of such anomalies - which cannot generically 
be expressed as exterior products of curvatures and ghosts
- was pointed out in
\cite{DixonRa1} for Yang-Mills gauge models with $U(1)$
factors.

\subsubsection{Counterterms of type B} The solutions that lead
to a non trivial descent can be assumed to live in the small
algebra, i.e., can be expressed in terms of exterior
products of the fields, the ghosts (which are all exterior
forms) and their exterior derivatives (modulo solutions of
type A). If $a$ is a non-trivial solution of the
Wess-Zumino consistency condition with ghost number zero,
then $da
\not= 0$ (in the universal algebra). Since
$a$ has ghost number zero, it  is the top of the descent
and $da$ is the obstruction to a further lift. Because $da$
is a
$\gamma$-cocycle, it is a gauge-invariant polynomial.  It
must, in addition, be $d$-closed but not $d$-exact in the
space of gauge-invariant polynomials since otherwise, $a$
could be redefined to be of type A.  Therefore, $da$ is an
element of the invariant cohomology of $d$ and it will be
easier to determine
$a$ directly from the obstruction
$da$ rather than from the bottom of the descent because one
knows  the invariant cohomology of $d$.

Thus we may assume that,
\begin{equation} da = P(H) = dQ(H,B)
\end{equation} where $Q$ is linear in the forms $B^a$,
and so up to trivial terms,
\begin{equation} a = R_a(H^b) B^a.\label{79}
\end{equation} Note that up to a $d$-exact term, one may in fact
assume that
$a$ involves only the potentials
$B^a$ of the curvatures of smaller form-degree present in
$P$.  
These are the familiar
Chern-Simons terms, which exist provided one can match the
spacetime dimension $n$ with a polynomial in the curvatures
$H^a$ and the forms $B^a$, linear in $B^a$.

The whole descent associated with $a$ is generated through
the "Russian formula" \cite{Stora1,Stora2,Zumino1,Baulieu1}
\begin{eqnarray} P &=& \tilde{\gamma} a(H, \tilde{B})
\label{descentCS} \\
\tilde{\gamma} &=& d + \gamma \\
\tilde{B}^a &=& B^a + C^a_1 + \cdots
C^a_{p_a},
\end{eqnarray} which follows from the ``horizontability
condition" (Eq. \eqref{formulerusse}),
\begin{equation}
\tilde{\gamma} \tilde{B}^a = H^a.
\end{equation} By expanding (\ref{descentCS}) according to the ghost
number, one gets the whole tower of descent equations.  The
bottom takes the form
$R_a(H^b) C^a_{p_a}$ and is linear in the last ghosts of
ghosts associated with the forms of smaller form degree
involved in $P$. That the bottoms should take this form
might have been anticipated since these are the only
bottoms with the right degrees that can be lifted all the
way to form-degree $n$.  The non-triviality of the bottom
implies also the non-triviality of the whole tower.

It is rather obvious that the Chern-Simons terms are
solutions of the Wess-Zumino consistency condition.  The
main result here is that these are the only solutions that
descend non trivially (up to solutions of type A).

\subsubsection{Anomalies of type B}

The anomalies $a$ of type B can themselves be of two types.
They can arise from an obstruction that lives one dimension
higher or from an obstruction that lives two dimensions
higher.  In the first case, the obstruction
$da$ has form degree $n+1$ and ghost number $1$.  This occurs
only for models with $1$-forms since other models have no
$\gamma$-cohomology in ghost number one. In the other case, the
anomaly can be lifted once,
$da + \gamma b =0$.  The obstruction
$db$ to a further lift is then a $(n+2)$-form of ghost
number
$0$.

In the first case, the obstruction $da$ reads
\begin{equation} da + \gamma(\hbox{something}) = P_A(H) C^A_1
\label{obstruction1}
\end{equation} where $A$ runs over the $1$-forms.  The right-hand
side of (\ref{obstruction1}) is necessarily the $d_1$ of something.
Indeed, it cannot be the $d_k$ ($k>1$) of something, say
$m$, since this would make $m$ trivial: the first
obstruction to the lift of $m$ would have to vanish and $m$
involves explicitly the variables of the $1$-form sector
(see Theorem {\bf \ref{triviality}} above). This implies
\begin{equation} P_A(H) C^A_1 = C_{AB}(H) H^A C^B_1, \; \;
C_{AB}(H) = - C_{BA}(H)
\label{obstruction2}
\end{equation} so that
$P_A(H) C^A_1 = d_1(\frac{1}{2}
C_{AB}(H)C^A_1 C^A_1$. One thus needs at least
two $1$-forms to construct such solutions. If
$C_{AB}(H)$ involves the curvatures $H^A$ of the $1$-forms,
it must be such that (\ref{obstruction2}) is not zero.  
The anomaly following from (\ref{obstruction1}) is
\begin{equation} a = C_{AB}(H) B^A C^A_1\label{722}
\end{equation} and the associated descent is generated through
\begin{equation} C_{AB}(H) H^A C^A_1 = \tilde{\gamma} (\frac{1}{2}
C_{AB}(H)
\tilde{B^A} \tilde{B^B})
\end{equation}

In the second case, the obstruction
$P\in H^{inv}(d)$ is a polynomial in $H^a$ of form-degree
$n+2$, which can be written $P = dQ$ with $Q$  linear in
the potentials associated to the curvatures of lowest
degree present in $P$.  The solution $a$ and the descent
are obtained from the Russian formula (\ref{descentCS}),
exactly as for the counterterms,
\begin{equation}
a = R_a(H^b) C^a_1.\label{724}
\end{equation}

They are linear in the
ghosts and exist only for models with forms of degree $>1$
which are the only ones that can occur in $P$ since
otherwise $a$ is either trivial or of type A. Indeed, if
variables from the
$1$-form sector are present in
$P$, then $P = d_1 a$ (if $P$ is non trivial) and the
descent has only two steps.  But this really means that $a$ is the
bottom of the descent and is actually of type A.

\subsection{Conclusions}
In this section, we have derived the general solution of the
antifield-independ\-ent Wess-Zumino consistency condition
for models involving $p$-forms.  We have justified in
particular why one can assume that the solutions can be
expressed in terms of exterior products of the fields, the
ghosts (which are all exterior forms) and their exterior
derivatives, {\em when these solutions appear in non trivial
descents}.  This is not obvious to begin with since there
are solutions that are not expressible in terms of forms
(those that descend trivially) and justify the usual
calculations made for determining the anomalies.  Once one
knows that the solutions involved in non trivial descents
can be expressed in terms of forms (up to solutions that
descend trivially), one can straightforwardly determine
their explicit form in ghost numbers zero and one.  This has been
done in the last section where all counterterms and
anomalies have been classified.  The
counterterms are either strictly gauge
invariant and given by (\ref{71}) or of the Chern-Simons
type (when available) and given by (\ref{79}). 

The anomalies are also either strictly annihilated by
$\gamma$, or lead to a non-trivial descent. The first type
generalizes the anomalies of Dixon and Ramon Medrano
\cite{DixonRa1} and are given by \eqref{454}. The more familiar
anomalies with a non trivial descent are listed in Eqs.
(\ref{722}) and (\ref{724}).

The method applies also
to other values of the ghost number, which are relevant in
the analysis of the antifield-dependent cohomology.

This result is also valid for more general lagrangians than
\eqref{Lagrangian}. Indeed, if one adds to the lagrangian of free
$p$-forms interactions which are gauge invariant (\eqref{71}
or \eqref{79}), then the gauge transformations of the resulting
theory are identical to those of the free theory. Therefore,
the definition of the longitudinal exterior derivative $\gamma$ is
unchanged and the results are not modified.

As we will
show in the next section, the natural appearance of exterior forms
holds also for the characteristic cohomology: all higher order
conservation laws are naturally expressed in terms of exterior
products of field strengths and duals to the field strengths.  It
is this property that makes the gauge symmetry-deforming
consistent interactions for $p$-form gauge fields expressible also
in terms of exterior forms and exterior products.

\newpage

\section{Antifield dependent solutions}
\label{adssec}
We now turn our attention to the antifield dependent solutions of
the Wess-Zumino consistency condition.

\subsection{Preliminary results}
\label{prelimres}

Any antifield dependent solution of
equation \eqref{WZ2} may be decomposed according to the
antighost number $a_g^n=a_{g,0}^n+
a_{g,1}^n+\ldots+ a_{g,q}^n$ with $q\not =0$.\footnote{In this
section we limit our attention to solutions of the Wess-Zumino
consistency condition in form degree $n$. In particular, this
applies to consistent interactions, counter\-terms, anomalies and
solutions related to the analysis of the gauge invariance of
conserved currents. Other values of the form-degree are treated
along the same lines.} When this is done, the Wess-Zumino
consistency condition,
\begin{equation}
s a^k_g + db^{k-1}_{g+1}=0,
\end{equation}
splits as,
\begin{align}
\gamma a_0 + \delta a_1 +db_0 &=0, \label{highanti0}\\
\gamma a_1 + \delta a_2 +db_1 &=0, \\
&\vdots \nonumber \\
\gamma a_{q-1} + \delta a_q +db_{q-1} &=0,\label{highanti2}\\
\gamma a_q + d b_q&=0,\label{highanti}
\end{align}
where we have dropped the indices labeling the ghost numbers and
form degrees (which are fixed) and we have performed the
decomposition
$b^{k-1}_{g+1}=b^{k-1}_{g+1,0}+\ldots +b^{k-1}_{g+1,q}$ (because
of the Algebraic Poincar\'e Lemma, we may assume that up to a
$d$-exact term, the expansion of
$b^{k-1}_{g+1}$ according to the antighost number stops at order
$q$). 

These equations resemble the descent equations used in the
analysis of antifield independent BRST cocycles.
The bottom equation now defines an element of $H(\gamma \vert d)$
which we denote ${\cal E}_0$. In order to `lift' this cocycle, one
needs to analyze whether or not its $\delta$-variation is trivial in
$H(\gamma\vert d)$. To address this question we define the map
$\delta_0:{\cal E}_0\rightarrow {\cal E}_0$ as follows,
\begin{equation}
\delta_0 [a_q] =[\delta a_q],
\end{equation}
where $[\ ]$ denotes the class in $H(\gamma \vert d)$. This map is
well-defined because: 1) $\gamma \delta a_q=-\delta \gamma
a_q=\delta d b_q=-d\delta b_q$ (so $\delta a_q$ is a $\gamma$ mod
$d$ cocycle; 2) $\delta (\gamma r_q + d c_q)=-\gamma (\delta
r_q)-d(\delta c_q)$ (so $\delta$ maps a coboundary on a
coboundary). Furthermore
$\delta_0^2=0$ so $\delta_0$ is a differential and one can define
its cohomology. Cocycles of
$\delta_0$ are elements of $H(\gamma\vert d)$ which can be lifted
at least once. By contrast, if $\delta_0 [a_q]\not =0$ then one
cannot lift $a_q$ to construct the component of antighost number
$q-1$ of $a^n_{g}$. Furthermore, if $[a_q]=\delta_0 [c_{q+1}]$ then
one can eliminate $a_q$ from $a^n_g$ by the addition of trivial
terms and the redefinition of the terms of lower antighost
numbers. The interesting $a_q$ involved in the construction of
BRST-cocycles are therefore the representatives of $H({\cal
E}_0,\delta_0)$.

To analyze the next lifts, one can define a sequence of spaces
${\cal E}_r$ and differentials $\delta_r$ which satisfy similar
properties as those encountered in the standard descent equations:
\begin{enumerate}
\item 
${\cal E}_r \equiv H({\cal E}_{r-1},\delta_{r-1})$.
\item 
There exist a map $\delta_r : {\cal E}_r \rightarrow {\cal
E}_r$ which is defined by $\delta_r [[\ldots [a_q]\ldots ]]=
[[\ldots
[\delta a_{q-r}]\ldots ]]$, for $[[\ldots [a_q]\ldots ]]\in {\cal
E}_r$  where $[[\ldots
[\delta a_{q-r}]\ldots ]]$ is the class of the $\gamma$ mod $d$
cocycle $\delta a_{q-r}$ and where $a_{q-r}$ is defined by the
equations \eqref{highanti0}-\eqref{highanti}.
\item $\delta_r^2=0.$
\item
A $\gamma$ mod $d$ cocycle $a_q$ can be `lifted' $q$ times if and
only if $\delta_0 [a_q]=0,\ 
\delta_1 [[a_q]]=0,
\ldots,
\delta_{q-1}[[\ldots [a_q]\ldots ]]=0.$ Such a $\gamma$ mod $d$
cocycle and its successive lifts constitute the components of a
BRST cocycle. If
$\delta_r [[\ldots [a_q]\ldots]]\not =0\ (r<q)$, the $\gamma$ mod
$d$ cocycle cannot be lifted $r+1$ times and it is not an $s$ mod
$d$ coboundary up to terms of lower antighost number.
\item One can eliminate $a_q$ from $a^n_g$ by the addition of
trivial terms and the redefinition of terms of lower antighost
numbers if and only if there exists a $k$ such $\delta_i [[\ldots
[a_q]\ldots ]]=0, (i=0,\ldots,k-1)$ and $[[\ldots
[a_q]\ldots]]=\delta_k [[\ldots [a_{q+k+1}]\ldots ]]$.
\end{enumerate}
The proof of these properties proceed as in the case of the
standard descent equations.

According to these properties, the $\gamma$ mod $d$ cocycles which
yield non-trivial BRST cocycles belong to ${\cal E}_r
\equiv H({\cal E}_{r-1},\delta_r),\ \forall r$. This condition
seems awkward to work with since it implies the
calculation of an infinite number of cohomologies ${\cal E}_r$. The
reason for this is that the antighost number is not bounded.
However, for a large class of theories such as linear theories and
normal theories \cite{BBH1} the cohomologies ${\cal E}_r$ coincide
for
$r>k$ where $k$ is a finite number. This result is a consequence
of the fact that their characteristic cohomology (to be defined
below) vanishes above a certain value of the
antighost number.

There is thus a strong analogy between the analysis of antifield
independent and antifield dependent solutions of the Wess-Zumino
consistency condition. However, compared to the $d_r$, the
operators
$\delta_r$ are not antiderivations.
Indeed they are maps in vector spaces ${\cal E}_r$ and not in
algebras. For instance, if $a_q$ and $b_q$ are representatives of
$H(\gamma\vert d)\equiv {\cal E}_0$ their product is not
necessarily in ${\cal E}_0$. The product of two elements of ${\cal
E}_r$ is therefore not an internal operation and one cannot even
speak of derivations. Because of this particularity, we cannot
calculate the spaces
${\cal E}_k$ as we did for the $E_k$. Instead we will have to lift
the possible $\gamma$ mod $d$ cocycles $a_q$ ``by hand" and find
those which can be lifted all the way up to produce BRST cocycles.

Before we do so, we begin by characterizing the ``bottoms"
$a_q$ as much as possible. To this end, we first prove that
 without changing the class of
$a^k_g$ in
$H(s\vert d)$ we can replace \eqref{highanti} by the simpler Eq.
$\gamma a_q=0$. To prove this, we use
the following theorem:
\begin{theorem}
Let $a^k$ be a $\gamma$-closed solution of $da^k=\gamma
b^{k+1}$ of form degree $k<n$ and antighost $>0$. Then,
$a^k =dc^{k-1} + \gamma b^k$, with $c^{k-1}$ in
$H(\gamma)$. In other words,  the cohomology in antighost
$>0$ and form degree $<n$ of
$d$ in
$H(\gamma)$ vanishes.
\label{dingamma}
\end{theorem}
\proof{Since $a^k$ is $\gamma$-closed we have
$a^k=P_J(\chi)\omega^J+ \gamma m^k$. Equation $da^k=\gamma
b^{k+1}$ then implies $dP_J \omega^J 
=\gamma (b^{k+1}+dm^k +P_J {\hat{\omega}}^J)$, where
$d\omega^J+\gamma {\hat{\omega}}^J=0$. Therefore we have $dP_J=0$
and using Theorem {\bf{\ref{invpoincare}}}, $P_J=dQ_J$. Thus,
$a^k=dQ_J\omega^J+\gamma m^k=d(Q_J \omega^J)+\gamma (Q_J
{\hat{\omega}}^J+m^k)$ which proves the theorem.}
Theorem {\bf{\ref{dingamma}}} now  allows us 
assume 
$\gamma a_q=0$. Indeed, if we act  on \eqref{highanti} with
$\gamma$ we can build a descent equation. However, the bottom of
this descent satisfies the condition of Theorem
{\bf{\ref{dingamma}}} which implies that it is trivial in
$H(\gamma \vert d)$ and thus that
$a_q$ cannot descend non-trivially. Therefore we have $a_q =
P_J(\chi)\omega^J + \gamma m_q + dr_q = P_J(\chi)\omega^J + sm_q
+dr_q -\delta m_q$. Because $\delta m_q$ is of antighost number
$q-1$ we see that in each class of $H(s\vert d)$ there is a
representative satisfying $a_q=P_J(\chi)\omega^J$.

The next equation we have to examine is \eqref{highanti2}. If we
substitute $a_q$ in this equation we obtain,
\begin{equation}
\delta P_J \omega^J +\gamma a_{q-1}+db_{q-1}=0.\label{charinvdef4}
\end{equation}
By acting on this equation with $\gamma$ we can once more build a
descent equation for $b_{q-1}$. If $q-1>0$ then again the bottom
satisfies the conditions of Theorem {\bf{\ref{dingamma}}} and
therefore
$b_{q-1}=N_J\omega^J + \gamma h_{q-1}+ dl_{q-1}$. If we substitute
this form in \eqref{charinvdef4} we obtain the condition,
\begin{equation}
\delta P_J + dN_J=0.\label{char1}
\end{equation}
If $q=1$ then we have for the descent,
\begin{align}
\delta P_J \omega^J +\gamma a_{0}+db_{0}&=0, \\
\gamma b_0 + dr_0&=0,\\
&\vdots\\
\gamma m_0&=0.
\end{align}
According to our analysis of $H(\gamma\vert d)$, we know that
$b_0$ can be written as $b_0=N_J\omega^J+ \overline{b}_0$, where
$\overline{b}_0$ is an element of $H(\gamma \vert d)$ which does
not descend trivially and which is in the small algebra. Thus we
have
$db_0= dN_J\omega^J -
\gamma(N_J{\hat{\omega}}^J)+ R_J(H^a)\omega^J+\gamma c_0$, where
$R_J(H^a)\omega^J+\gamma c_0=d\overline{b}_0$. The condition
\eqref{charinvdef4} therefore becomes,
\begin{equation}\label{Rj}
\delta P_J  +dN_J+R_J(H^a)=0.
\end{equation}
In the algebra of $x$-independent
forms ${\cal P}_{-}$, one easily obtains $R_J=0$ by counting the
number of derivatives of
$H_{\mu\nu\rho}$. The above equation therefore reduces to
\eqref{char1}. In the algebra of $x$-dependent forms ${\cal P}$,
this is no longer true and we shall comment later on the
consequences of this.

A solution of \eqref{char1} is called trivial when it is of the
form, $P_J = \delta M_J + d R_J$ where $M_J$ and $R_J$ are
polynomials in the $\chi$. Such solutions are irrelevant in the
study of BRST cocycles because the corresponding $a_q$ can be
eliminated from $a^n_g$. Indeed we have, $a_q=(\delta M_J + d
R_J)\omega^J= \delta (M_J \omega^J)+ d (R_J \omega^J) + \gamma
(R_J {\hat \omega}^J)= s (M_J \omega^J + R_J {\hat\omega}^J) +
d(R_J\omega^J) -\delta R_J {\hat\omega}^J$ so we may assume that
the expansion of
$a^n_g$ stops at order $q-1$ in the antighost number.

We therefore need to solve equation \eqref{char1} for invariant
polynomials $P_J$ and identify two solutions which differ by
trivial terms of the form $\delta M_J + d R_J$. 

These equivalence
classes define the invariant $\delta$ mod $d$ cohomology denoted
$H^{inv}(\delta \vert d)$. This cohomology is related to the
so-called characteristic cohomology and both will be calculated in
the next section.

\vspace{.5cm}
To summarize, we have shown in this section that the antifield
dependent solutions of the Wess-Zumino consistency condition
can be chosen of the form $a_q= P_J\omega^J$ where $P_J$ has to be a
non-trivial element of
$H^{inv}(\delta \vert d)$ (for $q=1$ this is only valid in the
algebra of $x$-independent forms).

In the next Chapter we study the characteristic cohomology in
detail and return to the calculation of $H(s\vert d)$ in Chapter
{\bf\ref{Chap6}}.

\chapter{Characteristic Cohomology}
\label{CharaCohosec}

\section{Introduction}
The characteristic cohomology \cite{BryantGriffiths}
plays a central role in the analysis of
any local field theory. The easiest way to define this
cohomology, which is contained in the so-called Vinogradov
$C$-spectral sequence
\cite{Vinogr1,Vinogr2,Vinogr3,Vinogr4,Tsujishita1,Tsujishita2}, 
is to start with the more familiar notion of conserved current. 
Consider a dynamical theory with field variables $\phi^i$
($i=1,\dots ,M$) and Lagrangian ${\cal L}(\phi^i,
\pp_\mu \phi^i,\dots, \pp_{\mu_1 \dots \mu_k} \phi^i)$.  The field
equations read
\begin{equation}
{\cal L}_i = 0, \label{FE0}
\end{equation}
with
\begin{equation}
{\cal L}_i = \frac{\delta {\cal L}}{\delta \phi^i} =
\frac{\pp {\cal L}}{\pp \phi^i} - \pp_\mu
\big( \frac{\pp {\cal L}}{\pp (\pp_\mu \phi^i)} \big) +
\dots + (-1)^k \pp_{\mu_1 \dots \mu_k} \big( \frac{\pp {\cal L}}{\pp 
(\pp_{\mu_1 \dots \mu_k} \phi^i)} \big).
\end{equation}
A (local) conserved current $j^\mu$ is a vector-density which
involves the fields and their derivatives up to some
finite order and which is conserved modulo the field 
equations, i.e., which fulfills
\begin{equation}
\pp_\mu j^\mu \approx 0.\label{ConsCurr}
\end{equation} 
Here and in the sequel, $\approx $ means ``equal 
when the equations of motion hold" or, as one also says
equal ``on-shell".  Thus, (\ref{ConsCurr}) is equivalent to
\begin{equation}
\pp_\mu j^\mu = \lambda^i {\cal L}_i + \lambda^{i \mu}
\pp_\mu {\cal L}_i + \dots + \lambda^{i \mu_1 \dots \mu_s}
\pp_{\mu_1 \dots \mu_s} {\cal L}_i
\label{ConsCurr2}
\end{equation}
for some $\lambda^{i \mu_1 \dots \mu_j}$, $j=0,\dots,s$.
A conserved current is said to be trivial if it can be written as
\begin{equation}
j^\mu \approx \pp _\nu S^{\mu \nu} \label{TrivCurr} 
\end{equation}
for some local antisymmetric tensor density $S^{\mu \nu} = -
S^{\nu \mu}$.  The terminology does not mean that trivial
currents are devoid of physical interest, but rather, that they
are easy to construct and that they are trivially conserved. 
Two conserved currents are said to be equivalent if they differ
by a trivial one.  The characteristic cohomology in degree $n-1$ is
defined to be the quotient space of equivalence classes of
conserved currents.  One assigns the degree $n-1$ because
the equations (\ref{ConsCurr}) and (\ref{TrivCurr}) can be
rewritten as
$d \omega \approx 0$
and
$\omega \approx d \psi$
in terms of the ($n-1$)-form $\omega$ and ($n-2$)-form
$\psi$ respectively dual to $j^\mu$ and $S^{\mu \nu}$.

One defines the characteristic cohomology in degree $k$
($k<n$) along exactly the same lines, by simply considering other
values of the form degree.  So, one says that a local
$k$-form $\omega$ is a cocycle of the characteristic 
cohomology in degree $k$ if it is weakly closed,
\begin{equation}
d \omega \approx 0; \;\; \hbox{``cocycle condition"}
\label{cocycleCC}
\end{equation}
and that it is a coboundary if it is weakly exact,
\begin{equation}
\omega\approx d \psi, \;\; \hbox{``coboundary condition"}
\label{coboundaryCC}
\end{equation}
just as it is done for $k=n-1$.
For instance, the characteristic cohomology in form degree
$n-2$ is defined, in dual notations, as the quotient space of
equivalence classes of weakly conserved antisymmetric tensors,
\begin{equation}
\pp _\nu S^{\mu \nu} \approx 0, \; S^{\mu \nu} = S^{[\mu \nu]},
\label{ConsLaw}
\end{equation}
where two such tensors are regarded as equivalent iff
\begin{equation}
S^{\mu \nu} - S^{'\mu \nu} \approx  \pp_\rho R^{\rho\mu \nu}, \; 
R^{\rho\mu \nu} = R^{[\rho\mu \nu]}.  
\end{equation}
We shall denote the characteristic cohomological 
groups by $H^k_{char}(d)$.

Higher order conservation laws 
involving antisymmetric tensors
of degree $2$ or higher 
are quite interesting in their own right. 
In particular, conservation laws of the form (\ref{ConsLaw}), 
involving an antisymmetric tensor $S^{\mu \nu}$ 
have attracted a great deal of interest in the past
\cite{Unruh1} as well as recently \cite{BBH3,Torre1} in the
context of the ``charge without charge" mechanism developed by
Wheeler 
\cite{MisnerWheeler1}. Higher order conservation laws also enter
the analysis of the BRST field-antifield
formalism extension incorporating global symmetries
\cite{BHW1,BHW3}.

But as we have seen in section {\bf \eqref{adssec}} the
characteristic cohomology is also important because it appears as
an  important step in the calculation of the local BRST
cohomology. 

In this section we will carry out the calculation of the
characteristic cohomology for a system of free $p$-form gauge
fields.  We give complete results in
degree
$<n-1$; that is, we determine all the
solutions to the equation $\pp_{\mu} S^{\mu \nu_1 \dots \nu_s} 
\approx 0$ with $s>0$. 
Although we do not solve the characteristic cohomology
in degree $n-1$, we comment at the end of the section
on the gauge invariance properties of the
conserved currents and provide an infinite number of them,
generalizing earlier results of the Maxwell case
\cite{Lipkin1,Morgan1,Kibble1,OConnell1}. The 
determination
of all the conserved currents is of course also an interesting
question, but it is not systematically 
pursued here for two reasons.  First, for the free theories
considered here,
the characteristic cohomology $H^{n-1}_{char}(d)$ is 
infinite-dimensional and does not appear to be completely known
even in the Maxwell case in an arbitrary number of dimensions. 
By contrast, the cohomological groups
$H^{k}_{char}(d)$, $k<n-1$, are all finite-dimensional
and can explicitly be  computed. 
Second, 
the group
$H^{n-1}_{char}(d)$ plays no role either in the analysis of the
consistent interactions of antisymmetric tensor fields of degree
$>1$, nor in the analysis of candidate anomalies if the
antisymmetric tensor fields all have degree $>2$.

An essential feature of theories involving $p$-form gauge fields
is that their gauge symmetries
are {\em reducible} (see \eqref{GaugeSym}).
General vanishing theorems have been established in
\cite{BryantGriffiths,Vinogr1,Vinogr2,BBH1} showing that the
characteristic  cohomology of reducible theories of reducibility
order
$p-1$  vanishes in form degree
strictly smaller than $n-p-1$.  Accordingly,  in the case of
$p$-form gauge theories, 
there can be a priori non-vanishing
characteristic cohomology only in form degree $n-p_M-1$, $n-p_M$, etc,
up to form degree $n-1$ (conserved currents). 
In the $1$-form case,
these are the best vanishing theorems one can prove, since a set of
free gauge fields $A^a_\mu$ has characteristic cohomology
both in form degree $n-1$ and $n-2$ \cite{BBH1}. Representatives
of the cohomology classes in form degree $n-2$ are given
by the duals to the field strengths, which are indeed closed
on-shell due to Maxwell equations.

Our main result is that the general vanishing theorems of
\cite{BryantGriffiths,Vinogr1,Vinogr2,Vinogr3,Vinogr4,BBH1} can be
considerably strengthened when $p >1$. 
For instance, if there
is a single $p$-form gauge field and if $n-p-1$ is odd,
there is only one non-vanishing group of
the characteristic cohomology in degree $<n-1$.  This is 
$H^{n-p-1}_{char} (d)$, which is one-dimensional. 
All the other groups 
$H^{k}_{char}(d)$  with  $n-p-1<k<n-1$ happen to be
zero, even though the general theorems of 
\cite{BryantGriffiths,Vinogr1,Vinogr2,Vinogr3,Vinogr4,BBH1}
left open the possibility that they might not vanish. The presence
of these additional zeros give $p$-form gauge fields and their
gauge transformations a strong rigidity. 
 
Besides the standard characteristic cohomology, one may consider
the invariant characteristic cohomology, in which the local forms
$\omega$ and $\psi$
occurring in (\ref{cocycleCC}) and (\ref{coboundaryCC}) are
required to be invariant under the
gauge transformations (\ref{GaugeSym}).  This is the
relevant cohomology for the resolution of the
Wess-Zumino consistency condition. We also completely determine in
this section the invariant characteristic cohomology in form degree
$<n-1$.

Our method for
computing the characteristic cohomology is based
on the reformulation performed in \cite{BBH1}
of the characteristic cohomology
in form degree $k$ in terms of the cohomology $H^n_{n-k}(\delta|d)$
of the Koszul-Tate differential
$\delta$ modulo the spacetime exterior derivative $d$.
Here, $n$ is the form degree and
$n-k$ is the antighost number.  

This section is organized as follows.
First, since the calculation of the characteristic cohomology is
rather long and intricate, we begin by formulating precisely our
main results, which state (i) that the characteristic cohomology
$H^{k}_{char}(d)$ with  $k<n-1$ 
is generated (in the exterior product) by the exterior forms
${\overline H}^{a}$ dual to the field strengths $H^a$; these are 
forms of
degree $n-p_a-1$; and (ii) that the invariant characteristic
cohomology $H^{k, inv}_{char}(d)$ with  $k<n-1$
is generated (again in the exterior product) by the exterior forms
$H^a$  and ${\overline H}^{a}$.
Then, we recall the relation between the characteristic cohomology
and the Koszul-Tate complex and show how they relate to 
the cohomology of the differential
$\Delta=\delta +d$.  After that we analyze the gauge invariance
properties of
$\delta$-boundaries modulo $d$.  We then determine the
characteristic  cohomology for a single
$p$-form gauge field and afterwards extend the results
to an arbitrary system of $p$-forms.
Next, we calculate the invariant cohomology and use
the results to obtain the  cohomological groups
$H^{*,inv}(\delta
\vert d)$. Finally, we show that the existence of
representatives  expressible in terms of the ${\overline H}^{a}$'s
does not hold for the characteristic cohomology in form degree
$n-1$, by exhibiting an infinite number of (inequivalent)
conserved currents which are not of that form.  We will also
comment on how the results on the free
characteristic cohomology in degree $<n-1$ 
generalize straightforwardly
if one adds to the free Lagrangian (\ref{Lagrangian})
gauge invariant interaction terms that involve the fields 
$B^a_{\mu_1 \dots \mu_{p_a}}$ and their derivatives only
through the gauge invariant field strength components
and their derivatives.

\section{Results}

\subsection{Characteristic cohomology}

Remember that the equations of motion (\ref{FE1}) can be written
as,
\begin{equation}
d {\overline H}^a \approx 0, \label{FE2}
\end{equation}
in terms of the ($n-p_a-1)$-forms ${\overline H}^a$
dual to the field strengths.  It follows that any polynomial
in the ${\overline H}^a$'s is closed on-shell and thus defines
a cocycle of the characteristic cohomology.

The remarkable feature is that these polynomials are
not only inequivalent in cohomology, but
also that they {\em completely
exhaust the characteristic cohomology in form degree
strictly smaller than $n-1$}.  Indeed, one has:
\begin{theorem}
Let ${\cal {\overline H}}$ be the algebra generated by
the ${\overline H}^a$'s and let ${\cal V}$ be the
subspace containing the polynomials in the ${\overline H}^a$'s
with no term of form degree exceeding $n-2$.
The subspace ${\cal V}$ is isomorphic to
the characteristic cohomology in form degree
$<n-1$.
\label{MainResult}
\end{theorem}
We stress
again that this theorem does not hold in degree $n-1$ 
because there exist conserved
currents not expressible in terms of the ${\overline H}^a$'s.

Since the form degree is limited by the spacetime dimension
$n$, and since ${\overline H}^a$ has strictly positive
form degree
$n-p_a-1$ (as explained previously, we
assume
$n-p_a-1>0$ for each $a$), the algebra ${\cal {\overline H}}$ 
is finite-dimensional.  In this algebra, the
${\overline H}^a$ with $n-p_a-1$ even  commute with all the
other generators, while the ${\overline H}^a$ with $n-p_a-1$ odd 
are anticommuting objects.

\subsection{Invariant characteristic cohomology}
While the cocycles of Theorem {\bf \ref{MainResult}} are all gauge
invariant, there exists co\-boundaries of the characteristic
cohomology that are gauge invariant, i.e., that involve only the
field strength components and their derivatives, but which cannot,
nevertheless, be written as coboundaries of  gauge invariant
local forms, even weakly. Examples are given by the field
strengths $H^a = dB^a$ themselves. For this reason, the
invariant characteristic cohomology and the characteristic
cohomology do not coincide. We shall denote by
${\cal H}$ the finite-dimensional algebra generated by the
$(p_a+1)$-forms $H^a$, and by ${\cal J}$ the finite-dimensional
algebra generated by the field strengths $H^a$ and their
duals ${\overline H}^a$.  One has,
 
\begin{theorem}
Let ${\cal W}$ be the
subspace of  ${\cal J}$  containing the polynomials
in the $H^a$'s and the ${\overline H}^a$'s
with no term of form degree exceeding $n-2$.
The subspace ${\cal W}$ is isomorphic to
the invariant  characteristic cohomology in form degree
$<n-1$.
\label{MainResult2}
\end{theorem}
\noindent This chapter is devoted to proving these theorems. 

\subsection{Cohomologies in the algebra of $x$-independent forms}

The previous theorems hold as such
in the algebra of local forms
that are allowed to have an explicit $x$-dependence. If one
restricts one's attention to the algebra of local forms with no
explicit dependence on the spacetime coordinates, then, one must
replace in the above theorems the polynomials in the curvatures
and their duals with coefficients that are {\em numbers} by 
the polynomials in the curvatures
and their duals with coefficients that are {\em constant exterior
forms}.

\section{Characteristic Co\-ho\-mo\-lo\-gy and
Koszul-Tate Complex}

Our analysis of the characteristic cohomology relies upon the
isomorphism established in \cite{BBH1} between 
$H^*_{char}(d)$ and the cohomology $H^*_*(\delta \vert d)$
of $\delta$ modulo $d$.
The cohomology $H^k_i(\delta \vert d)$ in form
degree $k$ and antighost number $i$ is obtained by solving in
the algebra ${\cal P}$ of local exterior forms the equation,
\begin{equation}
\delta a^k_i + db^{k-1}_{i-1}=0,
\end{equation}
and by identifying solutions which differ by $\delta$-exact and
$d$-exact terms, i.e,
\begin{equation}
a^k_i \sim {a'}^k_i = a^k_i +\delta n^k_{i+1}+dm^{k-1}_i.
\end{equation}
One has,
\begin{theorem} \label{CharAnddelta}
\begin{eqnarray}
H^{k}_{char}(d) &\simeq& H^n_{n-k}(\delta \vert d), \; 0<k<n 
\label{CharAndDelta1}\\
\frac{H^{0}_{char}(d)}{R}  &\simeq& H^n_{n}(\delta \vert d).
\label{CharAndDelta2}\\
0 &\simeq& H^n_{n+k}(\delta \vert d),\; k>0
\label{CharAndDelta3}
\end{eqnarray}
\end{theorem}
\proof{Although the proof is standard and can be 
found in \cite{DVHTV1,BBH1}, we shall repeat 
it explicitly here because it involves 
ingredients which will be needed below.
Let $\alpha$ be a class of $H^k_{char}(d)$ ($k<n$) and let
$a^k_{0}$ be a representative of $\alpha$, $\alpha = [a^k_{0}]$.
One has,
\begin{equation}
\delta a^{k+1}_{1} + da^{k}_{0} =0,
\label{MapdDelta1}
\end{equation}
for some
$a^{k+1}_{1}$ since any antifield-independent
form that is zero on-shell
can be written as the $\delta$ of something.  By acting with
$d$ on this equation, one finds that $d a^{k+1}_{1}$
is $\delta$-closed and thus, by  Theorem {\bf \ref{propkoszul}},
that it is $\delta$-exact,
$\delta a^{k+2}_{2}+ da^{k+1}_{1} =0$ for some
$a^{k+2}_{2}$.  One can repeat the procedure until one reaches
degree $n$, the last term $a^n_{n-k}$ fulfilling
\begin{equation}
\delta a^n_{n-k} + d a^{n-1}_{n-1-k} = 0, 
\label{MapdDelta2}
\end{equation}
and, of course,
$d  a^n_{n-k}= 0$ (it is a $n$-form).  For future reference we 
collect all the terms appearing in this tower of equations
as
\begin{equation}
a^k = a^n_{n-k} + a^{n-1}_{n-1-k} + \dots + a^{k+1}_1 + a^k_0.
\label{Tower}
\end{equation}

Eq. (\ref{MapdDelta2}) shows that $a^n_{n-k}$
is a cocycle of the cohomology of $\delta$ modulo $d$, in form-%
degree $n$ and antighost number $n-k$.  Now, given the
cohomological class $\alpha$ of $H^k_{char}(d)$,
it is easy to see, using again 
Theorem {\bf \ref{propkoszul}}, that the corresponding 
element $a^n_{n-k}$ is well-defined in $H^n_{n-k}(\delta \vert d)$.
Consequently, the above procedure
defines an non-ambiguous map $m$ from $H^k_{char}(d)$
to $H^n_{n-k}(\delta \vert d)$.
 
This map is surjective.  Indeed, let $a^n_{n-k}$ be
a cocycle of $H^n_{n-k}(\delta \vert d)$. 
By acting with $d$ on Eq. (\ref{MapdDelta2})
and using the second form of the
Poincar\'e lemma (Theorem
{\bf \ref{poincarebis}}), one finds that $a^{n-1}_{n-1-k}$
is also $\delta$-closed modulo $d$. 
Repeating the procedure
all the way down to antighost number zero, one sees that there
exists a cocycle $a^k_0$ of the characteristic 
cohomology such that $m([a^k_0]) = [a^n_{n-k}]$.

The map $m$ is not quite injective, however, because
of the constants.  Assume that $a^k_0$ is mapped on zero.
This means that the corresponding $a^n_{n-k}$ is trivial in
$H^n_{n-k}(\delta \vert d)$, i.e., $a^n_{n-k}
= \delta b^n_{n-k+1} + d b^{n-1}_{n-k}$.  Using the Poincar\'e
lemma (in the second form) one then successively
finds  that all $a^{n-1}_{n-k-1}$
$\dots$ up to $a^{k+1}_1$ are  trivial.  The last
term $a^k_0$ fulfills $da^k_0 + \delta db^k_1 = 0$ and
thus, by the Poincar\'e lemma (Theorem {\bf \ref{Poincare}}),
 $a^k_0 = \delta b^k_1 + db^{k-1}_0 + c^k$.  In the algebra of
$x$-dependent local forms, the constant
$k$-form
$c^k$ is present only if $k=0$. 
This establishes (\ref{CharAndDelta1}) and (\ref{CharAndDelta2}).

That $H^n_m(\delta \vert d)$ vanishes for $m>n$
is proved as follows. If $a^n_m$
is a solution of $\delta a^n_m + da^{n-1}_{m-1}$ then one can
build a descent by acting with $\delta$ on this equation.
The bottom of this descent satisfies, $\delta
a^{n-j}_{m-j}=0$. Since $j\leq n$ and $m>n$, Theorem
{\bf{\ref{propkoszul}}}  implies that the bottom of the
descent and all the cocycles above him are trivial.} 
The proof of the theorem also shows
that (\ref{CharAndDelta1}) holds as
such because one allows for an explicit $x$-dependence of the
local forms.  Otherwise, one must take into account the constant forms
$c^k$ which appear in the analysis of injectivity
and which are no longer exact even when $k>0$, so 
that (\ref{CharAndDelta1}) becomes,
\begin{equation}
\frac{H^{k}_{char}(d)}{\Lambda^k} \simeq H^n_{n-k}(\delta \vert d),
\end{equation}
while (\ref{CharAndDelta2}) and (\ref{CharAndDelta3}) remain
unchanged.

\section{Characteristic Cohomology and Co\-ho\-mo\-lo\-gy of
$\Delta=\delta+d$}
We have seen in Section {\bf\ref{mixedforms}} that in form
notation, the action on the antifields of the Koszul-Tate
differential can be written in the compact form, 
\begin{equation}
\Delta {\tilde H}^a = 0,
\end{equation}
with,
\begin{equation}
\Delta = \delta + d,
\end{equation}
and
\begin{equation}
{\tilde H}^a = {\overline H}^a + \sum_{j=1}^{p_a+1} {\overline
B}^{*a}_j.
\end{equation}
The parity of the mixed form ${\tilde H}^a$ is equal to
$n-p_a-1$. Quite generally, it should be
noted that the dual ${\overline H}^a$ to the field
strength $H^a$ is the term of lowest form degree in ${\tilde H}^a$.
It is also the term of lowest antighost number,
namely, zero. At the other end, the term of highest form degree in
${\tilde H}^a$ is ${\overline B}^{*a}_{p_a+1}$, which has form
degree $n$ and antighost number $p_a+1$. If we
call the difference between the form degree and
the antighost number
the ``$\Delta$-degree", all the terms present in
the expansion of ${\tilde H}^a$ have same $\Delta$-degree,
namely $n-p_a-1$.   

The differential $\Delta=\delta+d$ enables
one to reformulate the characteristic cohomology as the cohomology
of $\Delta$. Indeed one has,

\begin{theorem}
The cohomology of $\Delta$ is isomorphic to the
characteristic cohomology,
\begin{eqnarray}
H^k(\Delta) \simeq H^k_{char}(d), \; 0 \leq k \leq n
\end{eqnarray}
where $k$ in $H^k(\Delta)$ means the
$\Delta$-degree, and in $H^k_{char}(d)$ k is
the form degree.
\label{ISO}
\end{theorem}
\proof{Let $a^k_0$ ($k<n$) be a cocycle of the characteristic
cohomology. Construct $a^k$ as in the proof of
Theorem {\bf \ref{CharAnddelta}}, formula (\ref{Tower}).
The form $a^k$ is easily seen to be a cocycle of $\Delta$,
$\Delta a^k= 0$, and furthermore, to be uniquely
defined in cohomology given the class of $a^k_0$.
It is also immediate to check that the map so defined
is both injective and surjective.  This proves
the theorem for $k<n$.  For $k=n$, the isomorphism
of $H^n(\Delta)$ and $H^n_{char}(d)$ is
even more direct ($da^n_0 = 0 $ is equivalent to
$\Delta a^n_0 = 0 $ and $a^n_0 = db^{n-1}_0 +
\delta b^n_1$ is equivalent to $a^n_0 =
\Delta (b^{n-1}_0 +  b^n_1)$).}
Our discussion has also established the following useful
rule:  the term of lowest form degree in a $\Delta$-cocycle
$a$ is a cocycle of the characteristic cohomology.  Its
form degree is equal to the $\Delta$-degree $k$ of
$a$.  For $a={\tilde H}^a$, this reproduce the rule
discussed above Theorem {\bf{\ref{ISO}}}. Similarly, the term of
highest form degree in
$a$ has always form degree equal to
$n$ if
$a$ is  not a $\Delta$-coboundary (up to a constant),  and
defines an element of $H^n_{n-k}(\delta \vert d)$.
 
Because $\Delta$ is a derivation, its cocycles form an algebra.
Therefore, any polynomial in the
${\tilde H}^a$ 
is also a $\Delta$-cocycle.
Since the form degree is limited by the spacetime dimension
$n$, and since the term ${\overline H}^a$ with
minimum form degree in ${\tilde H}^a$ has strictly positive form
degree
$n-p_a-1$,
the algebra generated by the ${\tilde H}^a$
is finite-dimensional.  
We shall show below that these $\Delta$-cocycles are not exact and
that any cocycle of form degree $<n-1$ is a
polynomial in the ${\tilde H}^a$ modulo trivial
terms.  According to the isomorphism expressed by Theorem
{\bf \ref{ISO}}, this is equivalent to proving  Theorem
{\bf \ref{MainResult}}.

\vspace{.3cm}

\noindent
{\bf Remarks: } (i) The $\Delta$-cocycle associated with a
conserved current contains only two terms,
\begin{equation}
a=a^n_1 + a^{n-1}_0,
\end{equation}
where $a^{n-1}_0$ is the dual of the conserved current.  The
product of such a $\Delta$-cocycle with a $\Delta$-cocycle of
$\Delta$-degree $k$ has
$\Delta$-degree $n-1+k$ and therefore vanishes unless
$k=0$ or $1$.

\noindent
(ii) It will be useful below to introduce another degree $N$ as 
follows. One assigns $N$-degree $0$ to the undifferentiated fields
and $N$-degree $1$ to all the antifields irrespective of their
antighost number.  One then extends the $N$-degree to the 
differentiated variables according to the rule $N(\pp_\mu \Phi) =
N(\Phi) +1$. Thus, $N$ counts the number of derivatives and of
antifields. Explicitly,
\begin{equation}
N = \sum_a N_a
\end{equation}
with
\begin{equation}
N_a = \sum_J \big[ (|J| \sum_i \partial_{J} B^a_i
\frac{\partial}{\partial_{J} B^a_i}
+ (|J|+1) \sum_\alpha \partial_{J} \phi^{*a}_\alpha
\frac{\partial}{\partial_{J}\phi^{*a}_\alpha} \big].
\label{Na}
\end{equation}
where (i) the sum over $J$ is a sum over all possible derivatives
including the zeroth order one;
(ii) $|J|$ is the differential order of the derivative $\pp_J$
(i.e.,  $|J|=k$ for $\pp_{\mu_1 \dots \mu_k}$); (iii) the sum
over $i$ stands for the sum over the independent components of $B^a$;
and (iv) the sum over $\alpha$ is a sum over the independent
components of all the antifields appearing in the tower associated
with $B^a$ (but there is {\em no} sum over the $p$-form
species $a$ in (\ref{Na})).
The differential $\delta$ 
increases $N$ by one unit.  The 
differentials $d$ and
$\Delta$  have in addition an inhomogeneous piece
not changing the $N$-degree, namely
$dx^\mu (\pp^{explicit} / \pp x^\mu)$, where
$\pp^{explicit} / \pp x^\mu$ sees only
the explicit $x^\mu$-dependence.
The forms ${\tilde H}^a$
have $N$-degree equal to one.  

\section{Acyclicity and Gauge Invariance}

\subsection{Preliminary results}

In the sequel, we shall denote by ${\cal I}_{Small}$
the algebra of local exterior forms with coefficients
$\omega_{\mu_1 \dots \mu_J}$ that depend only on the field strength
components and their derivatives (and possibly $x^\mu$); these are
the only invariant objects that can be formed out of the 
``potentials" $B^a_{\mu_1 \dots \mu_{p_a}}$ and their
derivatives (c.f. Theorem {\bf{\ref{hgamma}}}).  The
algebras ${\cal H}$, ${\cal {\overline H}}$ and ${\cal J}$ respectively
generated by the ($p_a+1$)-forms $H^a$, ($n-p_a-1$)-forms
${\overline H}^a$ and ($H^a$, ${\overline H}^a$) are subalgebras
of ${\cal I}_{Small}$. Remember that we have also defined the
algebra ${\cal I}$ as the algebra of invariant polynomials 
in the field strength components, the antifield components and
their derivatives and that in pureghost number $0$ the cohomology
$H(\gamma)$ is given by ${\cal I}$.

Since the field equations are gauge invariant and since
$d$ maps ${\cal I}_{Small}$ on ${\cal I}_{Small}$,
one can consider the cohomological problem  
(\ref{cocycleCC}), (\ref{coboundaryCC}) in the algebra 
${\cal I}_{Small}$.  This defines the invariant 
characteristic cohomology $H^{*,inv}_{char}(d)$.
Furthermore, the differentials $\delta$,
$d$  and $\Delta$ map  the algebra ${\cal I}$ on itself. Clearly,
${\cal I}_{Small} \subset {\cal I}$. 
The invariant cohomologies $H^{*,inv}(\Delta)$
and  $H^{n,inv}_j(\delta \vert d)$
are defined by considering only local exterior forms that belong
to ${\cal I}$.

In order to analyze the invariant characteristic cohomology 
and to prove the non triviality of the cocycles listed
in Theorem {\bf \ref{MainResult}}, we shall need some
preliminary results on the invariant cohomology of the
Koszul-Tate differential $\delta$ and on the gauge invariant
$\delta$-boundaries modulo $d$.

We define ${\cal A}$ and ${\cal A}_{-}$ respectively as the
restrictions of ${\cal P}$ and ${\cal P}_{-}$ in pureghost
number $0$. 

The variables generating 
${\cal A}$ are, together
with
$x^\mu$ and
$dx^\mu$,
$$B_{a\mu_1
\ldots
\mu_{p_a}},
\partial_\rho B_{a\mu_1\ldots \mu_{p_a}},\ldots,B^{*a\mu_1 \ldots
\mu_{p_a-m}}, \partial_\rho B^{*a \mu_1 \ldots
\mu_{p_a-m}},\ldots,B^{*a},\partial_\rho B^{*a},...\ .$$ These
variables can be conveniently split into two subsets. The first
subset consists of the variables $\chi$ defined below Theorem
{\bf\ref{hgamma}} (field strengths, antifields and their
derivatives). Note that the field strengths and their derivatives
which are present in
$\chi$ are not independent, since they are constrained by the
identity
$dH^a = 0$ and its differential consequences, but this is not a
difficulty for the considerations of this section.  The
$\chi$'s are invariant under the gauge transformations and they
generate the algebra ${\cal I}$ of invariant
polynomials.
In order to generate the full algebra ${\cal A}$ we need to add to
the $\chi$'s some extra variables that will be
collectively denoted
$\Psi$. The $\Psi$'s contain the field
components $B^{a \mu_1\ldots \mu_{p_a}}$ and their
appropriate derivatives not already present in the  $\chi$'s.
The explicit form of the  $\Psi$'s may be found in the proof of
Theorem {\bf\ref{hgamma}} and have the property that they
are algebraically independent from the
$\chi$'s and that, together with
the
$\chi$'s, they generate ${\cal A}$.
\begin{theorem}
Let $a$ be a polynomial in the $\chi$: $a=a(\chi)$. If $a=\delta
b$ then we can choose $b$ such that $b=b(\chi)$.  In particular,
\begin{equation}
H^{inv}_j(\delta) \simeq 0 \; \hbox{for } j>0.
\end{equation}
\label{deltachi}
\end{theorem}
\proof{We can decompose $b$ into two parts: $b={\overline
b}+ {\overline {\overline b}}$, with ${\overline b} = {\overline
b}(\chi)=b(\Psi=0)$ and ${\overline {\overline b}} =\sum_m
R_m(\chi)S_m(\Psi)$, where $S_m(\Psi)$ contains at least one
$\Psi$. Because $\delta \Psi =0$, we
have, $\delta ({\overline b} + {\overline {\overline b}})=
\delta {\overline b}(\chi) + \sum_m \delta R_m(\chi)S_m(\Psi).$
Furthermore if
$M=M(\chi)$ then $\delta M(\chi) = (\delta M)(\chi).$ We thus
get,
$$a(\chi)=(\delta {\overline b})(\chi) + \sum_m (\delta
R_m)(\chi)S_m(\Psi).$$ The above equation has to be satisfied
for all the values of the $\Psi$'s and in particular for
$\Psi=0$. This means that $a(\chi)=
(\delta {\overline b})(\chi) =\delta {\overline b}(\chi)$.}

\subsection{Gauge invariant $\delta$-boundary modulo $d$}

For the following theorem we assume  that the antisymmetric tensors
$B^a_{\mu_1 \mu_2 \dots \mu_p}$ all have the same degree $p$.
This covers, in particular, the case of a single $p$-form. 
\begin{theorem}
(Valid when the $B^a_{\mu_1 \mu_2 \dots \mu_p}$'s
have all the same form degree $p$).  Let $a_q^k=a_q^n(\chi) \in 
{\cal I}$ be an invariant local $k$-form of
antighost number $q>0$.
If $a_q^k$ is $\delta$-exact modulo
$d$, $a_q^k=\delta \mu_{q+1}^k + d\mu_q^{k-1}$, then
one can assume that $\mu_{q+1}^k$ and $\mu_q^{k-1}$ only depend
on the $\chi$'s, i.e., are invariant ($\mu_{q+1}^k$ and
$\mu_q^{k-1}$ $\in {\cal I}$).
\label{invardelta}
\end{theorem}
\proof{The proof goes along exactly the
same lines as the proof of a similar
statement made in \cite{BBH2} (theorem 6.1) for $1$-form gauge
fields.

By acting with $d$ and $\delta$ on the equation
$a_q^k=\delta \mu_{q+1}^k + d\mu_q^{k-1}$ one can construct a
ladder of equations:
\begin{eqnarray}
a^n_{q+n-k}=\delta \mu^n_{q+n-k+1}+d\mu^{n-1}_{q+n-k}\\
\vdots \nonumber \\
a_q^k=\delta \mu^k_{q+1} + d\mu^{k-1}_q \\
\vdots \nonumber \\
\left\{
\begin{array}{l}
a^{k-q+1}_1 = \delta \mu_2^{k-q+1} +d\mu_1^{k-q} \\
\text{or} \\
a^0_{q-k}=\delta \mu^0_{q-k+1}, 
\end{array}
\right.
\end{eqnarray}
where the $a^i_j$ only depend on the variables $\chi$. To go up
the ladder one acts with $d$ on one of the equations and uses
Theorem {\bf{\ref{deltachi}}} and the fact that $d$ maps a
$\chi$ on a
$\chi$. To go down the ladder one acts with
$\delta$ on one of the equations and uses Theorem
{\bf{\ref{invpoincare}}} and the fact that $\delta$ also maps a
$\chi$ on a $\chi$. 

Using the same theorems, it is easy to
see that if one of the $\mu^i_j$ may be assumed to depend only
on the
$\chi$, then one can suppose that it is also true for the
other $\mu^i_j$. Therefore, the theorem will hold if we can
prove it for the equation at the top of the ladder and thus for
$a^n_{q}=\delta
\mu^n_{q+1}+d\mu^{n-1}_{q}$ with $q>0$. 

Let us first treat the case where $q>n$ in which the
bottom of the ladder is $a^0_{q-n}=\delta \mu_{q-n+1}^0$. A
direct application of Theorem {\bf{\ref{deltachi}}} tells us that
we may assume that
$\mu_{q-n+1}^0$ only depends on the $\chi$ and therefore that the
property also holds for the other $\mu^i_j$. This ends the
proof for $q>n$.

Let us now examine the more difficult case $q \leq n$. We shall
work in dual notations where the equation reads,
\begin{equation}
a_q =\delta b_{q+1}+\partial_\mu j^{\mu}_{q}.\label{aqdual}
\end{equation}
All we need to prove is that one can assume
that $b_{q+1}$ only depends on the $\chi$. To this end, let us
take the Euler-Lagrange derivatives of \eqref{aqdual}
with respect to all the fields and the antifields. Using Theorem
{\bf{\ref{propkoszul}}} and the fact that the variational
derivative of a divergence vanishes, we obtain:
\begin{eqnarray}
{\delta^L a_q \over \delta B^{*a}}&=& \delta Y_a, \\ &\vdots&
\nonumber
\\
{\delta^L a_q \over \delta B^{*a\mu_1 \ldots
\mu_{p-m}}}&=&\delta Y_{a\mu_1
\ldots \mu_{p-m}} + (-1)^{m+1}\partial_{[\mu_1}Y_{a\mu_2
\ldots \mu_{p-m}]},\\ &\vdots& \nonumber \\
{\delta^L a_q \over \delta B^{*a\mu_1 \ldots
\mu_{p}}}&=&\delta Y_{a\mu_1
\ldots \mu_{p}}  -\partial_{[\mu_1}Y_{a\mu_2
\ldots \mu_{p}]},\\ &\vdots& \nonumber \\ \label{bot}
{\delta^L a_q \over \delta B^{a\mu_1 \ldots
\mu_{p}}}&=&\delta X_{a\mu_1 \ldots \mu_{p}} 
-(p+1)\partial^\rho\partial_{[\rho}Y_{a\mu_1 \ldots \mu_{p} ]},
\end{eqnarray}
where $[\ ]$ denotes complete antisymmetrization (factorial
included), $Y_{a\mu_1 \ldots \mu_{p-m}}$ $= (-1)^{m+1}{\delta^L
b_{q+1}
\over \delta B^{a*\mu_1 \ldots \mu_{p-m}}}$ and $X={\delta^L 
b_{q+1} \over \delta B^{q\mu_1 \ldots \mu_{p}}}$. Using
Theorem {\bf{\ref{deltachi}}} successively in all the above
equations we conclude that all the $Y_{a\mu_1 \ldots \mu_{p-m}}$
may be assumed to depend only on the $\chi$. Furthermore, since
$a_q$ depends on
$B^{a\mu_1 \ldots \mu_{p}}$ only through $H^{a\mu_1 \ldots
\mu_{p+1}}$, its Euler-Lagrange derivatives with respect to
$B^{a\mu_1 \ldots \mu_p}$ are of the form $\partial^\rho Z_{a\rho
\mu_1 \ldots \mu_{p}}$ where $Z_{a\rho
\mu_1 \ldots \mu_{p}}$ only depends on $H^{a\mu_1 \ldots
\mu_{p+1}}$ and is completely antisymmetric.
This implies that we may also assume that
$X_{a\mu_1 \ldots \mu_{p}}$ only depends on the $\chi$. 

Eq.
\eqref{bot} means that $X_{a\mu_1 \ldots \mu_{p}}$ is dual
to an element of $H^{n-p}_{q+1}(\delta \vert d)$  $\simeq
H^n_{q+p+1}(\delta \vert d) \equiv 0$ according to Theorem
{\bf{\ref{trivlin}}} (see below). We thus have $X_{a\mu_1 \ldots
\mu_{p}}=\partial^\rho S_{a\rho \mu_1 \ldots \mu_{p}} + \delta
N_{a\mu_1
\ldots
\mu_{p}}.$ Let us make the recurrence hypothesis that the theorem
holds for $q'=q+p+1$, so
that
$S_{a\rho \mu_1 \ldots \mu_{p}}$ and $N_{a\mu_1 \ldots
\mu_{p}}$ can be chosen  in such a way that they only depend on
the
$\chi$. Setting $a_q(t)=a_q(tB^{\mu_1 \ldots \mu_p},t\partial_\rho
B^{\mu_1
\ldots \mu_p},\ldots , tB^{*\mu_1 \ldots \mu_p},$ $t\partial_\rho
B^{*\mu_1 \ldots \mu_p},\ldots )$ we can reconstruct
$a_q$ from its Euler-Lagrange derivatives as follows:

\begin{align}
a_q&=\int_0^1 dt {d\over dt}a_q(t)\\ 
&=\int_0^1 dt (B^{a\mu_1
\ldots
\mu_p}{\delta^L a \over
\delta B^{a\mu_1
\ldots \mu_p}}(t) + B^{*a\mu_1 \ldots \mu_p}{\delta^L a
\over
\delta B^{*a\mu_1
\ldots \mu_p}}(t) + \ldots \nonumber \\
& \hspace{3cm} +
B^{*a\mu_1}{\delta^L a \over \delta B^{*a\mu_1}}(t) +
B^{*a} {\delta^L a
\over \delta B^{a*}}(t)) +\partial_\mu j^{'\mu}
\\&=\delta (  B^{a\mu_1 \ldots \mu_p} \int_0^1 dt 
\partial^{\rho}S_{a\rho \mu_1 \ldots \mu_p}(t)- B^{a*\mu_1
\ldots
\mu_p}\int_0^1 dt
Y_{a\mu_1 \ldots \mu_p}(t) \nonumber\\ 
& \hspace{2cm}+
(-1)^{p+1}B^{a*}\int_0^1 dt Y_a(t))+\partial_\mu
j^{''\mu} 
\\
&=\delta ( {1\over p+1} H^{a\rho \mu_1 \ldots \mu_p} \int_0^1 dt 
S_{a\rho \mu_1 \ldots \mu_p}(t) - B^{a*\mu_1 \ldots
\mu_p}\int_0^1 dt
Y_{a\mu_1 \ldots \mu_p}(t) \nonumber \\ &\hspace{1cm}  + \ldots 
 + (-1)^{p+1}B^{a*}\int_0^1 dt
Y_a(t))+\partial_\mu j^{''\mu}
= \delta m_{p+1}+\partial_\mu j^{'''\mu},
\end{align}
where $m_{p+1}$ manifestly only depends on the $\chi$.

The theorem will thus be proven if the recurrence hypothesis is
correct. This is the case since we have proved that
the theorem holds for $q>n$ and therefore for $q^{''}=q+r(p+1)$ for
a sufficiently great $r$.}
{\bf Remark :} The theorem does not hold if the forms have various
form degrees (see Theorem {\bf \ref{xx}} below).

\section{Characteristic Cohomology for a Single $p$-Form
Gauge Field}

Our strategy for computing the characteristic cohomology
is as follows.  First, we compute  $H^n_*(\delta \vert d)$ 
(cocycle condition, coboundary condition) for a single
$p$-form.  We then use the isomorphism theorems to infer
$H^*_{char}(d)$. Finally, we solve the case of a system involving 
an arbitrary (but finite) number of $p$-forms of various
form degrees.

\subsection{General theorems}
Before we compute $H^n_*(\delta \vert d)$ for a single abelian
$p$-form gauge field $B_{\mu_1 \dots \mu_p}$,
we will recall some general results which will be
needed in the sequel.  These theorems hold for an arbitrary
linear theory of reducibility order $p-1$.

\begin{theorem}
For a linear gauge theory of reducibility order p-1, one has,

\begin{equation}
H^n_j(\delta \vert d)=0,\ \ \ \ j > p+1.
\end{equation}
\label{trivlin}
\end{theorem}
\proof{The technic of the proof is the same as the one used
in the previous theorem. Although it is valid for any linear
theory \cite{BBH1} we only prove it for free $p$-forms. Suppose
there exists some
$a$ with antighost$(a)>p+1$, satisfying,
\begin{equation}
\delta a + \partial_\rho b^\rho =0.
\label{aderiver1}
\end{equation}
If we take the Euler-Lagrange
derivatives of (\ref{aderiver1}) with respect to all the fields
and antifields we get:
\begin{eqnarray}
\delta ( {\delta^L a \over \delta B^*})&=&0, \\ \nonumber
&\vdots& \\ \delta( {\delta^L a \over \delta B^{*\mu_1 \ldots
\mu_{p-m}}})&=&(-1)^{m+1}\partial_{[\mu_1} {\delta^L a
\over \delta B^{*\mu_2 \ldots \mu_{p-m}]}},\\ 
\nonumber &\vdots& \\
\delta( {\delta^L a \over \delta B^{\mu_1 \ldots
\mu_{p}}})&=&-{(p+1)}\partial^\rho\partial_{[\rho} {\delta^L a
\over \delta B^{*\mu_1 \ldots \mu_{p}]}}.
\end{eqnarray}
Because antighost($a$)$>p+1$ we have 
antighost(${\delta^L a \over \delta B^*}$)$>1$; by using Theorem
{\bf{\ref{propkoszul}}} successively in all the above equations we
obtain:
\begin{eqnarray}
{\delta^L a \over \delta B^*}&=& \delta f, \\ &\vdots& \nonumber \\
{\delta^L a \over \delta B^{*\mu_1 \ldots
\mu_{p-m}}}&=&\delta f_{\mu_1
\ldots \mu_{p-m}} + (-1)^{m+1}\partial_{[\mu_1}f_{\mu_2
\ldots \mu_{p-m}]},\\ &\vdots& \nonumber \\
 {\delta^L a \over \delta B^{\mu_1 \ldots
\mu_{p}}}&=&\delta g_{\mu_1 \ldots \mu_{p}} 
-(p+1)\partial^\rho\partial_{[\rho}f_{\mu_1 \ldots \mu_p ]}.
\end{eqnarray}
Using those relations we can reconstruct $a$ as in Theorem
{\bf{\ref{invardelta}}}:
\begin{eqnarray}
a&=&\int_0^1 dt {d\over dt}a(t) \label{reconstruction} \\
&=&\delta (  B^{\mu_1 \ldots \mu_p} \int_0^1 dt 
g_{\mu_1 \ldots
\mu_p}(t) -  B^{*\mu_1 \ldots \mu_p}\int_0^1 dt
f_{\mu_1 \ldots \mu_p}(t) + \ldots \nonumber \\
&&\hspace{0.3cm} + (-1)^{p+1}B^*\int_0^1 dt f(t))+\partial_\mu
j^{'\mu}.\label{tr}
\end{eqnarray}
Equation (\ref{tr}) shows
explicitly that
$a$ is trivial.}
Theorem 
{\bf \ref{trivlin}} is particularly useful because it limits the
number of possible non-vanishing cohomologies. 
The calculation of the characteristic cohomology is further
simplified by the following theorem:
\begin{theorem}
Any solution of  $\delta a +\partial_\rho b^\rho =0$ that is at
least bilinear in the antifields is necessarily trivial.
\label{trivmu}
\end{theorem}
\proof{the proof proceed along the same lines of Theorem
{\bf{\ref{trivlin}}}. One takes the Euler-Lagrange derivatives of
the equation $\delta a +\partial_\rho b^\rho=0$ with respect to
all the fields and antifields keeping in mind that since $a$ is
at least bilinear in the antifields it cannot depend on $B^*$. One
then uses the reconstruction formula (\ref{tr}) but
with
$f=0$.}
Both theorems hold whether  the local forms
are assumed to have an explicit $x$-dependence
or not.

\subsection{Cocycles of $H^n_{p+1}(\delta \vert d)$}

We have just seen that the first possible
non-vanishing cohomological group is $H^n_{p+1}(\delta
\vert d)$.  We show
in this section that this group is one-dimensional
and provide  explicit representatives.  
We systematically use the dual notation involving
divergences of antisymmetric tensor densities.

\begin{theorem}
$H^n_{p+1}(\delta \vert d)$ is one-dimensional. One can take as
representatives of the cohomological
classes $a=kB^*$ where $B^*$ is the last antifield, of antighost
number $p+1$ and where $k$ is a 
number.\label{Cohop+1}
\end{theorem}
\proof{Any polynomial of antighost number $p+1$ can be
written $a = f B^*+f^\rho\partial_\rho B^*+\ldots+\mu$,
where $f$ does not involve the antifields and where
$\mu$ is at least bilinear in the antifields. 
By adding a divergence to $a$, one can remove the derivatives
of $B^*$, i.e., one can assume $f^\rho = f^{\rho\sigma} = \dots =0$.
The cocycle condition $\delta a +\partial_\rho
b^\rho=0$ then implies $-\partial_\rho f B^{*\rho} 
+\delta \mu +\partial_\rho (b^\rho + f
B^{*\rho})=0$. By taking the Euler-Lagrange derivative of 
this equation with respect to $B^{*\rho}$, one gets
\begin{equation}
-\partial_\rho f + \delta ((-1)^p {\delta^L \mu \over \delta
B^{*\rho}})=0.
\end{equation}
This shows that $f$ is a cocycle of the characteristic cohomology
in degree zero since $\delta$(anything of antighost number one) $
\approx 0$.
Furthermore, if $f$ is trivial in $H^0_{char}(d)$,
then $a$ can be redefined so as to be at least bilinear
in the antifields and thus is also trivial in the cohomology
of $\delta$ modulo $d$. 
Now, the isomorphism
of $H^0_{char}(d)/R$ with $H^n_n(\delta \vert d)$
implies 
$f=k+\delta g$ with $k$ a constant ($H^n_n(\delta \vert d)=0$
because $n>p+1$). As we already pointed out, the second term can be
removed by adding a trivial term, 
so we may assume $f=k$.
Writing $a=kB^* +\mu$, we see that $\mu$ has to be a solution
of
$\delta \mu +\partial_\rho {b'}^\rho=0$ by itself and is therefore
trivial by Theorem {\bf \ref{trivmu}}. So
$H^n_{p+1}(\delta \vert d)$ can indeed be represented by $a=kB^*$.
In form notations, this is just the $n$-form $k {\overline B}^*_{p+1}$.
Note that the calculations are true both in the $x$-dependent and
$x$-independent cases.

To complete the proof of the theorem, it remains to show that
the cocycles $a=kB^*$, which belong to the invariant algebra ${\cal I}$
and which contain the undifferentiated antifields, are non trivial.  
If they were trivial,
one would have according to 
Theorem {\bf \ref{invardelta}},
that ${\overline B}^*_{p+1}  = \delta u + dv$ for some $u$, $v$ also in
${\cal I}$.  But this is impossible, because both $\delta$
and $d$ bring in one derivative of the invariant generators
$\chi$
while ${\overline B}^*_{p+1}$ does not contain derivatives of $\chi$.  
[This derivative counting argument is direct if $u$ and
$v$ do not involve explicitly the spacetime coordinates
$x^\mu$.  If they do, one must expand $u$, $v$ and the equation
${\overline B}^*_{p+1} = \delta u + dv$ according to the number of
derivatives of the fields in order to reach the conclusion.
Explicitly, one sets 
$u = u_0 + \dots + u_k$, $v = v_0 + \dots + v_k$, where $k$
counts the number of derivatives of the $H_{\mu_1 \dots \mu_{p+1}}$
and of the antifields.  The condition ${\overline B}^*_{p+1}=\delta u
+dv$ implies in degree
$k+1$ in the  derivatives that $\delta u_k + d'v_k = 0$, where
$d'$ does not differentiate with respect to the explicit dependence
on $x^\mu$. This relation implies in turn that $u_k$ is
$\delta$-trivial modulo $d'$ since there is no cohomology in
antighost number $p+2$. Thus, one can remove
$u_k$ by adding trivial terms.  Repeating the argument for $u_{k-1}$,
and then for $u_{k-2}$ etc, leads to the desired conclusion].}

\subsection{Cocycles of $H^n_{i}(\delta \vert d)$ with $i \leq p$}

We now solve the cocycle condition for the remaining
degrees. 
First we prove
\begin{theorem}
Let $K$ be the greatest integer such that
$n -K(n-p-1)>1$. 
The cohomological groups $H^n_j(\delta \vert d)$
($j>1$) vanish unless $j = n - k(n-p-1)$,
$k = 1, 2,\dots,K$.  Furthermore, for those
values of $j$,
$H^n_j(\delta \vert d)$
is at most one-dimensional.
\label{Cohop}
\end{theorem}
\proof{We already know that $H^n_j(\delta \vert d)$
is zero for $j>p+1$ and that $H^n_{p+1}(\delta \vert d)$  
is one-dimensional.  Assume thus that the theorem has
been proved for all $j$'s strictly greater than $J<p+1$ and
let us extend it to $J$. In a manner analogous to what
we did in the proof of Theorem {\bf \ref{Cohop+1}}, we
can assume that 
the cocycles of $H^n_J(\delta \vert d)$
take the form,
\begin{equation}
f_{\mu_1 \ldots \mu_{p+1-J}}B^{*\mu_1 \ldots \mu_{p+1-J}}+\mu,
\label{particular}
\end{equation}
where $f_{\mu_1 \ldots \mu_{p+1-J}}$ does not
involve the antifields and defines an 
element of $H^{p+1-J}_{char}(d)$.
Furthermore, if $f_{\mu_1 \ldots \mu_{p+1-J}}$ is trivial, then the
cocycle (\ref{particular}) is also trivial. Now, using the
isomorphism $H^{p+1-J}_{char}(d) \simeq H^n_{n-p-1+J}(\delta \vert d)$
($p+1-J >0$),
we see that $f$ is trivial unless $j'=n-p-1+J$, which is strictly
greater than $J$ and is of the form $j' = n - k(n-p-1)$.
In this case, $H^n_{j'}$ is at most one-dimensional. 
Since $J= j' -(n-p-1)=  n -(k+1)(n-p-1)$ is of the
required form, 
the property extends to $J$. This proves the theorem.}
Because we explicitly used the isomorphism 
$H^{p+1-J}_{char}(d) \simeq H^n_{n-p-1+J}(\delta \vert d)$,
which holds only if the local forms are allowed to involve explicitly
the coordinates $x^\mu$, the theorem must be amended for
$x$-independent local forms.  This will be done in section
{\bf \eqref{xmudep}}. 

Theorem {\bf \ref{Cohop}} 
goes beyond the vanishing theorems of 
\cite{BryantGriffiths,Vinogr1,Vinogr2,BBH1}
since it sets further cohomological groups
equal to zero,  in antighost number smaller
than $p+1$.  This is done by viewing the cohomological group
$H^n_i(\delta \vert d)$ as a subset of 
$H^n_{n-p-1+i}(\delta \vert d)$ at a 
{\em higher} value of the antighost number, through
the form (\ref{particular}) of the cocycle
and the isomorphism between $H^{p+1-i}_{char}(d)$
and $H^n_{n-p-1+i}(\delta \vert d)$.  In that manner, the known
zeros at values of the ghost number greater than
$p+1$ are ``propagated" down to values of the ghost number
smaller than $p+1$.
 
To proceed with the analysis, we have to consider two cases:

\noindent
(i) Case I: $n-p-1$ is even. 

\noindent
(ii) Case II: $n-p-1$ is odd. 

\vspace{.1cm}

We start with the simplest case, namely, case I.
Here, ${\tilde H}$ is a commuting object and we can
consider its various powers $({\tilde H})^k$,
$k=1,2,\dots,K$ with $K$  as in Theorem {\bf \ref{Cohop}}.
These powers have $\Delta$-degree
$k(n-p-1)$. By Theorem {\bf \ref{ISO}},
the term of form
degree $n$ in $({\tilde H})^k$ defines a cocycle of
$H^n_{n-k(n-p-1)}(\delta \vert d)$, which is non trivial as is
indicated by the same invariance argument used in the previous
subsection. Thus, 
$H^n_{n-k(n-p-1)}(\delta \vert d)$, which
we know is at most one-dimensional,  is
actually exactly one-dimensional and one
may take as representative the term 
of form degree $n$
in $({\tilde H})^k$.  This settles the case when $n-p-1$ is
even. 

In the case when $n-p-1$ is odd, ${\tilde H}$
is an anticommuting object and its
powers $({\tilde H})^k$, $k>0$ all vanish unless $k=1$.
We want to show that $H^n_{n-k(n-p-1)}(\delta \vert d)$ similarly 
vanishes
unless $k=1$.  To that end, it is enough
to prove that $H^n_{n-2(n-p-1)}(\delta \vert d)
 = H^n_{2p+2-n}(\delta \vert d)=0$ 
as the proof of Theorem {\bf \ref{Cohop}} indicates (we
assume, as before, that $2p+2-n>1$ since we only
investigate here the cohomological groups
$H^n_i(\delta \vert d)$ with $i>1$).
Now, as we have seen, the most general cocycle in 
$H^n_{2p+2-n}(\delta \vert d)$
may be assumed to take the form
$a = f_{\mu_{p+2} \ldots \mu_n} B^{*\mu_{p+2} \ldots \mu_n} +\mu$,
where $\mu$ is at least quadratic in the antifields and where
$f_{\mu_{p+2} \ldots \mu_n}$ does not involve the antifields
and defines an element of $H^{n-p-1}_{char}(d)$. But 
$H^{n-p-1}_{char}(d) \simeq H^n_{p+1}(\delta \vert d)$ is
one-dimensional and one may take as representative
of $H^{n-p-1}_{char}(d)$ the dual
$k\epsilon_{\mu_1\ldots \mu_n}H^{\mu_1 \ldots
\mu_{p+1}}$ of the field strength.  
This means that $a$ is necessarily of the form,
\begin{equation}
a=k\epsilon_{\mu_1\ldots \mu_n}H^{\mu_1 \ldots
\mu_{p+1}}B^{*\mu_{p+2} \ldots \mu_n} +\mu,
\label{a}
\end{equation}
so the question to be answered is: for which
value of $k$ can one adjust $\mu$
in (\ref{a}) so that,
\begin{equation}
k \epsilon_{\mu_1\ldots \mu_n}H^{\mu_1 \ldots
\mu_{p+1}}\delta B^{*\mu_{p+2} \ldots \mu_n}
+ \delta \mu +\partial_\rho b^\rho =0\; ?
\label{cohomo}
\end{equation}
In (\ref{cohomo}), $\mu$ does not contain $B^{*\mu_{p+2} \ldots 
\mu_n}$
and is at least quadratic in the antifields.  
Without loss of generality, we can assume that it is
exactly quadratic in the antifields and that it
does not contain derivatives, since $\delta$ and $\pp$
are both linear and bring in one derivative. [That
$\mu$ can be assumed to be quadratic is obvious.  That it
can also be assumed not to contain the derivatives
of the antifields is a little
bit less obvious since we allow for explicit $x$-dependence, but
can be easily checked by expanding $\mu$ and $b^\rho$ according
to the number of derivatives of the variables and using
the triviality of the cohomology of $\delta$ modulo
$d$ in the space of local forms that are at least quadratic in
the fields and the antifields].
Thus we can write,
$$\mu = \sigma_{\mu_1 \ldots \mu_n} B^{*\mu_1 \ldots
\mu_{p}}_{(1)} B^{*\mu_{p+1} \ldots \mu_n}_{(2p+1-n)}
+ \mu', $$ where $\mu'$ neither involves $B^{*\mu_{p+2}
\ldots
\mu_n}_{(2p+2-n)}$
nor $B^{*\mu_{p+1} \ldots \mu_n}_{(2p+1-n)}$. We have 
explicitly indicated the
antighost number in parentheses in order to keep
track of it.   
Inserting this form of $\mu$ in (\ref{cohomo}) one finds by taking
the Euler-Lagrange derivatives of the result by $B^{*\mu_{p+1}
\ldots \mu_n}_{(2p+1-n)}$ that
$\sigma_{\mu_1
\ldots
\mu_n}$ is equal to $k \epsilon_{\mu_1\ldots \mu_n}$ if 
$2p+1-n>1$ (if $2p+1-n=1$, see below).  One can
then successively eliminate $B^{*\mu_{p} \ldots \mu_n}_{(2p-n)}$,
$B^{*\mu_{p-1} \ldots \mu_n}_{(2p-n-1)}$ etc 
from $\mu$. 

So the question ultimately boils down to: is
$$k \epsilon_{\mu_1\ldots \mu_{2j}} B^{*\mu_1 \dots \mu_j}_{(p+1-j)}
\delta
B^{*\mu_{j+1} \dots \mu_{2j}}_{(p+1-j)}\quad  \text{for $n$ even
$=2j$}$$ or
$$k \epsilon_{\mu_1\ldots \mu_{2j+1}} B^{*\mu_1 \dots 
\mu_{j+1}}_{(p-j)}
\delta
B^{*\mu_{j+2} \dots \mu_{2j+1}}_{(p+1-j)}\quad \text{for $n$ odd
$=2j+1$}$$
$\delta$-exact modulo $d$, i.e., of the form
$\delta \nu + \pp_\rho c^\rho$, where $\nu$
does not involve the antifields $B^*_s$ for
$s>p+1-j$ ($n$ even) or $s>p-j$ ($n$ odd)?

That the answer to this question
is negative unless $k=0$ and accordingly that $a$ is
trivial, is easily seen by trying to construct
explicitly $\nu$.  We treat for
definiteness the case $n$ even ($n=2j$). 

One
has $$\nu = \lambda_{\mu_1\ldots \mu_{2j}} 
B^{*\mu_1 \dots \mu_j}_{(p+1-j)}
B^{*\mu_{j+1} \dots \mu_{2j}}_{(p+1-j)}$$ where 
$\lambda_{\mu_1\ldots \mu_{2j}}$ is antisymmetric (respectively
symmetric) for the exchange of $(\mu_1\ldots \mu_{j})$ with
$(\mu_{j+1} \dots \mu_{2j})$ if $j$ is even (respectively
odd) (the $j$-form ${\overline B}^*_{(p+1-j)}$ is odd 
by assumption and this can happen only if the components
$B^{*\mu_1 \dots \mu_j}_{(p+1-j)}$ are odd for $j$ even, or
even for $j$ odd).  From
the equation
\begin{equation}
k \epsilon_{\mu_1\ldots \mu_{2j}} B^{*\mu_1 \dots \mu_j}_{(p+1-j)}
\delta
B^{*\mu_{j+1} \dots \mu_{2j}}_{(p+1-j)} = \delta \nu + \pp_\rho c^\rho
\end{equation}
one gets
\begin{equation}
k \epsilon_{\mu_1\ldots \mu_{2j}} B^{*\mu_1 \dots \mu_j}_{(p+1-j)}
\pp_\rho B^{* \rho \mu_{j+1} \dots \mu_{2j}}_{(p-j)} =\pm
2 \lambda_{\mu_1\ldots \mu_{2j}} B^{*\mu_1 \dots \mu_j}_{(p+1-j)}
\pp_\rho B^{* \rho \mu_{j+1} \dots \mu_{2j}}_{(p-j)}
+ \pp_\rho c^\rho
\end{equation}
Taking the Euler-Lagrange derivative of this equation
with respect to $B^{*\mu_1 \dots \mu_j}_{(p+1-j)}$ yields next
$$ \big( k\epsilon_{\mu_1\ldots \mu_{2j}}\mp
2 \lambda_{\mu_1\ldots \mu_{2j}} \big)
\pp_\rho B^{* \rho \mu_{j+1} \dots \mu_{2j}}_{(p-j)} = 0,$$ which
implies $k \epsilon_{\mu_1\ldots \mu_{2j}}=\pm
2 \lambda_{\mu_1\ldots \mu_{2j}}$.  This contradicts the symmetry
properties of $\lambda_{\mu_1\ldots \mu_{2j}}$, unless $k=0$,
as we wanted to prove.

\subsection{Characteristic Cohomology}

By means of the isomorphism Theorems
{\bf\ref{CharAnddelta}} and {\bf{\ref{ISO}}}, our
results on
$H^n_*(\delta \vert d)$ can be translated in terms of the
characteristic cohomology as follows.

\noindent
(i) If $n-p-1$ is odd, the only non-vanishing group of the 
characteristic
cohomology in form degree $<n-1$ is $H^{n-p-1}_{char}(d)$, which is
one-dimensional.  All the other groups vanish.  One may take as
representatives for $H^{n-p-1}_{char}(d)$ the cocycles $k
{\overline H}$.  Similarly, the only non-vanishing group 
$H^{j}(\Delta)$ with
$j<n-1$ is $H^{n-p-1}(\Delta)$ with representatives $k {\tilde H}$
and the only non-vanishing group $H^n_i(\delta \vert d)$ with $i>1$
is
$H^n_{p+1} (\delta \vert d)$ with representatives $k {\overline
B}^*_{p+1}$.

\noindent
(ii) If $n-p-1$ is even, there is further cohomology.
The degrees in which there is non trivial cohomology are
multiples of
$n-p-1$ (considering  again values of  the
form degree strictly smaller than $n-1$).
Thus, there is characteristic
cohomology only in degrees $n-p-1$, $2(n-p-1)$, $3(n-p-1)$ etc.
The corresponding groups are one-dimensional and one may take as
representatives $k{\overline H}$, $k({\overline H})^2$,
$k({\overline H})^3$ etc. There is also non-vanishing
$\Delta$-cohomology for the same values of the
$\Delta$-degree, with representative cocycles given
by $k {\tilde H}$, $k ({\tilde H})^2$, $k ({\tilde H})^3$, etc.
By expanding these cocycles according to the form degree
and keeping the terms of form degree $n$, one gets
representatives for the only non-vanishing groups $H^n_i(\delta \vert 
d)$
( with $i>1$), which are respectively $H^n_{p+1}(\delta \vert d)$,
$H^n_{p+1 -(n-p-1)}$, $H^n_{p+1 -2(n-p-1)}$ etc.

An immediate consequence of our analysis is the following
useful theorem:
\begin{theorem} \label{useful}
If the polynomial $P^k(H)$ of form degree $k<n$ in the
curvature $(p+1)$-form $H$ is $\delta$-exact
modulo $d$ in the invariant algebra ${\cal I}$, then $P^k(H)=0$.
\end{theorem}
\proof{The theorem is straightforward in the
algebra of $x$-independent local forms; it follows from a direct
derivative counting argument.  To prove it when an explicit
$x$-dependence is allowed, one proceeds as follows.
If $P^k(H) = \delta a^k_1 + d a^{k-1}_0$ where $a^k_1$ and $a^{k-1}_0
\in {\cal I}$,
then $da^k_1 + \delta a^{k+1}_2 = 0$ for some invariant $a^{k+1}_2$.
Using the results on the cohomology of $\delta$ modulo $d$ that
we have just established, we know that $a^k_1$
differs from the component of form degree $k$ and antighost
number $1$ of a polynomial $Q({\tilde H})$ by a term of the
form $\delta \rho
+ d \sigma$, where $\rho$ and $\sigma$ are both invariant.
But then, $\delta a^k_1$ has the form $d \big([Q({\tilde H})]^{k-1}_0
+ \delta \sigma \big)$, which implies $P^k(H) = d 
\big(-[Q({\tilde H})]^{k-1}_0
-\delta \sigma + a^{k-1}_0 \big)$, i.e., $P^k(H) = d$(invariant).
According to the theorem on the invariant cohomology of $d$, this
can  only occur if $P^k(H) = 0$.}

\subsection{Characteristic cohomology in the algebra of 
 $x$-inde\-pen\-dent local forms} \label{xmudep}

Let us denote $({\tilde H})^m$ by $P_m$ ($m = 0, \dots, K$).
We have just shown (i) that the most general cocycles of the
$\Delta$-cohomology are given, up to trivial terms,
by the linear combinations $\lambda_m P_m$ with
$\lambda_m$ real or complex numbers; and (ii)
that if $\lambda_m P_m$ is
$\Delta$-exact, then the $\lambda_m$ are all
zero.  In establishing these results, we allowed
for an explicit $x$-dependence of the local
forms (see comments after the proof of
Theorem {\bf \ref{Cohop}}).  How are our results affected if we work
exclusively with local forms with no explicit $x$-dependence?

In the above analysis, it is in calculating the 
cocycles that arise in anti\-ghost number
$<p+1$ that we used the $x$-dependence of the local forms,
through the isomorphism 
$H^{p+1-J}_{char}(d) \simeq H^n_{n-p-1+J}(\delta \vert d)$.
If the local exterior forms are not allowed to depend
explicitly on $x$, one must take into account the constant $k$-forms
($k>0$).  The derivation goes otherwise
unchanged and one finds that
the cohomology of $\Delta$
in the space of $x$-independent local forms 
is given by the polynomials
in the $P_m$ with coefficients $\lambda_m$ that
are constant forms, $\lambda_m = \lambda_m(dx)$.
In addition, if $\lambda_m P_m$ is $\Delta$-exact,
then, $\lambda_m P_m = 0$ for each $m$.  One cannot
infer from this equation that $\lambda_m$ vanishes, because it is
an exterior form.  One can simply assert that the
components of $\lambda_m$ of form degree $n-m(n-p-1)$ or
lower are zero (when multiplied by $P_m$, the other
components of $\lambda_m$ yields forms of degree $>n$
that identically vanish, no matter what these other
components are).

It will be also useful in the sequel to know the
cohomology of $\Delta'$, where $\Delta'$ is the part of
$\Delta$ that acts only on the fields and antifields,
and not on the explicit $x$-dependence.  One has
$\Delta = \Delta' + d_x$, where 
$d_x \equiv \pp^{explicit}/\pp x^\mu$ sees
only the explicit $x$-dependence.  By the
above result, the cohomology
of $\Delta'$ is clearly given by the polynomials in the
$P_m$ with coefficients $\lambda_m$ that are now
arbitrary spacetime forms, $\lambda_m = \lambda_m(x, dx)$.

\section{Characteristic Cohomology in the General Case}

To compute the cohomology $H^n_i(\delta \vert d)$ 
for an arbitrary set of $p$-forms, 
one proceeds along the lines of the Kunneth theorem.
Let us illustrate explicitly the procedure
for two fields $B^1_{\mu_1 \dots \mu_{p_1}}$  and
$B^2_{\mu_1 \dots \mu_{p_2}}$. One may split
the  differential $\Delta$ as a sum of terms
with definite $N_a$-degrees,
\begin{equation}
\Delta = \Delta_1 + \Delta_2 +d_x
\label{split}
\end{equation}
(see  (\ref{Na})).  In (\ref{split}), $d_x$ 
leaves both
$N_1$ and $N_2$ unchanged.  By contrast, $\Delta_1$
increases $N_1$ by one unit without changing $N_2$,
while $\Delta_2$ increases $N_2$  by one unit without changing
$N_1$. The differential $\Delta_1$ acts only on the
fields $B^1$ and its associated antifields (``fields and
antifields of the first set"), whereas the differential $\Delta_2$
acts only on the fields $B^2$ 
and its associated antifields (``fields and
antifields of the second set"). Note that $\Delta_1 +
\Delta_2 = \Delta'$. 

Let $a$ be a cocycle of $\Delta$ with $\Delta$-degree
$< n-1$.  Expand $a$ according to the $N_1$-degree,
\begin{equation}
a= a_0 + a_1 + a_2 + \dots + a_m, \; N_1(a_j) = j.
\end{equation}
The equation $\Delta a = 0$ implies $\Delta_1 a_m =0$ for the term
$a_m$ of highest $N_1$-degree. Our
analysis of the $\Delta'$-cohomology for a single $p$-form
then yields $a_m = c_m ({\tilde H}^1)^k + \Delta_1$(something),
where $c_m$ involves only the fields and antifields of the
second set, as well as $dx^\mu$ and
possibly $x^\mu$.
There can be no conserved current in $a_m$ since we assume the
$\Delta$-degree of $a$ - and thus of each $a_j$ - to be strictly
smaller than $n-1$.  Now, the exact term in $a_m$ can be
absorbed by adding to $a_m$ a $\Delta$-exact term, through a
redefinition of $a_{m-1}$.  Once this is done, one finds
that the next equation for $a_m$ and $a_{m-1}$ following
from $\Delta a = 0$ reads
\begin{equation}
[(\Delta_2 + d_x) c_m] ({\tilde H}^1)^k + \Delta_1 a_{m-1} = 0.
\end{equation}
But we have seen that $\lambda_m ({\tilde H}^1)^k$ cannot be exact
unless it is zero,
and thus this last equation implies both
\begin{equation}
[(\Delta_2 + d_x) c_m] ({\tilde H}^1)^k = 0
\label{vanish}
\end{equation}
and 
\begin{equation}
\Delta_1 a_{m-1} = 0.
\end{equation}
Since $({\tilde H}^1)^k$ has independent
form components in degrees
$k(n-p-1)$, $k(n-p-1) +1$ up to degree $n$, we
infer from (\ref{vanish})
that the form components of $(\Delta_2 + d_x) c_m$
of degrees $0$ up to degree $n- k(n-p-1)$ are zero. If we
expand $c_m$ itself according to the form degree,
$c_m = \sum c_{m}^i$, we obtain the equations
\begin{equation}
\delta c_m^i + d c_m^{i-1} = 0, \; i= 1, \dots, n- k(n-p-1),
\end{equation}
and
\begin{equation}
\delta c_m^0 = 0.
\end{equation}
Our analysis of the relationship between the $\Delta$-cohomology
and the cohomology of $\delta$ modulo $d$ indicates then that one
can redefine the terms of form degree $>n- k(n-p-1)$ of $c_m$
in such a way that $\Delta c_m = 0$.
This does not affect the product $c_m ({\tilde H}^1)^k$. 
We shall assume that the 
(irrelevant) higher order terms in $c_m$ have been chosen in that manner.
With that choice, $c_m$ is given, up to trivial terms that can
be reabsorbed, by $\lambda_m ({\tilde H}^2)^l$,
with $\lambda_m$ a number (or a constant form in the case of
${\cal P}_{-}$), so that
$a_m =
\lambda_m ({\tilde H}^2)^l ({\tilde H}^1)^k$ is a $\Delta$-cocycle
by itself. One next successively repeats the analysis for
$a_{m-1}$,
$a_{m-2}$ until one reaches the desired conclusion that
$a$ may indeed be assumed to be a polynomial in the
${\tilde H}^a$'s, as claimed above.

The non-triviality of the polynomials in the ${\tilde H}^a$'s
is also easy to prove.  If $P({\tilde H}) = \Delta \rho$, with
$\rho = \rho_0 + \rho_1 + \dots + \rho_m$, $N_1(\rho_k) = k$, then
one gets at $N_1$-degree $m+1$ the condition $(P({\tilde H}))_{m+1}
= \Delta_1 \rho_m$, which implies $(P({\tilde H}))_{m+1}=0$ and
$\Delta_1 \rho_m=0$, since no polynomial in ${\tilde H}^1$ is
$\Delta_1$-trivial, except zero.  It follows that $\rho_m =
u ({\tilde H}^1)^m$ up to trivial terms that play no role, where
$u$ is a function of the variables of the second set as well as of
$x^\mu$ and $dx^\mu$.  The equation of order $m$ implies then
$(P({\tilde H}))_m = \big( (\Delta_2 + d_x)u \big) ({\tilde H}^1)^m
+ \Delta_1 \rho_{m-1}$.  The non-triviality of the polynomials in
${\tilde H}^1$ in $\Delta_1$-cohomology yields next $\Delta_1 
\rho_{m-1}
=0$ and $(P({\tilde H}))_m = \big( (\Delta_2 + d_x)u \big) 
({\tilde H}^1)^m$.
Since the coefficient of $({\tilde H}^1)^m$ in $(P({\tilde H}))_m$ is
a polynomial in ${\tilde H}^2$, which cannot be 
$(\Delta_2 + d_x)$-exact,
one gets in fact $(P({\tilde H}))_m= 0$ and $(\Delta_2 + d_x)u =0$.
It follows that $\rho_m$ fulfills $\Delta \rho_m=0$ and can be dropped.
The analysis goes on in the same way at the lower values of the
$\Delta_1$-degree, until one reaches the desired conclusion
that the exact polynomial $P({\tilde H})$ indeed vanishes.

In view of the isomorphism between the characteristic cohomology 
and $H^*(\Delta)$, 
this completes the proof of Theorem {\bf \ref{MainResult}}
in the case of two $p$-forms.
The case of more $p$-forms is treated similarly.

\section{Invariant Characteristic Cohomology}

\subsection{Isomorphism theorems for the invariant
co\-ho\-mo\-lo\-gies} To compute the invariant characteristic
cohomology, we proceed as follows.  First, we establish
isomorphism theorems between $H^{k,inv}_{char}(d)$, 
$H^{n,inv}_{n-k}(\delta \vert d)$ and $H^{k,inv}(\Delta)$.
Then, we compute $H^{k,inv}(\Delta)$ for a single $p$-form.
Finally, we extend the calculation to an arbitrary systems 
of $p$-forms.

\begin{theorem} \label{CharAnddeltaInv}
\begin{eqnarray}
\frac{H^{k,inv}_{char}(d)}{{\cal H}^k} &
\simeq& H^{n,inv}_{n-k}(\delta \vert d), \; 0\leq k<n
\label{CharAndDeltaInv1}\\
0 &\simeq& H^{n,inv}_{n+k}(\delta \vert d),\; k>0
\label{CharAndDeltaInv2}
\end{eqnarray}
\end{theorem}

\begin{theorem}
The invariant cohomology of $\Delta$ is isomorphic to the
invariant characteristic cohomology,
\begin{eqnarray}
H^{k,inv}(\Delta)
\simeq H^{k,inv}_{char}(d), \; 0 \leq k \leq n.
\end{eqnarray}
\label{ISObis}
\end{theorem}

\proof{First we prove (\ref{CharAndDeltaInv1}).  To
that end we observe that the map $m$ introduced in the
demonstration of Theorem {\bf \ref{CharAnddelta}} associates
$H^{k,inv}_{char}(d)$ and $ H^{n,inv}_{n-k}(\delta \vert d)$.
Indeed, in the expansion (\ref{Tower}) for $a$, all the
terms can be assumed to be invariant on account of Theorem
{\bf \ref{deltachi}}.  The surjectivity of $m$ is also
direct, provided that the polynomials in the
curvature $P(H)$ are not trivial in $H^*(\delta \vert d)$,
which is certainly the case if there is a single $p$-form 
(Theorem {\bf
\ref{useful}}).  We shall thus use Theorem {\bf \ref{CharAnddeltaInv}}
first only in the case of a single $p$-form.  We shall then prove that
Theorem {\bf \ref{useful}} extends to an arbitrary system of
forms of various form degrees, so that the proof  of
Theorem {\bf \ref{CharAnddeltaInv}} will be completed.

To compute the kernel of $m$, consider an element 
$a^k_0 \in {\cal I}$ such that the corresponding
$a^n_{n-k}$ is trivial in  $ H^{n,inv}_{n-k}(\delta \vert d)$.
Then, again as in the proof of Theorem {\bf \ref{CharAnddelta}},
one finds that all the terms in the expansion (\ref{Tower})
are trivial, except perhaps $a^k_0$, which fulfills
$da^k_0 + \delta d b^k_1 = 0$, where $b^k_1 \in {\cal I}$
is the $k$-form appearing in the equation expressing the
triviality of $a^{k+1}_1$, $a^{k+1}_1 =
d b^k_1 + \delta b^{k+1}_2$. This implies
$d(a^k_0 - \delta  b^k_1) = 0$, and thus, by
Theorem {\bf \ref{invpoincare}}, $a^k_0 = P + db^{k-1}_0
+\delta b^k_1$
with $P \in {\cal H}^k$ and $b^{k-1}_0 \in {\cal I}$.
This proves (\ref{CharAndDeltaInv1}), since $P$ is not trivial
in $H^*(\delta \vert d)$ (Theorem {\bf \ref{useful}}). [Again,
we are entitled to use this fact only for a single $p$-form
until we have proved the non-triviality of $P$ in the general
case].

The proof of (\ref{CharAndDeltaInv2}) is a direct consequence of
Theorem {\bf \ref{deltachi}} and parallels step by step
the proof of a similar statement demonstrated for
1-forms in \cite{BBH2} (lemma 6.1).  It will not be
repeated here.  Finally, the proof of Theorem
{\bf \ref{ISObis}}  amounts to observing that the map
$m'$ that sends $[a^k_0]$ on $[a]$ (Equation (\ref{Tower}))
is indeed well defined in cohomology, and
is injective as well as surjective (independently of
whether $P(H)$ is trivial in the invariant cohomology
of $\delta$ modulo $d$).}
Note that if the forms do not depend explicitly on $x$, on
must replace (\ref{CharAndDeltaInv1}) by
\begin{equation}
\frac{H^{k,inv}_{char}(d)}{(\Lambda \otimes {\cal H})^k}
\simeq H^{n,inv}_{n-k}(\delta \vert d).
\end{equation}

\subsection{Case of a single $p$-form gauge field}

Theorem {\bf \ref{invardelta}} enables one to compute also
the invariant characteristic cohomology for a single
$p$-form gauge field.  Indeed,
this theorem implies that $H^{n,inv}_{n-k}(\delta \vert d)$
and $H^n_{n-k}(\delta \vert d)$ actually coincide since
the cocycles of $H^n_{n-k}(\delta \vert d)$ are invariant
and the coboundary conditions are equivalent.  The isomorphism
of Theorem {\bf \ref{CharAnddeltaInv}} 
shows then that the invariant characteristic
cohomology for a single $p$-form gauge field in form degree $<n-1$
is isomorphic to  the subspace of form degree $<n-1$ of
the direct sum  ${\cal H} \oplus {\cal {\overline H}}$.  Since
the product $H \wedge {\overline H}$ has form degree $n$, 
which exceeds $n-1$, this
is the same as the subspace ${\cal W}$ of Theorem 
{\bf \ref{MainResult2}}. The invariant characteristic cohomology
in form degree $k<n-1$ is thus given by $({\cal H} \otimes
{\cal {\overline H}})^k$, i.e., by the invariant polynomials
in the curvature $H$ and its dual ${\overline H}$ with 
form degree $<n-1$.  Similarly, by the isomorphism of
Theorem {\bf \ref{ISObis}}, the invariant cohomology
$H^{k,inv}(\Delta)$ of $\Delta$ is given by the polynomials
in ${\tilde H}$ and $H$ with $\Delta$-degree smaller 
than $n-1$.

\subsection{Invariant cohomology of $\Delta$ in the general case}

The invariant $\Delta$-cohomology for an arbitrary
system of $p$-form gauge fields follows again from a straightforward
application of the Kunneth formula and is thus
given by the polynomials in the ${\tilde H}^a$'s
and $H^a$'s  with $\Delta$-degree smaller 
than $n-1$.  The explicit proof of this statement works as in
the non-invariant case (for that matter, it is actually more
convenient to use as degrees not $N_1$ and $N_2$, but rather,
degrees counting the number of derivatives of the invariant
variables $\chi$'s. These degrees have the advantage that the
cohomology is entirely in degree zero).  In particular, none of the
polynomials in the ${\tilde H}^a$'s and $H^a$'s is trivial.

The isomorphism of Theorem
{\bf \ref{ISObis}} implies next that the invariant characteristic
cohomology $H^{k,inv}_{char} (d)$ ($k<n-1$)
is given by the polynomials in the 
curvatures $H^a$ and their duals ${\overline H}^a$, restricted to
form degree smaller than $n-1$.  Among these, those
that involve the curvatures $H^a$ are weakly exact, but not
invariantly so. The property of Theorem {\bf \ref{useful}}
thus extends as announced to an arbitrary system of dynamical gauge
forms of various form degrees.

Because the forms have now different form degrees, one may have
elements in $H^{k,inv}_{char} (d)$ ($k<n-1$) that involve
both the curvatures and their duals.  For instance,
if $B^1$ is a $2$-form and $B^2$ is a $4$-form, the
cocycle $H^1 \wedge {\overline H}^2$ is a ($n-2$)-form.
It is trivial in $H^k_{char} (d)$, but not in
$H^{k,inv}_{char} (d)$.

\section{Invariant cohomology of $\delta$ mod $d$}
 
The easiest way to work out explicitly $H^{n,inv}_{n-k}(\delta \vert d)$
in the general case is to use the above isomorphism
theorems, which we are now entitled to do.  Thus, one
starts from $H^{k,inv}(\Delta)$
and one works out the component of form degree $n$ in the
associated cocycles. 

Because one has elements in $H^{k,inv}(\Delta)$
that involve simultaneously both the curvatures and their 
$\Delta$-invariant
duals ${\tilde H}^a$, the property that
$H^{n,inv}_{n-k}(\delta \vert d)$
and $H^n_{n-k}(\delta \vert d)$ coincide can no longer hold.
In the previous example, one would find that $H^{(1)}_{\lambda
\mu \nu} B^{*(2)\lambda \mu \nu}$, which has antighost number
two, is a $\delta$-cocycle modulo $d$, but it cannot
be written invariantly so.
An important case where the isomorphism
$H^{n,inv}_{n-k}(\delta \vert d) \simeq
H^n_{n-k}(\delta \vert d)$ ($k>1$) does hold, however, is when the 
forms
have all the same degrees.

To write down the generalization of Theorem {\bf \ref{invardelta}}
in the case of $p$-forms of different degrees, let
$P(H^a, {\tilde H}^a)$ be a polynomial in the curvatures
$(p_a +1)$-forms $H^a$ and their $\Delta$-invariant
duals ${\tilde H}^a$.  One has $\Delta P = 0$.  We shall be
interested in polynomials of $\Delta$-degree $<n$ that are
of degree $>0$ in both $H^a$ and ${\tilde H}^a$.  The
condition that $P$ be of degree $>0$ in $H^a$ implies that
it is trivial (but not invariantly so), while the condition
that it be of degree $>0$ in ${\tilde H}^a$ guarantees that
when expanded according to the antighost number, $P$ has
non-vanishing components of antighost number $>0$,
\begin{equation}
P= \sum_{j=k}^n [P]^j_{j-k}.
\end{equation}
{}From $\Delta P = 0$, one has $\delta [P]^n_{n-k} + d
[P]^{n-1}_{n-k-1} =0$.

There is no polynomial in $H^a$ and ${\tilde H}^a$ with the required
properties if all the antisymmetric tensors $B^a_{\mu_1 \dots 
\mu_{p_a}}$
have the same form degree ($p_a = p$ for all $a$'s) since the 
product $H^a {\tilde H}^b$ has necessarily
$\Delta$-degree $n$.  When there are tensors of different form degrees,
one can construct, however, polynomials $P$ with the given
features.

The analysis of the previous subsection implies straightforwardly:
\begin{theorem} \label{xx}
Let $a_q^n=a_q^n(\chi) \in {\cal I}$
be an invariant local $n$-form of
antighost number $q>0$.
If $a_q^n$ is $\delta$-exact modulo
$d$, $a_q^n=\delta \mu_{q+1}^n + d\mu_q^{n-1}$, then
one has
\begin{equation}
a_q^n= [P]_q^n + \delta \mu_{q+1}^{'n} + d\mu_q^{'n-1}
\end{equation}
for some polynomial $P(H^a, {\tilde H}^a)$ of degree
at least one in $H^a$ and at least one in ${\tilde H}^a$, and
where $\mu_{q+1}^{'n}$ and $\mu_q^{'n-1}$ can be assumed to
depend only on the $\chi$'s, i.e.,
to be invariant.
In particular, if all the $p$-form gauge fields have the same
form degree, $[P]_q^n$ is absent and one has
\begin{equation}
a_q^n= \delta \mu_{q+1}^{'n} + d\mu_q^{'n-1}
\end{equation}
where one can assume that $\mu_{q+1}^{'n}$ and $\mu_q^{'n-1}$
are invariant ($\mu_{q+1}^{'n}$ and
$\mu_q^{'n-1}$ $\in {\cal I}$).
\end{theorem}

\section{Remarks on Conserved Currents}
\label{Currentssec}

That the characteristic cohomology  is
finite-dimensional and that it is entirely
generated by the duals ${\overline H}^a$'s 
to the field strengths holds only in form degree
$k<n-1$. 
This property is not true
in form degree equal to $n-1$, where there are conserved
currents that cannot be expressed in terms of the forms
${\overline H}^a$,
even up to trivial terms. 

An infinite number of conserved currents that cannot be expressible
in terms of the forms ${\overline H}^a$
are given by,
\begin{eqnarray}
T_{\mu \nu \alpha_1 \ldots \alpha_s \beta_1
\ldots \beta_r}= -{1\over 2}({1\over p!}H_{ \mu \rho_1 \ldots
\rho_p, \alpha_1 \ldots \alpha_s} H^{\ \rho_1 \ldots
\rho_p}_{\nu\hspace{0.8cm},\beta_1 \ldots \beta_r}
\nonumber \\ -{1\over (n-p-2)!}H^*_{ \mu \rho_2 \ldots
\rho_{n-p-1}, \alpha_1 \ldots \alpha_s} H^{*\rho_2 \ldots
\rho_{n-p-1}}_{\nu\hspace{1.6cm},\beta_1 \ldots \beta_r}).
\label{gencurr}
\end{eqnarray}
These quantities are easily checked to be conserved,
\begin{equation}
T^{\mu}_{\nu \alpha_1 \ldots \alpha_s \beta_1\ldots \beta_r ,\mu}
\equiv 0,
\end{equation}
and generalize the conserved currents given in
\cite{Lipkin1,Morgan1,Kibble1,OConnell1} for free
electromagnetism.  They are symmetric for the exchange of $\mu$
and $\nu$ and are duality invariant in the critical dimension
$n = 2p+2$ where the field strength and its dual have
same form degree $p+1$.  In this critical dimension, there
are further conserved currents which generalize the ``zilches",
\begin{eqnarray}
Z^{\mu\nu\alpha_1\ldots\alpha_r\beta_1\ldots\beta_s}&=&H^{\mu\sigma_1
\ldots\sigma_p,\alpha_1\ldots\alpha_r}
H_{\ \ \sigma_1\ldots\sigma_p}^{*\nu \hspace{0.9cm} ,\beta_1\ldots
\beta_s} \nonumber \\&&-
H^{*\mu\sigma_1
\ldots\sigma_p,\alpha_1\ldots\alpha_r}
H_{\ \ \sigma_1\ldots\sigma_p}^{\nu \hspace{0.9cm} ,\beta_1\ldots
\beta_s}. 
\end{eqnarray}

Let us prove that the conserved currents (\ref{gencurr}) which contain
an
even total number of derivatives are  not trivial in the space of
$x$-independent local forms. To avoid cumbersome notations we will 
only
look at the currents with no $\beta$ indices.  One may reexpress
(\ref{gencurr}) in terms of the field strengths as
\begin{eqnarray}
T^{\mu \nu \alpha_1 \ldots \alpha_m}&=& -{1\over 2
p!}(H^{ \mu \sigma_1 \ldots
\sigma_p, \alpha_1 \ldots \alpha_m} H^{\nu}_{\ \sigma_1 \ldots
\sigma_p}+
H^{ \mu \sigma_1 \ldots
\sigma_p} H^{\nu\hspace{.8cm}, \alpha_1 \ldots \alpha_m}_{\
\sigma_1 \ldots
\sigma_p})
\nonumber \\
&&+\eta^{\mu\nu}{1\over 2(p+1)!}H_{\sigma_1 \ldots
\sigma_{p+1}}H^{\sigma_1 \ldots
\sigma_{p+1},\alpha_1 \ldots \alpha_m}.
\end{eqnarray}
If one takes the divergence of this expression one gets,
\begin{eqnarray}
T^{\mu \nu \alpha_1 \ldots
\alpha_m}_{\hspace{1.4cm},\mu}&=& \delta K^{\nu \alpha_1 \ldots
\alpha_m}
\end{eqnarray}
where $K^{\nu \alpha_1 \ldots\alpha_m}$ differs from
$k H^{\nu\hspace{.8cm} ,\alpha_1\ldots\alpha_m}_{\
\sigma_1\ldots\sigma_p}B^{*\sigma_1\ldots\sigma_p}$
by a divergence.
It is easy to see that $T^{\mu\nu \alpha_1\ldots\alpha_m}$ is
trivial if and only if
$H^{\nu\hspace{.8cm} ,\alpha_1\ldots\alpha_m}_{\
\sigma_1\ldots\sigma_p}$ $B^{*\sigma_1\ldots\sigma_p}$ is
trivial. So the question is: can we write,
\begin{equation}
H^{\nu\hspace{.8cm} ,\alpha_1\ldots\alpha_m}_{\
\sigma_1\ldots\sigma_p}B^{*\sigma_1\ldots\sigma_p}=\delta
M^{\nu\alpha_1 \ldots \alpha_m}+\partial_\rho N^{\rho
\nu\alpha_1 \ldots \alpha_m},\label{testreef}
\end{equation}
for some $M^{\nu\alpha_1 \ldots \alpha_m}$ and
$N^{\rho \nu\alpha_1 \ldots \alpha_m}$?
Without loss of generality, one can assume that $M$ and $N$
have the Lorentz transformation properties indicated
by their indices type (the parts of
$M$ and
$N$ transforming in other representations would cancel by
themselves). 
We can also decompose both sides of
Eq. \eqref{testreef} according to tensors of given symmetry type
(under the permutations of $\nu \alpha_1\ldots \alpha_m$); in
particular \eqref{testreef} implies:
\begin{equation}
H^{(\nu\hspace{.9cm} ,\alpha_1\ldots\alpha_m)}_{\ \ 
\sigma_1\ldots\sigma_p}B^{*\sigma_1\ldots\sigma_p}=\delta
M^{(\nu\alpha_1 \ldots \alpha_m)}+\partial_\rho N^{\rho
(\nu\alpha_1 \ldots \alpha_m)}.\label{testreef2}
\end{equation}

Moreover, according to Theorem {\bf \ref{invardelta}}, one can also
assume that $M$ and $N$ are gauge invariant, i.e., belong to
${\cal I}$. If one takes into account all the symmetries of the
left-hand side and use the identity $dH=0$, Eq. \eqref{testreef2} 
reduces to,
\begin{equation}
H^{(\nu\hspace{.9cm} ,\alpha_1\ldots\alpha_m)}_{\ \ 
\sigma_1\ldots\sigma_p}B^{*\sigma_1\ldots\sigma_p}=\partial_\rho
N^{\rho (\
\nu\alpha_1 \ldots \alpha_m)}+ {\hbox{terms that vanish
on-shell}}.
\end{equation}
This is a consequence of the fact that there is no
suitable polynomial in
$(\partial)H_{\mu_1\ldots\mu_r}$ and
$(\partial)B^{*}_{\mu_1\ldots\mu_s}$ which can be present in
$M^{(\nu\alpha_1 \ldots \alpha_m)}$. 

If one takes the
Euler-Lagrange derivative of this equation with respect to
$B^{*\sigma_1\ldots\sigma_p}$ one gets,
\begin{equation}
H_{(\nu
\vert \sigma_1\ldots\sigma_p\vert ,\alpha_1\ldots\alpha_m)}
\approx 0,
\end{equation}
which is not the case. 

This shows that
$T^{\mu \nu \alpha_1 \ldots \alpha_m}$ (with $m$ even) is not trivial 
in
the algebra of $x$-independent local forms.  It then follows, by
a mere counting of derivative argument, that the
$T^{\mu \nu \alpha_1 \ldots \alpha_m}$ define independent
cohomological classes and cannot be expressed as polynomials
in the undifferentiated dual to the field strengths 
${\overline H}$ with coefficients that are constant forms.

The fact that the conserved currents are not
always expressible in terms of the forms
${\overline H}^a$ makes the validity
of this property for higher
order conservation laws more striking.
In that respect, it should be indicated
that the computation of the characteristic
cohomology in the 
algebra generated by the ${\overline H}^a$
is clearly a trivial question.  The non trivial issue is
to demonstrate that this computation does not miss other cohomological
classes in degree $k<n-1$.

Finally, we point out that the 
conserved currents can all be redefined so as to be
strictly gauge-invariant, apart from a few of them 
whose complete list can be systematically
determined for each given system of $p$-forms.
This point will be fully established in Section
{\bf{\ref{conscurent}}}; it extends to higher degree antisymmetric
tensors a property established in \cite{BBH4} for one-forms
(see also \cite{Torre2} in this context).

\section{Introduction of Gauge Invariant Interactions}

The analysis of the characteristic cohomology proceeds in the same
fashion if one adds to the Lagrangian (\ref{Lagrangian}) interactions
that involve gauge invariant terms of higher dimensionality. These
interactions may increase the derivative order of the field equations.
The resulting theories should be regarded as effective theories and
can be handled through a systematic perturbation expansion
\cite{Weinberg2}.

The new equations of motion read,
\begin{equation}
\partial_\mu {\cal L}^{a\mu \mu_1 \mu_2 \dots \mu_{p_a}} = 0,
\end{equation}
where ${\cal L}^{a\mu \mu_1 \mu_2 \dots \mu_{p_a}}$ are the 
Euler-Lagrange
derivatives of the Lagrangian with respect to the field strengths
(by gauge invariance, ${\cal L}$ involves only the field strength
components and their derivatives).  These equations can be rewritten 
as,
\begin{equation}
d {\overline {\cal L}}^a \approx 0,
\end{equation}
where ${\overline {\cal L}}^a$ is the $(n-p_a-1)$-form dual to the
Euler-Lagrange derivatives.

The Euler-Lagrange equations obey the same Noether identities as in
the free case, so that the Koszul-Tate differential takes
the same form, with ${\overline H}^a$ everywhere replaced by
${\overline {\cal L}}^a$.  It then follows that 
\begin{equation}
{\tilde {\cal L}}^a = {\overline {\cal L}}^a + 
\sum_{j=1}^{p+1} {\overline B}^{*a}_j
\end{equation}
fulfills
\begin{equation}
\Delta {\tilde {\cal L}}^a = 0.
\end{equation}
This implies, in turn, that any polynomial in the
${\tilde {\cal L}}^a$ is $\Delta$-closed. It is also clear that
any polynomial in the ${\overline {\cal L}}^a$ is
weakly $d$-closed.  By making the regularity
assumptions on the higher order terms in the
Lagrangian explained in \cite{BBH1}, one
easily verifies that these are the only cocycles
in form degree $<n-1$,
and that they are non-trivial.  The characteristic
cohomology of the free theory possesses therefore
some amount of ``robustness" since it survives
deformations.  By contrast, the infinite number
of non-trivial conserved currents is not expected
to survive interactions (even gauge-invariant ones).

[In certain dimensions, one may add Chern-Simons
terms to the Lagrangian. These interactions are not strictly
gauge invariant, but only gauge-invariant
up to a surface term.  The equations of motion still
take the form $d$(something) $ \approx 0$, but
now, that ``something" is not gauge invariant.  Accordingly,
with such interactions, some of the cocycles of the characteristic
cohomology are no longer gauge invariant.
These cocycles are removed from the invariant
cohomology].

\section{Summary of Results and Conclusions}

In this section, we have completely worked
out the characteristic cohomology $H^k_{char}(d)$
in form degree $k<n-1$ for an arbitrary
collection of free, antisymmetric tensor theories.
We have shown in particular that the
cohomological groups $H^k_{char}(d)$
are finite-dimensional and take a simple form,
in sharp contrast with $H^{n-1}_{char}(d)$,
which  is
infinite-dimensional and appears to be quite
complex.  Thus,  
even in free field theories with  an infinite number of conserved
local currents,
the existence
of higher degree local conservation laws
is quite constrained.  For instance,
in ten dimensions, there is one 
and only one (non trivial) higher
degree conservation law for
a single 2-, 3-, 4-, 6-, or 8-form
gauge field, in respective form degrees
7, 6, 5, 3 and 1.  It is $d {\overline H} \approx 0$.
For a 5-form, there are two higher degree 
conservation laws, namely
$d {\overline H} \approx 0$ and
$d ({\overline H})^2 \approx 0$, in form degrees
4 and 8.  For a 7-form, there are four  higher degree 
conservation laws, namely
$d {\overline H} \approx 0$, $d ({\overline H})^2 \approx 0$,
$d ({\overline H})^3 \approx 0$ and
$d ({\overline H})^4 \approx 0$,  in form degrees
2, 4, 6 and 8. 

Our results provide at the same
time the complete list of the  
isomorphic groups $H^k(\Delta)$, as well as of $H^n_{n-k}(\delta
\vert d)$.  We have also worked out the
invariant characteristic cohomology,
which is central in the investigation of
the BRST cohomology since it controls the antifield
dependence of BRST cohomological classes.

An interesting feature of the characteristic cohomology in
form degree $<n-1$
is its ``robustness" to the introduction of strictly
gauge invariant interactions, in contrast to the conserved
currents.

\chapter{Free theory: BRST cohomology (Part II)}\label{Chap6}
\section{Main theorems}\label{matho}

Having studied the characteristic cohomology, we can
return to our analysis of the antifield
dependent solutions of the Wess-Zumino consistency condition. In
Section {\bf\ref{prelimres}} we have shown that the most general
form for the component of highest antighost number of a BRST
cocycle is, $a^n_{g,q}=P_J \omega^J$ where $P_J$ is an element of
the invariant characteristic cohomology. According to the results
of the previous section and in particular Theorem {\bf\ref{xx}}, we
are able to specify further $a^n_{g,q}$. These terms fall into
two categories:

\begin{enumerate}
\item
If $q=1$ we have,
\begin{equation}
a^n_{g,q=1}=k_{\Delta a_1\ldots a_r}a^\Delta
{\cal Q}^{a_1\ldots a_r}_{0,g+1},
\end{equation}
where the $a^\Delta$ form a
complete set of non-trivial gauge invariant global symmetries
\cite{BBH1} of the action (\ref{Lagrangian});  they
satisfy,
$\delta a_\Delta + \partial_\mu j_\Delta^\mu =0,$
where the $j_\Delta^\mu$ form a complete set of non-trivial
conserved currents \cite{BBH1}. These belong to an infinite
dimensional space and are not all known explicitly. In the
previous section we have exhibited an infinite number of them.
\item
If $q\geq 2$ we have,
\begin{align}
a_{g,q}^n &= [P_J]^n_{-q} \omega^J \\
&=[P_{a_1 \ldots a_m}]^n_{-q}
{\cal Q}^{a_1 \ldots a_m}_{0,g+q},
\label{fakgq}
\end{align}
 where $P_{a_1 \ldots a_m}$ is a polynomial in the variables $H^a$
and $\tilde H^a$ and thus $[P_{a_1 \ldots a_m}]^n_{-q}$ is a
representative of the cohomology $H^{inv}(\delta\vert d)$. 
\end{enumerate} 
These
results clearly exhibits the fact that the existence of antifield
dependent solutions is closely tied to the presence of a
non-vanishing invariant characteristic cohomology. 
Furthermore, the component of form degree
$n$ of
$P_{a_1 \ldots a_m}$ is necessarily of antighost number
$q\leq p_M+1$. This means that the expansion of $a^n_g$ according
to the antighost number cannot stop after degree $p_M+1$.

To find
elements of the BRST cohomology we must try to complete
the possible
$a_{g,q}^n$ with components of lower antighost number in order to
have $sa^n_g + da^{n-1}_{g+1}=0$.

When $q=1$, this construction is immediate and we have,

\begin{theorem}
\label{bigtheo1}
The term $a^n_{g,q=1}=k_{\Delta a_1\ldots a_r}a^\Delta
{\cal Q}^{a_1\ldots a_r}_{0,g+1}$ can be completed in a solution
of the Wess-Zumino consistency condition $a^n_g$ given by,
\begin{equation}
a^n_g= k_{\Delta a_1\ldots a_r}(j^\Delta {\cal Q}^{a_1\ldots
a_r}_{1,g}+ a^\Delta {\cal Q}^{a_1\ldots a_r}_{0,g+1}),
\end{equation}
where $\delta a_\Delta + d j_\Delta=0.$
\end{theorem}

When $q\geq 2$ the situation is more complicated.
The next theorem classifies a first set of solutions:
\begin{theorem}
\label{bigtheo2}
If $a_{g,q}^n = [P_J]^n_{-q} \omega^J$ 
only involves the ghosts of ghosts corresponding to $p$-forms of
degree
$\geq q$, then it can be completed in a solution of the Wess-Zumino
consistency condition $a^n_g$ given by,
\begin{equation}
a^n_g= [P_{a_1 \ldots a_m}
{\cal Q}^{a_1 \ldots a_m}]_g^n.
\end{equation}
\end{theorem}

\proof{The $\Delta$-degree of $P_J$ is $n-q$. This implies
that it is a sum of terms of form
degrees $\geq n-q$. 
Furthermore, since ${\cal Q}^{a_1\ldots a_m}$ only involves the
ghosts of ghosts corresponding to $p$-forms of degrees $\geq q$,
the term of lowest degree occurring in
$\tilde s {\cal Q}^{a_1\ldots a_m}$ is at least of form degree
$\geq q+1$. Therefore $\tilde s P_{a_1 \ldots a_m}{\cal
Q}^{a_1\ldots a_m}=0$ and $a^n_g= [P_{a_1 \ldots a_m}
{\cal Q}^{a_1 \ldots a_m}]_g^n$ is a solution of the Wess-Zumino
consistency condition.}
We now investigate what happens when $a^n_{g,q}$ involves ghosts
of ghosts corresponding to $p$-forms of degrees $< q$. 

\begin{theorem}
\label{bigtheo3}
If $a_{g,q}^n = [P_J]^n_{-q} \omega^J$ 
involves ghosts of ghosts corresponding to $p$-forms of
degree
$p < q$, then it can only be completed in a solution of the
Wess-Zumino consistency condition $a^n_g$ if it is of the form
$a^n_{g,q}=k_{A_1 A_2 \ldots A_r b_1\ldots b_s}{\overline
B}^{*A_1}_{p+1} {\cal Q}^{A_2\ldots A_r b_1\ldots b_s}_{0,g+p+1}$,
where the labels $A_i$ (resp. $b_j$) correspond to the variables of
forms of degree $p$ (resp. $r>p$) and
$k_{A_1 A_2
\ldots A_r b_1\ldots b_s}=-k_{A_2 A_1
\ldots A_r b_1\ldots b_s}$. 

The corresponding solution of the
Wess-Zumino consistency condition is 
\begin{align}
a^n_g=&[k_{A_1 A_2 \ldots A_r b_1\ldots b_s}{\tilde 
H}^{A_1}_{p+1} {\cal Q}^{A_2\ldots A_r b_1\ldots b_s}]^n_g,\\
&\text{{\rm with}}\quad  k_{A_1 A_2
\ldots A_r b_1\ldots b_s}=-k_{A_2 A_1
\ldots A_r b_1\ldots b_s}.
\end{align}
\end{theorem}

\proof{Let
$p$ be the lowest form degree appearing
in
${\cal Q}^{a_1
\ldots a_m}_{0,g+q}$. We can then write $a^n_{g,q}$ as,
\begin{equation}
a^n_{g,q}=[P_{a_1 \ldots a_r b_1 \ldots b_s}]^n_{-q}[{\tilde
B}^{a_1}_1
\ldots {\tilde B}^{a_r}_1 {\tilde B}^{b_1}_2 \ldots {\tilde
B}^{b_s}_2]^0_{g+q},
\end{equation}
where the ${\tilde B}^{a_i}_1$ correspond to forms of
degree $p$ and ${\tilde B}^{b_j}_2$ correspond to forms of
any higher degree. A direct calculation then shows that we
have,
\begin{eqnarray}
a^n_{g,q-j}&=&[P_{a_1 \ldots a_r b_1 \ldots
b_s}]^{n-j}_{-q+j}[{\tilde B}^{a_1}_1
\ldots {\tilde B}^{a_r}_1 {\tilde B}^{b_1}_2 \ldots {\tilde
B}^{b_s}_2]^j_{g+q-j},\label{recu1} \\ && \hspace{7cm} for\ 
0\leq j \leq p
\nonumber
\label{recurP}.
\end{eqnarray}

The ambiguity in $a^n_{g,q-j}$ is $a^n_{g,q-j}\rightarrow
a^n_{g,q-j} + m_0 + m_1 +\ldots +m_{j-1}$ where $m_0$ satisfies
$\gamma m_0=0$, $m_1$ satisfies $\gamma m_1 +\delta n_1 + db_1=0,\
\gamma n_1=0$, $m_2$ satisfies $\gamma m_2+\delta n_2+db_2=0,\
\gamma n_2 +dl_2+dc_2=0,\ \gamma l_2=0$, etc. Using the vocabulary
of Section {\bf{\ref{Sec:Antidep1}}}, we will say that the
ambiguity in $a^n_{g,q-j}$ is the sum of a $\gamma$-cocycle
$(m_0)$, the first ``lift" $(m_1)$ of a $\gamma$-cocycle, the
second ``lift"
$(m_2)$ of a $\gamma$-cocycle, etc. 

However, none of these
ambiguities except $m_0$ in $a_{q-p}$ can play a role in the
construction of a non-trivial solution. To see this, we note that
$\delta$,
$\gamma$ and $d$ conserve the polynomial degree of the variables
of any given sector. We can therefore work at fixed polynomial
degree in the variables of all the different $p$-forms. Since
$n_1$, $l_2$, etc. are $\gamma$-closed terms which have to be
lifted at least once, they have the generic form $R[H,\tilde
H]{\cal Q}$ where
${\cal Q}$ has to contain a ghost of ghost of degree $p_A<p$.
Because we work at fixed polynomial degree, the presence of such
terms imply that
$P_{a_1\ldots a_r b_1\ldots b_s}$ depends  on either $H^A$ or on
$\tilde H^A$. However, if $P$ depends on $\tilde H^A$ then its
component of degree $n$ is of antighost $q\leq p_A+1<p+1$ which is
in contradiction with our assumption $q>p$. This means that $P$
has to depend on $H^A$ so that,
\begin{multline}
\nonumber [P_{a_1 \ldots b_s}]^n_{-q}
{\cal Q}^{a_1 \ldots b_s}_{0,g+q}= [M_{A_1 a_1\ldots
b_s}]^{n-p_A-1}_{-q} H^{A_1}{\cal Q}^{a_1 \ldots
b_s}_{0,g+q}\\ =  s
((-)^{\epsilon_{M}}[M_{A_1 a_1\ldots b_s} 
{\cal Q}^{A_1  a_1
\ldots b_s}]^n_{g-1})\\  \quad +
d((-)^{\epsilon_{M}}[M_{A_1 a_1\ldots b_s}
{\cal Q}^{A_1  a_1
\ldots b_s}]^{n-1}_{g})\\
+ \text{terms of lower antighost number}.
\label{triviadeP}
\end{multline}
Thus, if $P_{a_1\ldots a_m}$ depends on $H^A$, one can 
eliminate $a_q$ from $a$ by the addition of trivial terms and the
redefinition of the terms of antighost number $<q$. Therefore we
may now assume that $a_q$ does not contain $H^A$ and that the only
ambiguity in the definitions of the $a_{q-j}$ is $m_0$ in
$a_{q-p}$.

Since $p<q$, we have to substitute
$a^n_{g,q-p}$ in the equation $\gamma a^n_{g,q-p-1}+\delta
a^n_{g,q-p}+ db^{n-1}_{g+1,q-p-1}=0$. We then get,
\begin{multline}
\gamma a^n_{g,q-p-1}+\delta [P_{a_1 \ldots a_r b_1 \ldots
b_s}]^{n-p}_{-q+p}[{\tilde B}^{a_1}_1
\ldots {\tilde B}^{a_r}_1 {\tilde B}^{b_1}_2 \ldots {\tilde
B}^{b_s}_2]^p_{g+q-p}\\ +\delta m_0+db^{n-1}_{g+1,q-p-1} =0,
\end{multline}
which can be written as,
\begin{multline}
\gamma a^{'n}_{g,q-p-1} + db^{' n-1}_{g+1,q-p-1} 
+\delta m_0 \hspace{2cm}
\\
 + (-)^{\epsilon_P}r [P_{a_1 \ldots a_r b_1
\ldots b_s}]^{n-p-1}_{-q+p+1}H^{a_1}_1{\cal Q}^{a_2
\ldots  a_r b_1 \ldots b_s}
_{0,g+q-p} =0.\label{eqdelobs}
\end{multline}
If we act with $\gamma$ on the above equation we obtain $d\gamma
b^{' n-1}_{g+1,q-p-1}=0 \Rightarrow \gamma b^{'
n-1}_{g+1,q-p-1}+db^{'
n-2}_{g+2,q-p-1}=0$ which means that $\gamma b^{'
n-1}_{g+1,q-p-1}$ is a $\gamma$ mod $d$ cocycle. There are two
possibilities according to whether $q-p-1>0$ or $q-p-1=0$. In the
first case we may assume that $b^{'n-1}_{g+1,q-p-1}$ is strictly
annihilated by $\gamma$ so that 
$db^{'n-1}_{g+1,q-p-1}=[d\beta_{a_2\ldots a_r b_1\ldots
b_s}(\chi)]{\cal Q}^{a_2
\ldots  a_r b_1 \ldots b_s}_{0,g+q-p} +\gamma l^{n}_{g,q-p-1}$.
Equation \eqref{eqdelobs} then reads,
\begin{align}
&(-)^{\epsilon_P} r [P_{a_1 \ldots a_r b_1
\ldots b_s}]^{n-p-1}_{-q+p+1}H^{a_1}_1 \nonumber \\ &\hspace{4cm}
+\delta \alpha_{a_2\ldots a_r b_1\ldots b_s}(\chi) + d
\beta_{a_2\ldots a_r b_1\ldots b_s}(\chi)=0,\label{xindeppourP}
\end{align}
where we have set $m_0=\alpha_{a_2\ldots a_r b_1 \ldots
b_s}(\chi) {\cal Q}^{a_2
\ldots  a_r b_1 \ldots b_s}
_{0,g+q-p}$. If we restrict ourselves to the algebra of
$x$-independent forms, Eq. \eqref{xindeppourP} implies,
\begin{align}
[P_{a_1 \ldots a_r b_1
\ldots b_s}]^{n-p-1}_{-q+p+1}H^{a_1}_1=0,\label{leqasa}
\end{align}
since $\delta$ and $d$ both increase the number of derivatives of
the $\chi$. The situation in the algebra of $x$-dependent forms is
more complicated and we shall discuss it at the end of the section.
Let us first note that $P_{a_1 \ldots a_r b_1 \ldots b_s}$ cannot
depend on $\tilde H^{c}_1$ because in that case we would have
$q-p-1\leq 0$ which contradicts our assumption. This means that
$P_{a_1 \ldots a_r b_1 \ldots b_s}$ will satisfy 
\eqref{leqasa} only if it is of the form, $P_{a_1 \ldots a_r b_1
\ldots b_s}=R_{c a_1\ldots a_r b_1\ldots b_s}H_1^{c}$ with $R_{c
a_1\ldots a_r b_1\ldots b_s}$ symmetric in $c \leftrightarrow a_1$
(resp. antisymmetric) if $H_1$ is anticommuting (resp. commuting).
However the same calculation as in \eqref{triviadeP} shows that
$a^n_{g,q}$ can then be absorbed by the addition of trivial terms
and a redefinition of the components of lower antighost number of
$a^n_g$. 

We now turn to the case $q-p-1=0$. According to our analysis of
Section {\bf\ref{Sec:Antidep1}} the obstruction to writing
$db^{'n-1}_{g+1,q-p-1}$ as a $\gamma$-exact term is of the form,
$[d\beta_{a_2\ldots a_r b_1\ldots
b_s}(\chi)+V_{a_2\ldots a_r b_1\ldots b_s}(H^a)]{\cal Q}^{a_2
\ldots  a_r b_1 \ldots b_s}_{0,g+q-p}$. Equation \eqref{eqdelobs} 
then reads,
\begin{align}\label{VFA}
&(-)^{\epsilon_P} r [P_{a_1 \ldots a_r b_1
\ldots b_s}]^{n-p-1}_{0}H^{a_1}_1 + V_{a_2\ldots a_r
b_1\ldots b_s}(H^a)\nonumber \\ &\hspace{4cm} +\delta
\alpha_{a_2\ldots a_r b_1\ldots b_s}(\chi) + d
\beta_{a_2\ldots a_r b_1\ldots b_s}(\chi)=0,
\end{align}
which becomes,
\begin{align}
(-)^{\epsilon_P} r [P_{a_1 \ldots a_r b_1
\ldots b_s}]^{n-p-1}_{0}H^{a_1}_1+V_{a_2\ldots a_r
b_1\ldots b_s}(H^a)=0,\label{EulercondpourP}
\end{align}
in the algebra of $x$-independent local forms. The fact that
$V_{a_2\ldots a_r b_1\ldots b_s}$ is $d$-exact implies that 
the variational derivatives with respect to all the
fields of 
$[P_{a_1 \ldots a_r b_1
\ldots b_s}]^{n-p-1}_{0}H^{a_1}_1$ must vanish. If $P_{a_1 \ldots
a_r b_1
\ldots b_s}$ depends on $\tilde H_1^c$ then the condition $q=p+1$
implies that 
\begin{equation}
P_{a_1 \ldots a_r b_1 \ldots b_s}=k_{c a_1 \ldots a_r b_1 \ldots
b_s}\tilde H^c_1,
\end{equation}
where the $k_{c a_1 \ldots a_r b_1 \ldots
b_s}$ are constants. If we take the Euler-Lagrange derivative of
\eqref{EulercondpourP} with respect to $B^b_{\mu_1\ldots \mu_{p}}$
we obtain,
\begin{equation}
k_{c a_1 \ldots a_r b_1 \ldots b_s}=-k_{a_1 c \ldots a_r b_1 \ldots
b_s}.
\end{equation}
In that case,
\begin{equation}
a^n_g=[k_{c a_1 \ldots a_r b_1\ldots b_s}\tilde H_1^c
{\tilde
B}^{a_1}_1
\ldots {\tilde B}^{a_r}_1 {\tilde B}^{b_1}_2 \ldots {\tilde
B}^{b_s}_2]^n_{g}
\end{equation}
is a solution of the Wess-Zumino consistency condition.

If $P_{a_1 \ldots a_r b_1
\ldots b_s}$ does not depend on $\tilde H_1^c$ it is of the form,
\begin{equation}
P_{a_1 \ldots a_r b_1 \ldots b_s} \sim 
(H^{a}_{p_{a}+1})^k \ldots
(H^{b}_{p_{b}+1})^i (\tilde H^{c}_{n-p_{c}-1})^u \ldots (\tilde
H^{d}_{n-p_{d}-1})^v,
\end{equation}
with $p\leq p_a<\ldots <p_b<p_c\ldots < p_d$. If we insert this
expression in \eqref{EulercondpourP} and take the Euler-Lagrange
derivative with respect to $B^b_{\mu_1\ldots \mu_{p}}$ we obtain
identically $0$ only if,
\begin{align}
[P_{a_1 \ldots a_r b_1
\ldots b_s}]^{n-p-1}_{-q+p+1}H^{a_1}_1=0,\label{leqasa2}
\end{align}
as in \eqref{leqasa}. By repeating exactly the discussion
following \eqref{leqasa} we reach the conclusion that
\eqref{leqasa2} implies that
$a^n_{g,q}$ can be absorbed by the addition of trivial terms and a
redefinition of the components of lower antighost number of
$a^n_g$.  All the results stated in the theorem have now been
proved.}

\subsubsection{Coboundary condition for antifield dependent
solutions} 

In this section we analyze the coboundary condition for the
antifield dependent solutions of the Wess-Zumino consistency
condition.

Let $a^n_g=a^n_{g,0}+\ldots +a^n_{g,q}$ be a BRST cocycle. From
our general analysis we know that $a^n_{g,q}$ is of the form (for
$q>1$),  $a^n_{q,q}=[P_{a_1 \ldots a_m}(H^a,{\tilde
H}^a)]^n_{-q} {\cal Q}^{a_1 \ldots a_m}_{0,g+q}$.

If
$a^n_g$ is trivial then there exist 
$c^n_{g-1}=c^n_{g-1,0}+\ldots+c^n_{g-1,l}$ and
$e^{n-1}_{g}=e^{n-1}_{g,0}+\ldots+e^{n-1}_{g,l}$ such that $a^n_g
= sc^{n}_{g-1}+ de^{n-1}_g$. Decomposing this equation according
to the antighost number we get:
\begin{align}
a_0&=\delta c_1+\gamma c_0 +d  e_0, \\
a_1&=\delta c_2+\gamma c_1 +d  e_1, \\
&\ \vdots\nonumber \\
a_q&=\delta c_{q+1}+\gamma c_q +d  e_q, \\
0&=\delta c_{q+2}+\gamma c_{q+1} +d  e_{q+1}, \\
&\ \vdots\nonumber \\
0&=\delta c_{l}+\gamma c_{l-1} +d  e_{l-1}, \\
0&= \gamma c_{l} +d  e_{l}.\label{trucmus}
\end{align}
We have dropped the indices labeling the ghost numbers and
form degrees which are fixed. In the above equations, $a^n_{g,q}$
appears as the obstruction to lifting $l-q$ times the term $c_l$.

From our analysis of Section {\bf\ref{prelimres}} we know that
\eqref{trucmus} implies $c_l=Q_J(\chi)\omega^J$. If $l=q+1$ we then
have,
\begin{equation}
a^n_{g,q}=\delta Q_J(\chi)\omega^J +\gamma c_q +de_q.
\end{equation}
Because $q>1$ we may assume that $de_q=dS_J(\chi)\omega^J + \gamma
v_q$ and therefore,
\begin{equation}
[P_{a_1 \ldots a_m}(H^a,{\tilde
H}^a)]^n_{-q}=\delta Q_J(\chi)+dS_J(\chi),
\end{equation}
which is not the case since $[P]^n_{-q}$ defines a non-trivial
class of
$H^{inv}(\delta\vert d)$.

Thus we now assume that $l>q+1$ and in that case $c_l$ has to be
lifted at least once. According to the analysis of Section
{\bf\ref{prelimres}}, the most general expression for
$c_l$ is then
\begin{equation}\label{cslwp}
c_l= [R_{a_1
\ldots a_r}(H^a,{\tilde H}^a)]^n_{-l} {\cal Q}^{a_1 \ldots
a_r}_{0,g+l}.
\end{equation}
By examining the calculations performed in the
proof of Theorem {\bf\ref{bigtheo3}} we conclude that $a^n_g$ is
trivial if and only if 
$a^n_{g,q}$ is given by, 
\begin{equation}\label{lamort}
[M_{a_1 \ldots a_r b_1
\ldots b_s}(H^a,{\tilde H}^a)]^{n-p-1}_{-q+p+1} H^{a_1}{\cal
Q}^{a_2
\ldots  a_r b_1 \ldots b_s}
_{0,g+q-p},
\end{equation}
since this is the obstruction arising when one tries to
lift a term of the form \eqref{cslwp}. In \eqref{lamort} the $a_i$
label
$p$-forms of the same degree while the $b_j$ label forms of higher
degree; furthermore
$M_{a_1
\ldots a_r b_1
\ldots b_s}$ is symmetric (resp. antisymmetric) in $(a_1\ldots
a_r)$ if $H^{a_j}$ is anticommuting (resp. commuting). 

In particular, the BRST cocycles described in Theorem
{\bf\ref{bigtheo3}} are not trivial.

\subsubsection{Results in the algebra of $x$-dependent forms}

The proofs of the previous theorems hold because we have limited
our attention to the algebra of $x$-independent local forms in
which the exterior derivative $d$ maps a polynomial containing $i$
derivatives of the fields onto a polynomial containing $i+1$
derivatives of the fields. This observation allowed us in
particular to obtain Eq. \eqref{leqasa} from Eq.
\eqref{xindeppourP} and Eq. \eqref{EulercondpourP} from Eq.
\eqref{VFA}.
The complete calculations in the algebra of $x$-dependent forms
of the BRST cocycles will not be done here. Instead we give an
example to illustrate the problem.

Let us examine a
system of
$1$-forms for which we want to construct in ghost number $-1$ the
solutions corresponding to
$a^n_{-1,2}= f_{ab}{\overline A}^{*a}_2 C^b$. By examining the BRST
cocycle condition at antighost number $1$ one easily gets
$a_1=f_{ab}{\overline A}^{*a}_1 A^b + m_1$ where $m_1$ can be
assumed of the form $m_1=\alpha(\chi)$. At antighost number $0$ we
then have the condition,
$f_{ab}{\overline F}^a F^b + V(F^a) + \delta \alpha(\chi)+
d\beta(\chi)=0$ which is just Eq. \eqref{VFA} for our particular
example. In the algebra of $x$-independent variables we have seen
that this condition implies that the symmetric part of $f_{ab}$
vanishes. In ${\cal P}$ this is no longer true. Indeed for the
symmetric part of $f_{ab}$ one finds the BRST cocycles \cite{BBH2},
\begin{equation}
a=a_1+a_2= f_{(ab)}\left( \frac{n-4}{2}C^{a*}C^{b} +
A^{*a\mu}\left[ x^\nu F^{b}_{\nu\mu}+\frac{n-4}{2}A^b_\mu \right]
\right).
\end{equation}
These are available in all dimensions except $n =4$  where they are
actually described by Theorem {\bf\ref{bigtheo1}}.

\section{Counterterms, first order vertices and a\-n\-o\-malies}
\label{cfova}

Using the results of the previous section, we obtain the
counterterms, first-order vertices and the anomalies which depend
non-trivially on the antifields. 

\subsection{Counterterms and first order vertices}

In ghost number $0$, Theorems {\bf{\ref{bigtheo1}}},
{\bf{\ref{bigtheo2}}} and {\bf{\ref{bigtheo3}}} imply the
existence of the following BRST cocycles:

\begin{enumerate}
\item
$a^n_0 = k_{\Delta a}(j^{\Delta}B^{a}+a^{\Delta}C^a)$ 
\hspace{5.55cm}(Theorem {\bf{\ref{bigtheo1}}}).
\item
$a^n_0= [P_c(H^a, \tilde H^b)Q^c]^n_0,\label{solantidep}$
\hspace{.1cm} 
where $P_c$ has $\Delta$-degree $n-p_c$.
(Theorem {\bf{\ref{bigtheo2}}}).
\item
$a^n_0 =[f_{abc}{\tilde H}^a {\tilde B}^b {\tilde B}^c]^n_0,$
with $f_{abc}$ completely antisymmetric.

\vspace{-.2cm}\hspace{9.95cm} (Theorem {\bf{\ref{bigtheo3}}}).
\end{enumerate}
{\em The cocycles of the first and third type are only
available for
$1$-forms.}
\vspace{.5cm}

To obtain the first order vertices and counterterms 
one isolates from the above cocycles the
component of antighost number
$0$. In particular, we see that a vertex of the Yang-Mills type is
only available for $1$-forms. In the absence of $1$-forms we have:
\begin{theorem}
For a system of free $p$-form gauge fields with $p\geq 2$, the
counterterms and the first order vertices are given by,
\begin{equation}
{\cal{V}}= P_c(H^a,\overline{H}^b)B^c.
\label{fov}
\end{equation}
where in \eqref{fov} the form degrees of the various forms present
must add up to $n$.
\end{theorem}
For such systems, the first order vertex
$P_c(H^a,\overline{H}^b)B^c$ is as announced a generalized Noether
coupling since
$P_c(H^a,\overline{H}^b)$ is a higher order conserved
current. All these vertices have the remarkable
property to be linear in the gauge potentials $B^a$; we stress
again that this is in sharp contrast to the situation where
$1$-forms are present since in those cases couplings of the
Yang-Mills type exist.

To first order in the coupling constants, the 
classical action corresponding to \eqref{fov} reads,
\begin{equation}
I_{int}=\int d^n x\{\sum_a \big({-1 \over 2(p_{a}
+1)!}H^a_{\mu_1
\ldots 
\mu_{p_{a}+1}} H^{a \mu_1
\ldots \mu_{p_{a}+1}}+g_aS^{a\mu_1 \ldots
\mu_{p_a}}B^a_{\mu_1 \ldots \mu_{p_a}}\big)\},
\end{equation}
where $S^{c\mu_1 \ldots
\mu_{p_a}}$ are the components of the form dual to
$P^c(H^a,\overline{H}^b)$. 

This action is no longer gauge
invariant under the original gauge transformations. Since $S^{a\mu_1 \ldots
\mu_{p_a}}$ is gauge invariant and has an on-shell vanishing
divergence, we have,
$\partial_\mu S^{a\mu_1
\ldots
\mu_{p_a}}=k_{b\nu_1\ldots\nu_k}^{a\mu_2 \ldots \mu_{p_a}}
\partial_\rho H^{b\rho\nu_1\ldots\nu_k}$, where the
$k_{b\nu_1\ldots\nu_k}^{a\mu_2 \ldots \mu_{p_a}}$ are gauge
invariant. It is then easy to check that up to
terms of order $g^2$, $I_{int}$ is invariant under,
\begin{equation}
\delta^{New}B^a_{\mu_1\ldots\mu_{p_a}}=\partial_{[\mu_1}
\Lambda^a_{\mu_2\ldots\mu_p ]} +g_bp_a!p_b
k_{b\mu_1\ldots\mu_{p_a}}^{a\nu_2 \ldots
\nu_{p_b}}\Lambda^b_{\nu_2\ldots\nu_{p_b}}.
\end{equation}
Since the $k_{b\mu_1\ldots\mu_{p_a}}^{a\nu_2 \ldots
\nu_{p_b}}$ are gauge invariant, the new gauge algebra
remains abelian up to order $g^2$. Another way to see this is
to remember from the general theory that in order to  deform the
gauge algebra to order $g$ one needs a term in $a^n_0$ which is
quadratic in the ghosts.

\subsection{Anomalies}

Using our analysis we can compute the
``anomalies" of the theory. Taking into account Theorems 
{\bf{\ref{bigtheo1}}}, {\bf{\ref{bigtheo2}}} and
{\bf{\ref{bigtheo3}}} we obtain in ghost number $1$ the following
BRST cocycles:
\begin{enumerate}
\item
$a^n_1 = k_{\Delta ab}(2 j^{\Delta} B^{a}C^b+a^{\Delta}C^a C^b)$,
\ for $1$-forms
\hspace{1.5cm}(Theorem {\bf{\ref{bigtheo1}}}).\\ \vspace{-.2cm}

$a^n_1 = k_{\Delta a}(j^{\Delta}B^{a}+a^{\Delta}C^a)$,\ for
$2$-forms
\hspace{2.95cm}(Theorem {\bf{\ref{bigtheo1}}}).
\item
$a^n_1= [P_c(H^a, \tilde H^b)Q^c]^n_1,$
\ 
where $P_c$ has $\Delta$-degree $n-p_c-1$.

\vspace{-.2cm}\hspace{9.57cm}
(Theorem {\bf{\ref{bigtheo2}}}).
\item
$a^n_1 =[f_{ABc}{\tilde H}^A {\tilde B}^B {\tilde B}^c]^n_1,$ with
$f_{ABc}=-f_{ABc}$ and where $B^A$ is a $1$-form and $B^c$ a
$2$-form.
\hspace{6.26cm} (Theorem {\bf{\ref{bigtheo3}}}).\\ \vspace{-.5cm}

$a^n_1 =[f_{ABCD}{\tilde H}^A {\tilde B}^B {\tilde B}^C {\tilde
B}^D]^n_1,$\ for $1$-forms and where $f_{ABCD}$ is completely
antisymmetric \hspace{6.855cm} (Theorem {\bf{\ref{bigtheo3}}}).
\end{enumerate}

Notice that anomalies of type 1 and 2 only exist in the
presence of
$1$-forms or $2$-forms. Therefore,
\begin{theorem}
In the absence of $1$-forms and $2$-forms, the
antifield dependent candidate anomalies are given by,
\begin{equation}
a^n_1= [P_c(H^a, \tilde H^b)Q^c]^n_1.
\end{equation}
\end{theorem}

\section{Gauge invariance of con\-ser\-ved
cur\-rents}\label{conscurent}

In this section we list all the conserved currents which can not
be covariantized. As we recalled in Section {\bf\ref{cocudef}}
these are related to the representatives of $H^n_{-1}(s\vert d)$
for which $a=a_1+\ldots +a_q$ with $q>1$.

The results of Section {\bf\ref{matho}} imply that these BRST
cocycles are necessarily of the form,
\begin{align}
a^n_{-1}=&[k_{ab}{\tilde 
H}^{a}_{p+1} {\tilde B}^{b}]^n_{-1}\ \ \text{{\rm with}}\quad  k_{a
b} =-k_{ba},\label{consc}
\end{align}
where the ${\tilde H}^a$ and ${\tilde B}^b$ are mixed forms
associated to exterior forms of the same degree. 

The global symmetries associated with the component
$a_1=k_{ab}{\overline B}^{*a}_1 B^b$ of the BRST cocycles
\eqref{consc} are,
\begin{equation}
\delta B^a_{\mu_1\ldots\mu_p}=k^a_{\
b}B^b_{\mu_1\ldots\mu_p},\label{globsymng}
\end{equation}
and correspond to ``rotations" of the forms $B^a$ among themselves
since the generators of these symmetries are antisymmetric matrices
$k_{ab}$.

Let us now prove that a conserved current $j$ associated to a
{\em gauge invariant} global symmetry through the relation $\delta
a_1+ dj=0$ can be assumed to be gauge invariant as well. 
\proof{
If
$a_1$ is gauge invariant, then this is also true of $\delta a_1$.
By assumption we have $d\delta a_1=0$ since $\delta
a_1+ dj=0$. Therefore, according to Theorem
{\bf{\ref{invpoincare}}} we have $\delta a_1 = d R$,
where $R$ is a gauge invariant polynomial (there can be no
polynomial in $H^a$ present in $\delta a_1$ because $\delta$ and
$d$ bring one derivatives of the $\chi$). Thus we conclude that up
to a $d$-exact term,
$j=R$.}

Conversely, using Theorem {\bf\ref{deltachi}} it is immediate to see
that any global symmetry associated to a gauge invariant 
conserved current may be assumed gauge invariant.
Therefore, we have:
\begin{theorem}
The only conserved currents which cannot be
assumed to be gauge invariant are associated to the global
symmetries
$\delta B^a_{\mu_1\ldots\mu_p}=k^a_{\
b}B^b_{\mu_1\ldots\mu_p}$ and given by,
\begin{equation}
j= k_{ab}{\overline H}^{a} B^b \text{ with $k_{ab}=-k_{ba}$}.
\end{equation}
\end{theorem}

\section{Conclusions}

In this section we have computed all the solutions of the
Wess-Zumino consistency condition which depend on the antifields.
Apart from those described by Theorem {\bf\ref{bigtheo1}}
which related to the conserved currents of the theory,
they are all explicitly known and given by Theorems
{\bf{\ref{bigtheo2}}} and {\bf{\ref{bigtheo3}}}. The latter have the
property to be expressible in terms of the forms $B^a, H^a,
{\overline H}^a, {\overline B}^{*a}_j$ and
$C^a_j$. This extends to antifield dependent solutions the
similar property we established in Section {\bf\ref{Sec:Antidep1}}
concerning the antifield independent BRST cocycles belonging to
non-trivial descents.

From the BRST cocycles in ghost number $0$, we obtained all the
first-order vertices, counterterms and ``anomalies" of the theory.
We also determined all the Noether conserved currents which are not
equivalent to gauge invariant currents.

In the absence of $1$-forms, all the first order vertices were shown
to be of the Noether type (conserved current x
potential) and exist only in particular spacetime dimensions. As a
consequence there is in that case no vertex of the Yang-Mills type
and accordingly no interaction which deforms the algebra of the
gauge transformations at first order. If
$1$-forms are present in the system, one finds additional
interactions which are also of the Noether form
$j^\mu A_\mu$. However, the conserved current $j^\mu$ which couples
to the
$1$-form need not be gauge-invariant.  There is actually only one non
gauge-invariant current that is available and it leads to the
Yang-Mills cubic vertex, which deforms the gauge algebra to order
$g$.  All other currents $j^\mu$ may be assumed to be gauge-invariant
and thus do not lead to algebra-deforming interactions.  There is in
particular no vertex of the form $\overline{H} B A$ where A are
1-forms and B are
$p$-forms ($p>1$) with  curvature $H$, which excludes charged
$p$-forms (i.e. $p$-forms transforming in some representation of a
Lie algebra minimally coupled to a Yang-Mills potential).

The ``anomalies" have also been computed and we have shown that for
systems of $p$-forms of degree $\geq 3$, they are all linear in the
ghosts variables. When $1$-forms and $2$-forms are included in
the system, one finds additional anomalies related to the
conserved currents of the theory. Among these, there are only two
for which the corresponding conserved currents cannot be made gauge
invariant.

Some of the above conclusions are based on our analysis of the gauge
invariant nature of the global symmetries and conserved currents.
Our results are:  1) the only global symmetries which are not gauge
invariant up to trivial terms are the rotation of forms among
themselves; 2) the only conserved currents which cannot be
improved to become gauge invariant are those related to these
global symmetries.

\section{Higher order vertices}
\label{highorver}

Once all the first order vertices are known, one can pursue
the analysis of the consistent deformations. This is done by
requiring that the equation $(S,S)=0$ should be satisfied to all
orders in the coupling constant. One then obtains a succession of
equations which read,
\begin{eqnarray}
(S_0,S_0)&=&0,\\
(S_0,S_1)&=&0,\label{hsmodd} \\
2(S_0,S_2) +(S_1,S_1)&=&0, \label{brack1} \\
(S_0,S_3) + (S_1,S_2) &=&0\label{brack2}, \\
&\vdots& \nonumber
\end{eqnarray}
The construction of the full interacting action
$S=S_0+gS_1+g^2S_2+\ldots$, consistent to all orders, can be
obstructed if one equation of the tower fails to be
satisfied. In this section we 
investigate this problem for a
system of exterior forms with degrees limited to two values
$p$ and $q$ such that $2 \leq p < q \leq n-2$ and we construct
lagrangians which are consistent to all orders in the coupling
constant.

The $p$-forms are denoted $A^a$ $(a=1,\ldots\,m)$ and the
$q$-forms $B^A$
$(A=1,\ldots,M)$; their curvatures $F^a=dA^a$ and
$H^A=dB^A$ are respectively $(p+1)$- and $q+1$-forms.
Their duals,
\begin{eqnarray}
{\overline F}^a = \frac{1}{(n-p-1)!} \epsilon_{\mu_1 \ldots
\mu_n} F^{a\mu_1\ldots \mu_p+1}dx^{\mu_{p+2}}\ldots
dx^{\mu_n},\\ 
{\overline H}^A = \frac{1}{(n-q-1)!} \epsilon_{\mu_1 \ldots
\mu_n} H^{A\mu_1\ldots \mu_{q+1}}dx^{\mu_{q+2}}\ldots
dx^{\mu_n},
\end{eqnarray}
are respectively $(n-p-1)$- and $(n-q-1)$-forms. In form
notation, the free lagrangian can be written as,
\begin{equation}
{\cal L}= -\frac{1}{2(p+1)!} F^a {\overline F}_a -
\frac{1}{2(q+1)!} H^A {\overline H}_A.
\label{acavecpetq}
\end{equation}

From the previous section we know that the first order
vertices are exterior products of form degree $n$ of one of
the forms, the curvatures and their duals and are thus given by,
\begin{equation}
(H^A)^k(F^a)^l ({\overline H}^A)^m({\overline F}^a)^rA^a,
\end{equation}
or 
\begin{equation}
(H^A)^k(F^a)^l ({\overline H}^A)^m({\overline F}^a)^r B^A,
\end{equation}
where the form degrees must add up to $n$. This imposes the
condition $k(q+1)+l(p+1)+m(n-q-1)+r(n-p-1)+p=n$ in the first case
and $k(q+1)+l(p+1)+m(n-q-1)+r(n-p-1)+q=n$ in the second case.
Furthermore, in order for these vertices to truly deform the gauge
transformations they must contain at least one dual so that
$m+r>1$. Using those conditions in addition to $n>q+1>p+1$ one
finds only three types of first order couplings which
truly deform the gauge transformations:

\vskip .5cm

\noindent (i) Chapline-Manton couplings, which are linear in
the duals
\cite{ChaplineManton,Nito,Cham1,Cham2,BergRooWitNieu},
\begin{eqnarray} V_1 &=& \int f^a_A \overline{F}_a  B^A, \;
\; (q=p+1),
\label{CM1} \\ V_2 &=& \int f^A_{a_1 \dots a_{k+1}} 
\overline{H}_A F^{a_1}  \dots F^{a_{k}} A^{a_{k+1}},
\nonumber \\ && (k(p+1) +p = q+1).
\label{CM2}
\end{eqnarray} Here, $f^a_A$ and $f^A_{a_1\dots a_{k+1}}$ are
arbitrary constants.  The
$f_{Aa_1\dots a_{k+1}}$ may be assumed to be completely
symmetric (antisymmetric) in the $a$'s if $p$ is odd (even).
The Chapline-Manton coupling (\ref{CM1}) only exists for
$q=p+1$ while the Chapline-Manton coupling (\ref{CM2}) only exists
if there is some integer $k$ such that 
$k(p+1) +p = q+1$.

\vskip .5cm

\noindent (ii) Freedman-Townsend couplings, which are
quadratic in the duals \cite{FT1},
\begin{eqnarray} V_3 &=& \int f^A_{BC} \overline{H}^B 
\overline{H}^C B_A ,
\label{FT1} \\ V_4 &=& \int t^a_{Ab} \overline{H}^A 
\overline{F}_a A^b .
\label{FT2}
\end{eqnarray} Here, $f^A_{BC}$ and $t^a_{Ab}$ are constants
arbitrary at first order but restricted at
second order. The Freedman-Townsend vertices (\ref{FT1}) and
(\ref{FT2}) only exist for
$q=n-2$.

\vskip .5cm

\noindent (iii) Generalized couplings, which are at least
quadratic in the duals $\overline{H}^A$,
\begin{eqnarray} V_5 = \int k^{A_1 \dots A_l}_{a_1 \dots
a_{k+1}}
\overline{H}_{A_1} ...  
\overline{H}_{A_l} F^{a_1} ... F^{a_k}  A^{a_{k+1}}
\label{GenInt}
\end{eqnarray} where $k^{A_1 \dots A_l}_{a_1 \dots
a_{k+1}}$ are arbitrary constants with the obvious
symmetries.  These interactions exist only if there are
integers $k$, $l$ (with $l \geq 2$) such that
$l(n-q-1) + k(p+1) +p = n$.

We next show how the above first order vertices can be extended to
higher orders in order to obtain a theory which is consistent to
all orders.

\subsection{Chapline-Manton couplings}

Instead of using the master equation $(S,S)=0$ to get the
higher order vertices, it is sometimes easier to try to
guess the full interacting action and then show that without
imposing any conditions on the arbitrary parameters of the
first order vertex, one has a consistent deformation in the sense
recalled in the introduction. The antibracket analysis is then
facultative.
This is how we proceed for the Chapline-Manton couplings
(\ref{CM1}), (\ref{CM2}) and the generalized couplings
(\ref{GenInt}).

For $V_1$, the complete lagrangian which reduces to ${\cal
L}+V_1$  at order $1$ in the coupling constant is:
\begin{equation}
{\cal L}_{I,1}= -\frac{1}{(p+1)!} F_I^a {\overline F}_{Ia} -
\frac{1}{(q+1)!} H^A {\overline H}_A,
\end{equation}
where $F_I^a=F^a + g' f^a_A B^A$ and
$g'=-\frac{(p+1)!}{2}(-)^{q(n-q)}g$. The ``improved" field
strengths are invariant under the gauge transformations:
\begin{equation}
A^a \rightarrow A^a + d\epsilon^a -g'f^a_A\eta^A,\quad
B^A\rightarrow B^A + d\eta^A,
\label{ginvl1}
\end{equation}
from which it follows that ${\cal L}_{I,1}$ is gauge invariant
under \eqref{ginvl1} to all orders.

For $V_2$, the role of the $p$-form and the $q$-form are in a
certain sense exchanged. The complete lagrangian is:
\begin{equation}
{\cal L}_{I,2}= -\frac{1}{(p+1)!} F^a {\overline F}_{a} -
\frac{1}{(q+1)!} H_I^A {\overline H}_{IA},
\end{equation}
with $H_I^A=H^a + g' f^A_{a_1\ldots a_{k+1}} F^{a_1}\ldots
F^{a_{k}}A^{a_{k+1}}$ and
$g'=-\frac{(q+1)!}{2}(-)^{(q+1)(n-q-1)}g$. The ``improved" field
strengths are invariant under the gauge transformations:
\begin{equation}
A^a \rightarrow A^a + d\epsilon^a ,\quad
B^A\rightarrow B^A + d\eta^A - (-)^{k(q+1)}g'f^A_{a_1\ldots
a_{k+1}} F^{a_1}\ldots F^{a_{k}}\epsilon^{a_{k+1}},\label{ginvl2}
\end{equation} 
from which it follows that ${\cal L}_{I,2}$ is gauge invariant
under \eqref{ginvl2} to all orders. 

In both cases, the number of
fields, gauge invariances and order of reducibility are the
same as for the free theory. Let us check this
explicitly for $V_1$. The fields are $A^a$ and $B^A$; their gauge
transformations \eqref{ginvl2} are not all independent since they
vanish when $\epsilon^a$ and $\eta^A$ are of the form,
\begin{equation}
\epsilon^a=d\mu^a+gf^a_A\Lambda^A,\quad
\eta^A=d\Lambda^A.\label{rdug}
\end{equation}
From \eqref{rdug} we see that the number of parameters in
the reducibility identities of order 1 are the same for the free
theory and ${\cal L}_{I,1}$. In the same way, one shows that this
property also holds for the reducibility identities at all orders.

Furthermore, no restrictions on the
coefficients
$f^a_A$ and $f^A_{a_1\ldots a_{k+1}}$ are imposed so the
antibracket analysis is
not required. Therefore, ${\cal L}_{I,1}$ and ${\cal
L}_{I,2}$ are the most general consistent interactions
corresponding to $V_1$ and $V_2$.

\subsection{Generalized couplings}

Before examining the Freedman-Townsend couplings, we study
the ``generalized couplings" because as in the
Chapline-Manton case, no antibracket analysis is required. 

\paragraph{First-order formulation}

A convenient way to analyze the Freedman-Townsend couplings and
the generalized couplings is to reformulate those
theories by introducing auxiliary fields. This has the advantage
that the full interacting theories are then polynomial.

In the first-order formulation, the lagrangian \eqref{acavecpetq}
is replaced by,
\begin{equation}
{\cal L}=-\frac{a}{2}(2B^Ad\beta_A +
\overline{\beta}^A\beta_A)-\frac{b}{2}(2A^ad\alpha_a +
\overline{\alpha}^a\alpha_a),
\label{1storder}
\end{equation}
with $a=-(n-q-1)!(-1)^{(q+1)(n-q-1)}$ and
$b=-(n-p-1)!(-1)^{(p+1)(n-p-1)}$. 

The equations of motion are,
\begin{equation}
d\beta^A =0,\quad d\alpha^a=0, \quad
\beta^A=c\overline{H}^A,\quad \alpha^a
=k\overline{F}^a,\label{eqpouraux}
\end{equation}
where $c=\frac{1}
{(n-q-1)!(q+1)!}(-)^{(n-q-1)(q+1)+q}$ and $k=c(q\rightarrow
p)$. The original lagrangian \eqref{acavecpetq} is recovered by
inserting \eqref{eqpouraux} in \eqref{1storder}.

The 
gauge transformations of the first-order lagrangian
\eqref{1storder} are:
\begin{equation}
\delta_\Lambda B^A = d\Lambda^A,\quad \delta_\Lambda A^a=d\Lambda^a,\quad
\delta_\Lambda \beta^A =0,\quad \delta_\Lambda \alpha^a=0.
\end{equation}
According to the rules of the BRST formalism, the differential $s$
can then be written as
$s=\delta +\gamma$ with:
\begin{eqnarray}
\delta {\overline B}^{*A}_1 +d\beta^A =0,&\gamma B^A +
dC^A_1 =0,
\nonumber \\
\delta {\overline B}^{*A}_2 +d{\overline B}^{*A}_1
=0,&\gamma C^A_{q-1} + dC^A_{q}=0, \nonumber \\
{\vdots\quad \quad \quad \quad},&\vdots \nonumber
\\
\delta {\overline B}^{*A}_{q+1}+d{\overline
B}^{*a}_{q} =0,&\gamma C^A_{q}=0,\nonumber \\ 
\delta \beta^{*A} + \beta^A -c\overline{H}^A=0,&\gamma
\beta^A=0.
\nonumber
\end{eqnarray}
along with similar expressions for the $p$-form sector.

It
has been shown in
\cite{BBH1} that the cohomology
$H(s\vert d)$ for the theory with auxiliary fields and for the
original theory are isomorphic. In our
case, the mapping between the BRST cohomologies of the two
formulations is implemented through the replacement
of
$\overline{H}^A$ by $\beta^A$ and $\overline{F}^a$ by
$\alpha^a$ in the BRST cocycles of the theory without auxiliary
fields. This is easily seen from the above definitions
of $\delta$ and $\gamma$.

\vspace{.7cm}

\paragraph{Generalized couplings} In the first-order formulation,
the full interacting lagrangian corresponding to the generalized
couplings can be written as,
\begin{eqnarray}
{\cal L}_{I,5}&=& -\frac{a}{2}(2B^A d\beta_A + {\overline
\beta}^A \beta_A) - \frac{1}{2(p+1)!} F^a {\overline F}_a 
\nonumber \\&&+g' k^{A_1 \ldots A_l}_{a_1 \ldots a_{k+1}}
\beta_{A_1}\ldots \beta_{A_l} F^{a_1} \ldots F^{a_k}
A^{a_{k+1}},
\label{lag5}
\end{eqnarray}
with $g'=c^{-l}g$.
To first order in $g$, the action
$\int {\cal L}_{I,5}$ reduce to $I+g\int V_5$ upon
elimination of the auxiliary fields (the auxiliary fields
$\alpha^a$ have already been eliminated). 

This lagrangian is
invariant to all orders in the coupling constant under the
following gauge transformations:
\begin{multline}
\delta_\Lambda A^a = d\Lambda^a, \ \delta_\Lambda \beta_A =0, \\
\delta_\Lambda B^{A_1}= d\Lambda^{A_1} -  \alpha k^{A_1 \ldots A_l}_{a_1
\ldots  a_{k+1}} \beta_{A_2}\ldots \beta_{A_l} F^{a_1}
\ldots F^{a_k}
\Lambda^{a_{k+1}},
\end{multline}
where $\alpha=\frac{l}{a}g'(-)^{(n-q-1)[k(p-1)-l]-(n-q)(p-1)}$ and 
no restriction on the $k^{A_1 \ldots A_l}_{a_1 
\ldots  a_{k+1}}$ is needed to achieve gauge invariance.
Furthermore, as in the case of $V_1$ it is easy to show that the
number of fields, gauge parameters and the order of reducibility
after elimination of
$\beta_A$ are the same as for the free theory (\ref{1storder}).

\subsection{Freedman-Townsend couplings}

For the Freedman-Townsend couplings the situation is more
complicated. However, it is easy to see that 
\begin{eqnarray}
{\cal L}_I=-\frac{a}{2}(2B^Ad\beta_A +
\overline{\beta}^A\beta_A)-\frac{b}{2}(2A^ad\alpha_a +
\overline{\alpha}^a\alpha_a) \nonumber \\ +g f^A_{BC}\beta^B
\beta^C B_A +g t^a_{Ab}\beta^A \alpha_a A^b 
\label{fulllagft}
\end{eqnarray}
is to all orders a consistent
interacting lagrangian invariant under the following
gauge transformations,
\begin{align}
\delta_\Lambda B_A=d\Lambda_A + \frac{2}{a}g
f^B_{AC}(-)^q\Lambda_B\beta^C +
\frac{1}{a}(-)^{p+1}g
t^a_{Ab}\Lambda^b \alpha_a,\\
\delta_\Lambda A^a =d\Lambda^a +
(-)^{p+1}\frac{1}{b}gt^a_{Ab}\beta^A\Lambda^b,\quad
\delta_\Lambda \beta^A=0,\quad \delta_\Lambda \alpha_a =0,
\end{align} 
provided we impose the conditions,
\begin{eqnarray}
f^B_{AC}f^A_{DE}+f^B_{AD}f^A_{EC}+f^B_{AE}f^A_{CD}=0,
\label{alglie} \\
 t^d_{Ab}t^a_{Bd}-
t^d_{Bb}t^a_{Ad}=t^a_{Cb}f^C_{AB}.
\label{repalglie}
\end{eqnarray}
The first condition expresses that the $f^A_{BC}$ are the
structure constants of a Lie algebra while the second
condition states that the $t^a_{Bc}$ are the matrices of a
representation of that Lie algebra.

Upon elimination of the auxiliary fields, \eqref{fulllagft}
reduces to ${\cal L}+g V_3+g V_4$ to first order in the
coupling constant.

Notice that we are not sure at this stage that the conditions
\eqref{alglie} and \eqref{repalglie} are mandatory and that they
cannot be dropped by adding to 
\eqref{fulllagft} higher order terms and by further modifying the
gauge transformations. To answer this question we use the
antibracket analysis recalled at the beginning of the section.

The solution of the Wess-Zumino consistency condition
corresponding to the Freedman-Townsend vertex $g f^A_{BC}\beta^B
\beta^C B_A +g t^a_{Ab}\beta^A \alpha_a A^b$ is,
\begin{equation}
S_1= [f^A_{BC} \tilde{H}^B 
\tilde{H}^C \tilde{B}_A + t^a_{Ab} \tilde{H}^A 
\tilde{F}_a \tilde{A}^b]^n_0.
\end{equation}
The construction of the second order vertex will be possible only
if the antibracket of $S_1$ with itself is $s$-exact. A direct
calculation yields,
\begin{eqnarray}
(S_1,S_1)&=&\int  [\{2 f^B_{AC}f^A_{DE}{\tilde H}^C
{\tilde H}^D {\tilde H}^E {\tilde B}_B \nonumber \\ &&\quad+
(t^a_{Cb}f^C_{AB}- t^d_{Ab}t^a_{Bd}){\tilde H}^A
{\tilde H}^B {\tilde F}_a {\tilde A}^b\}]^n_1.
\label{antibracketS1}
\end{eqnarray}
Using our results on the antifield dependent anomalies, we see
that the r.h.s. of \eqref{antibracketS1} is the sum of two
\emph{non-trivial} BRST cocycles. Therefore $(S_1,S_1)$ can only
be $s$-exact if these two terms vanish which means that the
conditions
\eqref{alglie} and \eqref{repalglie} are needed\footnote{In a
recent paper \cite{BrandtD1}, F. Brandt and N. Dragon have described
an interaction between two $1$-forms $A^a,\  a=1,2$ and one $2$-form
$B^1$. Their example is a particular case of (\ref{fulllagft})
with $f^A_{BC}=0$, $t^2_{11}=1$ and other components of
$t^a_{1b}=0$. It is clear that this choice of $f^A_{BC}$ and
$t^a_{Bc}$ satisfies the requirements
(\ref{alglie},\ref{repalglie}). One recovers the action they
obtain upon elimination of the auxiliary fields. Their approach
is complementary to ours and is based on the gauging of a global
symmetry of the free lagrangian (\ref{Lagrangian}).}.

It is interesting to notice that (\ref{fulllagft}) remains
polynomial after the elimination of the auxiliary fields
$\alpha^a$. Indeed, by using their equation of motions one
gets:
\begin{equation}
{\cal L}_I=-\frac{a}{2}(2B^A\Phi_A +
\overline{\beta}^A\beta_A)-\frac{1}{2(p+1)!}F^{'a}{\overline
{F^{'}}}_a,\label{consisa}
\end{equation}
where $\Phi^A=d\beta^A -\frac{1}{a}g_1f^A_{BC}\beta^B
\beta^C$ and
$F^{'a}=dA^a+ \frac{(-)^{(n-p)}}{b}g_2 t^a_{Ab}\beta^A A^b$.
The gauge invariances for (\ref{consisa}) read:
\begin{eqnarray}
\delta_\Lambda B_A=d\Lambda_A + \frac{2}{a}g_1
f^B_{AC}(-)^q\Lambda_B\beta^C +
\frac{1}{ab(p+1)!}(-)^{(n-p+q)}g_2
t^a_{Ab}\Lambda^b{\overline F}_a,\\
\delta_\Lambda A^a =d\Lambda^a +
(-)^{(n-p)}\frac{1}{b}g_2t^a_{Ab}\beta^A\Lambda^b,\quad
\delta_\Lambda \beta^A=0.
\end{eqnarray}

\subsection{Remarks} 

\indent

1) For the three types of couplings described above, the algebra of
the gauge transformations remains abelian on-shell to all orders in
the coupling constant. This is a particularity of systems which
contain exterior forms of degrees limited to two values. Indeed,
with three (or more) degrees, one can modify the algebra at order
$g^2$.  For instance, if $A$, $B$ and $C$ are respectively 3-, 4-
and 7-forms, the Lagrangian $ \sim
\overline{F} \wedge F +
\overline{H} \wedge H + \overline{G} \wedge G$ is invariant under
the gauge transformations $\delta A = d \epsilon + g \Lambda$,
$\delta B = d \Lambda$ and $\delta C = d \mu + g \epsilon dB - g^2
\Lambda B$, where $\epsilon$, $\Lambda$ and $\mu$ are respectively
2-, 3- and 6-forms.  Here, $F = dA -g B$, $H = dB$ and $G = dC - g A
dB + (1/2) g^2 B^2$.  The commutator of two
$\Lambda$-transformations is a $\mu$-transformation with $\mu = g^2
\Lambda_1 \Lambda_2$. This model will be studied in detail in
Chapter {\bf\ref{chapCM}} where we discuss the BRST cohomology
for Chapline-Manton models.

2) The interaction vertices of this
section are still available in the presence of $1$-forms. They just
fail to exhaust all the possible vertices.

3) The basic interaction vertices described above can of
course be combined, or can be combined with Chern-Simons terms.  This
leads, in general, to additional constraints on their coefficients
(which may actually have no non trivial solutions in some cases). 
One example of a non-trivial combination is the description of 
massive vector fields worked out in \cite{FT1,Thierry1}, which
combines the Freedman-Townsend vertex with a Chern-Simons term
\cite{HenLemesetAl}.  Another example is given in \cite{Anco1},
where both the Freedman-Townsend vertex and the Yang-Mills vertex
are introduced simultaneously.

\chapter{Chapline-Manton models}
\label{chapCM}
\section{Introduction}
In the previous chapter we studied the BRST cohomology for a
system of free $p$-forms. From this analysis we were
able to get all the first order vertices that could be added to
the free lagrangian. We have proved that except for those which are
strictly gauge invariant, the interactions are quite constrained,
i.e., for a fixed spacetime dimension, a given system of
$p$-forms only allows for a finite number of consistent vertices
(even to first order).
This result complements the geometric
analysis performed in \cite{Teitelboim1} where it is shown that the
non-abelian Yang-Mills construction cannot be generalized to
$p$-forms viewed as connections for extended objects.
[Topological field theory offers ways to bypass some of the
difficulties
\cite{Baulieu3}, but will not be discussed here].

In this chapter we present the BRST cohomology for
different Chapline-Manton couplings. These are
particularly interesting because their gauge algebra remains closed
off-shell and the reducibility identities hold strongly even after
the interactions are switched on.  Using those properties one
can study their cohomologies essentially along the lines drawn for
the system of free $p$-forms. 

For the other couplings, the
new gauge algebra generally closes only on-shell and the
reducibility identities become on-shell relations.  This occurs for
the Freedman-Townsend interaction \eqref{fulllagft} and also for the
generalized couplings \eqref{lag5}. In those cases,
$\gamma^2 \approx 0$ and it becomes meaningless to consider the
cohomology of
$\gamma$ since it is no longer a differential. One cannot
therefore study antifield independent BRST cocycles as we did in the
free case; the antifields must be incorporated from the beginning. 

In section {\bf{\ref{highorver}}} we have limited our 
study of Chapline-Manton couplings to a system of forms
of only two different degrees. However, 
Chapline-Manton type interactions can be constructed for more
general systems of
$p$-forms
along the lines discussed in \cite{Baulieu1}.  Rather
than facing the general case, which would lead to non
informative and uncluttered formulas, we shall illustrate the
general construction through four particular example.

Chapline-Manton models are characterized by
gauge-invariant curvatures
$H^a$ which differ from the free ones by terms proportional to the
coupling constant $g$,
\begin{equation} H^a = dB^a + g \mu^a + O(g^2).
\end{equation} The gauge transformations are,
\begin{equation}
\delta_\epsilon B^a = d \epsilon^a + g \rho^a + O(g^2).
\end{equation} Here, $\mu^a$ is  a sum of exterior products of
$B$'s and
$dB$'s -- which must match the form degree of $dB^a$ --
while $\rho^a$ is a sum of exterior products of
$B$'s, $dB$'s and
$\epsilon$'s (linear in the $\epsilon$'s). The modified
curvatures and gauge transformations must fulfill the
consistency condition,
\begin{equation}
\delta_\epsilon H^a = 0,
\label{invar}
\end{equation} which means that the modified curvatures should be
invariant under the modified gauge transformations. Furthermore,
off-shell reducibility must be preserved, i.e.,
$\delta_\epsilon B^a$ should identically vanish for
$\epsilon^a = d
\lambda^a +  \theta^a $ for some appropriate
$\theta^a(\epsilon, B, dB,g)$. The Lagrangian being a function
of the curvatures and their derivatives, ${\cal L}= {\cal
L}([H^a_{\mu_1 ...
\mu_{p_a+1}}])$ is automatically gauge-invariant.
To completely define the model, it is thus necessary to
specify, besides the field spectrum, the modified curvatures
and the gauge transformations leaving them invariant. In many
cases, the curvatures are modified by the addition of
Chern-Simons forms of same degree, but this is not the only
possibility as Example 3 below indicates. In the sequel we set the
coupling constant $g$ equal to one.

\section{The models}

\subsubsection{Model 1} The first example contains one
$p$-form, denoted $A \equiv A^p_0$, and one
$(p+1)$-form, denoted $B
\equiv B^{p+1}_0$. The new field strengths are,
\begin{equation} F = dA + B, \; H = d B,
\label{CM1a}
\end{equation} while the modified gauge transformations take the
form,
\begin{eqnarray}
\delta_{\epsilon,\eta} A &=& d \epsilon - \eta,
\label{CM1b}\\
\delta_{\epsilon,\eta} B &=& d \eta,
\label{CM1c}
\end{eqnarray} where $\epsilon$ is a $(p-1)$-form and
$\eta$ a
$p$-form. The gauge transformations are abelian and remain
reducible off-shell since the particular gauge parameters
$\epsilon = d \rho +\sigma$,
$\eta = d \sigma$ clearly do not affect the
fields. The BRST transformations of the undifferentiated
fields and ghosts are,
\begin{equation}
s A^{p-k}_k + d A^{p-k-1}_{k+1} + B^{p-k}_{k+1} = 0,
\end{equation} for the $A$-variables, and
\begin{eqnarray}
s B^{p+1-k}_k + d B^{p-k}_{k+1} &=& 0, \\
s B^0_{p+1} &=& 0,
\end{eqnarray} ($k= 0, \dots, p$) for the $B$-ones. One has,
\begin{equation}
s F = 0 = s H.
\end{equation} 

This model describes in fact a massive $(p+1)$-form.
One can indeed use the gauge freedom of $B$ to set $A
=0$.  Once this is done, one is left with the Lagrangian
of a massive $(p+1)$-form.

The BRST action of this model can be written as,
\begin{align}
S&=\int d^n x ( {\cal L}+(-)^{\epsilon_A}\phi^*_A s\phi^A
) \\ &= \int d^n x( {\cal L} +
\frac{1}{p!}B^{*\mu_1\ldots
\mu_{p+1}}\partial_{\mu_1}C^B_{\mu_2\ldots\mu_{p+1}}+\ldots+
B^{*\mu}\partial_\mu C^B\nonumber
\\ &+ \frac{1}{p!}A^{*\mu_1\ldots
\mu_{p}}(p\partial_{\mu_1}C^A_{\mu_2\ldots\mu_{p}}-
C^B_{\mu_1\ldots\mu_p})+\ldots +A^{*\mu}(\partial_\mu C^A
+(-)^{p}C^B_\mu)).\label{Sformod1}
\end{align}

Using, $s\phi^*_A = \frac{\delta^R S}{\delta \phi^A}$, we
extract from
\eqref{Sformod1} the BRST transformations of the
antifields:
\begin{align}
&sA^{*\mu_1\ldots\mu_p}=\partial_\nu
F^{\nu\mu_1\ldots\mu_p},\\ &s
A^{*\mu_1\ldots\mu_{p-j}}= -
\partial_\nu A^{*\nu\mu_1\ldots\mu_{p-j}},\\
&sB^{*\mu_1\ldots\mu_{p+1}}=\partial_\nu
H^{\nu\mu_1\ldots\mu_{p+1}}-F^{\mu_1\ldots\mu_{p+1}},
\\ &sB^{*\mu_1\ldots \mu_{p+1-j}}=-\partial_\nu
B^{*\nu\mu_1\ldots\mu_{p+1-j}}+(-)^{j+1}A^{*\mu_1\ldots\mu_{p+1-j}}.
\end{align}
The action of $s$ on the fields, antifields and ghosts decomposes
(as in the free theory) as the sum of the Koszul-Tate differential
and the longitudinal exterior derivative: $s=\delta + \gamma$ with
$s\phi^*_A=\delta \phi^*_A$ and $s\phi^A=\gamma\phi^A$.

\subsubsection{Model 2} The second example contains an
abelian
$1$-form $A \equiv A^1_0$ and a $2r$-form $B \equiv
B^{2r}_0$ ($r>0$). The field strengths are,
\begin{equation} F = dA , \; H = d B +  F^r A,
\label{CM2a}
\end{equation} with $F^r \equiv FF \cdots F$ ($r$ times). The gauge
transformations read,
\begin{eqnarray}
\delta_{\epsilon,\eta} A &=&  d\epsilon,
\label{CM2b}\\
\delta_{\epsilon,\eta} B &=& d \eta -
 F^r\epsilon,
\label{CM2c}
\end{eqnarray} and they clearly leave the curvatures invariant.

The BRST transformations of the fields and the ghosts are,
\begin{eqnarray}
s A^1_0 + d A^0_1 &=& 0, \\
s A^0_1 &=& 0, \\
s B^{2r}_0 + d B^{2r -1}_1 + F^r A^0_1 &=& 0,
\\
s B^{2r-k}_k + d B^{2r -k -1}_{k+1} &=& 0, \\
s B^0_{2r} &=& 0,
\end{eqnarray}
$(k=1, ..., 2r-1)$.

The BRST action of this model can be written as,
\begin{align}
S&=\int d^n x ( {\cal L}+(-)^{\epsilon_A}\phi^*_A s\phi^A
) \\ 
&= \int d^n x( {\cal L} +
\frac{1}{(2r)!}B^{*\mu_1\ldots
\mu_{2r}}(2r\partial_{\mu_1}C^B_{\mu_2\ldots\mu_{2r}}\\ & \quad
\quad -
\frac{(2r)!}{2^r}F_{\mu_1\mu_2}\ldots
F_{\mu_{2r-1}\mu_{2r}}C^A) +
\ldots+ B^{*\mu}\partial_\mu C^B\nonumber
 + A^{*\mu}\partial_{\mu}C^A.\label{Sformod2}
\end{align}
Therefore, the BRST transformations of the antifields are:
\begin{align}
&sA^{*\mu}=\partial_\nu F^{\nu\mu}-\frac{1}{2^r}
H^{\nu_1\ldots\nu_{2r}\mu}F_{\nu_1\nu_2}\ldots
F_{\nu_{2r-1}\nu_{2r}}\nonumber \\ \nonumber
&\quad\quad\quad -\frac{2r}{2^r}\partial_\rho( H^{\mu\rho\nu_1
\ldots
\nu_{2r-1}}F_{\nu_1\nu_2}\ldots
F_{\nu_{2r-3}\nu_{2r-2}}A_{\nu_{2r-1}}) 
\\  &\quad\quad\quad
-\frac{2r}{2^r}\partial_\rho(B^{*\mu\rho\nu_1\ldots\nu_{2r-2}}F_{\nu_1\nu_2}\ldots
F_{\nu_{2r-3}\nu_{2r-2}}C^A),\\
&sA^*= -\partial_\mu A^{*\mu} -
\frac{1}{2^r}  B^{*\mu_1\ldots
\mu_{2r}}F_{\mu_1\mu_2}\ldots
F_{\mu_{2r-1}\mu_{2r}},\\ \label{cfCM21}
&sB^{*\mu_1\ldots \mu_{2r}}=\partial_\nu
H^{\nu\mu_1\ldots\mu_{2r}},\\
&sB^{*\mu_1\ldots \mu_{2r-j}}=-\partial_\nu
B^{*\nu\mu_1\ldots\mu_{2r-j}}.\label{cfCM22}
\end{align}
The action  of $s$ on the fields, antifields and ghosts decomposes
as in the free theory as the sum of the Koszul-Tate differential
and the longitudinal exterior derivative: $s=\delta+\gamma$.
However, here there are two undesirable features. First, the
variation of
$A^{*\mu}$ involves contributions of antighost number $0$ so that
$\gamma A^{*\mu}\not = 0$. Furthermore, $sA^{*\mu}$ also
contains the undifferentiated fields
$A_\mu$  which are not invariant under the gauge transformations. 
This is in contrast with the free theory where all the antifields
were in $H(\gamma)$ and their variations were invariant. Both
defects can be corrected by making the following invertible
transformations,
\begin{align}
&A^{*\mu}\rightarrow A^{*\mu}-
\frac{2r}{2^r}F_{\nu_1\nu_2}\ldots F_{\nu_{2r-3}\nu_{2r-2}}
(B^{*\mu\nu_1\ldots\nu_{2r-1}}A_{\nu_{2r-1}}-
B^{*\mu\nu_1\ldots\nu_{2r-2}}C^A),\\
&A^*\rightarrow A^* -
\frac{2r}{2^r}F_{\nu_1\nu_2}\ldots F_{\nu_{2r-3}\nu_{2r-2}}
(B^{*\nu_1\ldots\nu_{2r-1}}A_{\nu_{2r-1}}-
B^{*\nu_1\ldots\nu_{2r-2}}C^A),
\end{align}
which cast $s$ into the form,
\begin{align}
&sA^{*\mu}=\partial_\nu F^{\nu\mu}-\frac{r+1}{2^r}
H^{\nu_1\ldots\nu_{2r}\mu}F_{\nu_1\nu_2}\ldots
F_{\nu_{2r-1}\nu_{2r}},\label{cfCM23}\\
&sA^* =-\partial_\mu A^{*\mu} -
\frac{r+1}{2^r}  B^{*\mu_1\ldots
\mu_{2r}}F_{\mu_1\mu_2}\ldots
F_{\mu_{2r-1}\mu_{2r}}.\label{cfCM24}
\end{align}
The $s$-variations of the new antifields are now gauge invariant and
$\gamma \phi^*_A=0,\ \forall \phi^*_A$.
\subsubsection{Model 3}
Let $A$, $B$ and $C$ be
respectively 1-, 2- and 3-forms. The curvatures are defined
through,
\begin{equation} F =dA + B, \; H = dB , \;G = dC +  A dB + (1/2)  B^2.
\label{CM3a}
\end{equation} The gauge transformations are
\begin{align}
\delta_{\epsilon,\Lambda,\mu} A &= d \epsilon -  \Lambda,\\
\delta_{\epsilon,\Lambda,\mu} B &= d \Lambda,\\
\delta_{\epsilon,\Lambda,\mu} C &= d \mu -  \epsilon dB -
\Lambda B,
\end{align}
where $\epsilon$, $\Lambda$ and $\mu$ are
respectively 0-, 1- and 2-forms.  Their gauge algebra is
non-abelian and the gauge
transformations are off-shell reducible, as is easily verified.

The BRST differential on the fields and ghosts is defined by
\begin{align}
s A^1_0 + dA^0_1 + B^1_1 &= 0, \\
s A^0_1 + B^0_2 &= 0, \\
s B^2_0 + d B^1_1 &= 0, \\
s B^1_1 + d B^0_2 &= 0, \\
s B^0_2 &= 0, \\
s C^3_0 + d C^2_1 + A^0_1 H + B^1_1
B^2_0&= 0, \\
s C^2_1  + d C^1_2 +\frac{1}{2}
B^1_1B^1_1 + B^0_2B^2_0&= 0, \\
s C^1_2 + d C^0_3 +  B^1_1 B^0_2 &=
0, \\
s C^0_3 + \frac{1}{2} B^0_2B^0_2 &= 0.
\end{align} This example arises in some formulations of
massive supergravity in $10$ dimensions
\cite{Romans1,BergRooPaGrTo}. 

The BRST action can again be written as,
\begin{align}
S=\int &d^n x ( {\cal L}+(-)^{\epsilon_A}\phi^*_A s\phi^A
)\\ \int &d^n x ({\cal L} + A^{*\mu}(\partial_\mu C^A-C^B_\mu) +
A^*C^B \nonumber \\
&+ B^{*\mu\nu}\partial_\mu C^B_\nu +B^{*\mu}\partial_\mu C^B
\nonumber \\
&+\frac{1}{3!}C^{*\mu\nu\rho}(3\partial_\mu
C^C_{\nu\rho}-3C^A\partial_\mu B_{\nu\rho}-3 C^B_\mu
B_{\nu\rho})\nonumber \\
&-\frac{1}{2!}C^{*\mu\nu}(-2\partial_\mu C^C_\nu + C^B_\mu
C^B_\nu -C^B B_{\mu\nu}) \nonumber \\
&+C^{*\mu}(\partial_\mu C^C- C^B_\mu C^B) \nonumber \\
&+\frac{1}{2}C^*(C^B)^2).
\end{align}
Therefore, the variations of the antifields are,
\begin{align}
&sA^{*\mu}=\partial_\nu F^{\nu\mu}-\frac{1}{6}
H_{\nu\rho\alpha}G^{\mu\nu\rho\alpha}, \\
&sA^*=-\partial_\mu
A^{*\mu}-\frac{1}{6}C^{*\mu\nu\rho}H_{\mu\nu\rho},\\
&sB^{*\mu\nu}=\partial_\rho H^{\rho\mu\nu} -F^{\mu\nu}
-\frac{1}{2} G^{\mu\nu\rho\alpha}B_{\rho\alpha}
+\partial_\alpha ( G^{\rho\alpha\mu\nu}A_\rho) \nonumber \\
&\quad\quad\quad\quad\partial_\rho(C^{*\mu\nu\rho} C^A) -
C^{*\mu\nu\rho}C^B_\rho -C^{*\mu\nu}C^B, \\
&sB^{*\mu}=-\partial_\nu
B^{*\nu\mu}-A^{*\mu}-\frac{1}{2}C^{*\mu\nu\rho}B_{\nu\rho} +
C^{*\mu\nu}C^B_{\nu}- C^{*\mu} C^B,\\
&sB^*=-\partial_\mu B^{*\mu} +A^*+\frac{1}{2}
C^{*\mu\nu}B_{\mu\nu}-C^{*\mu}C^B_\mu+ C^* C^B,\\
&sC^{*\mu\nu\rho}=\partial_\alpha G^{\alpha\mu\nu\rho},\\
&sC^{*\mu\nu}=-\partial_\rho C^{*\rho\mu\nu},\\
&sC^{*\mu}=-\partial_\rho C^{*\rho\mu},\\
&sC^*=-\partial_\rho C^{*\rho}.
\end{align}
The definition BRST differential's action on the
antifields of the $B$-sector suffers from defects similar to those
encountered in Chapline-Manton model 2 with the
original antifields. For example, $sB^{*\mu}$ involves
components of antighost number $0$ and $1$ and the
undifferentiated fields $B_{\mu\nu}$ and $C_\mu^B$. This justifies
the replacements,
\begin{align}
&B^{*\mu\nu}\rightarrow B^{*\mu\nu}+C^{*\rho\mu\nu}C^A_\rho
+C^{*\mu\nu}C^A,\\
&B^{*\mu} \rightarrow B^{*\mu}+ C^{*\mu\rho}C^A_\rho
+C^{*\mu}C^A,\\
&B^*\rightarrow B^* + C^{*\rho}C^A_\rho+ C^*C^A.
\end{align}
In terms of these modified antifields, the action of $s$ reads,
\begin{align}
&sB^{*\mu\nu}=\partial_\rho H^{\rho\mu\nu} -F^{\mu\nu}
-\frac{1}{2} G^{\mu\nu\rho\alpha}F_{\rho\alpha}, \nonumber\\
&sB^{*\mu}=-\partial_\nu
B^{*\nu\mu}-A^{*\mu}-\frac{1}{2}C^{*\mu\nu\rho}F_{\nu\rho},\\
&sB^*=-\partial_\mu B^{*\mu} +A^*+\frac{1}{2}
C^{*\mu\nu}F_{\mu\nu}.
\end{align}
The BRST differential now decomposes as $s=\delta+\gamma$ and the
new antifields are gauge invariant,
$\gamma \phi^*_A=0,\ \forall \phi^*_A$.

\subsubsection{Model 4} Our last example is the original
Chapline-Manton model, coupling a Yang-Mills connection
$A^a$ with a
$2$-form
$B$.  We will assume the gauge group to be $SU(N)$ for
definiteness although the analysis holds 
for any other compact group.  The curvatures are,
\begin{align} F &= dA + A^2, \label{CM4a} \\ H  &= dB +
\omega_3,
\label{CM4b}
\end{align} where $\omega_3 (A, dA)$ is the
Chern-Simons
$3$-form,
\begin{equation}
\omega_3 = \frac{1}{2} [tr(A dA + \frac{2}{3} A^3)].
\end{equation} The BRST differential's action on the fields and
ghosts reads,
\begin{align}
s A+DC &= 0, \\
s C-C^2 &= 0, \\
s B+  \omega_2 + d\eta &= 0, \\
s\eta +  \omega_1 + d\rho &= 0, \\
s\rho + \frac{1}{3} tr C^3 &= 0.
\end{align} Here, the one-form $\omega_1$ and the
two-form
$\omega_2$ are related to the Chern-Simons form $\omega_3$ 
through the descent,
\begin{align}
s\omega_3 + d \omega_2&=0, \ \ \omega_2= tr (CdA), \\
s \omega_2 +d\omega_1&=0, \ \ \omega_1= tr (C^2 A),\\
s \omega_1 + d(\frac{1}{3} tr C^3)&=0.
\end{align} 
The BRST action for this model is,
\begin{align}
S=\int &d^n x ( {\cal L}+(-)^{\epsilon_A}\phi^*_A s\phi^A
)\\ \int &d^n x ( {\cal L} + A^{*\mu}_a(\partial_\mu C^a
-C^a_{bc}A^b_\mu C^c) -\frac{1}{2}C^* C^a_{bc}C^b C^c\\
&+\frac{1}{2} B^{*\mu\nu}(2C_a(\partial_\mu
A_\nu^a-\partial_\nu A^a_\mu)+(\partial_\nu
\eta_\nu-\partial_\nu\partial_\mu))\\
&-\eta^{*\mu}(C_{abc}C^aC^bA^c_\mu-\partial_\mu
\rho)+\frac{1}{3}\rho^*C_{abc}C^aC^bC^c).
\end{align}
Therefore, the variation of the antifields are,
\begin{align}
&sB^{*\mu\nu}=\partial_\rho H^{\rho\mu\nu},\\
&s\eta^{*\mu}=-\partial_\nu B^{*\nu\mu},\\
&s\rho^*=-\partial_\mu \eta^{*\mu},\\
&sA^{*\mu}_a=D_\nu
F^{\nu\mu}_a+2H^{\mu\nu\rho}F_{a\nu\rho}-2\partial_\rho
H^{\rho\mu\nu}A_{a\nu}\label{sastar}\\
&\quad\quad\quad -2\partial_\nu
(B^{*\nu\mu}C_a)-\eta^{*\mu}C_{abc}C^b C^c +C_{abc}A^{*b\mu}C^c,\\
&sC_a^* =-D_\mu A^{*\mu}_a +2B^{*\mu\nu}\partial_\mu A_{a\nu}+2
C_{abc}\eta^{*\nu}C^b A^c_\mu \\&\quad\quad\quad + C_{abc}\rho^* C^b
C^c -C_{abc}C^{*b}C^c.
\end{align}
In terms of the above variables,
the BRST differential suffers once more from some defects. First it
has a component of antighost number $1$, e.g.
$\eta^{*\mu}C_{abc} C^a C^b$ in (\ref{sastar}), as a consequence of
which the BRST differential
does not split
as the sum of the Koszul-Tate differential and the
longitudinal exterior derivative. 
The second undesirable feature is that the
BRST variations of the antifields of the Yang-Mills sector
contain contributions which are not covariant  under the gauge
transformations, e.g. $\partial_\rho H^{\rho\mu\nu}
A_{a\nu}$ in (\ref{sastar}). One can remedy
both problems by redefining the
antifields of the Yang-Mills sector according to the
following invertible transformations:
\begin{eqnarray}
A^{*\mu}_a \rightarrow A^{*\mu}_a + 2 B^{*\mu\nu}A_{a\nu} -
2 \eta^{*\mu}C_a,\label{var1} \\ \label{var2}
C^*_a \rightarrow C^*_a + 2 \eta^{*\mu}A_{a\mu}-2\rho^*C_a.
\end{eqnarray}
In terms of the new variables, the BRST differential
now takes the familiar form, $s=\delta + \gamma$
with:
\begin{eqnarray}
\delta B^{*\mu\nu}=\partial_\rho H^{\rho\mu\nu};\ \
\delta \eta^{*\mu}=-\partial_\nu B^{*\nu\mu};\ \
\delta\rho^* = -\partial_\mu \eta^{*\mu}; \label{delta1} \\
\delta A^{*\mu}_a = D_\nu F_a^{\nu\mu}+2\lambda
H^{\mu\nu\rho}F_{a\nu\rho};\ \
\delta C_a^* =2\lambda B^{*\mu\nu}F_{a\mu\nu} - D_\mu
A^{*\mu}_a, \label{defdelta}
\end{eqnarray}
and
\begin{equation}
\gamma B^{*\mu\nu}=\gamma \eta^{*\mu}=\gamma \rho^* = 0;\ \
\gamma A^{*\mu}_a = C_{abc}A^{*b\mu}C^c;\ \ \gamma C_a^* =
-C_{abc}C^{*b}C^c \label{delta2};
\end{equation}
\begin{equation}
\gamma \ (fields) = s \ (fields).
\end{equation}
The
$\gamma$ variations of the Yang-Mills variables are
now identical to those of the uncoupled theory and $A^{*a}_\mu$
and $C^*_a$ transform according to the  adjoint
representation.

\section{Cohomology of $\gamma$}

The cohomology $H^*(\gamma)$ of the Chapline-Manton model
can be worked out as in the free case, by exhibiting
explicitly the contractible part of the algebra.  This
contractible part typically gets larger with the coupling:
some cocycles are removed from
$H^*(\gamma)$. This happens for the models 1, 3 and 4.

\subsubsection{Chapline-Manton model 1}
 
In the absence of couplings, the $\gamma$-cohomology for
the first model is given, according to Theorem {\bf{\ref{hgamma}}},
by the polynomials in $(dA)_{\mu_1 ...\mu_{p+1}}$,
$(dB)_{\mu_1 ...
\mu_{p+2}}$, the antifields, their derivatives and the
last ghosts
$A^0_p$, $B^0_{p+1}$.
When the coupling is turned on, however, some of these
``$x$"-variables become contractible pairs and get canceled in
cohomology. Specifically, it is the last ghosts of ghosts that
disappear.  

Indeed, as in the proof of Theorem {\bf{\ref{hgamma}}}	one can
replace the original variables by,
\begin{equation}
\partial_{(
{\alpha_1\ldots\alpha_k}}A_{[\mu_1{)}_1\ldots
\mu_l{]}_2},
\partial_
{\alpha_2\ldots\alpha_{k}}F^0_{\alpha_1\mu_1\ldots
\mu_l},\partial_{(
{\alpha_1\ldots\alpha_k}}B_{[\mu_1{)}_1\ldots
\mu_l{]}_2},\text{ and } \partial_
{\alpha_2\ldots\alpha_{k}}H_{\alpha_1\mu_1\ldots
\mu_l},
\end{equation}
with $F^0_{\mu_1\ldots
\mu_l}=\partial_{[\mu_1}A_{\mu_2\ldots\mu_r]}$ ($2\leq r\leq p$)
and
$H_{\mu_1\ldots
\mu_l}=\partial_{[\mu_1}B_{\mu_2\ldots\mu_l]}$ ($2\leq l\leq p+1$).
One then makes a further change of coordinates by replacing
$F^0_{\mu_1\ldots
\mu_l}$ with $F_{\mu_1\ldots
\mu_l}=\partial_{[\mu_1}A_{\mu_2\ldots\mu_r]} -
B_{\mu_1\ldots\mu_r}$, which is obviously invertible. The
variables can now be associated as in Theorem {\bf{\ref{hgamma}}}
except that a new contractible pair appears, i.e., $\gamma
A^0_p=B^0_{p+1}$; the $x$-variables are now made of the
$H_{\mu_1\ldots
\mu_l}$,  $F_{\mu_1\ldots
\mu_l}$ and their derivatives. 

The
important point which makes the argument correct is that
$F^0_{\mu_1\ldots
\mu_l}$ and $F_{\mu_1\ldots
\mu_l}$ only differ by terms which are of lower order in the
derivatives of the $A$-sector (here they even differ only by
variables of the $B$-sector). As this property also holds for the
other Chapline-Manton models, the same change of variables will
also be possible.

Note that the Bianchi identities for the new field strengths
read
\begin{equation} dF = H, \; \; dH=0.
\end{equation} They can be used to express the $H$-components and their
derivatives in terms of the components $F_{\mu_1 ...
\mu_{p+1}}$ and their derivatives, which thus completely
generate the cohomology.

To summarize, we have proved:
\begin{theorem}
\label{gammaCohoCM1} For the Chapline-Manton model
1, the cohomology
$H(\gamma)$ is given by the polynomials in the improved
field strength components
$F_{\mu_1 ...\mu_{p+1}}$, the antifields and their derivatives,
\begin{equation}
\gamma \omega = 0 \Leftrightarrow \omega = \frac{1}{q!}
\omega_{\nu_1 ... \nu_q}([F_{\mu_1 ... \mu_{p+1}}],[\phi^*_A])
dx^{\nu_1} \dots dx^{\nu_q}.
\end{equation} In particular, there is {\em no} cohomology at
non-vanishing pureghost number.
\end{theorem}

The situation is very similar to the discussion of the
gauged principal $U(1)$ sigma model
\cite{HenneauxW1} (see also \cite{BizSa1} in this context).

\subsubsection{Chapline-Manton model 2} 

In this case, the change of variables proceeds as for model 1 so
that the
$\gamma$-cohomology is unchanged compared with the free
case except that the free curvatures are replaced by the
improved, gauge invariant curvatures \eqref{CM2a}.  The last ghosts
remain in cohomology because
$A^{(0,1)}$ is still $\gamma$-closed, so the mechanism of
the previous subsection is not operative. We thus have,
\begin{theorem} 
The cohomology of $\gamma$ for the
Chapline-Manton model 2, is given by
\begin{equation} 
H(\gamma)= \tilde{{\cal I}} \otimes {\cal C}
\end{equation} 
where ${\cal C}$ is the algebra generated by the last,
undifferentiated ghosts $A^0_1$ and $B^0_{2r}$, and
where
$\tilde{{\cal I}}$ is the algebra generated by the gauge
invariant field strength components
$F_{\mu \nu}$, $H_{\mu_1 \dots \mu_{2r+1}}$, the antifields and
their derivatives.
\end{theorem} 
Note the new form of the Bianchi identities
on the curvatures,
\begin{equation} dF=0,\; \; dH=F^{r+1}.
\end{equation}

\subsubsection{Chapline-Manton model 3} 
The discussion of the
third example proceeds to a large extent like that of the
first one.  The last ghosts of ghosts $A^0_1$ and
$B^0_2$ form a contractible pair and disappear in
cohomology; the improved last ghost of ghost
\begin{equation}
\tilde{C}^0_3=C^0_3-\frac{1}{2} A^0_1 B^0_2
\label{improvedGhost}
\end{equation} remains. 
Thus one has,
\begin{theorem} 
The cohomology of $\gamma$ for the
Chapline-Manton model 3 is given by,
\begin{equation} 
H(\gamma) = \tilde{\tilde{{\cal I}}} \otimes
\tilde{{\cal C}},
\end{equation} 
where $\tilde{\tilde{{\cal I}}}$ is the algebra
generated by the gauge invariant field strength components
$F_{\mu \nu}$, $G_{\mu \nu
\rho
\sigma}$, the antifields and their derivatives, and where
$\tilde{{\cal C}}$ is the algebra generated by the last, improved
ghost of ghost
$\tilde{C}^0_3=C^0_3-\frac{1}{2} A^0_1 B^0_2$.
\end{theorem} 
Again, note the new form of the Bianchi
identities,
\begin{equation} dF = -H, \; dH = 0, \; dG = -F H,
\label{Bianchi5}
\end{equation} 
which express $H$ in terms
of the derivatives of $F$.

\subsubsection{Chapline-Manton model 4}

In the absence of coupling, the cohomology of
$\gamma$ is given by the tensor product of
the pure Yang-Mills cohomology
\cite{Dixon1,Dixon2,Bandelloni1,Bandelloni3,BDK1,BDK2,BDK3,
HenneauxPL1} and
of the free 2-form cohomology.
We collectively denote by 
 $\chi_0$
(i) the Yang-Mills field
strengths, their covariant derivatives
$D_{\alpha_1}\ldots D_{\alpha_k} F^a_{\mu\nu}$, the
antifields
and their co\-va\-riant
de\-ri\-va\-tives $D_{\alpha_1}\ldots  D_{\alpha_k}
A^{*\mu}_a$,$D_{\alpha_1}\ldots D_{\alpha_k}
C^{*}_a$; these transform according to 
the adjoint representation; and
(ii) the free 2-form field strengths
$H^0_{\mu\nu\rho}=(dB)_{\mu\nu\rho}$, their derivatives,
the antifields $B^{* \mu \nu}$,  $\eta^{* \mu}$,
$\rho^*$, their derivatives and the undifferentiated
ghost of ghost $\rho$. 
Then the representatives of $H(\gamma)$ in the uncoupled case can
be written as
$a=\sum_J \alpha_J(\chi_0){\omega}^J(C^a)$, where
the $\alpha_J(\chi_0)$ are invariant polynomials (under $SU(N)$)
in the $\chi_0$ 
and where the ${\omega}^J(C^a)$
form a basis of the Lie algebra
cohomology of the Lie algebra of the gauge group.  The
$\omega^J$ are
polynomials in the so-called ``primitive forms", i.e $tr
C^3, tr C^5 \hbox{ if } tr C^5 \not = 0$, etc.

When the Chern-Simons coupling is turned on, the results are very
similar but with two modifications: (i) one must
replace in the above  cocycles
the free
field strengths $H^0_{\mu\nu\rho}$ and their derivatives by
the improved invariant field strengths $H_{\mu\nu\rho}$
and their derivatives (we shall denote 
the new set 
of improved variables defined in this manner by $\chi$); 
(ii) the
ghost of ghost $\rho$ and the primitive form
$tr C^3$ now drop from the cohomology since these elements are
related by
$\gamma
\rho=\frac{\lambda}{3} tr C^3$, which indicates that $tr
C^3$ is exact, while $\rho$ is no longer closed. This last
feature underlies the Green-Schwarz anomaly cancellation
mechanism. 
We thus have:
\begin{theorem}\label{gamCM4}The cohomology of $\gamma$ for the
Chapline-Manton model 4 is given by,
\begin{equation} H(\gamma) = {\cal J} \otimes {\cal D},
\end{equation} where (i) ${\cal J}$ is the algebra of the invariant
polynomials in the Yang-Mills curvature components, the antifields
and their covariant derivatives, as well as in the components
of the gauge invariant curvature $H$ and their derivatives;
and (ii) ${\cal D}$ is the algebra generated by the
``primitive forms" $trC^5$, $trC^7$, ...,
$trC^{2N-1}$.
\end{theorem} We recall that the Lie algebra cohomology for
$SU(N)$ is generated by the primitive forms $trC^3$,
$trC^5$, ... up to
$trC^{2N-1}$
\cite{GreubHalVan1,Koszul1}. 
The Bianchi identities read,
\begin{equation} DF = 0, \; \; dH=trF^2.
\end{equation}

\section{$H(s\vert d)$ - Antifield independent solutions}

\subsection{Covariant Poincar\'e Lemma}
In the free case, we have shown that the various spaces $E_k, k>0$
used in the lifts of elements of $H(\gamma)$ might
be calculated in the so-called ``small algebra". This result
relies upon the invariant Poincar\'e lemma which states that
each class of the invariant cohomology of $d$  has a
representative in ${\cal A}$. We now show that it is also the case
for the   Chapline-Manton models and therefore that ${\cal A}$ is
again the relevant space in which to calculate the spaces $E_k
,k>0$. One has,

\begin{theorem} Let $P$ be a gauge invariant polynomial. If
$P$ is closed, then $P$ is the sum of a closed, gauge
invariant polynomial belonging to the small algebra and of
the exterior derivative of an invariant polynomial,
\begin{equation} dP = 0  \Leftrightarrow P=Q+dR, \; Q \in {\cal A}, \;
dQ=0,
\label{invariantpol}
\end{equation} (with $P$, $Q$ and $R$ all gauge-invariant). 
Furthermore, if
$Q$ is $d$-exact in the algebra of gauge-invariant
polynomials, $Q= dS$ with $S$ gauge-invariant, one may
assume that $S$ is in the small algebra (and
gauge-invariant). Therefore, the invariant cohomology of
$d$ can be evaluated in ${\cal A}$ instead of the bigger algebra
${\cal P}$.
\end{theorem}

Note that while in the free case the conditions $Q \in {\cal A}$ and
$Q=dS$ (with $S$ gauge-invariant) imply
$Q=0$, this is no longer true here.

\vspace{.5cm}

\proof{We shall prove the theorem for the specific case of the
second model.  The proof proceeds in the same way for the
other models. We introduce a grading $N$ that counts the
number of derivatives of the
$B$ field.  According to this grading $P$ and $d$ split as,
\begin{equation} P = P_k + P_{k-1} + \cdots + P_1 + P_0, \; \; d = D_1 +
D_0,
\end{equation} with,
\begin{equation} N(P_i) = i, \; \; N(D_i)=i.
\end{equation} The differential $D_1$ takes derivatives only of the
$B$-field, the differential $D_0$ takes derivatives only of
the $A$-field. Because $P$ is gauge-invariant, the
$B$-field enters $P$ only through the components of $dB$
and their derivatives.  Furthermore, even though the
$P_i$'s with $i<k$ may involve the components
$A_\mu$'s and their symmetrized derivatives, $P_k$ depends
on $A$ only through the $F_{\mu \nu}$ and their derivatives.

The equation $dP = 0$ yields $D_1 P_k =0$ at the highest
value of the
$N$-degree. According to the results for the free case,
this implies
$P_k$ = $D_1 R_{k-1} + m_k$ where $R_{k-1}$ is a polynomial
in the components of $dB$ and their derivatives, while
$m_k$ is a polynomial in the form $dB$, both with
coefficients made of the components of $F$ and their derivatives
(which fulfill
$D_1 F_{\mu \nu} =0$). One then covariantize $R_{k-1}$ and $m_k$
by completing
$dB$ into $H$.  This only introduces terms of lower
$N$-degree.  We denote the covariant objects by $r$ and
$m$, respectively. One has
$P_k = [dr + m]_k$ and $P = dR_{k-1} + m_k + \; more$,
where ``$more$" is an invariant polynomial of maximum
$N$-degree strictly smaller than $k$. The invariant
polynomial $m$ - which exists only if $k=1$ or $0$ since
$H^2 = 0$ - is of order $k$ in the exterior form $H$.  It
must be closed by itself since there can be no compensation
between $D_0 m$ and $D_1(more)$ which is necessarily of
lower degree in the components of $H$ and their
derivatives.  It follows from $D_0 m = 0$ that $m =
\mu(F,H) + ds$, where $\mu$ is a polynomial in the forms
$F$ and $H$ and where $s$ is an invariant polynomial (using
again the results for the free case and
$H dp = -d(Hp)+ more$).  Thus one can get rid of $P_k$ by adding
to $P$ terms of the form (\ref{invariantpol}) of the
theorem.   By repeating the argument at the successive
lower degrees, one reaches the desired conclusion.

To prove the second part of the theorem, one first observes
that if
$dQ=0$, then $Q(F,H)$ does in fact not involve $H$,
$Q=Q(F)$ (because $d(\alpha(F)+\beta(F)H)=0\Rightarrow
\beta(F)F^{r+1}=0\Rightarrow \beta(F)=0$).  Assume then that
$Q=dU$, where $U$ is a gauge-invariant polynomial,
$U=U([H],[F])$. By expanding $U$ according to the
$N$-degree, $U=U_0 + U_1 +... + U_l$, one finds at higher
order $D_1 U_l = 0$, which implies as above
$U_l = D_1 R_{l-1} + m_l$ where $m_l$ is a polynomial in
the form
$dB$.  One can remove $D_1 R_{l-1}$ from $U_l$ by
subtracting
$dR_{l-1}$ from $U$, which does not affect $Q$.  Thus, only
$m_l$, which is present for
$l=1$ or $l=0$, is relevant.  By repeating the argument,
one finally arrives at,
\begin{equation} U = H a([F]) + b([F]).
\end{equation} The condition $Q = dU$ further implies $da=0$ and
thus
$a = d
\nu([F]) +
\rho(F)$ where $\rho(F)$ is a polynomial in the form $F$.
The term $H d \nu([F])$ is irrelevant since it can be
absorbed into $b([F])$ with a $d$-exact term.  Thus, $U =
H \rho(F) + b'([F])$.  The condition
$Q(F) = dU$ now reads $Q(F) = k(F) + db'([F])$ where $k(F)$
is a polynomial in $F$ and implies $db' = 0$ (invariant Poincar\'e
Lemma in the free case). But then, again, one can drop
$b'$ from $U$, which proves the second assertion.}
It follows from this theorem that there is no restriction in
investigating the invariant $d$-cohomology in the small
algebra.  Elements of $H(\gamma)$ that can be lifted at
least once necessarily belong to ${\cal A}$ up to trivial
terms. There is no restriction in the investigation of the
next lifts either because again $E_1^{small} \simeq E_1$. If a
$\gamma$-cocycle $a \in  {\cal A}$ can be written as $a =
du + \gamma v$ where
$u$ and $v$ are in the big algebra and $\gamma u = 0$, then
one may find $u'$ and $v'$ in ${\cal A}$ such that
$a=du'+\gamma v'$ (with
$\gamma u'=0$).  This follows from the second part of the
theorem. Obstructions to lifts within ${\cal A}$ cannot be
removed by going to the big algebra.

\subsubsection{Chapline-Manton model 1} For the first
Chapline-Manton model discussed above, the invariant cohomology of
$d$ is trivial. Indeed, in the algebra generated by $F$ and $H$,
the differential $d$ takes the contractible form $dF = H$,
$dH= 0$. Thus
\begin{equation} E_1 \equiv H(d_0, E_0^{small}) = 0
\end{equation} where $E_0^{small}$ is the algebra generated by $F$ and
$H$.

\subsubsection{Chapline-Manton model 2} In the algebra generated
by the gauge-invariant curvatures,
$d$ takes the form
\begin{equation} dF = 0, \; dH = F^{r+1}.
\end{equation} Since $H^2=0$, any element in this algebra is of the
form
\begin{equation} a = \alpha(F) + \beta(F) H,
\end{equation} where $\alpha(F)$ and $\beta(F)$ are polynomials in F.
The condition that $a$ is closed implies $\beta(F)F^{k+1}
=0$, which forces $\beta (F)$ to vanish.  Furthermore $a
\equiv \alpha(F)$ is exact if it is in the ideal generated
by $F^{r+1}$. Thus, we have the theorem:
\begin{theorem} The invariant cohomology of $d$ for the
Chapline-Manton mo\-del 2 is the quotient of the algebra
generated by the $F$'s by the ideal generated by
$F^{r+1}$.\label{invardmod2}
\end{theorem}

\subsubsection{Chapline-Manton model 3} For the third model, $d$ is
given by (\ref{Bianchi5}). By redefining the curvature $G$ as
\begin{equation} G_M = G- \frac{F^2}{2},
\end{equation} 
the algebra is brought to the form
\begin{equation} dF = -H, \; dH=0, \; dG_M = 0,
\end{equation} from which it follows that:
\begin{theorem} For the third model, the invariant
cohomology of $d$ is given by the polynomials in the
variable $G_M=G-F^2/2$.\label{invardmod3}
\end{theorem}

\subsubsection{Chapline-Manton model 4}

The invariant polynomials in the small algebra are the
polynomials in the gauge-invariant curvature $H$ of the
$2$-form and in the ``fundamental" invariants
$tr F^2$, $tr F^3$, ... $tr F^N$ for $SU(N)$ (this is a
basis for the
$SU(N)$ symmetric polynomials). These polynomials are all
closed, except
$H$, which obeys
$dH = tr F^2$.  Hence,
$H$ and  $tr F^2$ do not appear in the cohomology.

\begin{theorem} For the fourth model, the invariant
cohomology of $d$ is given by the polynomials in $tr F^3$,
$tr F^4$, ... $tr F^N$.
\end{theorem}

\subsection{Results}
\label{rescmgen}

We can now compute the various $E_k$ for the Chapline-Manton
models.

\subsubsection{Chapline-Manton model 1} 

The analysis is obvious in
this case since there is no non trivial descent.  All solutions
of the Wess-Zumino consistency condition can be taken to be
strictly annihilated by
$\gamma$, i.e., can be taken to be in $E_0$ ($E_1 = 0$). 
They are thus completely described by Theorem {\bf
\ref{gammaCohoCM1}} (from which one must remove the
$d$-exact terms $d\alpha([F])$).

\subsubsection{Chapline-Manton model 2} 

The second model is more
interesting.  Using Theorem \textbf{\ref{invardmod2}} we know
that $E_1\simeq E^{small}_1$ is 
isomorphic to the algebra generated by $F$, $A^0_1$ and
$B^0_{2r}$, with the relation $F^{r+1} = 0$.  This is no longer a
free algebra contrary to the situation encountered in the free case.

The differential $d_1$ is non trivial and
given by,
\begin{equation} d_1 A^0_1 = F, \; d_1 F = 0, \; d_1 B^0_{2r} = 0,
\end{equation} when $r>1$, which we shall assume at first.  
Because 
$F$ is subject to the relation $F^{r+1} = 0$, the
cohomological space
$E_2 \equiv H(d_1,E_1)$ is isomorphic to the algebra
generated by
$B^0_{2r}$ and $\mu(A,F)$ with
\begin{equation}
\mu(A,F) = -A^0_1 F^r.
\end{equation} 
One can take for $F_1$ the space  of polynomials of the form
$(B^0_{2r})^l Q_l(F) A^0_1$ where $Q_l$ is a
polynomial in
$F$ of degree strictly less than
$r$. To obtain the lifts of these cocycles, one can use
modified Russian formulas:
\begin{equation}
\tilde\gamma \tilde B=H-F\tilde A,\; \tilde\gamma \tilde A=F,
\end{equation}
with $\tilde B=B^0_2+B^1_1+B^2_0$ and $\tilde A=A^0_1+A^1_0$. The
lifts are therefore, $l(B^0_{2r})^{l-1}B^1_{2r-1} Q_l(F) A^0_1
+(B^0_{2r})^l Q_l(F) A^1_0$.

The next differentials $d_2$, $d_3$ ... vanish up to
$d_{2r-1}$. So, $E_2 = E_3 = \dots = E_{2r-1}$. 
One has
\begin{equation} 
d_{2r-1} B^0_{2r} = \mu(A,F), \; d_{2r-1} \mu(A,F) =0.
\end{equation} 
Thus $E_{2r} = 0$.

One may take for  $F_{2r-1}$ the space of
polynomials in
$B^0_{2r}$ (with no constant piece). The $k$-th lift of the
monomial $(B^0_{2r})^l=[(\tilde B)^l)]^0_{2rl}$ is $[(\tilde
B)^l)]^k_{2rl-k}$. 

Note in particular that $\mu(A,F)$ does not
appear in any of the spaces
$F_k$, because it is now trivial.  In the free
case,
$\mu(A,F)$ was an element of $F_1$ and the bottom of a
non-trivial descent of length two.  The coupling to the
$2r$-form makes it disappear from the cohomology. At the
same time, the cocycle
$F^{r+1}$, which was in the invariant cohomology of $d$ in
the free case, is now $d$-exact in the space of
invariant polynomials.  Also, while $B^0_{2r}$ could be
transgressed all the way up to $H$ in the free case, its
lift now stops at ghost number one with $\mu$.

The situation for $r=1$ is similar. The two steps
corresponding to the differentials $d_1$ and $d_{2r-1}$
are now combined in a single one so that the space $E_2$ vanishes. 
The easiest way to see this is to observe that $H(d_1,E_1)$
(with $d_1 A^{ (0,1)} = F$, $d_1 F = 0$ and $d_1 B^{(0,2)}
= \mu(A,F)$ for $r=1$) is isomorphic to
$H(D,E_0)$ with $DA^{ (0,1)} = F$, $DF = 0$,
$DH = F^{r+1}$, $D B^{(0,2)} = \mu(A,F) + H$.  Indeed, one
may view the generator $H$ as Koszul generator for the
equation
$F^{r+1} = 0$.   The change of variable $H \rightarrow H' =
H + \mu$ brings then $D$ to the manifestly contractible
form.

\subsubsection{Chapline-Manton model 3}

The third model is essentially a combination of the first
model in the ($A$, $B$)-sector and of the free model for
the improved $3$-form
$C_M = C - AB- \frac{1}{2} A dA$, with curvature $G_M =
dC_M$ and improved last ghost of ghost
$\tilde{C}^{(0,3)}$ (\ref{improvedGhost}).  
Using Theorem \textbf{\ref{invardmod3}} we know that
$E_1\simeq E^{small}_1$  has generators
$G_M$ and
$\tilde C^0_3$. 
The differentials $d_1$ and $d_2$ vanish so
$E_1=E_2=E_3$. One next finds that the differential $d_3$ acts
as,
\begin{equation}
d_3 \tilde C^0_3=G_M,\; d_3 G_M=0,
\end{equation}
so that $E_4=0$.

For $F_3$, one can take as representatives polynomials of the
form $P(G_M)\tilde C^0_3$. Their successive lifts are obtained
by using the Russian formula,
\begin{equation}
\tilde\gamma\tilde C_M=G_M,\label{defCm}
\end{equation}
with $\tilde{C}_M = C_M + E_M + L_M + {\tilde C}^0_3
$, $C_M = C-AB-\frac{1}{2} A dA$, $E_M= C^2_1-\frac12
A^1_0B^1_1-\frac12 dA A^0_1-BA^0_1$ and $L_M
= C^1_2-\frac12 A B^0_2-\frac12
A^0_1B^1_1$. The successive lifts
of $P(G_M)\tilde C^0_3$ are therefore $P(G_M)L_M$, $P(G_M)E_M$ and
$P(G_M)C_M$.

\subsubsection{Chapline-Manton model 4}

The cohomology $H(\gamma\vert d)$ for the $2$-form has been
studied in Section \textbf{\ref{results}} while $H(\gamma\vert d)$
for Yang-Mills theory has been extensively studied in the
literature \cite{DVTV1,DVTV2}. Here we only highlight the points
which are relevant when the two models interact.

In the absence of coupling, the non trivial differentials are,
\begin{equation} d_2 B^0_2 = H, \; d_2 H =0,
\label{d2CM4}
\end{equation} ($B^0_2 \equiv \rho$) and
\begin{eqnarray} d_3 trC^3 &=& tr F^2, \; d_3 tr F^2 =0,
\label{d3CM4}\\ d_5 trC^5 &=& tr F^3, \; d_5 tr F^3 = 0,
\label{d5CM4}\\ &\vdots& \\ d_{2N-1} tr C^{2N-1} &=& tr
F^{N},
\label{dNCM4}
\end{eqnarray} (see \cite{DVTV1}).  We have here only written
down explicitly the actions of the non trivial $d_k$'s on
the contractible pairs. The last ghost of ghost $B^0_2$
is non trivial and can be lifted twice;
$trC^3$ is non trivial and can be lifted three times;
$trC^5$ is non trivial and can be lifted five times; more
generally, $tr C^{2k+1}$ is non trivial and can be lifted
$(2k+1)$ times.

When the coupling is turned on, the variables
$\rho$ and $trC^3$ disappear from the $\gamma$-cohomology. It
follows that all the solutions of the Wess-Zumino
consistency condition that previously were above a
polynomial in $\rho$ and $trC^3$  disappear or
become trivial.  This last feature is known as the Green-Schwarz
anomaly cancellation mechanism
\cite{GS1}. At the same time, the differential $d_0$ becomes
non trivial, as for the previous Chapline-Manton models.
One has
\begin{equation} d_0 H = tr F^2, \; d_0 tr F^2 = 0
\end{equation} which explicitly shows that $tr F^2$ disappears from
the invariant cohomology.  The other
differentials (\ref{d5CM4}) through (\ref{dNCM4}) remain
unchanged.

\subsection{Counterterms and anomalies}

As in the free case, we summarize the previous results by producing
explicitly the antifield-inde\-pen\-dent counterterms and
anomalies, i.e.,
$H^n_0(\gamma \vert d)$ and $H^n_1(\gamma \vert
d)$. 

\subsubsection{Counterterms and anomalies of type A}

The counterterms that lead to a trivial descent involve in
general the individual components of the gauge-invariant
field strengths and their derivatives and cannot generically
be expressed as exterior products of the forms $F$
or $H$. They are the gauge-invariant polynomials  and read
explicitly,
 \begin{equation}
a = a([F]) d^n x, 
\end{equation}
for the Chapline-Manton model 1,
\begin{equation}
a = a([F],[H]) d^n x,
\end{equation}
for the Chapline-Manton model 2,
\begin{equation}
a = a([F],[G]) d^n x,
\end{equation}
 for the Chapline-Manton model 3 and,
\begin{equation}
a=P_I ([F],[H])d^nx,
\end{equation}
for the Chapline-Manton model 4,
where $P_I$ is an invariant function of $F^a_{\mu
\nu}$ and their covariant derivatives, as well as of $[H]$.

In order for those counterterms to be non-trivial $a$ should
satisfy 
$a \not= db$ in all cases which is equivalent to the
condition that its variational derivatives with
respect to the fields do not identically vanish.

We have assumed that the spacetime forms
$dx^\mu$ could occur only through the product $dx^0 dx^1 \cdots
dx^{n-1}
\equiv d^n x$ as is required by Lorentz-invariance.

The anomalies that lead to a trivial descent are sums of
terms of the form
$a = P \, C \, d^n x$ where $P$ is a gauge-invariant
polynomial and
$C$ is a last ghost of ghost of ghost number one, which
must be non trivial in $H(\gamma)$.  These anomalies exist
only  in the
Chapline-Manton model 2 which has last ghost of ghosts with ghost
number one.  One has explicitly,
\begin{equation} a = P([F],[H]) A^0_1 d^n x \text{ \quad (second
CM models)}.
\label{sccm}
\end{equation}
$a$ will be trivial if and only if $P=dR([F],[H])$. Indeed, if $a$
is trivial then it is of the form, $a=\gamma c+de$ with 
$\gamma e + dm=0$, where $e$ is of ghost number one and form
degree $n-1$. Using the results of Section {\bf{\ref{rescmgen}}}
we see that no element of an $F_k$ can be lifted on a
solution in ghost number one and form degree $n-1$. Therefore, up
to irrelevant terms,  $e$ is of the form $e=R([F],[H])A^0_1$ which
implies $P=dR([F],[H])$.

\subsection{Counterterms of type B} As in the free case (Section
{\bf{\ref{cafree}}}),  we determine the solutions $a$ which descend
non-trivially starting directly from the obstruction
$P=da$ since the invariant cohomology of $d$ is known.

\subsubsection{Chapline-Manton model 1} There is in this
case no non trivial solution of type B since there is no
non trivial descent.

\subsubsection{Chapline-Manton model 2} One may proceed as
for the free theory.  The polynomial
$P=da$ must be taken in the invariant cohomology of $d$ and so
is a polynomial in the curvatures $F$ with
$F^{r+1}$ identified with zero.  
This implies,
\begin{equation}
da=F^{m+1}=d(F^mA) \text{ with $m<r$}.
\end{equation}
As in the free
theory,this leads to the Chern-Simons terms,
\begin{equation}
a=F^m A,
\end{equation}
except that
$F^r A$ is now absent because it can be brought into class A
by the addition of exact terms. However, due to the restriction
$m<r$, these Chern-Simons terms are never of form degree $n$ and
therefore, they do not contribute to the counterterms.

\subsubsection{Chapline-Manton model 3}
\label{above}

In this case, the obstruction $P=da$ is a polynomial in the
improved field strength $G_M$.  Therefore, one has $P=dQ(G_M, C_M)$
and so up to trivial terms $a=Q(G_M, C_M) = R(G_M) C_M$ and $a$ is
linear in the improved potential $C_M$.  The Chern-Simons
solution $Q$ exists only in spacetime dimension $4k -1$.

\subsubsection{Chapline-Manton model 4} Again, one finds as
solutions the familiar higher order Yang-Mills Chern-Simons
not involving $tr F^2$ or $\omega_3$. These are only available
in odd dimensions $>3$.

\subsection{Anomalies of type B}

As in the free theory, the anomalies $a$ of type B can 
be of two types. They can arise from an obstruction that lives one
dimension higher or from an obstruction that lives two dimensions
higher.  

In the first case, the obstruction
$da$ has form degree $n+1$ and ghost number $1$.  This case
is only possible in the 
second Chapline-Manton model 2, since there is no
$\gamma$-cohomology in ghost number one for the other
models.  
The obstruction $da$ reads,
\begin{equation} da + \gamma(\hbox{something}) = P(F) A^0_1.
\label{obstructioncm1}
\end{equation}
The right-hand side of
(\ref{obstructioncm1}) is necessarily the $d_k$ of some element in
$F_{k-1}$ of ghost number $>1$. According to the results of
Section {\bf{\ref{rescmgen}}} we see that the only term which
has as obstruction a polynomial of the form $P(F)A^0_1$ is
$B^0_{2r}$. However, the corresponding lift $a$ is $B^{2r-1}_1$
which is not of form degree $n$. There is thus no anomaly in
this case.

In the second case, the anomaly can be lifted once,
$da + \gamma b =0$.  The obstruction
$db$ to a further lift is then a $(n+2)$-form of ghost
number
$0$.

There is no solution of this type for the Chapline-Manton
model 1 because of the lack of a non-trivial descent. 

For the Chapline-Manton model 2, there is again no anomaly
that can be lifted once since the obstruction $db=k F^m \in
H^{inv}(d)$ cannot be of form-degree $n+2$ due to the restriction
$m<r+1$.

For the Chapline-Manton model 3, solutions
descending from polynomials
$P(G_M)$ in two dimensions higher exist only
in spacetime dimensions equal to
$4k-2$. They are given by $a=Q(G_M)L_M$ with $L_M$ defined
in Section {\bf{\ref{rescmgen}}} above. 

Finally, for the Chapline-Manton model 4, one has all the
anomalies of the $SU(N)$ pure Yang-Mills theory, except those
involving the cocycle $trC^3$ and its lifts which are now trivial.

\section{$H(s\vert d)$ - Antifield dependent solutions}

The calculation of the antifield dependent solutions of the
Wess-Zumino consistency condition for the Chapline-Manton models
proceeds in very much the same way as for the free theory. 

To begin with, one repeats the analysis of Section
{\bf\ref{prelimres}}. In particular, Theorem {\bf\ref{dingamma}}
which states that there can be no non-trivial descents in
$H(\gamma\vert d)$ involving the antifields remains valid. Using
this result, it is again easy to prove that up to allowed
redefinitions, the component of highest antighost number of a BRST 
cocycle can be written as,
\begin{equation}
a^n_{g,q}=P_J \omega^J.\label{ilesttard}
\end{equation}
In \eqref{ilesttard}, $P_J$ is in the invariant
cohomology $H^{inv}(\delta\vert d)$ while the $\omega^J$ are a
basis of the polynomials in the ghosts belonging to
$H(\gamma)$.

To make use of this result we must calculate
$H^{inv}(\delta\vert d)$ for the Chapline-Manton models. This is
the subject of the next section. Afterwards, we will study which of
the terms \eqref{ilesttard} can be completed by components of lower
antighost numbers to produce solutions of the Wess-Zumino
consistency condition.

\subsection{Invariant characteristic cohomology}

The calculation of the cohomology $H^{inv}(\delta\vert d)$ proceeds
virtually identically for the four Chapline-Manton models
considered.  One
decomposes the representatives of 
$H^{inv}(\delta\vert d)$ and the Koszul-Tate differential $\delta$
according to specific degrees in order to use the results on the
invariant characteristic cohomology for the free theory. To avoid
repetition, the method will only be explicited for the first
Chapline-Manton model.

\subsubsection{Chapline-Manton model 1}

\begin{theorem}
For the Chapline-Manton model 1, the invariant characteristic
cohomology $H^{inv}(\delta \vert d)$ in antighost $>1$ and form
degree $n$ vanishes.
\label{thinCM1}\end{theorem}

\proof{Let us first recall for this CM model the action
of the Koszul-Tate differential on the antifields,
\begin{align}
&\delta A^{*\mu_1\ldots\mu_p}=\partial_\nu
F^{\nu\mu_1\ldots\mu_p},\label{cfCM1}\\ &\delta
A^{*\mu_1\ldots\mu_{p-j}}= -
\partial_\nu A^{*\nu\mu_1\ldots\mu_{p-j}},\\
&\delta B^{*\mu_1\ldots\mu_{p+1}}=\partial_\nu
H^{\nu\mu_1\ldots\mu_{p+1}}-F^{\mu_1\ldots\mu_{p+1}},\label{cfCM13}
\\ &\delta B^{*\mu_1\ldots \mu_{p+1-j}}=-\partial_\nu
B^{*\nu\mu_1\ldots\mu_{p+1-j}}+(-)^{j+1}A^{*\mu_1\ldots\mu_{p+1-j}}.
\label{cfCM14}
\end{align}

In the absence of coupling, Theorem {\bf{\ref{xx}}} indicates
that the invariant characteristic cohomology in antighost $>1$ and
form degree $n$ is given by the linear combinations of the monomials
$[\tilde H^m (\tilde {F^0})^l]^n_q$ with $F^0=dA$.

When the coupling is turned on, there are two modifications in the
definition of the Koszul-Tate differential: in \eqref{cfCM1} the
curvature $F^0$ is replaced by the improved curvature $F=dA+B$; in 
\eqref{cfCM13} and \eqref{cfCM14} there are some extra
(invariant) terms in the variations of the antifields of the
$B$-sector.

To obtain the elements $a$ of $H^{inv}(\delta \vert d)$ in the
interacting case, we first decompose $a$ according to the number
of derivatives of the invariant variables (field strengths,
antifields):
$a=a_0+\ldots +a_k$. According to this degree
$\delta$ splits as $\delta_1+\delta_0$; $\delta_1$ increases by one
the number of derivatives of the invariants while
$\delta_0$ leaves it unchanged and has a non-vanishing action only
on the antifields of the
$B$-sector. 

At highest degree in the derivatives, Eq. $\delta
a+db=0$ implies,
\begin{equation}
\delta_1 a_k + db_k=0.\label{gkfk}
\end{equation}
The differential
$\delta_1$ is identical to the Koszul-Tate differential of the
free theory except for the substitution $F^0\rightarrow F$ with
the consequence that $F$ is now subject to the Bianchi identity
$dF=H$ instead of
$dF^0=0$.
To properly take this into account we decompose the solutions $c$
of
\begin{equation}
\delta_1 c +
dm=0,\label{indienA}
\end{equation}
according to the polynomial degree of the $A$-sector. According to
this degree we have $\delta_1=\delta_f +\delta'$ where:
1) $\delta_f$ has the same action on the antifields as the
Koszul-Tate differential of the free theory; 2) $\delta'$
decreases the polynomial degree in $A$ and has a vanishing action on
all the antifields except $A^{*\mu_1\ldots\mu_p}$ for which
$\delta'A^{*\mu_1\ldots\mu_p}=\partial_\nu B^{\nu\mu_1\ldots\mu_p}$.
If we set $c=c_0+\ldots +c_l$, \eqref{indienA} implies
$\delta_f c_l+ dm_l=0$ where $c_l$ is now a polynomial in
$F^0_{\mu_1\ldots \mu_{p+1}}$ and $H_{\mu_1\ldots\mu_{p+2}}$.
Using the results on $H^{inv}(\delta\vert d)$ for the free
case we then have, 
$c_l=\lambda_l [\tilde H^{r_l}(\tilde {F^0})^l]^n_q +\delta_f
\mu([F^0],[H]) +d
\nu([F^0],[H])$ for $l>0$ with $\lambda_l$ a constant. By a
redefinition of the terms of lower polynomial degree in the
$A$-sector and the addition of trivial terms we conclude that
$c=c_0+\ldots c_{l-1} +
\lambda_l [\tilde H^{r_l}\tilde F^l]^n_q$, where
$c_0 +\ldots +c_{m-1}$ is of maximal order $m-1$ in the variables of
the
$A$-sector and has to satisfy
\eqref{indienA} on its own. By recurrence we thus 
have up to trivial terms
$c=\lambda_0 [\tilde H^{r_0}]^n_q +\sum_l \lambda_l [\tilde H^{r_l}
\tilde F^l]^n_q$.

Using this result, we deduce that unless $k=0$ in
\eqref{gkfk} $a_k$ can
be removed from $a$ so we necessarily have $a=a_0=
\lambda_0 [\tilde H^{r_0}]^n_q +\sum_l \lambda_l [\tilde H^{m_l}
\tilde F^l]^n_q$.

Finally, the last condition which $a$ has to satisfy is $\delta_0
a_0=0$ which immediately implies
$k_0= k_l=0$ and therefore $a=\sum_l \lambda_l [\tilde F^l]^n_q$.
However, because $(\delta + d)\tilde H +\tilde F=0$, these
cocycles are all trivial.}

\subsubsection{Chapline-Manton model 2}

\begin{theorem}\label{thinCM2}
For the Chapline-Manton model 2, the invariant characteristic
cohomology $H^{inv}(\delta \vert d)$ in antighost $>1$ and form
degree $n$ is given by linear combinations of the monomials
$[F^l \tilde H^k]^n_{q(l,k)}$, with $l$ and $k$ such that
$q(k,l)=n-2l-k(n-p-1)>1$.
\end{theorem}

\subsubsection{Chapline-Manton model 3}

\begin{theorem}
For the Chapline-Manton model 3, the invariant characteristic
cohomology $H^{inv}(\delta \vert d)$ in antighost $>1$ and form
degree $n$ is given by linear combinations of the monomials
$[\tilde G^k]^n_{q(k)}$, with $k$ such that
$q(k)=n-k(n-4)>1$.\label{thinCM3}
\end{theorem}

\subsubsection{Chapline-Manton model 4}

\begin{theorem}
For the Chapline-Manton model 4, the invariant characteristic
cohomology $H^{inv}(\delta \vert d)$ in antighost $>1$ and form
degree $n$ is given by linear combinations of the monomials
$[\tilde H^k]^n_{q(k)}$, with $k$ such that
$q(k)=n-k(n-3)>1$.\label{thinCM4}
\end{theorem} 

\subsection{Results}
\label{cmres}

Using the above four theorems we can continue our
construction of the antifield dependent solutions of the Wess-Zumino
consistency condition. Here, we will focus our attention on the
counterterms and the anomalies. The other values of the ghost
number are analyzed similarly.

\subsubsection{Counterterms and anomalies of type I}

As in the free case, the representatives of $H(s\vert d)$ for
which the expansion according to the antighost number stops at order
1 are related to the gauge invariant conserved currents of the
theory. These solutions exist a priori for the four CM models and
are given by,
\begin{equation}
a^n_g= k_{\Delta a_1\ldots a_r}(j^\Delta {\cal Q}^{a_1\ldots
a_r}_{1,g}+ a^\Delta {\cal Q}^{a_1\ldots
a_r}_{0,g+1}),\label{solcurCM}
\end{equation}
where the $k_{\Delta a_1\ldots a_r}$ are constants and the
$a^\Delta$ form a complete set of non-trivial gauge invariant
global symmetries of the model and satisfy $\delta
a_\Delta + d j_\Delta=0.$

Counterterms and anomalies of the form \eqref{solcurCM} exist only
if the cohomology $H(\gamma)$ has non-trivial elements in pureghost
number $1$ or $2$. This occurs only in the
Chapline-Manton model 2. If $r=1$, the counterterms and anomalies
are given respectively by,
\begin{align}
&a^n_g= k_{\Delta}(j^\Delta A^1_0
+ a^\Delta A_1^0)\quad \text{(counterterm),}\label{countCM2}
\\
&a^n_g=  k_{\Delta}(j^\Delta B^1_1
+ a^\Delta B^0_2)\quad \text{(anomaly).}
\end{align}
If $r>1$ then only the counterterms \eqref{countCM2} are present.

\subsubsection{Counterterms and anomalies of type II}

The counterterms and anomalies of this type correspond to
solutions of the Wess-Zumino consistency condition for which the
expansion according to the antighost number stops at order
$\geq 2$.

\paragraph{Chapline-Manton model 1}
There is  no counterterm or anomaly of this type because the
cohomology $H(\gamma)$ vanishes at pureghost number $>0$.
Alternatively, one can view the absence of solutions of type II as a
consequence of Theorem {\bf\ref{thinCM1}}.

\paragraph{Chapline-Manton model 2} For this model, the ghosts of
ghosts available to construct the component of highest antighost
number of a BRST cocycle are $A^0_1$ and $B^0_{2r}$. Combining this
with Theorem {\bf{\ref{thinCM2}}}, we obtain for the
counterterms $a^n_{g,q}=a^n_{0,2r+1}=k[\tilde H]^n_{2r+1} B^0_{2r}
A^0_1$ and for the anomalies
$a^n_{g,q}=a^n_{1,2r-1}=k[F\tilde H]^n_{2r-1} B^0_{2r}$ or
$a^n_{g,q}=a^{n}_{1,2r-1}=k[\tilde H^2]^n_{2r-1} B^0_{2r}$ (the
last term is only available in spacetime dimension $n=2r+3$).

The $a^n_{g,q}$ corresponding to the anomalies are easily
completed into solutions of the Wess-Zumino consistency
conditions. They yield the following representatives of $H(s\vert
d)$,
\begin{align}
&a^n_{1}=k[F\tilde H \tilde B]^n_1 \quad \label{anoC21}
\text{(anomaly)},\\
&a^{n}_{1}=k[\tilde H^2 \tilde B]^n_1 \quad \label{anoC22}
\text{(anomaly) \em{in spacetime dimension $n=2r+3$}.}
\end{align}
Note that in \eqref{anoC21} and \eqref{anoC22} we suppose $r\geq
2$ otherwise the corresponding anomalies are of type I.

For the counterterms, the situation is more complicated. Indeed,
$a^n_{g,q}=a^n_{0,2r+1}=k[\tilde H]^n_{2r+1} B^0_{2r} A^0_1$ cannot
be completed in a BRST cocycle. This implies that for the second CM
model there are no counterterms of type II. The proof is the
following: 
\begin{list}{{\hspace{\blength}}}
{\listparindent=\parindent\parsep=0pt
\labelwidth=0cm
\labelsep=\parindent
\addtolength{\labelsep}{-\blength}
\addtolength{\labelsep}{1.2cm}
\itemindent=-\blength
\addtolength{\itemindent}{\parindent}
\leftmargin=1.2cm}
\item 
We have, 
\begin{multline}
\delta (k[\tilde H]^n_{2r+1}
B^0_{2r} A^0_1)= -d(k[\tilde H]^{n-1}_{2r}B^0_{2r} A^0_1)\\
-\gamma (k[\tilde H]^{n-1}_{2r} (B^1_{2r-1} A^0_1 +
B^0_{2r} A^1_0)),
\end{multline}
and thus 
\begin{equation}
a^n_{0,2r}=k[\tilde H]^{n-1}_{2r} (B^1_{2r-1} A^0_1 +
B^0_{2r} A^1_0)+ m_{2r}B^0_{2r},
\end{equation}
where $m_{2r}$ is a polynomial in the invariants. At order $2r-1$
in the antighost number we then have,
\begin{multline}
\gamma (a_{2r-1}-k[\tilde H]^{n-2}_{2r-1} (B^2_{2r-2} A^0_1 +
B^1_{2r-1} A^1_0)) \\ +d(b_{2r-1}-k[\tilde
H]^{n-2}_{2r-1} (B^1_{2r-1} A^0_1 + B^0_{2r} A^1_0))+(\delta
m_{2r})B^0_{2r}\\  +(-)^{n-2r-1}k[\tilde
H]^{n-2}_{2r-1}B^0_{2r}F=0.\label{choixR}
\end{multline}
Acting with $\gamma$ on this equation we see
that $b'_{2r-1}=b_{2r-1}-k[\tilde H]^{n-2}_{2r-1}$ $(B^1_{2r-1}
A^0_1 + B^0_{2r} A^1_0)$ is an element of $H(\gamma
\vert d)$. Because $2r-1>0$ we have up to
irrelevant terms $b'_{2r-1}= u_{2r-1}B^0_{2r}$ where $u_{2r-1}$
only depends on the invariants. Eq.
\eqref{choixR} then implies,
\begin{equation}
(-)^{n-2r-1}k[\tilde
H]^{n-2}_{2r-1}F+\delta m_{2r}+du_{2r-1}=0.\label{condipourCM2}
\end{equation}
If $r>1$, we must have $k=0$ because according to Theorem
{\bf\ref{thinCM2}},
$[\tilde
H]^{n-2}_{2r-1}F$ defines a non-trivial class of
$H^{inv}(\delta
\vert d)$. Our statement is thus proved for $r>1$.

If $r=1$, Eq. \eqref{condipourCM2} admits solutions. Indeed in form
notation, the action  of $\delta$ on
the antifields  of the
$1$-form (Eq. \eqref{cfCM23}-\eqref{cfCM24}) reads,
\begin{align}
\delta {\overline A}^*_{1}+ d{\overline F} +\alpha F^r {\overline
H}=0, \\
\delta {\overline A}^*_{2}+ d{\overline A}^*_1 +\alpha F^r
{\overline B}^*_{1}=0,
\end{align}
where $\alpha=\frac{2(r+1)}{(2r+1)!}$. Up to trivial terms, the
solutions of
\eqref{condipourCM2} are therefore,
\begin{equation}
m_{2r}=\frac{1}{\alpha}(-)^{n-3}k{\overline A}^*_2; \quad
u_{1}=\frac{1}{\alpha}(-)^{n-3}k{\overline A}^*_1.
\end{equation}
Returning to Eq. \eqref{choixR} we thus have,
\begin{equation}
a_{1}=k[\tilde H]^{n-2}_{1} (B^2_{0} A^0_1 +
B^1_{1} A^1_0)+\frac{1}{\alpha}(-)^{n-3}k{\overline A}^*_1 B^1_1 +
m_1 A^0_1,
\end{equation}
where $m_1$ only depends on the invariants. It thus appears that for
some Chapline-Manton models, it is possible to eliminate
the first obstruction which is met in the
construction of BRST cocycles. This is in contrast with the free
case where those obstructions cannot be eliminated without imposing
constraints on the arbitrary parameters present in $a^n_{g,q}$. 

However, we now show that for the second CM Model, the construction
of the counterterm is obstructed at the next step. Indeed, at order
$0$ in the antighost number we have,
\begin{multline}
\gamma a'_0 + db'_0  \\ 
 +k(-)^{n-3}{\overline H} H A^0_1
+\frac{1}{\alpha}(-)^{n-3}k{\overline F} F A^0_1 +(\delta m_1)
A^0_1=0.\label{cmcmreduc}
\end{multline}
Acting with $\gamma$ on this equation and using our results on
$H(\gamma\vert d)$ we see that the obstruction for
$db'_0$ to be $\gamma$-exact  is of the form
$P(F)+du_0$ where $u_0$ only depends on the invariants. Therefore,
\eqref{cmcmreduc} reduces to,
\begin{align}
k(-)^{n-3}{\overline H} H 
+\frac{1}{\alpha}(-)^{n-3}k{\overline F} F  + P(F) +\delta
m_1 + du_0=0.\label{frez}
\end{align}
Since $m_1$ is
linear in the antifields of antighost number $1$, it can be
written as $m_1= A^{*\mu}I_{\mu}d^n x + B^{*\mu_1\mu_{2}}
G_{\mu_1\mu_{2}}d^n x$ where $I_{\mu}$ and
$G_{\mu\nu}$ are functions of the fields strength and their
derivatives (terms containing derivatives of the antifields are
absorbed in a redefinition of
$u_0$ in \eqref{frez}). We thus have the condition,
\begin{multline}
k(-)^{n-3}{\overline H} H 
+\frac{1}{\alpha}(-)^{n-3}k{\overline F} F  + P(F)+ \partial_\rho
H^{\rho\mu\nu} G_{\mu\nu}d^n x  \\
 +(\partial_\nu F^{\nu\mu}-
H^{\nu\alpha\mu}F_{\nu\alpha})I_\mu d^n x  + du_0=0.\label{frez2}
\end{multline}
Because $d$ is a linear operator which increases by one the number
of derivatives, Eq. \eqref{frez2} reads at
order 2 in the derivatives of the fields and polynomial degree 2 in
the invariants,
\begin{equation}
k(-)^{n-3}{\overline H} H 
+\frac{1}{\alpha}(-)^{n-3}k{\overline F} F  + fF^2+
H^{\nu\alpha\mu}F_{\nu\alpha}f_\mu d^n x=0  \label{frez3},
\end{equation}
where $f_\mu$ and $f$ are constants. To obtain this condition one
use the fact that $u_0$, $I_\mu$ and $G_{\mu\nu}$ only depend on
the field strengths and their derivatives. 

If we now take the
Euler-Lagrange derivative of \eqref{frez3} with respect to
$B_{\mu\nu}$ we reach the conclusion that
$k=0$ and this proves our statement.
$\qedsymbol$
\end{list} 
\paragraph{Chapline-Manton model 3} For this model, the
ghost part of the cohomology $H(\gamma)$ 
is generated by the improved anticommuting ghost $\tilde
C^0_3-\frac12 A^0_1B^0_2$. 

In order to
construct counterterms we thus need
elements of
$H^{inv}(\delta\vert d)$ in antighost number $3$. However, using
Theorem {\bf\ref{thinCM3}} we conclude that there are no such
elements (the only candidates arise in spacetime dimension $n=5$
and are of the form $[\tilde G^2]^6_3$ but vanish because
$\tilde G$ is anticommuting for $n=5$.)

To construct an anomaly we need elements of the
$H^{inv}(\delta\vert d)$ in antighost number $2$. Again using
Theorem {\bf\ref{thinCM3}} we see that such terms exist only in
spacetime dimension $6$ and are of the form $[k\tilde G^2]^6_2$.
The corresponding anomalies are given by,
\begin{equation}
a^n_1=k [\tilde G^2 \tilde C_M]^6_1,
\end{equation}
where $\tilde C_M$ is defined below \eqref{defCm}.

\paragraph{Chapline-Manton model 4}
According to Theorem {\bf\ref{gamCM4}}, the $\omega^J$ are at least
of pureghost number $5$. In order to construct counterterms or
anomalies we therefore need elements of $H^{inv}(\delta\vert
d)$ in antighost number $\geq 4$. However, Theorem
{\bf\ref{thinCM4}} implies that $H^{inv}(\delta\vert d)$ vanishes
in antighost number $>3$. Therefore there are no
counterterms or anomalies of type II for the fourth CM model.

\subsubsection{Remarks} 

To summarize, we have shown that one can construct antifield
dependent candidate anomalies for the Chapline-Manton models 2 and
3. However, for the four models considered, there are no
antifield dependent BRST cocycles in ghost number $0$. This shows
that these models are quite rigid because it is impossible to
construct consistent interactions which deform their gauge
transformations.

\section{Conclusions}

In this section we have discussed the Wess-Zumino consistency
condition for Chapline-Manton models by explicitly analyzing four
examples. 

The cohomology $H(s\vert d)$ was worked out but using the same
procedure as for free $p$-forms. This is possible because
the four models considered share the
following properties:
\begin{enumerate}
\item
The gauge algebra is closed on-shell and
therefore the action of the longitudinal exterior derivative
$\gamma$ on the fields and ghosts is nilpotent: $\gamma^2=0$; 
\item
Although the action of the BRST differential $s$ on the antifields
contains components of antighost numbers $>-1$, it is possible to
redefine the antifields to eliminate those components. Furthermore,
the BRST variations of the new antifields only involve
combinations of the invariant variables (denoted $\chi$ in the
text).
\end{enumerate} 

The first property allows to calculate separately the BRST
cocycles which do not depend on the antifields. The analysis
explicitly shows that the ``number" of such solutions is reduced
compared to the free theory. This is due to the fact that the
cohomologies
$H(\gamma)$ and $H^{inv}(d)$ typically get smaller because of
the emergence of new contractible pairs. In the case of $H(\gamma)$,
those contractible pairs consist of ghosts of ghosts while for
$H(d)$ they are made up of curvatures since these obey new
Bianchi identities.

In the calculation of the antifield dependent BRST cocycles one
observes the same reduction in the ``number" of solutions. This is
a consequence of the fact that $H(\gamma)$ contains less elements
but also because the invariant characteristic cohomology
$H^{inv}(\delta\vert d)$ vanishes for more values of the antighost
number.

That the BRST cohomology contains less
elements for interacting theories than for the free
ones is easily understood by considering the following
argument. Let us decompose the BRST action of an interacting
theory in powers of the coupling constant, $S=S_0+gS_1+g^2
S_2+\ldots$ and let $A=A_0+gA_1+g^2 A_2+\ldots$ be a BRST cocycle:
$(S,A)=0$. This condition implies,
\begin{align}
(S_0, A_0)&=0, \label{trucmuch} \\
(S_0, A_1) +(S_1, A_0)&=0, \\
(S_0,A_2)+(S_1,A_1)+(S_2,A_0)&=0, \\
&\vdots 
\end{align}
The first equation tells us that $A_0$ is an elements of the
cohomology
$H(s\vert d)$ of the free theory. All the other equations beneath
\eqref{trucmuch} are conditions which the free BRST cocycle must
satisfy. It is therefore natural that for many of them the
construction of the higher order terms $A_1, A_2, \ldots$ gets
obstructed.

\chapter{Comments}

In this thesis we have solved the Wess-Zumino consistency condition
for an arbitrary system of free $p$-forms but also for models of
the Chapline-Manton type. Using this analysis we have listed for
each theory the first-order vertices, counterterms and anomalies;
for the free system we have also discussed the gauge structure of
the conserved currents.

We insist that our calculations were done in the algebra of forms
depending on the components of the antisymmetric tensors, the
ghost, the antifields and their derivatives up to an arbitrary high
order. However we have shown that:
\begin{enumerate}
\item
All the antifield independent BRST cocycles can be expressed in
terms of exterior products of the fields and the ghost {\em when
these solutions occur in non-trivial descents}; this justifies why
previous calculations made in the so-called ``small algebra" to
obtain counterterms and anomalies are nearly exhaustive.
\item 
The natural appearance of exterior forms also holds for antifield
dependent solutions of the Wess-Zumino consistency condition.
Indeed, except for those related to the conserved
currents of the theory, one may assume that the BRST cocycles only
depend $B^a_{p_a},H^a, C^a_{1},\ldots,C^a_{p_a},{\overline
H^a},$ ${\overline B}^{*a}_1,\ldots,$ ${\overline B}^{*a}_{p_a+1}$.
This is a direct consequence of our analysis of the characteristic
cohomology
\end{enumerate}

A second feature which deserves to be highlighted is the fact that
the calculation of the local BRST cohomology of interacting
theories such as the Chapline-Manton models is greatly simplified when
$H(s\vert d)$ is known in the free case. Indeed we have seen in
Chapter 7 how we could obtain the BRST cocycles in the
Chapline-Manton models from those of the free theory by
``perturbative arguments". This encourages future works on the BRST
cohomology in the context of supersymmetric theories and
supergravity theories where the $p$-forms are coupled to other
gauge fields.
\vspace{.9cm}
\begin{center}
\rule{5cm}{0.1mm}
\end{center}


\begin{thebibliography}{100}
\addcontentsline{toc}{chapter}{\numberline{}Bibliography}
\bibitem{GSW}
M.~Green, J.~Schwarz and E.~Witten, {\em Superstring Theory\/}, Cambridge
  University Press,  (1987).

\bibitem{Polchinskibook}
J.~Polchinski, {\em String Theory\/}, Cambridge University Press,  (1998).

\bibitem{FaddeevPopov}
L.~D. Faddeev and V.~N. Popov, {\em Phys. Lett.\/} {\bf B25} (1967) 29.

\bibitem{BRS1}
C.~Becchi, A.~Rouet and R.~Stora, {\em Commun. Math. Phys.\/} {\bf 42} (1975)
  127.

\bibitem{BRS2}
C.~Becchi, A.~Rouet and R.~Stora, {\em Ann. Phys.\/} {\bf 98} (1976) 287.

\bibitem{Tyutin}
I.V. Tyutin, {\em Gauge invariance in field theory and statistical
  mechanics\/}, Lebedev preprint FIAN, n.39 (1975).

\bibitem{Kallosh1}
R.~E. Kallosh, {\em Nucl. Phys.\/} {\bf B141} (1978) 141.

\bibitem{WitHol1}
B.~de~Wit and J.W. van Holten, {\em Phys. Lett.\/} {\bf B79} (1978) 389.

\bibitem{BV1}
I.A. Batalin and G.A. Vilkovisky, {\em Phys. Lett.\/} {\bf B102} (1981) 27.

\bibitem{BV2}
I.A. Batalin and G.A. Vilkovisky, {\em Phys. Rev.\/} {\bf D28} (1983) 2567.

\bibitem{BV3}
I.A. Batalin and G.A. Vilkovisky, {\em Phys. Rev.\/} {\bf D30} (1984) 508.

\bibitem{WZ}
J.~Wess and B.~Zumino, {\em Phys. Lett.\/} {\bf 37B} (1971) 95.

\bibitem{theseGlenn}
G.~Barnich, {\em {Local BRST cohomology in Yang-Mills theory}\/}, PhD thesis,
  Universit\'e Libre de Bruxelles, 1995.

\bibitem{BBHreport}
G.~Barnich, F.~Brandt and M.~Henneaux, {\em Phys. Rep.\/}, in preparation.

\bibitem{DeserJT}
S.~Deser, R.~Jackiw and S.~Templeton, {\em Phys. Rev. Lett.\/} {\bf 48} (1982)
  975.

\bibitem{FT1}
D.~Freedman and P.K. Townsend, {\em Nucl. Phys.\/} {\bf B177} (1981) 282.

\bibitem{Teitelboim1}
C.~Teitelboim, {\em Phys. Lett.\/} {\bf B167} (1986) 63.

\bibitem{BergRooWitNieu}
E.~Bergshoeff, M.~de~Roo, B.~de~Wit and P.~van Nieuwenhuizen, {\em Nucl.
  Phys.\/} {\bf B195} (1982) 97.

\bibitem{ChaplineManton}
G.F. Chapline and N.S. Manton, {\em Phys. Lett.\/} {\bf B120} (1983) 105.

\bibitem{Baulieu1}
L.~Baulieu, {\em {\rm in} Perspectives in Particles and Fields\/}, Carg\`ese
  1983, M. Levy, J.-L. Basdevant, D. Speiser, J. Weyers, M. Jacob and R.
  Gastmans eds, NATO ASI Series B126, Plenum Press, New York (1983).

\bibitem{BrandtD1}
F.~Brandt and N.~Dragon, {\em Nonpolynomial gauge invariant interactions of
  1-form and 2-form gauge potentials\/}, in {\em Theory of Elementary
  Particles}, pp. 149-154, H. Dorn, D. L\"ust, G. Weigt (eds.) (Wiley-VCH,
  Weinheim, 1998), hep-th/9709021.

\bibitem{HK1}
M.~Henneaux and B.~Knaepen, {\em Phys. Rev. D\/} {\bf 56} (1997) 6076,
  hep-th/9706119 (v3).

\bibitem{HenneauxTeitelboim}
M.~Henneaux and C.~Teitelboim, {\em Quantization of Gauge Systems\/}, Princeton
  University Press,  (1992).

\bibitem{henneauxSC}
M.~Henneaux, {\em Consistent Interactions Between Gauge Fields: The
  Cohomological Approach\/}, hep-th/9712226, International Conference on {\em
  Secondary calculus and cohomological Physics}, Moscow, August 1997.

\bibitem{BH1}
G.~Barnich and M.~Henneaux, {\em Phys. Lett.\/} {\bf B311} (1993) 123.

\bibitem{BBH4}
G.~Barnich, F.~Brandt and M.~Henneaux, {\em Phys. Lett.\/} {\bf B346} (1995)
  81.

\bibitem{BBH1}
G.~Barnich, F.~Brandt and M.~Henneaux, {\em Commun. Math. Phys.\/} {\bf 174}
  (1995) 57.

\bibitem{BBH2}
G.~Barnich, F.~Brandt and M.~Henneaux, {\em Commun. Math. Phys.\/} {\bf 174}
  (1995) 93.

\bibitem{PiguetSor1}
O.~Piguet and S.P. Sorella, {\em Algebraic Renormalization\/}, Lecture notes in
  Physics vol. m28, Springer-Verlag, Berlin,  (1995).

\bibitem{Siegel}
W.~Siegel, {\em Phys. Lett.\/} {\bf B93} (1980) 170.

\bibitem{Thierry1}
J.~Thierry-Mieg, {\em Nucl. Phys.\/} {\bf B335} (1990) 334.

\bibitem{Thierry2}
L.~Baulieu and J.~Thierry-Mieg, {\em Nucl. Phys.\/} {\bf B228} (1983) 259.

\bibitem{FHST}
J.~Fisch, M.~Henneaux, J.~Stasheff and C.~Teitelboim, {\em Commun. Math.
  Phys.\/} {\bf 120} (1989) 379.

\bibitem{FH}
J.~Fisch and M.~Henneaux, {\em Commun. Math. Phys.\/} {\bf 128} (1990) 627.

\bibitem{HenneauxCMP1}
M.~Henneaux, {\em Commun. Math. Phys.\/} {\bf 140} (1991) 1.

\bibitem{hamermesh}
M.~Hamermesh, {\em Group Theory\/}, Addison Wesley,  (1962).

\bibitem{Dragon1}
N.~V. Dragon, {\em Tensor Algebra and Young Tableaux\/}, HD-THEP-81-16.

\bibitem{Vinogr1}
A.M. Vinogradov, {\em Sov. Math. Dokl.\/} {\bf 18} (1977) 1200.

\bibitem{Vinogr2}
A.M. Vinogradov, {\em Sov. Math. Dokl.\/} {\bf 19} (1978) 1220.

\bibitem{DeWilde1}
M.~{De Wilde}, {\em Lett. Math. Phys.\/} {\bf 5} (1981) 351.

\bibitem{Tulczyjew1}
W.M. Tulczyjew, {\em Lecture Notes in Math.\/} {\bf 836} (1980) 22.

\bibitem{Tulczyjew2}
P.~Dedecker and W.M. Tulczyjew, {\em Lecture Notes in Math.\/} {\bf 836} (1980)
  498.

\bibitem{Tsujishita1}
T.~Tsujishita, {\em Osaka J. Math.\/} {\bf 19} (1982) 19.

\bibitem{BonoraCoRa1}
L.~Bonora and P.~Cotta-Ramusino, {\em Commun. Math. Phys.\/} {\bf 87} (1983)
  589.

\bibitem{Olver1}
P.J. Olver, {\em Appli\-ca\-tions of Lie Groups to Dif\-fe\-rential
  Equations\/}, {\rm Graduate Text in Mathematics, volume 107},
  Sprin\-ger-Verlag,  (1986).

\bibitem{Wald1}
R.M. Wald, {\em J. Math. Phys.\/} {\bf 31} (1990) 2378.

\bibitem{Dickey1}
L.A. Dickey, {\em Contemp. Math.\/} {\bf 132} (1992) 307.

\bibitem{BDK3}
F.~Brandt, N.~Dragon and M.~Kreuzer, {\em Nucl. Phys.\/} {\bf B332} (1990) 224.

\bibitem{DVHTV1}
M.~Dubois-Violette, M.~Henneaux, M.~Talon and C.~M. Viallet, {\em Phys.
  Lett.\/} {\bf B267} (1991) 81.

\bibitem{Stora2}
R.~Stora, {\em {\rm in} Recent Progress in Gauge Theories\/}, Lehmann G. et al
  eds, Plenum Press, New-York (1984).

\bibitem{Thorn1}
C.B. Thorn, {\em Phys. Rep.\/} {\bf 175} (1989) 1.

\bibitem{BauBergSezg1}
L.~Baulieu, E.~bergshoeff and E.~Sezgin, {\em Nucl. Phys.\/} {\bf B307} (1988)
  348.

\bibitem{CarVilSasSor1}
M.~Carvalho, L.C.Q. Vilar, C.A.G. Sasaki and S.P. Sorella, {\em BRS Cohomology
  of Zero Curvature Systems I. The Complete Ladder Case\/}, hep-th/9509047.

\bibitem{Baulieu4}
L.~Baulieu, {\em Field Anti-Field Duality, p-Form Gauge Fields and Topological
  Field Theories\/}, hep-th/9512026.

\bibitem{Stora1}
R.~Stora, {\em {\rm in} New Developments in Quantum Field Theory and
  Statistical Mechanics\/}, Carg\`ese 1976, M. Levy, P. Mitter eds, NATO ASI
  Series B26, Plenum Press, New York (1977).

\bibitem{Zumino1}
B.~Zumino, {\em {\rm in} Relativity, Groups and Topology II\/}, B. S. De Witt
  and R. Stora eds, North Holland, Amsterdam, (1984).

\bibitem{DVTV0}
M.~Dubois-Violette, M.~Talon and C.~Viallet, {\em Phys. Lett.\/} {\bf B158}
  (1985) 231.

\bibitem{DVTV1}
M.~Dubois-Violette, M.~Talon and C.~M. Viallet, {\em Commun. Math. Phys.\/}
  {\bf 102} (1985) 105.

\bibitem{Massey}
S.~Mac Lane, {\em Homology\/}, Springer,  (1963).

\bibitem{DVTV2}
M.~Dubois-Violette, M.~Talon and C.~Viallet, {\em Ann. Inst. Henri
  Poincar\'e\/} {\bf 44} (1986) 103.

\bibitem{Talon1}
M.~Talon, {\em Algebra of Anomalies\/}, Cargese Summer Inst. Jul 15-31 (1985)
  433.

\bibitem{BDK1}
F.~Brandt, N.~Dragon and M.~Kreuzer, {\em Phys. Lett.\/} {\bf B231} (1989) 263.

\bibitem{BDK2}
F.~Brandt, N.~Dragon and M.~Kreuzer, {\em Nucl. Phys.\/} {\bf B332} (1990) 250.

\bibitem{BonoraCoRaRiSta1}
L.~Bonora, P.~Cotta-Ramusino, M.~Rinaldi and J.~Stasheff, {\em Commun. Math.
  Phys.\/} {\bf 112} (1987) 237.

\bibitem{GreubHalVan1}
W.~Greub, S.~Halperin and R.~Vanstone, {\em Connections, curvature and
  cohomology, vol III\/}, Academic Press, New-York,  (1976).

\bibitem{Cartan}
H.~Cartan, {\em {\rm in Colloque de Topologie (Bruxelles 1950)}\/}, Masson
  (Paris: 1951).

\bibitem{DixonRa1}
J.~Dixon and M.~Ramon Medrano, {\em Phys. Rev.\/} {\bf D22} (1980) 429.

\bibitem{BryantGriffiths}
R.L. Bryant and P.A. Griffiths, {\em Characteristic Cohomology of
  Dif\-fe\-ren\-tial Systems (I): General Theory\/}, Duke University
  Mathematics Preprints Series, volume 1993 n$^0$1 (January 1993).

\bibitem{Vinogr3}
A.M. Vinogradov, {\em Sov. Math. Dokl.\/} {\bf 20} (1979) 985.

\bibitem{Vinogr4}
A.M. Vinogradov, {\em J. Math. Anal. Appl.\/} {\bf 100} (1984) 1.

\bibitem{Tsujishita2}
T.~Tsujishita, {\em Diff. Geom. Appl.\/} {\bf 1} (1991) 3.

\bibitem{Unruh1}
W.~Unruh, {\em Gen. Rel. Grav.\/} {\bf 2} (1971) 27.

\bibitem{BBH3}
G.~Barnich, F.~Brandt and M.~Henneaux, {\em Nucl. Phys.\/} {\bf B455} (1995)
  357.

\bibitem{Torre1}
C.G. Torre, {\em Class. Quant. Grav.\/} {\bf 12} (1995) L43.

\bibitem{MisnerWheeler1}
C.W. Misner and J.A. Wheeler, {\em Ann. Phys.\/} {\bf 2} (1957) 525.

\bibitem{BHW1}
F.~Brandt, M.~Henneaux and A.~Wilch, {\em Nucl. Phys.\/} {\bf B550} (1999) 495.

\bibitem{BHW3}
F.~Brandt, M.~Henneaux and A.~Wilch, {\em Phys. Lett.\/} {\bf B387} (1996) 320.

\bibitem{Lipkin1}
D.M. Lipkin, {\em J. Math. Phys.\/} {\bf 5} (1964) 698.

\bibitem{Morgan1}
T.A. Morgan, {\em J. Math. Phys.\/} {\bf 5} (1664) 1659.

\bibitem{Kibble1}
T.W. Kibble, {\em J. Math. Phys.\/} {\bf 6} (1965) 1022.

\bibitem{OConnell1}
R.F. O'Connell and D.R. Tompkins, {\em J. Math. Phys.\/} {\bf 6} (1965) 1952.

\bibitem{Torre2}
C.G. Torre, {\em J. Math. Phys.\/} {\bf 36} (1995) 2113.

\bibitem{Weinberg2}
S.~Weinberg, {\em Physica\/} {\bf 96A} (1979) 327.

\bibitem{Nito}
H.~Nicolai and P.K. Townsend, {\em Phys. Lett.\/} {\bf 98B} (1981) 257.

\bibitem{Cham1}
A.H. Chamseddine, {\em Nucl. Phys.\/} {\bf B185} (1981) 403.

\bibitem{Cham2}
A.H. Chamseddine, {\em Phys. Rev.\/} {\bf D24} (1981) 3065.

\bibitem{HenLemesetAl}
M.~Henneaux, V.E.R. Lemes, C.A.G. Sasaki, S.P. Sorella, O.S. Ventura and L.C.Q.
  Vilar, {\em Phys. Lett.\/} {\bf B410} (1997) 195.

\bibitem{Anco1}
S.~C. Anco, {\em J. Math. Phys.\/} {\bf 38} (1997) 3399.

\bibitem{Baulieu3}
L.~Baulieu, {\em Phys. Lett.\/} {\bf B441} (1998) 250.

\bibitem{Romans1}
L.~Romans, {\em Phys. Lett.\/} {\bf B169} (1986) 374.

\bibitem{BergRooPaGrTo}
E.~Bergshoeff, M.~de~Roo, G.~Papadopoulos, M.B. Green and P.K. Townsend, {\em
  Nucl. Phys.\/} {\bf B470} (1996) 113.

\bibitem{HenneauxW1}
M.~Henneaux and A.~Wilch, {\em Phys. Rev.\/} {\bf D58} (1998) 025017.

\bibitem{BizSa1}
C.~Bizdadea and S.O. Saliu, {\em Int. J. Mod. Phys.\/} {\bf A11} (1996) 3523.

\bibitem{Dixon1}
J.A. Dixon, {\em Cohomology and Renormalization of Gauge Theories I, II,
  III\/}, Unpublished preprints (1976-1979).

\bibitem{Dixon2}
J.A. Dixon, {\em Commun. Math. Phys.\/} {\bf 139} (1991) 495.

\bibitem{Bandelloni1}
G.~Bandelloni, {\em J. Math. Phys.\/} {\bf 27} (1986) 2551.

\bibitem{Bandelloni3}
G.~Bandelloni, {\em J. Math. Phys.\/} {\bf 28} (1987) 2775.

\bibitem{HenneauxPL1}
M.~Henneaux, {\em Phys. Lett.\/} {\bf B313} (1993) 35.

\bibitem{Koszul1}
J.L. Koszul, {\em Bull. Soc. Math. Fr.\/} {\bf 78} (1950) 65.

\bibitem{GS1}
M.B. Green and J.H. Schwarz, {\em Phys. Lett.\/} {\bf B149} (1984) 117.

\end{thebibliography}
\end{document}